%% file: garcms00.tex
\documentclass{article}
\usepackage{myllncs,amstext,harvard,epsfig}

\protect

\input{garcmacr}


\newcommand{\eps}[2]{\centerline{\epsfig{figure={#1},width={#2\textwidth}}}}

\begin{document}


\title{
\normalsize
\bf
ON THE STATICS AND DYNAMICS OF MAGNETO-ANISOTROPIC NANOPARTICLES
}

\author{
J.\ L.\ Garc\'{\i}a-Palacios%
\thanks{
{\em On leave from:}
Instituto de Ciencia de Materiales de Arag\'{o}n, Consejo Superior de
Investigaciones Cient\'{\i}ficas-Universidad de Zaragoza, 50015 Zaragoza,
Spain.
\protect\\
{\em Electronic addresses:}
{\tt jose.garcia@angstrom.uu.se}
and
{\tt jlgarcia@posta.unizar.es}
\protect\\
---------------------------------
\protect\\
{\em Advances in Chemical Physics, Volume 112}, Edited by I. Prigogine
and Stuart A. Rice
\protect\\
ISBN 0-471-38002-4. \copyright\, 2000 John Wiley \& Sons, Inc.
}
\protect\\
{\small
Department of Materials Science - Division of Solid State Physics, Uppsala
University,
}
\protect\\
{\small
P O Box 534, \mbox{SE-751 21} Uppsala, Sweden
}
}

\date{}

\maketitle

\pagestyle{myheadings}

\markboth
{\scriptsize
{\sc J. L. GARC\'IA-PALACIOS}}
{\scriptsize
{\sc ON THE STATICS AND DYNAMICS OF MAGNETOANISOTROPIC NANOPARTICLES}
}


\setcounter{tocdepth}{3}
{\footnotesize
\tableofcontents
}
\setcounter{secnumdepth}{4}


\renewcommand{\thesection}{\Roman{section}}
\renewcommand{\thesubsection}{\thesection.\Alph{subsection}}
\renewcommand{\thesubsubsection}{\arabic{subsubsection}.}
\renewcommand{\theparagraph}{\alph{paragraph}.}
\renewcommand{\thetable}{\Roman{table}}
\renewcommand{\theequation}{\arabic{section}.\arabic{equation}}
\renewcommand{\thefootnote}{\arabic{footnote}}
\renewcommand{\labelenumi}{\theenumi}
\renewcommand{\theenumi}{(\roman{enumi})}


\setcounter{equation}{0}
\input{garcms01}

\setcounter{equation}{0}
\input{garcms02}

\setcounter{equation}{0}
\input{garcms03}

\setcounter{equation}{0}
\input{garcms04}

\setcounter{equation}{0}
\input{garcms05}

\setcounter{equation}{0}
\input{garcms06}

\setcounter{equation}{0}
\input{garcms07}

\renewcommand{\theequation}{\Alph{section}.\arabic{equation}}
\renewcommand{\thesubsection}{\thesection.\arabic{subsection}}
\renewcommand{\thesubsubsection}{\alph{subsubsection}}
\renewcommand{\theparagraph}{\arabic{paragraph}.}
\bigskip
\bigskip
\centerline{\Large\bf APPENDICES}
\addcontentsline{toc}{section}{Appendices}
\input{garcms08}



{\small

\input{garcbi00}
}





\end{document}

%% file: garcmacr.tex
\newcommand{\Weq}{P_{{\rm e}}}
\newcommand{\W}{P}

\newcommand{\llg}{g}
\newcommand{\legunb}{\tilde{S}_{2}}
\newcommand{\legunbfour}{\tilde{S}_{4}}

\newcommand{\llangle}{\left\langle}
\newcommand{\rrangle}{\right\rangle}

\newcommand{\blangle}{\big\langle}
\newcommand{\brangle}{\big\rangle}

\newcommand{\eq}{{\rm e}}
\newcommand{\ran}{{\rm ran}}
\newcommand{\D}{{\rm d}}
\newcommand{\ym}{{\bf y}}

\newcommand{\half}{{\textstyle \frac{1}{2}}}
\newcommand{\third}{{\textstyle \frac{1}{3}}}
\newcommand{\threehalfs}{{\textstyle \frac{3}{2}}}

\def\tm{t_{{\rm m}}}

\def\F{R}

\def\lan{{\rm Lan}}
\def\ising{{\rm Ising}}
\def\rotator{{\rm rot}}
\def\unb{{\rm unb}}

\def\B{\vec{B}}

\newcommand{\Hs}{{\cal H}}
\newcommand{\Hsm}{{\cal H}^{({\rm m})}}

\newcommand{\Ham}{{\cal H}}
\def\dU{\Delta U}

\def\m{\vec{m}}
\def\mm{m}
\newcommand{\mx}{m_{x}}
\newcommand{\my}{m_{y}}
\newcommand{\mz}{m_{z}}
\newcommand{\mi}{m_{i}}
\newcommand{\mj}{m_{j}}
\newcommand{\mk}{m_{k}}
\newcommand{\ml}{m_{\ell}}
\newcommand{\mr}{m_{r}}
\newcommand{\ms}{m_{s}}

\newcommand{\Xo}{\frac{\mu_{0}m^{2}}{\T}}
\newcommand{\Kuni}{(B_{K}/m)}
\def\BK{B_{K}}
\def\gBK{\gmr B_{K}}
\def\Bred{B/B_{K}}

\def\Delt{\Delta t}
\def\DelW{\Delta W}
\def\dB{\Delta\vec{B}}
\def\dB{\Delta\vec{B}}
\def\dBo{\Delta B}
\newcommand{\bp}{\hat{b}}

\def\red{{\rm red}}

\def\tlo{\tau_{\|}}
\def\ttr{\tau_{\perp}}
\def\wto{\omega\tau_{0}}
\def\wtlo{\omega\tau_{\|}}
\def\lnwto{\ln(\omega\tau_{0})}

\def\Eb{E_{{\rm b}}}

\def\sb{\sigma_{{\rm b}}}

\def\Tb{T_{{\rm b}}}

\def\EM{E_{{\rm M}}}
\def\TM{T_{{\rm M}}}

\def\Min{{\rm min}}
\def\Max{{\rm max}}

\def\Beff{\vec{B}_{{\rm eff}}}
\def\weff{\omega_{{\rm eff}}}
\def\bfl{\vec{b}_{{\rm fl}}}

\def\eff{{\rm eff}}
\def\fl{{\rm fl}}

\def\tN{\tau_{{\rm N}}}
\def\tNalt{\tau_{{\rm N}}}
\def\tK{\tau_{K}}
\def\tB{\tau_{B}}
\def\tint{\tau_{\rm int}}
\def\Lamone{\Lambda_{1}}

\def\wB{\w_{B}}
\def\wK{\w_{K}}

\newcommand{\tp}{t'}

\newcommand{\Hfl}{{\cal H}_{{\rm fl}}}
\newcommand{\R}{\vec{R}}

\newcommand{\Ma}{}
\newcommand{\Mb}{}
\newcommand{\Mg}{}

\newcommand{\Mr}{}
\newcommand{\overMa}{}

\newcommand{\sP}{p}
\newcommand{\sQ}{q}
\newcommand{\eP}{P}
\newcommand{\eQ}{Q}
\newcommand{\ePm}{{\bf P}}
\newcommand{\eQm}{{\bf Q}}

\newcommand{\coupling}{\varepsilon}

\newcommand{\G}{{\cal B}}

\newcommand{\h}{{\rm h}}

\newcommand{\pbra}{\big\{}
\newcommand{\pket}{\big\}}

\newcommand{\lin}{\scriptscriptstyle{\rm (L)}}
\newcommand{\qua}{\scriptscriptstyle{\rm (Q)}}

\newcommand{\TGIP}{\hat{G}}
\newcommand{\TL}{\hat{\Lambda}^{\lin}}
\newcommand{\TQ}{\hat{\Lambda}^{\qua}}
\newcommand{\TLQ}{\hat{\Lambda}}

\newcommand{\tGIP}{G}
\newcommand{\tL}{\Lambda^{\lin}}
\newcommand{\tQ}{\Lambda^{\qua}}
\newcommand{\tLQ}{\Lambda}

\newcommand{\TLpre}{\hat{\Gamma}^{\lin}}
\newcommand{\TQpre}{\hat{\Gamma}^{\qua}}

\newcommand{\ffl}{f}
\newcommand{\Vint}{V}
\newcommand{\Fint}{F}
\newcommand{\Cint}{{\scriptstyle C}}

\newcommand{\K}{{\cal K}}

\newcommand{\bt}{b}
\newcommand{\kt}{\kappa}

\newcommand{\ila}{{\bf l}}
\newcommand{\ilb}{{\bf l'}}
\newcommand{\iqa}{{\bf q}}
\newcommand{\iqb}{{\bf q'}}

\newcommand{\drift}{A}
\newcommand{\diff}{B}
\def\Lan{L}

\def\T{k_{{\rm B}}T}
\def\kB{k_{{\rm B}}}

\def\Mss{\mu_{0}M_{s}^{2}}

\def\q{\tan(\vartheta/2)}
\def\qs{\tan^{2}(\vartheta/2)}
\def\qo{\tan(\vartheta_{0}/2)}

\def\hfpi{\frac{\pi}{2}}

\def\vecpro{\wedge}

\def\Z{{\cal Z}}
\def\Zp{{\cal Z}_{\|}}

\def\FE{{\cal F}}
\def\K{{\cal K}}
\def\E{{\cal U}}
\def\ent{{\cal S}}
\def\sh{c_{B}}

\def\qed{\quad{\rm Q.E.D.}}

\def\s{\sigma}
\def\spl{\sigma_{+}}
\def\smi{\sigma_{-}}
\def\spm{\sigma_{\pm}}
\def\la{\lambda}
\def\gmr{\gamma}

\def\xipara{\xi_{\|}}
\def\xiperp{\xi_{\perp}}
\def\w{\omega}

\def\cosal{\cos\alpha}
\def\cosqal{\cos^{2}\!\alpha}
\def\coscal{\cos^{4}\!\alpha}

\def\senal{\sin\alpha}
\def\senqal{\sin^{2}\!\alpha}
\def\sencal{\sin^{4}\!\alpha}

\def\avcosqal{\langle\cos^{2}\!\alpha\rangle}
\def\avsenqal{\langle\sin^{2}\!\alpha\rangle}
\def\avX{\llangle\chi\rrangle_{{\rm ran}}}
\def\avXr{\llangle\chi'\rrangle_{{\rm ran}}}
\def\avXi{\llangle\chi''\rrangle_{{\rm ran}}}

\def\phase{\phi}

\def\gsim{
\,\raisebox{0.35ex}{$>$}
\hspace{-1.7ex}\raisebox{-0.65ex}{$\sim$}\,
}

\def\lsim{
\,\raisebox{0.35ex}{$<$}
\hspace{-1.7ex}\raisebox{-0.65ex}{$\sim$}\,
}

%% file: garcms01.tex
\section{Introduction}
\label{sect:introduction}

Small, magnetically ordered particles, are ubiquitous both in
naturally occurring and manufactured forms. On the one hand, it is
remarkable the wide spectrum of applications of these systems, which
range from magnetic recording media, catalysts, magnetic fluids,
filtering and phase separation in mineral processing industry,
magnetic imaging and magnetic refrigeration, to numerous geophysical,
biological, and medical uses. On the other hand, the {\em
nanometric\/} magnetic particles can be considered as model systems
for the study of various basic physical phenomena.  Among others we
can mention: rotational Brownian motion and thermally activated
processes in multistable systems, mesoscopic quantum phenomena,
dipole-dipole interaction effects, and the dependence of the
properties of solids on their size.

Magnetically ordered particles of nanometric size generally consist of
a single domain, whose constituent spins, at temperatures well below
the Curie temperature, rotate in unison. The magnetic energy of a
nanometric particle is then determined by its magnetic moment
orientation, and has a number of stable directions separated by
potential barriers (associated with the magnetic anisotropy). As a
result of the coupling of the magnetic moment of the particle, $\m$,
with the microscopic degrees of freedom of its environment (phonons,
conducting electrons, nuclear spins, etc.), the magnetic moment is
subjected to thermal fluctuations and may undergo a Brownian-type
rotation surmounting the potential barriers. This solid-state
relaxation process was proposed and studied by N\'{e}el (1949), and
subsequently reexamined by Brown (1963), by dint of the theory of
stochastic processes.

In the high potential-barrier range, $\dU/\T\gg1$, the characteristic
time for the over-barrier rotation process, $\tlo$, can approximately
be written in the Arrhenius form $\tlo\simeq\tau_{0}\exp(\dU/\T)$,
where $\tau_{0}$ ($\sim10^{-10}$--$10^{-12}$\,s) is related with the
intra-potential-well dynamics. For $\tlo\ll\tm$ ($\tm$ is the
measurement or observation time), $\m$ maintains the equilibrium
distribution of orientations as in a classical paramagnet; because
$\mm=|\m|$ is much larger than a typical microscopic magnetic moment
($\mm\sim10^{3}$--$10^{5}\,\mu_{{\rm B}}$) this phenomenon is named
{\em superparamagnetism}. In contrast, when $\tlo\gg\tm$, the magnetic
moment rotates rapidly about a potential minimum whereas the
over-barrier relaxation mechanism is {\em blocked}. This corresponds
to the state of stable magnetization in a bulk magnet. Finally, under
intermediate conditions ($\tlo\sim\tm$) {\em non-equilibrium
phenomena}, accompanied by magnetic ``relaxation," are observed. It is
to be noted that, in the Arrhenius range mentioned, the system may
pass through all these regimes in a relatively narrow temperature
interval.

We shall describe a nanoparticle as a classical magnetic moment with
magnetic-anisotropy energy. This brings generality to the results and
the connection with other physical systems that can approximately be
described as ensembles of ``rotators" in certain orientational
potentials. Examples include: molecular magnetic clusters with high
spin in their ground state (in the ranges where a classical
description of their spins is reasonable); nematic liquid crystals
with uniaxial physical properties; relaxor ferroelectrics; certain
high-spin dilutely-doped glasses described by the
random-axial-anisotropy model; and superparamagnetic-like spin
glasses.

Indeed, the analogies between the macroscopic behavior of certain
electric and magnetic ``glassy" systems and that of ensembles of small
magnetic particles have received recurrent attention during the last
20 years. The magnetic nanoparticle systems exhibit glassy-like
phenomena associated with the distribution of particle parameters
(anisotropy constants, volumes, magnetic moments, etc.), which lead to
more or less wide distributions of relaxation times. On the other
hand, ensembles of interacting nanoparticles apparently exhibit
genuine glassy properties, mainly due to the extremely anisotropic
character of the dipole-dipole interaction.  Therefore, it is
important to determine which phenomena are intrinsically due to the
presence of interactions in the nanoparticle ensemble and which others
not. In this connection, owing to the lack of enough knowledge about
some of the properties of {\em independent\/} magnetic particles, it
is not always known from which ``laws" the corresponding quantities
depart as a consequence of the inter-particle interactions. Similar
considerations also apply to the study of the effects associated with
quantum phenomena in small magnetic particles; as complete a knowledge
as possible of the {\em classical\/} regime is the mandatory starting
point towards the study of, for example, quantum tunnelling and
coherence in these systems.

Finally, the study of the dynamics of non-interacting classical
magnetic moments is an interesting strand of research {\em per se\/},
which seems to be far from exhausted.  Indeed, relevant developments
of the pioneering works of the 1960s and 1970s have been performed
during the last 15 years.

The purpose of this Chapter is to gain a deeper insight into the
statical (thermal-equilibrium) and dynamical (non-equilibrium)
properties of {\em non-interacting\/} magnetically anisotropic
nanoparticles in the framework of {\em classical\/} physics.

The scheme followed in this work is as follows: In Sections
\ref{sect:gen} and \ref{sect:quantities} some thermal-equilibrium
properties of classical magnetic moments are studied. Section
\ref{sect:gen} is devoted to the obtainment of general results for the
basic thermodynamical functions (partition function and
thermodynamical potentials), some of which are subsequently used in
Section \ref{sect:quantities} to calculate various important
thermal-equilibrium quantities.  Some known results are reobtained
(presenting in some cases alternative expressions and/or derivations),
whereas the superparamagnetic theory is extended by calculating a
number of other quantities. The central issue along these first two
Sections is the study of the effects of the magnetic anisotropy on the
thermal-equilibrium properties of superparamagnetic systems. These
effects are sometimes ignored because superparamagnetism is {\em
restrictively\/} associated with the temperature range where the
anisotropy energy is smaller than the thermal energy.

In the remainder Sections we shall concentrate on the dynamical
properties of classical magnetic moments. The heuristic approach to
the dynamics of these systems is considered in Section
\ref{sect:heuristic}, where the analyses of the corresponding models
in order to extract certain parameters of ensembles of magnetic
nanoparticles are revised and developed. In Section
\ref{sect:stochastic} the dynamical properties of classical magnetic
moments are studied by the methods of the {\em theory of stochastic
processes}. The Brown--Kubo--Hashitsume stochastic model is presented
in a unified way and Langevin-dynamics simulations are performed to
study the non-zero temperature dynamical properties. Both the study of
individual stochastic trajectories and the response of ensembles of
magnetic moments are undertaken.  Finally, Section \ref{sect:gle} is
devoted to the foundation of the dynamical equations that are the
basis of Section \ref{sect:stochastic}. The techniques of the
formalism of the {\em independent-oscillator environment\/} are
employed to derive dynamical equations for the magnetic moment that
take the effects of its interaction with the surrounding medium into
account.

%% file: garcms02.tex
\section{Equilibrium properties: generalities and methodology}
\label{sect:gen}

\subsection{Introduction}

Throughout this Chapter we shall concentrate on the study of magnetic
moments whose physical support (the crystal lattice in magnetic
nanoparticles), to which they are linked by the magnetic anisotropy,
is fastened in space. In small-particle magnetism, this corresponds to
particles dispersed in a solid matrix. Although this apparently
excludes the so called ``magnetic fluids" (where the physical rotation
of the particles plays a fundamental r\^{o}le), these belong to the
class of solid dispersions when the liquid carrier is frozen (which is
besides the case of experimental interest when studying
low-temperature properties). On the other hand, we shall also restrict
our study to systems with axially symmetric magnetic anisotropy. This
choice makes the problem mathematically tractable and provides
valuable insight into more complex situations.

As was mentioned in Section \ref{sect:introduction}, the thermal-equilibrium
(superparamagnetic) behavior is observed when the measurement or observation
time, $\tm$, is much longer than the characteristic relaxation times of the system (this
is of course a general statement). In Table \ref{measurement_times:table} the
measurement times of various experimental techniques are displayed.

Note that the thermal-equilibrium range can extend down to temperatures where the
heights of the energy barriers (created by the magnetic anisotropy) are much larger
than the thermal energy. To illustrate, for a system with an axially symmetric
Hamiltonian and in the high-barrier range, the mean time for the over-barrier
rotation process, $\tlo$, can be written in the Arrhenius form
\begin{equation}
\label{arrhenius:tau:0}
\tlo
=
\tau_{0}\exp(\dU/\T)
\;.
\end{equation}
Besides, the ``high-barrier" range where this expression for the relaxation time holds,
extends down to $\dU/\T\gsim2$; moreover, for $\dU/\T\lsim2$, the relaxation time
$\tlo$ is of the order of $\tau_{0}$ ($\sim10^{-10}$--$10^{-12}$\,s for magnetic
nanoparticles). Therefore, the exponential decrease of $\tlo$ as $T$ increases, yields
the range
\[
\ln(\tm/\tau_{0})>\dU/\T\geq 0
\;,
\]
as the thermal-equilibrium range ($\tlo\ll\tm$) for a given measurement time $\tm$.
For instance, for magnetic measurements with $\tm\sim1$--$100$\,s, this range is
extremely wide ($25>\dU/\T\geq 0$). This entails that the frequently encountered
statement, ``superparamagnetism occurs when the thermal energy is comparable or
larger than the energy barriers", is unnecessarily restrictive. 
\begin{table}[t]
\caption[]
{
Characteristic measurement times of various experimental techniques.
\label{measurement_times:table}
}
\begin{center}
\begin{tabular}{|c|c|}
Experimental technique
&
Measurement time
\cr
\hline
magnetization
&
$1$--$100$\,s
\cr
\hline
ac susceptibility
&
$10^{-6}$--$100$\,s
\cr
\hline
M\"{o}ssbauer spectroscopy
&
$10^{-9}$--$10^{-7}$\,s
\cr
\hline
Ferromagnetic resonance
&
$10^{-9}$\,s
\cr
\hline
Neutron scattering
&
$10^{-12}$--$10^{-8}$\,s
\end{tabular}
\end{center}
\end{table}

Let us further illustrate this important point which rests essentially on the magnitude
of $\tau_{0}$ and the exponential dependence of $\tlo$ on $T$ in Eq.\
(\ref{arrhenius:tau:0}). For an experiment with measurement time $\tm$, the {\em
blocking temperature}, $\Tb$, defined as the temperature where $\tm=\tlo$, is given
by $\tm=\tau_{0}\exp(\dU/\kB\Tb)$. Accordingly, one has
$\ln(\tm/\tau_{0})=\dU/\kB\Tb$ so that, if $\tm=\tau_{0}10^{12}$ (a typical value for
standard magnetic measurements), it follows that
$\dU/\kB\Tb=\ln(10^{12})\simeq27.6$. However, for $\dU/(\kB 1.1\Tb)\simeq25$,
one already finds $\tlo=0.08\tm$ while for $\dU/(\kB 1.2\Tb)\simeq23$, one has
$\tlo=0.01\tm$, i.e., {\em the system is clearly in the thermal-equilibrium regime,
whereas $\dU$ is still much larger than $\T$}.

Thus, there exists an extremely wide range where superparamagnetism occurs
($\tlo\ll\tm$) and, simultaneously, the ``na\"{\i}ve condition of superparamagnetism"
$\dU/\T\lsim1$, is not necessarily obeyed. Consequently, in that range, the effects of
the anisotropy-energy on the equilibrium quantities can be sizable. Indeed, for any
thermal-equilibrium quantity, prior to the observation of the corresponding
``blocking" (departure from thermal-equilibrium behavior) when the temperature is
sufficiently lowered, one can clearly observe a crossover from the isotropic-type
behavior at high temperatures (where the anisotropy potential plays a minor r\^{o}le)
to either a discrete-orientation- or plane-rotator-type behavior at low temperatures
(where the magnetic moment stays most of the time in the potential-minima regions),
{\em without leaving the thermal-equilibrium range}.

The organization of the remainder of this Section is as follows. In Subsec.\
\ref{hamiltonian} we shall introduce and discuss the Hamiltonian for a small magnetic
particle. In Subsec.\ \ref{Z-F} the partition function and free energy are introduced.
In Subsec.\ \ref{series:Z} we shall carry out the expansion of the partition function in
powers of either the external field or the anisotropy constant, along with an
asymptotic expansion for strong anisotropy. Finally, in Subsec.\
\ref{series:FE}, we shall derive the corresponding expansions of the free energy.

\subsection{Hamiltonian}
\label{hamiltonian}

To begin with, we shall discuss the concept of effective Hamiltonian for a small,
magnetically ordered particle. Then we shall introduce the basic form of the
Hamiltonian that will be studied along this work, to conclude with the study of the
energy barriers in the longitudinal-field case.

\subsubsection{Effective Hamiltonian of a nanoparticle}

A basic assumption in small-particle magnetism is that a single-domain
particle, with a given physical orientation, is in {\em internal\/}
thermodynamical equilibrium at temperature $T$. Not too close to the
Curie temperature, its constituent spins rotate in unison (coherent
rotation), so the only relevant degree of freedom left is the
orientation of the net magnetic moment. With respect to this variable
the thermal equilibration can take place in a time scale that can be
considerably longer than that of the internal equilibration. Under
such conditions, the internal free energy (for a given instantaneous
orientation) can be considered as an effective energy (Hamiltonian)
for the orientational degrees of freedom.

The consideration of a internal free energy as an effective Hamiltonian for the
remainder degrees of freedom is indeed general, and it is founded in the very
statistical-mechanical definition of the free energy. Let $(\sP,\sQ)$ be the
canonical variables ``of interest" and $(\ePm,\eQm)$ the set of ``internal" variables.
The partition function, $\Z$, and the free energy, $\FE$, are defined in terms of the
total Hamiltonian of the system, $\Ham_{\rm T}$, as
\[
\Z
=
\int\!\!\D\sP \D\sQ \D\ePm \D\eQm\,
\exp[-\beta \Ham_{\rm T}(\sP,\sQ;\ePm,\eQm)]
\;,
\qquad
\FE
=
-\frac{1}{\beta}\ln\Z
\;,
\]
where $\beta=1/\T$.%
\footnote{
In these preliminary considerations, we omit in $\Z$ a factor $(2\pi\hbar)^{-s}$ where
$s$ is the number of degrees of freedom (Landau and Lifshitz, 1980, \S~31). This
factor, which renders $\Z$ dimensionless, when multiplied by the volume element in
the phase space
$\D\sP_{1}\cdots\D\sQ_{s}$ gives the semiclassical ``number of states" in this volume
element, providing in this way the proper link with the quantum-mechanical
expression for the partition function.
}
One can define {\em internal\/} quantities for given values of the variables $\sP$ and
$\sQ$ (marked by a tilde), as follows
\[
\tilde{\Z}(\sP,\sQ)
=
\int\!\!\D\ePm \D\eQm\,
\exp[-\beta \Ham_{\rm T}(\sP,\sQ;\ePm,\eQm)]
\;,
\qquad
\tilde{\FE}(\sP,\sQ)
=
-\frac{1}{\beta}\ln\tilde{\Z}(\sP,\sQ)
\;.
\]
Note that, by definition, the internal free energy obeys the relation
\[
\exp[-\beta\tilde{\FE}(\sP,\sQ)]
=
\int\!\!\D\ePm \D\eQm\,
\exp[-\beta \Ham_{\rm T}(\sP,\sQ;\ePm,\eQm)]
\;.
\]
Therefore, the total partition function $\Z$, from which all the equilibrium quantities
of the system can be derived, can be written as
\[
\Z
=
\int\!\!\D\sP \D\sQ \exp[-\beta\tilde{\FE}(\sP,\sQ)]
\;.
\]
This equation demonstrates the above statement: the so-defined internal free energy
$\tilde{\FE}(\sP,\sQ)$ plays the r\^{o}le of an effective Hamiltonian for the variables
$\sP$ and $\sQ$ when studying the {\em equilibrium\/} properties of the system. Note
that this effective Hamiltonian may have, by its very definition, terms dependent on
$T$.

Naturally, this approach is in principle applicable to any chosen pair of
variables $(\sP,\sQ)$. However, for this procedure to be useful, a time-scale separation
between some internal ``fast" variables and certain ``slow" ones must occur. In our
case, the orientation of the total magnetic moment plays the r\^{o}le of the latter and,
in what follows, we shall refer to the so-introduced internal free energy as the {\em
magnetic energy (Hamiltonian) of the nanoparticle}, and it will be simply denoted by
$\Hs(\m)$.

Similar considerations can, in principle, be applied to a magnetic domain in a bulk
magnet but, for such a macroscopic system, the time scale separation mentioned is so
huge that the probability of thermally activated magnetization reversal is almost zero
over astronomical time scales; the system is then effectively confined in a restricted
region of the phase space. Note finally that the separation procedure between
``internal" and ``relevant" variables would lead to exact results if one in fact uses the
above definitions to calculate $\tilde{\FE}(\sP,\sQ)$ by ``integrating out" the internal
variables. However, this is not possible in general, but one determines
$\tilde{\FE}(\sP,\sQ)$ on the basis of series truncations, symmetry arguments, etc.\
(Brown, 1979).

\subsubsection{Hamiltonian studied}

The magnetic energy of a nanoparticle has a number of different  contributions, e.g.,
magnetostatic self-energy (``demagnetization" or ``shape" energy), magneto-crystalline
energy, surface terms, magneto-elastic energy, etc. All these contributions give rise to
a dependence of the energy of the nanoparticle on the orientation of its magnetic
moment, i.e., in the absence of an external magnetic field the magnetic properties
of the system are anisotropic. We shall mainly consider systems where the {\em
magnetic-anisotropy energy} has the simplest axial symmetry. Then, if an external
field $\B$ is applied (assumed to be uniform over the volume of the system), the total
magnetic energy reads
\begin{equation}
\label{U0}
\Hs(\m)
=
-\frac{Kv}{\mm^{2}}(\m\cdot\hat{n})^{2}
-\m\cdot\B
\;,
\end{equation}
where $K$ is the magnetic-anisotropy energy constant, $v$ is the volume of the
nanoparticle, and $\hat{n}$ is a unit vector along the symmetry axis of the
magnetic-anisotropy term (hereafter referred to as the {\em anisotropy axis}).

On introducing the unit vectors $\vec{e}$, in the direction of the magnetic moment
($\vec{e}=\m/\mm$), and $\hat{b}$, in the direction of the external magnetic field
($\hat{b}=\B/B$), as well as the dimensionless anisotropy and field parameters
\begin{equation}
\label{sigma-xi}
\s
=
\frac{Kv}{\T}
\;,
\qquad
\xi
=
\frac{\mm B}{\T}
\;,
\end{equation}
the Hamiltonian (\ref{U0}) can be written as
\begin{equation}
\label{U1}
-\beta\Hs
=
\s(\vec{e}\cdot\hat{n})^{2}
+\xi(\vec{e}\cdot\hat{b})
\;.
\end{equation}
For $K>0$ the anisotropy is of ``easy-axis" type, since the two  existing minima of the
anisotropy term point along $\pm\hat{n}$ (the ``poles"). On the other hand, for $K<0$
the anisotropy is of ``easy-plane" type, the minima of the anisotropy term being then
continuously distributed over the plane perpendicular to $\hat{n}$ (the ``equatorial"
region).

The adopted expression for the magnetic anisotropy is the leading term in the
expansion of a general uniaxial magneto-crystalline anisotropy energy with respect to
the direction cosines of the magnetization.%
\footnote{
For instance, directions of easy magnetization in the equatorial plane would
be determined by higher-order terms in the expansion for $K<0$ (Landau and Lifshitz,
1984, \S~40).
}
On the other hand, such a form is also the appropriate one for shape
anisotropy (demagnetization self-energy) of an ellipsoid of revolution
\[
\Hs_{{\rm dem}}
=
\half v\Mss
\left(D_{a}\cos^{2}\!\vartheta
+D_{b}\sin^{2}\!\vartheta\right)
\;,
\]
where $\vartheta$ is the angle between the magnetic moment and the long (polar)
axis of the ellipsoid, $M_{s}=\mm/v$ is the spontaneous magnetization, $D_{a}$ the
demagnetization factor along the polar axis, and $D_{b}$ the demagnetization factor
along an equatorial axis. Indeed, we can write the above expression as
$\Hs_{{\rm dem}}
={\rm cte}-\half v\Mss\left(D_{b}-D_{a}\right)\cos^{2}\!\vartheta$, so that the
corresponding anisotropy constant reads
\begin{equation}
\label{Kdem} K_{{\rm dem}}
=
\half\Mss\left(D_{b}-D_{a}\right)
\;.
\end{equation}
In this case easy-axis and easy-plane anisotropy correspond, respectively, to prolate
and oblate ellipsoids of revolution.

For many materials, slight deviations from spherical shape make the shape  anisotropy
to dominate the remainder contributions to the magnetic anisotropy. On the other
hand, as was shown by Brown and Morrish (1957), a single-domain particle with an
{\em arbitrary\/} shape is equivalent to a suitably chosen general ellipsoid, as far as
the behavior of its magnetization in a uniform applied field is concerned. Therefore,
after these results, the seemingly specialized study of ellipsoids of revolution (i.e., of
uniaxial anisotropy) can be of great importance to account for the effects of a general
shape anisotropy.

In what follows we shall phrase our discussion in the language of classical magnetic
moments. Nevertheless, the results obtained will be applicable to systems consisting of
classical dipole moments that could approximately be described by Hamiltonians akin
to (\ref{U0}), i.e., Hamiltonians comprising a coupling term to an (electric or magnetic)
external field plus an axially symmetric orientational potential.

\subsubsection{Energy barriers in the longitudinal-field case}

We shall now study the behavior of the Hamiltonian in the illustrative
$\B\parallel\hat{n}$ case, determining its extrema and how they change as a function
of the several parameters in the Hamiltonian.

\begin{figure}[t!]
\vspace{-3.ex}
\eps{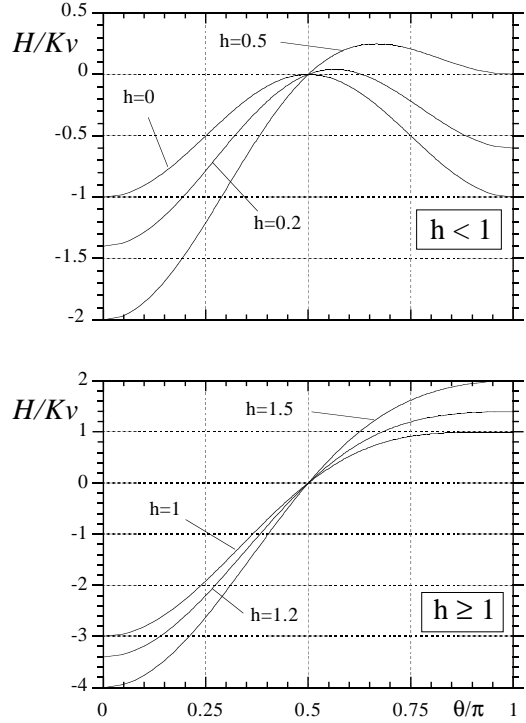}{0.7}
\vspace{-3.ex}
\caption[]
{
Magnetic energy in the longitudinal-field case for a number of values of
the reduced field parameter $h=B/\BK$. Upper panel: $0\leq h<1$, so that the potential
has two minima with an energy-barrier between them. Lower panel: $h\geq1$, so that
no potential barrier exists.
\label{barrier:plot}
}
\end{figure}

Before proceeding, let us introduce two useful quantities: the maximum {\em
anisotropy field}, $\BK$, and $h$, the external field measured in units of $\BK$,
\begin{equation}
\label{BK-h}
\BK
=
\frac{2Kv}{\mm}
\;,
\qquad h
=
\frac{B}{\BK}
=
\frac{\xi}{2\s}
\;.
\end{equation}
Let us now write the energy in terms of $\s$, the reduced field $h$, and the angle
$\vartheta$ between $\m$ and the anisotropy axis [cf.\ Eq.\ (\ref{U1})]
\begin{equation}
\label{Upara1}
\beta\Hs
=
-\s(\cos^{2}\!\vartheta+2h\cos\vartheta)
\;.
\end{equation}
To fix ideas, we shall assume $\s>0$, i.e., anisotropy of easy-axis type. The results for
$\s<0$, will be analogous but what is a maximum for $\s>0$, becomes a minimum for
$\s<0$, and vice-versa. The extrema of $\Hs$ are obtained by equating to zero the
$\vartheta$-derivative $\partial(\beta\Hs)/\partial\vartheta
=2\s\sin\vartheta(\cos\vartheta+h)$ getting
\[
\frac{\partial(\beta\Hs)}{\partial\vartheta}
=
0
\quad
\Longrightarrow
\quad
\left\{
\begin{array}{lcl}
\sin\vartheta
=
0
&
\Leftrightarrow
&
\vartheta
=
0,\pi\\
\cos\vartheta
=
-h
&
\mbox{if}
&
|h|\le 1
\end{array}
\right.
\;.
\]
The type of extrema is obtained by evaluating the second derivative at the extrema:
\[
\frac{\partial{}^{2}(\beta\Hs)}{\partial\vartheta^{2}}
=
\left\{
\begin{array}{ll}
\;\;\; 2\s(1+h)
&
\mbox{ for } \vartheta
=
0
\\
\;\;\; 2\s(1-h)
&
\mbox{ for } \vartheta
=
\pi
\\
-2\s(1-h^{2})
&
\mbox{ for } \cos\vartheta
=
-h \quad(\mbox{if~} |h|\le1)
\end{array}
\right.
\;,
\]
so that one gets the following results
\begin{center}
\begin{tabular}{c|c|c}
&
minima
&
maxima
\cr
\hline
$|h|<1$
&
$\vartheta=0,\pi$
&
$\vartheta=\arccos(-h)$
\cr
$h>1$
&
$\vartheta=0$
&
$\vartheta=\pi$
\cr
$h<-1$
&
$\vartheta=\pi$
&
$\vartheta=0$
\cr
\end{tabular}
\end{center}
Thus, for $|h|<1$ (i.e., for $|B|<\BK$), the energy has minima at
$\vartheta=0$ and $\vartheta=\pi$, with a maximum between them (see the upper
panel of Fig.\ \ref{barrier:plot}). On the other hand, for $|h|>1$ (that is, for
fields higher than the maximum anisotropy field $\BK$), the upper (shallower) energy
minimum ($\vartheta=\pi$ for $h>0$) turns into a maximum as it merges with the
intermediate maximum, which disappears (lower panel of Fig.\ \ref{barrier:plot}).

Finally, from the values of the energy at $\vartheta=0,\pi$, and, when
it exists, at the intermediate maximum $\vartheta_{\rm
M}=\arccos(-h)$, one gets the energy-barrier heights ($|h|<1$)
\[
\beta [\Hs(\vartheta_{\rm M})-\Hs(0)]
=
\spl
\;,
\qquad
\beta[\Hs(\vartheta_{\rm M})-\Hs(\pi)]
=
\smi
\;,
\]
where
\begin{equation}
\label{sigma:pm}
\spm
=
\s(1\pm h)^{2}
\;.
\end{equation}

\subsection{Partition function and free energy}
\label{Z-F}

\subsubsection{General definitions}

The statistical independence of non-interacting magnetic moments allows one to
express the thermodynamical quantities as sums over one-dipole contributions.
Consequently, we shall study these contributions and the results for the whole
system will be obtained by summation (or integration) of them over the ensemble
of dipoles, taking their different anisotropy constants, orientations about the external
field, magnitude of their dipole moments, etc. into account.

The {\em partition function\/} associated with a Hamiltonian
$\Hs(\vartheta,\varphi)$, where $\vartheta,\varphi$ are the angular coordinates of
$\m$ in a spherical coordinate system, can be defined as
\begin{equation}
\label{Z:def}
\Z
=
\frac{1}{2\pi}
\int_{0}^{\pi}\!\!\D{\vartheta}\,
\sin\vartheta
\int_{0}^{2\pi}\!\!\D{\varphi}\,
\exp[-\beta\Hs(\vartheta,\varphi)]
\;,
\end{equation}
while the associated free energy is then given by
\[
\FE
=
-\T\ln\Z
\;.
\]
The definition (\ref{Z:def}) deserves some discussion. First, as was
mentioned above, the definition of the partition function for a system
with one degree of freedom is $\Z=\int
(\D\sP\D\sQ/2\pi\hbar)\exp(-\beta\Hs)$ (Landau and Lifshitz, 1980,
\S~31). On the other hand, for a classical magnetic moment a
convenient pair of conjugate canonical variables is $\sP=\mz/\gmr$ and
$\sQ=\varphi$ [see Eq.\ (\ref{canonical_variables}) in Section
\ref{sect:gle}], where $\mz=\mm\cos\vartheta$ and $\gmr$ is the
gyromagnetic (or rather ``magnetogyric") ratio. Therefore
\[
\int
\frac{\D\sP \D\sQ}{2\pi\hbar} (\cdot)
=
\frac{\mm}{\gmr\hbar}\frac{1}{2\pi}
\int_{-1}^{1}\!\!\D{(\cos\vartheta)}\,
\int_{0}^{2\pi}\!\!\D{\varphi}\, (\cdot)
=
S\times\frac{1}{2\pi}
\int_{0}^{\pi}\!\!\D{\vartheta}\,
\sin\vartheta
\int_{0}^{2\pi}\!\!\D{\varphi}\, (\cdot)
\;,
\]
where $S=(\mm/\gmr)/\hbar$ is the quantum number associated with the
angular momentum $\mm/\gmr$. This expression yields $\Z=2S$ for
$\Hs\equiv0$, which is the correct semiclassical case ($S\gg1$) of the
corresponding quantum expression
$\Z=\sum_{S_{z}=-S}^{S}1=2S+1$. Therefore the definition (\ref{Z:def})
corresponds to the proper statistical-mechanical definition, except
for the factor $S$, which when required can be introduced by hand.

The {\em equilibrium probability distribution\/} of magnetic moment
orientations is given by the Boltzmann distribution
\[
\Weq(\cos\vartheta,\varphi)
=
\Z^{-1}\exp[-\beta\Hs(\vartheta,\varphi)]
\;,
\]
so that the {\em statistical-mechanical average\/} of any observable
$A=A(\m)=A(\vartheta,\varphi)$ reads
\begin{equation}
\label{average}
\llangle A\rrangle_{\eq}
=
\int\!\!\D{\Omega}\, A(\vartheta,\varphi)\Weq(\vartheta,\varphi)
=
\frac
{\int\!\D{\Omega}\,A(\vartheta,\varphi)\exp[-\beta\Hs(\vartheta,\varphi)]}
{\int\!\D{\Omega}\,\exp[-\beta\Hs(\vartheta,\varphi)]}
\;,
\end{equation}
where $\int\!\D{\Omega}\,(\cdot)\equiv(1/2\pi)
\int_{-1}^{1}\!\D{(\cos\vartheta)}\, \int_{0}^{2\pi}\!\D{\varphi}\,
(\cdot)$. The relevant thermodynamical quantities can be written as
the statistical-mechanical average of a certain function
$A=A(\vartheta,\varphi)$ as above. Besides, all of them can be
obtained as combinations of $\Z$ (or $\FE$) and its derivatives. Table
\ref{thermodyn:relations:table} summarizes some of these celebrated
relations, which illustrate the pivotal r\^{o}le that the calculation
of the partition function (or the free energy) plays in equilibrium
statistical mechanics.
\begin{table}
\caption[]
{
Definition of various thermodynamical quantities and their expressions in
terms of the partition function $\Z$, and of the free energy $\FE$.
\label{thermodyn:relations:table} }
\begin{center}
\begin{tabular}{|c||c|c|c|c|}
&
$\quad{\cal A}\quad$
&
def.
&
${\cal A}(\Z)$
&
${\cal A}(\FE)$
\cr
\hline
\hline
energy
&
$\E$
&
$\llangle\Hs\rrangle_{\eq}$
&
$-\frac{\partial {}}{\partial\beta}(\ln\Z)$
&
$\FE+\beta\frac{\partial {}}{\partial\beta}\FE$
\cr
\hline
entropy
&
$\ent$
&
$-\llangle\ln \Weq\rrangle_{\eq}$
&
$\ln\Z-\beta\frac{\partial {}}{\partial\beta}(\ln\Z)$
&
$\beta^{2}\frac{\partial {}}{\partial\beta}\FE$
\cr
\hline
\parbox{7em}{\center magnetization}
&
$M_{B}$
&
$\blangle\m\cdot\hat{b}\brangle_{\eq}$
&
$\mm\frac{\partial {}}{\partial\xi}(\ln \Z)$
&
$-\mm\beta\frac{\partial {}}{\partial\xi}\FE$
\end{tabular}
\end{center}
\end{table}

\subsubsection{Partition function for the simplest axially symmetric anisotropy
potential}

\begin{figure}[b!]
\vspace{-3.ex}
\eps{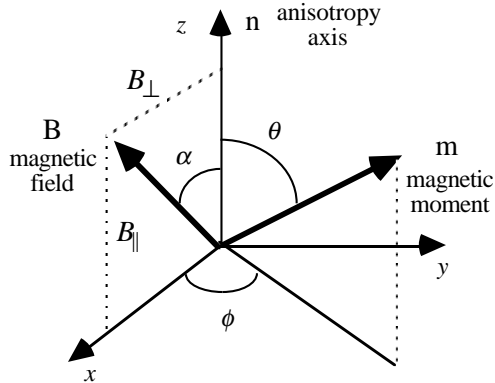}{0.56}
\vspace{-3.ex}
\caption[]
{
Coordinate system used in the calculation of the thermal-equilibrium
quantities. The plane determined by $\hat{n}$ and $\B$ is chosen as the $xz$-plane.
\label{coordinates:plot}
}
\end{figure}

We shall usually choose the anisotropy axis $\hat{n}$ as the polar
axis of a spherical coordinate system. Then, if $(\vartheta,\varphi)$
and $(\alpha,0)$ denote the angular coordinates of $\m$ and $\B$,
respectively (see Fig.\ \ref{coordinates:plot}), the Hamiltonian
(\ref{U1}) reads
\begin{equation}
\label{U2}
-\beta\Hs
=
\s\cos^{2}\!\vartheta
+\xipara\cos\vartheta+\xiperp\sin\vartheta\cos\varphi
\;,
\end{equation}
where we have introduced the longitudinal and transverse components (with respect
to the anisotropy-axis direction) of the dimensionless field $\vec{\xi}=\mm\B/\T$,
namely
\begin{equation}
\label{xiparaperp}
\xipara
=
\xi\cosal
\;,
\qquad
\xiperp
=
\xi\senal
\;.
\end{equation}

In order to analyze the partition function we, following Shcherbakova (1978), do
first the integral over $\varphi$ in the expression for $\Z$ associated with the
Hamiltonian (\ref{U2}), getting
\begin{equation}
\label{Z2}
\Z
=
\int_{0}^{\pi}\!\!\D{\vartheta}\,
\sin\vartheta
\exp(\s\cos^{2}\!\vartheta+\xipara\cos\vartheta)
     I_{0}(\xiperp\sin\vartheta)
\;,
\end{equation}
where
\begin{equation}
\label{mod:bessel:1st} I_{n}(y)
=
\frac{1}{\pi}
\int_{0}^{\pi}\!\!\D{t}\, e^{y\cos t}\cos nt
=
\sum_{k=0}^{\infty}\frac{1}{k!(k+n)!}
\left(\frac{y}{2}\right)^{2k+n},\quad n\ge 0
\;,
\end{equation}
is the modified Bessel function of the first kind of order $n$ (see, for example, Arfken,
1985, Sect.~11.5).

Equation (\ref{Z2}) gives the partition function in terms of an integral over $\vartheta$
only. Therefore, the integrand (divided by $\Z$) can be interpreted as an effective
probability distribution of the polar angle. Indeed, on introducing the substitution
$z=\cos\vartheta$ one can first write Eq.\ (\ref{Z2}) as
\begin{equation}
\label{Z_z}
\Z
=
\int_{-1}^{1}\!\!\D{z}\,
\exp(\s z^{2}+\xipara z)
    I_{0}(\xiperp\sqrt{1-z^{2}})
\;.
\end{equation}
Then, the thermal-equilibrium average of functions of $\cos\vartheta$ {\em only\/}
can be obtained through $\langle
A\rangle_{\eq}=\int_{-1}^{1}\!\D{z}\,A(z)\Weq^{\eff}(z)$ where
\begin{equation}
\label{pdfboltzmann:eff}
\Weq^{\eff}(z)
=
\frac{1}{\Z}
\exp(\s z^{2}+\xipara z)
       I_{0}(\xiperp\sqrt{1-z^{2}})
\;,
\end{equation}
is the effective or averaged (over the azimuthal angle), probability
distribution.  Naturally $\Weq^{\eff}(z)$ coincides with the actual
probability distribution when the total $\Hs(\m)$ is axially
symmetric.

\subsubsection{Particular cases and limiting regimes}

In various special cases, one can write down the partition function
and the free energy in a closed analytical form. Accordingly, along
with being relevant to get insight into the thermal-equilibrium
properties of the system, these expressions will be used as reference
for the general or approximate formulae derived along this Section.

\paragraph{Isotropic case.}

\begin{figure}[b!]
\vspace{-3.ex}
\eps{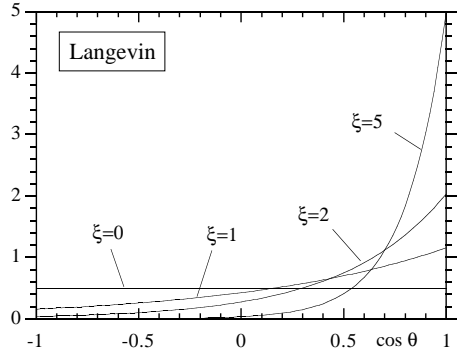}{0.7}
\vspace{-3.ex}
\caption[]
{
Probability distribution of the $z$ component of the magnetic moment for
$\s=Kv/\T=0$ in a magnetic field [Eq.\ (\ref{pdfboltzmann:eff:langevin})], for various
values of the dimensionless field parameter $\xi=\mm B/\T$. The value $0.5$
corresponds to the uniform probability distribution ($\s=\xi=0$).
\label{pdf:langevin:plot}
}
\end{figure}

We shall first consider the case $\s=0$. This {\em isotropic\/} or
{\em Langevin\/} regime will be attained if the anisotropy constant is
identically zero or at high temperatures where $|\s|\ll 1$. Then, the
partition function does not depend on $\alpha$
($\cosal=\hat{n}\cdot\hat{b}$), so we can choose $\alpha$ at will in
Eq.\ (\ref{Z_z}). On setting $\alpha=0$ (so that $\xiperp=0$ and
$\xipara=\xi$) and using $I_{0}(0)=1$, equation (\ref{Z_z}) reduces to
$\Z_{\lan}=\int_{-1}^{1}\!\D{z}\,\exp(\xi z)$. Therefore, the
partition function and free energy in the isotropic case can be
written as
\begin{equation}
\label{Z-F:langevin}
\Z_{\lan}
=
\frac{2}{\xi}\sinh\xi
\;,
\qquad
\FE_{\lan}
=
\T [\ln(\xi)-\ln(2\sinh\xi)]
\;.
\end{equation}
Similarly, the probability distribution (\ref{pdfboltzmann:eff}) reduces in this case to
\begin{equation}
\label{pdfboltzmann:eff:langevin}
\W_{\eq,\lan}(z)
=
\frac{\exp(\xi z)}{(2/\xi)\sinh\xi}
\;,
\end{equation}
which is is displayed in Fig.\ \ref{pdf:langevin:plot}

\paragraph{Zero-field case.}

\begin{figure}[b!]
\vspace{-3.ex}
\eps{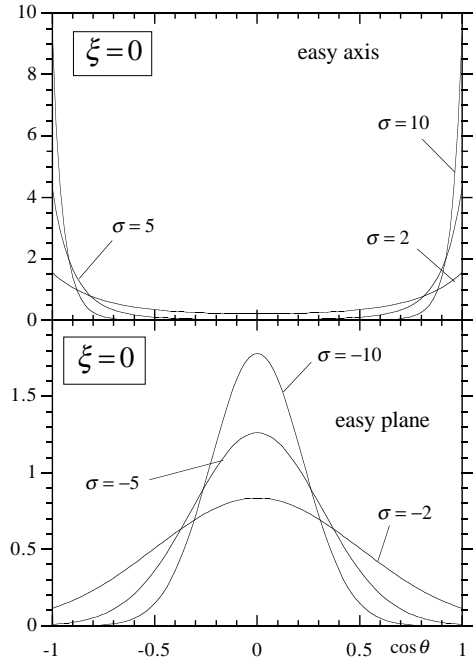}{0.7}
\vspace{-3.ex}
\caption[]
{
Probability distribution of the $z$ component of the magnetic moment in
zero field [Eq.\ (\ref{pdfboltzmann:eff:unb})], for different values of the dimensionless
anisotropy parameter $\s=Kv/\T$. The value $0.5$ corresponds to the uniform
probability distribution.
\label{pdf:unbiased:plot}
}
\end{figure}

In the absence of an external field (unbiased case), one can use again $I_{0}(0)=1$ in
Eq.\ (\ref{Z_z}), to get $\Z_{\unb}=2\int_{0}^{1}\!\D{z}\,\exp(\s z^{2})$. It will be very 
useful to introduce the function (Ra{\u{\i}}kher and Shliomis, 1975)
\begin{equation}
\label{F}
\F(\s)
\equiv
\int_{0}^{1}\!\!\D{z}\,\exp(\s z^{2})
\;,
\end{equation}
in terms on which one can simply write the partition function and the free energy in
the unbiased case as
\begin{equation}
\label{Z-F:unb}
\Z_{\unb}
=
2\F(\s) 
\;,
\qquad
\FE_{\unb}
=
-\T\ln[2\F(\s)]
\;.
\end{equation}
On the other hand, the probability distribution (\ref{pdfboltzmann:eff}) reduces in this
case to
\begin{equation}
\label{pdfboltzmann:eff:unb}
\W_{\eq,\unb}(z)
=
\frac{\exp(\s z^{2})}{2\F(\s)}
\;.
\end{equation}
In the easy-axis anisotropy case ($\s>0$), this probability
distribution evolves from uniform for $\s\ll 1$, to be quite
concentrated around the poles for $\s\gg1$ (see Fig.\
\ref{pdf:unbiased:plot}). Then the system approaches and effective
Ising spin, since the magnetic moment stays most of the time close to
the potential minima ($\m=\pm\mm\,\hat{n}$). For $\s<0$ (easy-plane
anisotropy), the probability distribution evolves from uniform for
$|\s|\ll1$, to be concentrated close to the equatorial circle for
$\s\ll-1$ (``plane-rotator" regime). Note that, in contrast to the
easy-axis anisotropy case, where for $\s\sim5$--$10$ the distribution
of magnetic moment orientations is rather concentrated around the
poles, for easy-plane anisotropy the corresponding shrink of the
probability distribution around the equatorial region is less steep as
a function of $|\s|$.

\paragraph{Ising regime.}

We shall now consider in more detail the $\s\gg 1$ range. Here, the function $\exp(\s
z^{2})$ in the integrand of Eq.\ (\ref{Z_z}) is sharply  peaked at the poles (see Fig.\
\ref{pdf:unbiased:plot}), so it can be approximated as a sum of two (non-normalized)
delta functions centered around $z=\pm1$. Consequently, one has
\begin{eqnarray*}
\Z
&
\simeq
&
\left[ e^{\xipara z}I_{0}(\xiperp\sqrt{1-z^{2}})
\right]_{z=1}
\int_{0}^{1}\!\!\D{z}\, e^{\s z^{2}}
+
\left[ e^{\xipara z}I_{0}(\xiperp\sqrt{1-z^{2}})
\right]_{z=-1}
\int_{-1}^{0}\!\!\D{z}\,e^{\s z^{2}}
\\
&
\lefteqn{\stackrel{I_{0}(0)=1}{=}}
&
\qquad
\F(\s)(e^{\xipara}+e^{-\xipara})
\;,
\qquad
\s\gg 1
\;.
\end{eqnarray*}
Then, on using the leading asymptotic result $\F(\s)\simeq e^{\s}/2\s$ (see Appendix
\ref{app:F}), the partition function and free energy in the ``Ising" regime, can be
written as
\begin{equation}
\label{Z-F:ising}
\Z_{\ising}
=
\frac{e^{\s}}{\s}\cosh\xipara
\;,
\qquad
\FE_{\ising}
=
-Kv+\T [\ln(\s)-\ln(\cosh\xipara)]
\;.
\end{equation}
Note however that for an Ising spin, the factor $e^{\s}/2\s$, is absent in the
corresponding $\Z$, which is equal to $e^{\xipara}+e^{-\xipara}=2\cosh\xipara$. This
factor does not alter quantities as the magnetization or the linear and non-linear 
susceptibilities, because they are obtained as $\xi$-derivatives of $\ln\Z$ (see Section
\ref{sect:quantities}). Nevertheless, the occurrence of the factor $e^{\s}/2\s$ moves
the ``thermal" quantities (thermodynamical energy, entropy, and specific heat) from
those of the archetypal Ising case.

Note finally that the employed replacement of the factor $\exp(\s z^{2})$ by a sum of
Dirac deltas will work if the remainder terms in the integrand vary slowly enough
with $z$. Naturally, this condition will not be obeyed for sufficiently high external
fields (specifically, for $\xi\gsim\s$).

\paragraph{Plane-rotator regime.}

For $\s\ll-1$ the term $\exp(\s z^{2})$ in the integrand of  Eq.\ (\ref{Z_z}) is peaked
at the equator (see Fig.\ \ref{pdf:unbiased:plot}). It can therefore be approximated by
a Dirac delta located at $z=0$, to get
\[
\Z
\simeq
\left[ e^{\xipara z} I_{0}(\xiperp\sqrt{1-z^{2}})
\right]_{z=0}
\int_{-1}^{1}\!\!\D{z}\, e^{\s z^{2}}
=
2\F(\s)I_{0}(\xiperp)
\;,
\quad
\s\ll-1
\;.
\]
Now, on employing the asymptotic ($\s\ll-1$) result $\F(\s)\simeq(-\pi/4\s)^{1/2}$
(Appendix \ref{app:F}), we obtain the following expressions for partition function and
free energy in the ``plane-rotator" regime
\begin{equation}
\label{Z-F:planerotator}
\Z_{\rotator}
=
\left(-\frac{\pi}{\s}\right)^{1/2} I_{0}(\xiperp)
\;,
\quad
\FE_{\rotator}
=
-\T
\left\{
\frac{1}{2}
\ln\left(-\frac{\pi}{\s}\right)
+\ln[I_{0}(\xiperp)]
\right\}
\;.
\end{equation}
The factor $(-\pi/\s)^{1/2}$ is absent in the partition function of the archetypal plane
rotator, which is merely given by $(1/2\pi)\int_{0}^{2\pi}\!\D{\varphi}\,
e^{\xiperp\cos\varphi}=I_{0}(\xiperp)$. Again,  this factor is irrelevant for the
quantities obtained as $\xi$-derivatives of $\ln\Z$, whereas is important for the
calculation of the thermal quantities. Similarly, the replacement of the factor
$\exp(\s z^{2})$ by a Dirac delta will only work for not very high external fields.

\paragraph{Longitudinal-field case.}

We shall finally consider the situation in which the external field points along the
anisotropy axis. In this case, without making assumptions concerning the magnitudes
of the anisotropy energy or the field, one can write down a closed analytical formula
for the partition function (and accordingly for all the thermodynamical quantities).

\begin{figure}[b!]
\vspace{-3.ex}
\eps{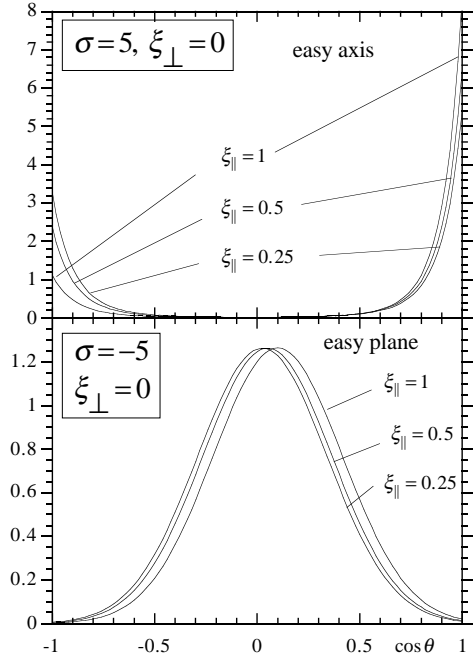}{0.7}
\vspace{-3.ex}
\caption[]
{
Probability distribution of the $z$ component of the magnetic moment [Eq.\
(\ref{distribution:para})] for $|\s|=|Kv/\T|=5$ and various values of the
longitudinal-field parameter $\xipara=\xi=\mm B/\T$.
\label{pdf:bpar:plot}
}
\end{figure}

When the external field is applied along the anisotropy axis one has $\xipara=\xi$ and
$\xiperp=0$, so that the general partition  function (\ref{Z_z}) reduces to
\begin{equation}
\label{Zpara1}
\Zp
=
\int_{-1}^{1}\!\!\D{z}\,
\exp(\s z^{2}+\xi z)
\;.
\end{equation}
Then, on completing the square in the argument of the exponential and taking the
definition (\ref{BK-h}) of $h$ into account, one gets $\Zp
=\exp(-\s h^{2})
\int_{-1}^{1}\!\D{z}\,
\exp[\s(z+h)^{2}]$. If we now introduce the substitution $t=z+h$, the partition function
reads
\[
\Zp
=
e^{-\s h^{2}}
\int_{h-1}^{h+1}\!\!\D{t}\, e^{\s t^{2}}
=
e^{-\s h^{2}}
\bigg[
\int_{0}^{h+1}\!\!\D{t}\, e^{\s t^{2}}-
\int_{0}^{h-1}\!\!\D{t}\, e^{\s t^{2}}
\bigg]
\;,
\]
so that, on using the substitutions $u=t/(h+1)$ in the first integral after the last equal
sign, and $u=t/(h-1)$ in the second one, we find
\[
\Zp
=
e^{-\s h^{2}}
\bigg\{ (1+h)
\int_{0}^{1}\!\!\D{u}\, e^{\s (1+h)^{2}u^{2}}
+(1-h)
\int_{0}^{1}\!\!\D{u}\, e^{\s (1-h)^{2}u^{2}}
\bigg\}
\;.
\]
However, the above integrals are merely the $\F$ function (\ref{F}) evaluated at
$\spm=\s(1\pm h)^{2}$ [the energy-barrier heights for $h<1$, Eq.\
(\ref{sigma:pm})], so that we can finally write the desired closed analytical formula for
$\Zp$ as
\begin{equation}
\label{Zpara}
\Zp
=
e^{-\s h^{2}}
\left[ (1+h)\F(\spl)+(1-h)\F(\smi)
\right]
\;.
\end{equation}
On the other hand, the probability distribution of $z=\cos\vartheta$ is in this case
given by
\begin{equation}
\label{distribution:para}
\W_{\eq,\|}(z)
=
\frac{\exp(\s z^{2}+\xi z)}{\Zp(\s,\xi)}
\;,
\end{equation}
which is displayed in Fig.\ \ref{pdf:bpar:plot} for various values of the longitudinal
field.

An alternative expression for $\Zp$ can be obtained by using the relation
(\ref{F-dawson}) between $\F(\s)$ and the Dawson integral $D(\cdot)$ [Eq.\
(\ref{dawson})], namely
\begin{equation}
\label{Zpara:dawson}
\Zp
=
\frac{e^{\s}}{\sqrt{\s}}
\left[ e^{\xi}D(\sqrt{\spl})+e^{-\xi}D(\sqrt{\smi})
\right]
\;.
\end{equation}
Note however that, since the relation employed only holds for $\s>0$, the above
formula for $\Zp$ is also subjected to the same restriction.

Let us finally consider some particular cases and approximations. On taking the
$h\to0$ limit in the expression (\ref{Zpara}), one again gets the unbiased partition
function $\Z_{\unb}=2\F(\s)$ [Eq.\ (\ref{Z-F:unb})]. The $\s\to0$ limit can also be
taken, but this should be done with some care. One must first realize that, since
$h=\xi/2\s$, the arguments of the $\F$ functions in Eq.\ (\ref{Zpara}) are large in this
case. Accordingly, on assuming for example $\s>0$ and using the leading term in the 
asymptotic expansion (\ref{Fderivatives:asympt1}) of $\F$, one has $\F(\spm)\simeq
e^{\spm}/2\spm$, whence (cf.\ Eq.\ (3.12) by Garanin, 1996)
\[
\Zp
\simeq e^{-\s h^{2}}
\left[ (1+h)\frac{e^{\spl}}{2\spl}+(1-h)\frac{e^{\smi}}{2\smi}
\right]
=
e^{\s}
\left[
\frac{e^{\xi}}{2\s+\xi}+\frac{e^{-\xi}}{2\s-\xi}
\right]
\;,
\]
where we have used $\spm=\s(1\pm h)^{2}$ and $\exp(\spm)
=\exp[\s(1+h^{2})]\exp(\pm\xi)$. On further manipulating the above expression, one
eventually gets the approximate result
\begin{equation}
\label{Zpara:approx}
\Zp
\simeq
\frac{2e^{\s}}{4\s^{2}-\xi^{2}}
(2\s\cosh\xi-\xi\sinh\xi)
\;,
\qquad
(K>0)
\;.
\end{equation}
Note that we have obtained more than we were initially looking for. Taking the limit
$\s\to0$ in this expression, we indeed get the isotropic partition function
$\Z_{\lan}=(2/\xi)\sinh\xi$ [Eq.\ (\ref{Z-F:langevin})]. However, on considering the
$\s\gg 1$ range of Eq.\ (\ref{Zpara:approx}), we get as a bonus the Ising partition
function $\Z_{\ising}=(e^{\s}/\s)\cosh\xi$ [Eq.\ (\ref{Z-F:ising})]. We have also
obtained this result since, for $\s\gg1$, the arguments of the functions $\F(\spm)$ in
$\Zp$ are also large and positive. Note finally that Eq.\ (\ref{Zpara:approx}) can also
be written in terms of $h=\xi/2\s$ as
\begin{equation}
\label{Zpara:approx:2}
\Zp\simeq\frac{e^{\s}}{2\s}\frac{1}{(1-h^{2})}
\left[ (1-h)e^{2\s h}+(1+h)e^{-2\s h}
\right]
\;,
\qquad
(K>0)
\;.
\end{equation}

\subsection{Series expansions of the partition function}
\label{series:Z}

We shall now carry out the expansion of the partition function  in powers of either the
external field or the anisotropy parameter, as well as an asymptotic expansion for
strong anisotropy. These expansions will enable us to derive the first few terms in the
corresponding expansions of the free energy in Subsec.\ \ref{series:FE}. From these
expressions one can obtain formulae for the linear and first non-linear susceptibilities,
as well as the deviations of the magnetization from the Langevin or Ising-type curves.

\subsubsection{Field expansion of the partition function}

Let us first consider the expansion of $\Z$ in powers of the external field
(Garc{\'{\i}}a-Palacios and L{\'{a}}zaro, 1997).

To begin with, we insert the power expansions of the functions $\exp(\xipara z)$ and
$I_{0}(\xiperp\sqrt{1-z^{2}})$ [see Eq.\ (\ref{mod:bessel:1st})], into the partition
function (\ref{Z_z}), to get
\begin{eqnarray*}
\Z
&
=
&
\sum_{i,k=0}^{\infty}
\frac{\xipara^{i}}{i!}
\left(\frac{\xiperp}{2}\right)^{2k}\frac{1}{(k!)^{2}}
\int_{-1}^{1}\!\!\D{z}\, z^{i}
\left(\sqrt{1-z^{2}}\right)^{2k}\exp(\s z^{2})
\\
&
=
&
2\sum_{i,k=0}^{\infty}
\frac{\xipara^{2i}\;\xiperp^{2k}}{(2i)!2^{2k}(k!)^{2}}
\int_{0}^{1}\!\!\D{z}\, z^{2i}(1-z^{2})^{k}\exp(\s z^{2})
\;.
\end{eqnarray*}
Note that the terms with odd powers of $z$ have vanished upon integration, while the
integration of the terms with even powers of $z$ has been reduced to the interval
$[0,1]$, by taking the symmetry of the corresponding integrand into account. Next, on
recalling the definitions (\ref{xiparaperp}) of $\xipara$ and $\xiperp$ and introducing
the angular coefficients
\begin{equation}
\label{bik} b_{i,k}(\alpha)
=
\frac{1}{(2i)!2^{2k}(k!)^{2}}
\cos^{2i}\!\alpha\sin^{2k}\!\alpha
\;,
\end{equation}
the partition function can be written as
\begin{equation}
\label{Z3}
\Z
=
2\sum_{i,k=0}^{\infty}b_{i,k}(\alpha)\xi^{2(i+k)}
\int_{0}^{1}\!\!\D{z}\, z^{2i}(1-z^{2})^{k}\exp(\s z^{2})
\;.
\end{equation}
Now, on expanding $(1-z^{2})^{k}$ by means of the binomial formula we obtain
\begin{equation}
\label{Z4}
\Z
=
2\sum_{i,k=0}^{\infty}b_{i,k}(\alpha)\xi^{2(i+k)}
\sum_{m=0}^{k}(-1)^{m}{k\choose m} \F^{(i+m)}(\s)
\;,
\end{equation}
where the ${k\choose m}=k!/[m!(k-m)!]$ are {\em binomial coefficients} and we have
used the derivatives $\F^{(\ell)}(\s)=\D^{\ell}\F/\D\s^{\ell}$ of the function $\F(\s)$
[Eq.\ (\ref{F})], namely
\begin{equation}
\label{F-Fderivatives}
\F^{(\ell)}(\s)
=
\int_{0}^{1}\!\!\D{z}\,z^{2\ell}\exp(\s z^{2})
\;,
\qquad
\ell
=
0,1,2,\ldots
\;,
\quad
\F^{(0)}\equiv\F
\;.
\end{equation}
Finally, on collecting the terms with the same power of $\xi$ by  means of the identity
\begin{equation}
\label{collecting_terms}
\sum_{i,k=0}^{\infty}A_{i,k}\,y^{(i+k)}
=
\sum_{j=0}^{\infty}
\bigg(
\sum_{\ell=0}^{j}A_{j-\ell,\ell}
\bigg) y^{j}
\;,
\end{equation}
the expansion (\ref{Z4}) can be rewritten as
\begin{equation}
\label{Zfinal}
\Z
=
2\F(\s)\sum_{i=0}^{\infty}\frac{C_{i}(\s,\alpha)}{i!}\xi^{2i}
\;,
\end{equation}
where the coefficients $C_{i}$ are given by
\begin{equation}
\label{Ci} C_{i}(\s,\alpha)
=
i!\sum_{k=0}^{i}b_{i-k,k}(\alpha)
\sum_{m=0}^{k}(-1)^{m}
{k\choose m}
\frac{\F^{(i-k+m)}(\s)}{\F(\s)}
\;.
\end{equation}
For the sake of later convenience, we have extracted the factor $\F(\s)$ in Eq.\
(\ref{Zfinal}) [recall that $2\F(\s)$ is the partition function at zero external field] and 
introduced the factor $i!$ in the definition of  the coefficients $C_{i}$.

\begin{figure}[t!]
\vspace{-3.ex}
\eps{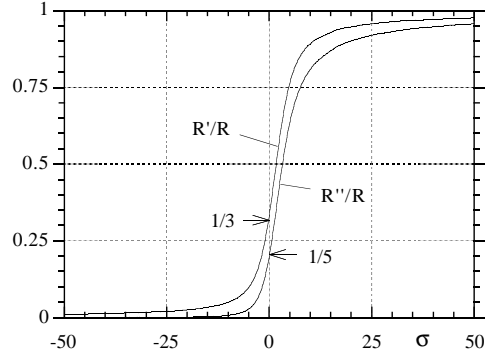}{0.7}
\vspace{-3.ex}
\caption[]
{
The functions $\F'/\F$ and $\F''/\F$.
\label{F-der:plot}
}
\end{figure}

The functions $\F^{(\ell)}$ are directly related with known special
functions ---confluent hypergeometric (Kummer) functions, error
functions, the Dawson integral, etc.--- and their properties are
summarized in Appendix \ref{app:F}.  All the combinations
$\F^{(\ell)}/\F$ occurring in the above coefficients are non-negative
and increase monotonically in the whole $\s$ range. $\F^{(\ell)}/\F$
tends to $0$ as $\s\to-\infty$, takes the value $1/(2\ell+1)$ at
$\s=0$ and tends to $1$ as $\s\to\infty$ [Eqs.\ (\ref{lim2}),
(\ref{F:zero}), and Eq.\ (\ref{lim1}), respectively]. The first two
quotients $\F^{(\ell)}/\F$ ($\F'/\F$ and $\F''/\F$) are shown in Fig.\
\ref{F-der:plot}. Note that we can write $\F'/\F
=\langle\cos^{2}\!\vartheta\rangle_{\eq}$, so that $\F'/\F$ is a
measure of the ``degree of polarization" of $\m$ along the anisotropy
axis in the absence of an external field.

\paragraph{Alternative expressions for the coefficients $C_{i}$.}

The coefficients $C_{i}$ can also be written in terms of the Kummer function
$M(a,c\,;x)$. First, on using the integral representation (\ref{kummer:integral}) for
$M(a,c\,;x)$, the integral occurring in the expression (\ref{Z3}) can be written as
\begin{equation}
\label{averages-kummer}
\int_{0}^{1}\!\!\D{z}\,z^{2i}(1-z^{2})^{k}e^{\s z^{2}}
=
\frac{\Gamma(i+\half)\Gamma(k+1)}{2\Gamma(i+k+\frac{3}{2})}
M(i+\half,i+k+\threehalfs;\s)
\;,
\end{equation}
where $\Gamma(z)$ is the gamma (factorial) function [Eq.\ (\ref{gamma:integral})]. If
we introduce this expression into the expansion (\ref{Z3}), we find the numerical
coefficient
\[
\frac{1}{(2i)!2^{2k}(k!)^{2}}
\frac{\Gamma(i+\half)\Gamma(k+1)}{2\Gamma(i+k+\frac{3}{2})}
=
\frac{1}{[2(i+k)+1]!}{i+k \choose k}
\;,
\]
where the basic property of the gamma function, $\Gamma(z+1)=z\Gamma(z)$, has
been used. Then, on gathering the terms with the same power of $\xi$ in the resulting
$\Z$ by dint of Eq.\ (\ref{collecting_terms}), we get
\[
\Z
=
2\sum_{i=0}^{\infty}\frac{\xi^{2i}}{(2i+1)!}
\left\{
\sum_{k=0}^{i}d_{i-k,k}(\alpha)M(i-k+\half,i+\threehalfs;\s)
\right\}
\;,
\]
where the angular coefficients $d_{i,k}(\alpha)$ are given by
\begin{equation}
\label{dik} d_{i,k}(\alpha)
=
{i+k\choose k}\cos^{2i}\!\alpha\sin^{2k}\!\alpha
\;.
\end{equation}
Consequently, on comparing with Eq.\ (\ref{Zfinal}), we can finally express the
coefficients $C_{i}$ as
\begin{equation}
\label{Ci:alt} C_{i}(\s,\alpha)
=
\frac{i!}{(2i+1)!}
\sum_{k=0}^{i}d_{i-k,k}(\alpha)
\frac{M(i-k+\half,i+\threehalfs;\s)}{M(\half,\threehalfs;\s)}
\;,
\end{equation}
where we have used (see Appendix \ref{app:F})
\begin{equation}
\label{F-kummer:particular}
\F(\s)
=
M(\half,\threehalfs;\s)
\;.
\end{equation}

Let us finally write in full the first few coefficients for future reference. If we
introduce the first few angular coefficients $d_{i,k}(\alpha)$
\[
\begin{array}{rclrclrcl} d_{0,0}
&
=
&
1
\;,
&
d_{1,0}
&
=
&
\cosqal
\;,
&
d_{0,1}
&
=
&
\senqal
\;,
\\
d_{2,0}
&
=
&
\coscal
\;,
&
d_{1,1}
&
=
&
2\cosqal\senqal
\;,
&
d_{0,2}
&
=
&
\sencal
\;,
\end{array}
\]
into Eq.\ (\ref{Ci:alt}), we get:
\begin{equation}
\label{C_1:alt} C_{1}
=
\frac{1}{3!}
\left[
\frac
{M(\threehalfs,{\textstyle \frac{5}{2}};\s)}
{M(\half,\threehalfs;\s)}
\cosqal
+\frac
{M(\half,{\textstyle \frac{5}{2}};\s)}
{M(\half,\threehalfs;\s)}
\senqal
\right]
\;,
\end{equation}
and
\[
C_{2}
=
\frac{1}{60}
\left[
\frac
{M({\textstyle \frac{5}{2}},{\textstyle \frac{7}{2}};\s)}
{M(\half,\threehalfs;\s)}
\cos^{4}\!\alpha
+\frac
{M(\threehalfs,{\textstyle \frac{7}{2}};\s)}
{M(\half,\threehalfs;\s)} 2\cos^{2}\!\alpha\sin^{2}\!\alpha
+\frac
{M(\half,{\textstyle\frac{7}{2};\s)}}
{M(\half,\threehalfs;\s)}
\sin^{4}\!\alpha
\right]
\;.
\]

The coefficients $C_{i}$ can also be expressed in terms of the averages of $\m$ in zero
field. To this end, let us begin from the definition of the partition function
\[
\Z
=
\int\!\!\D{\Omega}\,
\exp[\s(\vec{e}\cdot\hat{n})^{2}+\xi(\vec{e}\cdot\hat{b})]
\;,
\]
where $\int\!\D{\Omega}\,(\cdot)
=(1/2\pi)
\int_{-1}^{1}\!\D{(\cos\vartheta)}\,
\int_{0}^{2\pi}\!\D{\varphi}\, (\cdot)$ and the expression (\ref{U1}) for $-\beta\Hs$
have been used. Next, on expanding $\exp[\xi(\vec{e}\cdot\hat{b})]$ in powers of
$\xi$, we obtain
\[
\Z
=
\sum_{i=0}^{\infty}\frac{\xi^{2i}}{(2i)!}
\int\!\!\D{\Omega}\, (\vec{e}\cdot\hat{b})^{2i}\exp[\s(\vec{e}\cdot\hat{n})^{2}]
\]
where to eliminate the odd powers of $\xi$ we have merely considered that
$(\vec{e}\cdot\hat{b})^{2i+1}$ reverses its sign when the transformation
$\vec{e}\to-\vec{e}$ is applied, whereas the term
$\exp[\s(\vec{e}\cdot\hat{n})^{2}]$ is invariant against such transformation, whence
$\int\!\D{\Omega}\,
(\vec{e}\cdot\hat{b})^{2i+1}\exp[\s(\vec{e}\cdot\hat{n})^{2}]\equiv0$.
Finally, on comparing the above expansion of $\Z$ with
$\Z=2\F\sum_{i=0}^{\infty}(C_{i}/i!)\xi^{2i}$, noting that $\F(\s)$ can be written as
$\F(\s) =(1/2)\int\!\D{\Omega}\,\exp[\s(\vec{e}\cdot\hat{n})^{2}]$, and introducing
the thermal-equilibrium averages in zero field [cf.\ Eq.\ (\ref{average})]
\[
\left.
\blangle (\vec{e}\cdot\hat{b})^{n}
\brangle_{\eq}
\right|_{B=0}
=
\frac
{\int\!\D{\Omega}\,(\vec{e}\cdot\hat{b})^{n}\exp[\s(\vec{e}\cdot\hat{n})^{2}]}
{\int\!\D{\Omega}\,\exp[\s(\vec{e}\cdot\hat{n})^{2}]}
\;,
\]
we arrive at the desired relation
\begin{equation}
\label{Ci:averages}
\frac{C_{i}(\s,\alpha)}{i!}
=
\frac{1}{(2i)!}
\left.
\blangle (\vec{e}\cdot\hat{b})^{2i}
\brangle_{\eq}
\right|_{B=0}
\;.
\end{equation}

\paragraph{Particular cases of the coefficients $C_{i}$.}

Let us briefly consider the form that the coefficients appearing in the field expansion
of the partition function take in the particular cases considered in Subsec.\ \ref{Z-F}.
To this end, the alternative expression for those coefficients in terms of Kummer
functions [Eq.\ (\ref{Ci:alt})] results to be more convenient.

\begin{enumerate}
\item
On noting that $M(a,c\,;x=0)=1$ [see the definition (\ref{kummer:series})], one
gets for $C_{i}$ in the {\em isotropic\/} case
\[
\frac{1}{i!}
\left. C_{i}
\right|_{\s=0}
=
\frac{1}{(2i+1)!}
\sum_{k=0}^{i}
{i\choose k}
\cos^{2(i-k)}\!\alpha\sin^{2k}\!\alpha
=
\frac{1}{(2i+1)!}
\;,
\]
since the sum is equal to $(\cosqal+\senqal)^{i}=1$.
\item
In the $\s\to\infty$ limit, on employing the asymptotic expansion
(\ref{kummer:asympt1}) of $M(a,c\,;x)$ for large positive argument, one finds
\[
\left.
\frac{M(i-k+\half,i+\threehalfs;\s)}{M(\half,\threehalfs;\s)}
\right|_{\s\gg 1}
=
\frac{\Gamma(i+\frac{3}{2})}{\Gamma(i-k+\half)}
\frac{2}{\s^{k}}\stackrel{\s\to\infty}{\longrightarrow} (2i+1)\delta_{k,0}
\;,
\]
where we have used $\Gamma(i+3/2)=(i+1/2)\Gamma(i+1/2)$. Therefore, the general
expression (\ref{Ci:alt}) reduces in the {\em Ising\/} case to
\[
\frac{1}{i!}
\left. C_{i}
\right|_{\s\to\infty}
=
\frac{\cos^{2i}\!\alpha}{(2i)!}
\;.
\]

\item
To get the $\s\to-\infty$ limit of $C_{i}$, we can now use the asymptotic
expansion (\ref{kummer:asympt2}) of $M(a,c\,;x)$ for large negative argument. On
doing so, one first finds
\[
\left.
\frac{M(i-k+\half,i+\threehalfs;\s)}{M(\half,\threehalfs;\s)}
\right|_{\s\ll-1}
=
\frac{\Gamma(i+\frac{3}{2})}{\half\pi^{1/2}k!}
\frac{1}{(-\s)^{i-k}}\stackrel{\s\to-\infty}{\longrightarrow}
\frac{2\Gamma(i+\frac{3}{2})}{\pi^{1/2}i!}\delta_{i,k}
\;.
\]
Therefore, by using Eq.\ (\ref{gamma:halfinteger}) for the gamma function of
half-odd-integer argument, the {\em plane-rotator\/} $C_{i}$ reads
\[
\frac{1}{i!}
\left. C_{i}
\right|_{\s\to-\infty}
=
\left(\frac{\senal}{2}\right)^{2i}\frac{1}{(i!)^{2}}
\;.
\]

\item
The {\em longitudinal-field\/} case corresponds to set $\alpha=0$ in the
expression (\ref{Ci:alt}) for $C_{i}(\s,\alpha)$. On doing this and using
$d_{i-k,k}|_{\alpha=0}=\delta_{k,0}$ [see Eq.\ (\ref{dik})], one gets
\[
\frac{1}{i!}
\left. C_{i}
\right|_{\alpha=0}
=
\frac{1}{(2i+1)!}
\frac{M(i+\half,i+\threehalfs;\s)}{M(\half,\threehalfs;\s)}
=
\frac{1}{(2i)!}\frac{\F^{(i)}(\s)}{\F(\s)}
\]
where the relations (\ref{F-kummer}) between the functions $\F^{(\ell)}$ and Kummer
functions have been taken into account.
\end{enumerate}
All these particular cases of the coefficients $C_{i}$ are summarized in Table
\ref{Ci:limits:table}, while the first few ones are displayed in Table
\ref{C1C2C3:limits}.
\begin{table}[b!]
\caption[]
{
Expressions for the coefficients $C_{i}/i!$ of the field expansion of the
partition function in the isotropic, Ising, plane-rotator, and longitudinal-field cases.
\label{Ci:limits:table}
}
\begin{center}
\begin{tabular}{|c||c|c|c|c|}
&
$\s=0$
&
$\s\to\infty$
&
$\s\to-\infty$
&
$\B\parallel\hat{n}$
\cr
\hline
\hline
${\displaystyle \frac{C_{i}}{i!}}$
&
${\displaystyle \frac{1}{(2i+1)!}}$
&
${\displaystyle \frac{\cos^{2i}\!\alpha}{(2i)!}}$
&
${\displaystyle \frac{\sin^{2i}\!\alpha}{2^{2i}(i!)^{2}}}$
&
${\displaystyle \frac{1}{(2i)!}\frac{\F^{(i)}}{\F}}$
\end{tabular}
\end{center}
\end{table}

\begin{table}[t!]
\caption[]
{
The coefficients $C_{1}$, $C_{2}$, and $C_{3}$ in the isotropic, Ising,
plane-rotator, and longitudinal-field cases. \label{C1C2C3:limits}}
\begin{center}
\begin{tabular}{|c||c|c|c|c|}
&
$\s=0$
&
$\s\to\infty$
&
$\s\to-\infty$
&
$\B\parallel\hat{n}$
\cr
\hline
\hline
$C_{1}$
&
${\displaystyle \frac{1}{6}}$
&
${\displaystyle \frac{1}{2}\cosqal}$
&
${\displaystyle \frac{1}{4}\senqal}$
&
${\displaystyle \frac{1}{2}\frac{\F'}{\F}}$
\cr
\hline
$C_{2}$
&
${\displaystyle \frac{1}{60}}$
&
${\displaystyle \frac{1}{12}\coscal}$
&
${\displaystyle \frac{1}{32}\sencal}$
&
${\displaystyle \frac{1}{12}\frac{\F''}{\F}}$
\cr
\hline
$C_{3}$
&
${\displaystyle \frac{1}{840}}$
&
${\displaystyle \frac{1}{120}\cos^{6}\!\alpha}$
&
${\displaystyle \frac{1}{384}\sin^{6}\!\alpha}$
&
${\displaystyle \frac{1}{120}\frac{\F'''}{\F}}$
\end{tabular}
\end{center}
\end{table}

\subsubsection{Expansion of the partition function in powers of the anisotropy
parameter}

We shall now derive the first few terms in the expansion of $\Z$ in powers of
$\s=Kv/\T$. This expansion will be a suitable description of the
thermodynamical properties when the anisotropy energy is sufficiently small in
comparison with the thermal energy.

In order to perform this expansion, it is more convenient to rotate the spherical
coordinate system to set the polar axis pointing along the external field $\B$ (see Fig.\
\ref{coordinates:plot}; the anisotropy axis $\hat{n}$ is now in the $xz$-plane and
$\alpha$ is its polar angle). With this choice of coordinates, the partition function reads
\[
\Z
=
\frac{1}{2\pi}
\int_{0}^{\pi}\!\!\D{\vartheta}\,
\sin\vartheta
\exp(\xi\cos\vartheta)
\int_{0}^{2\pi}\!\!\D{\varphi}\,
\exp[\s(\cosal\cos\vartheta+\senal\sin\vartheta\cos\varphi)^{2}]
\;.
\]
If we now expand the second exponential, we get an expression of the form
\begin{equation}
\label{Zs2}
\Z
=
\sum_{i=0}^{\infty}\frac{\s^{i}}{i!}\Z_{i}
\;,
\end{equation}
where
\begin{equation}
\label{Zsubi}
\Z_{i}
=
\frac{1}{2\pi}
\int_{0}^{\pi}\!\!\D{\vartheta}\,
\sin\vartheta
\exp(\xi\cos\vartheta)
\int_{0}^{2\pi}\!\!\D{\varphi}\,
(\cosal\cos\vartheta+\senal\sin\vartheta\cos\varphi)^{2i}
\;.
\end{equation}
Note that the zeroth order coefficient is naturally the isotropic partition function
$\Z_{0}=(2/\xi)\sinh\xi$ [Eq.\ (\ref{Z-F:langevin})].

On using the binomial expansion in the second integrand of Eq.\ (\ref{Zsubi}), and
employing the following result (Arfken, 1985, p.~318),
\begin{equation}
\label{int:cosnphi}
\frac{1}{2\pi}\int_{0}^{2\pi}\!\!\D{\varphi}\,
\cos^{n}\!\varphi
=
\left\{
\begin{array}{cl} 0
&
\mbox{for~odd~} n
\\
{\displaystyle \frac{(2k)!}{2^{2k}(k!)^{2}}}
&
\mbox{ for } n
=
2k
\end{array}
\;,
\right.
\end{equation}
to do the integrals over the azimuthal angle, we see that only even  powers of $\cosal$
and $\senal$ appear in $\Z_{i}$. On the other hand, $\sin^{2k}\!\vartheta$ can always
be expressed as a sum of powers of the form $\cos^{2\ell}\!\vartheta$, with $\ell
\le k$, namely
\[
\sin^{2k}\!\vartheta
=
(1-\cos^{2}\!\vartheta)^{k}
=
\sum_{\ell=0}^{k}{k\choose \ell}(-1)^{\ell}\cos^{2\ell}\!\vartheta
\;.
\]
Accordingly, on introducing once more the substitution $z=\cos\vartheta$ and noting
that,
\begin{equation}
\label{dZ0}
\int_{-1}^{1}\!\!\D{z}\,z^{n}\exp(\xi z)
=
\frac{\D^{n}{}}{\D\xi^{n}}
\int_{-1}^{1}\!\!\D{z}\,\exp(\xi z)
=
\frac{\D^{n}{}}{\D\xi^{n}}\Z_{0}
\;,
\end{equation}
one realizes that all the functions $\Z_{i}$ can be expressed in terms of the isotropic
partition function, $\Z_{0}$, and its $\xi$-derivatives. For instance, $\Z_{1}$ reads
\begin{eqnarray}
\label{Zsub1}
\Z_{1}
&
=
&
\cosqal
\int_{-1}^{1}\!\!\D{z}\,z^{2}\exp(\xi z)
+\frac{1}{2}\senqal
\int_{-1}^{1}\!\!\D{z}\,(1-z^{2})\exp(\xi z)
\nonumber
\\
&
=
&
\Z''_{0}\cosqal+\frac{1}{2}(\Z_{0}-\Z''_{0})\senqal
\;,
\end{eqnarray}
where the prime denotes differentiation with respect to $\xi$. On the other hand,
since $\Z_{0}=(2/\xi)\sinh\xi$, the derivative $\Z'_{0}$ is given by
\begin{equation}
\label{Zsub0:derivative}
\Z'_{0}
=
L(\xi)\Z_{0}
\;,
\end{equation}
where
\begin{equation}
\label{langevin:function}
L(\xi)
=
\coth\xi-\frac{1}{\xi}
\;,
\end{equation}
is the celebrated Langevin function. On taking a further
$\xi$-derivative and using the relation between $L'$ and $L$, namely
\begin{equation}
\label{L-Lp}
L'
=
1-\frac{2}{\xi}L-L^{2}
\;,
\end{equation}
we get for the combinations of $\Z_{0}$ and $\Z''_{0}$ occurring in Eq.\ (\ref{Zsub1})
\begin{equation}
\label{Zsub0:derivatives2}
\Z''_{0}
=
\Z_{0}\left(1-\frac{2}{\xi}L\right)
\;,
\qquad
\frac{1}{2}(\Z_{0}-\Z''_{0})
=
\Z_{0}\frac{1}{\xi}L
\;.
\end{equation}
Therefore, on introducing these results in Eq.\ (\ref{Zsub1}), we finally get
\begin{equation}
\label{Zsub1:final}
\frac{\Z_{1}}{\Z_{0}}
=
\left(1-\frac{2}{\xi}L\right)\cosqal
+\frac{1}{\xi}L\senqal
\;.
\end{equation}

The calculation of $\Z_{2}$ proceeds similarly. On taking the definition (\ref{Zsubi})
into account and using $z=\cos\vartheta$, one obtains
\begin{eqnarray*}
\Z_{2}
&
=
&
\coscal
\int_{-1}^{1}\!\!\D{z}\,z^{4}e^{\xi z}
+\frac{6}{2}\cosqal\senqal
\int_{-1}^{1}\!\!\D{z}\,z^{2}(1-z^{2}) e^{\xi z}
\\
&
& {}+
\frac{3}{8}\sencal
\int_{-1}^{1}\!\!\D{z}\,(1-z^{2})^{2}e^{\xi z}
\;,
\end{eqnarray*}
where Eq.\ (\ref{int:cosnphi}) has been used for calculating the integrals over
$\varphi$. Consequently, in terms of $\Z_{0}$ and its derivatives, $\Z_{2}$ is given by
\begin{equation}
\label{Zsub2}
\Z_{2}
=
\Z''''_{0}\coscal
+3(\Z''_{0}-\Z''''_{0})\cosqal\senqal
+\frac{3}{8}(\Z_{0}-2\Z''_{0}+\Z''''_{0})\sencal
\;.
\end{equation}
In order to take the 4th-order derivative $\Z''''_{0}$, one can repeatedly use
Eqs.\ (\ref{Zsub0:derivative}) and (\ref{L-Lp}). However, it significantly simplifies the
calculations to obtain first the derivative $(L/\xi)'$, which can be written as
\begin{equation}
\label{fracxiL:derivative}
\left(\frac{1}{\xi}L\right)'
=
-\frac{1}{\xi}
\left[ L^{2}-\left(1-\frac{3}{\xi}L\right)
\right]
\;.
\end{equation}
Thus, after some manipulation, one gets the expression
\[
\Z''''_{0}
=
\Z_{0}
\left[ 1-\frac{4}{\xi}L+\frac{8}{\xi^{2}}\left(1-\frac{3}{\xi}L\right)
\right]
\;,
\]
which, along with Eqs.\ (\ref{Zsub0:derivatives2}), gives
\[
\Z''''_{0}-\Z''_{0}
=
2\Z_{0}
\left[
\frac{4}{\xi^{2}}\left(1-\frac{3}{\xi}L\right)-\frac{1}{\xi}L
\right]
\;,
\quad
\Z_{0}-2\Z''_{0}+\Z''''_{0}
=
\Z_{0}\frac{8}{\xi^{2}}\left(1-\frac{3}{\xi}L\right)
\;.
\]
On introducing all these results into Eq.\ (\ref{Zsub2}), we finally find for $\Z_{2}$:
\begin{eqnarray}
\label{Zsub2:final}
\frac{\Z_{2}}{\Z_{0}}
&
=
&
\left[ 1-\frac{4}{\xi}L+\frac{8}{\xi^{2}}\left(1-\frac{3}{\xi}L\right)
\right]
\coscal
\nonumber
\\
&
& {}+6
\left[
\frac{1}{\xi}L-\frac{4}{\xi^{2}}\left(1-\frac{3}{\xi}L\right)
\right]
\cosqal\senqal
+\left[
\frac{3}{\xi^{2}}\left(1-\frac{3}{\xi}L\right)
\right]
\sencal
\;.
\qquad
\quad
\end{eqnarray}
This formula completes the explicit expansion of the partition
function in powers of the anisotropy parameter up to second order.

\subsubsection{Asymptotic expansion of the partition function for strong anisotropy}

In order to complement the above derived weak-anisotropy expansion, we
shall now carry out an asymptotic expansion of the partition function
for strong anisotropy (easy-axis case only). As will be seen below,
the approximate thermal-equilibrium quantities obtained from the
combined use of those expansions, well approximate the exact results
in the whole temperature range. Therefore we shall be able to get
simple analytical expressions for the thermodynamical quantities that
reasonably avoid the necessity of their computation by numerical
methods.

In order to perform an expansion of the partition function for large $\s=Kv/\T$, we
shall start from the field expansion (\ref{Zfinal}) of $\Z$ and use the asymptotic
results for its coefficients. Then, we shall obtain a number of infinite series of powers
of $\xi=\mm B/\T$, which will be identified as certain elementary functions, obtaining
in this way a closed asymptotic expression for $\Z$.

We start by recalling that the whole coefficient of $\xi^{2i}$ in the general
$\xi$-expansion of $\Z$ reads [see Eqs.\ (\ref{Zfinal}) and (\ref{Ci:alt})]
\begin{equation}
\label{Ci:complete}
\frac{2\F(\s)C_{i}}{i!}
=
\frac{2}{(2i+1)!}
\sum_{k=0}^{i}d_{i-k,k}(\alpha) M(i-k+\half,i+\threehalfs;\s)
\;,
\end{equation}
where $\F(\s)=M(\half,\threehalfs;\s)$ has been used [Eq.\
(\ref{F-kummer:particular})], and $d_{i-k,k}(\alpha)$ is explicitly given by [see Eq.\
(\ref{dik})]
\begin{equation}
\label{dimkk} d_{i-k,k}(\alpha)
=
{i\choose k}\cos^{2(i-k)}\!\alpha\sin^{2k}\!\alpha
\;.
\end{equation}
On the other hand, the asymptotic expansion (\ref{kummer:asympt1}) of the confluent
hypergeometric functions yields for $\s\gg 1$
\begin{eqnarray*}
M(i-k+\half,i+\threehalfs;\s)
&
=
&
\frac{e^{\s}}{2\s}
\frac{2\Gamma(i+\threehalfs)}{\Gamma(i-k+\half)}
\frac{1}{\s^{k}}
\bigg[ 1+\frac{(2k-2i+1)(k+1)}{2\s}
\\
&
& {}+\frac{(2k-2i+3)(2k-2i+1)(k+1)(k+2)}{8\s^{2}}
+\cdots
\bigg]
\;.
\end{eqnarray*}
Considering that the sum in Eq.\ (\ref{Ci:complete}), begins at $k=0$, and that we 
shall carry out the expansion of $\Z$ through order $1/\s^{2}$, we write
\begin{eqnarray*}
\lefteqn{
\sum_{k=0}^{i}d_{i-k,k}M(i-k+\half,i+\threehalfs;\s)
\simeq\cos^{2i}\!\alpha
\, M(i+\half,i+\threehalfs;\s) }
\qquad
\qquad
\\
&
& {}+\frac{1}{2}(2i)\cos^{2(i-1)}\!\alpha\senqal
\, M(i-\half,i+\threehalfs;\s)
\\
&
& {}+\frac{1}{8}(2i)(2i-2)\cos^{2(i-2)}\!\alpha\sencal
\, M(i-\threehalfs,i+\threehalfs;\s)
\;,
\end{eqnarray*}
where we have taken Eq.\ (\ref{dimkk}) into account. Now, on using
$\Gamma(z+1)=z\Gamma(z)$, we get for the quotients of gamma functions occurring in
the above equation via the Kummer functions
\[
\frac{2\Gamma(i+\threehalfs)}{\Gamma(i-k+\half)}
=
\left\{
\begin{array}{cl}
\;\;\;(2i+1),&
\mbox{ for } k
=
0
\\
\half(2i+1)(2i-1),&
\mbox{ for } k
=
1
\\
\frac{1}{4}(2i+1)(2i-1)(2i-3),&
\mbox{ for } k
=
2
\end{array}
\right.
\;.
\]

On collecting all these intermediate results, we can approximately write  the $i$th
term in the $\xi$-expansion of $\Z$ in the form
\begin{eqnarray}
\label{Ci:complete:approx}
\frac{\s}{e^{\s}}\frac{2\F(\s)C_{i}}{i!}\;\xi^{2i}
&
\simeq
&
\frac{\xipara^{2i}}{(2i)!}
\left[ 1-\frac{(2i-1)}{2\s}+\frac{(2i-1)(2i-3)}{4\s^{2}}
\right]
\nonumber
\\
&
& {}+
\frac{\xipara^{2(i-1)}\xiperp^{2}}{[2(i-1)]!}
\left[
\frac{1}{4\s}-\frac{(2i-3)}{4\s^{2}}
\right]
+\frac{\xipara^{2(i-2)}\xiperp^{4}}{[2(i-2)]!}
\frac{1}{32\s^{2}}
\;,
\nonumber
\\
\end{eqnarray}
where we have multiplied across by $\s/e^{\s}$ to avoid writing $e^{\s}/\s$ in all the
right-hand sides of the subsequent equations. In addition, in the above equation we
have introduced the longitudinal and transverse components of the dimensionless
field: $\xipara=\xi\cosal$ and $\xiperp=\xi\senal$. Note however that Eq.\
(\ref{Ci:complete:approx}) only holds for the terms with $i\ge2$. For $i=1$, the sum in
$k$ in the expression (\ref{Ci:alt}) only runs over $k=0$ and $k=1$; therefore, the last
term on the right-hand side of Eq.\ (\ref{Ci:complete:approx}) is absent. Similarly, for
$i=0$, only the first term remains. Taking these considerations into account by
properly adjusting the summation limits in the following expression, we can already
write down the partition function
$\Z=2\F\sum_{i=0}^{\infty}(C_{i}/i!)\xi^{2i}$ as
\begin{eqnarray*}
\frac{\s}{e^{\s}}\Z
&
\simeq
&
\sum_{i=0}^{\infty}\frac{\xipara^{2i}}{(2i)!}
\left[ 1-\frac{(2i-1)}{2\s}+\frac{(2i-1)(2i-3)}{4\s^{2}}
\right]
\\
&
& {}+
\frac{1}{4\s}
\sum_{i=1}^{\infty}
\frac{\xipara^{2(i-1)}\xiperp^{2}}{[2(i-1)]!}
\left[ 1-\frac{(2i-3)}{\s}
\right]
+
\frac{1}{32\s^{2}}
\sum_{i=2}^{\infty}
\frac{\xipara^{2(i-2)}\xiperp^{4}}{[2(i-2)]!}
\;.
\end{eqnarray*}
If we now redefine the summation indices in order to force all the above series to
start at the value zero of the corresponding new index and gather the terms
multiplying the same type of series, we get
\begin{eqnarray}
\label{Z:asympt:1}
\frac{\s}{e^{\s}}\Z
&
\simeq
&
\left(1+\frac{1}{4\s}\xiperp^{2}+\frac{1}{32\s^{2}}\xiperp^{4}\right)
\sum_{i=0}^{\infty}\frac{\xipara^{2i}}{(2i)!}
\nonumber
\\
&
& {}-\left(\frac{1}{2\s}+\frac{1}{4\s^{2}}\xiperp^{2}\right)
\sum_{i=0}^{\infty}\frac{\xipara^{2i}}{(2i)!}(2i-1)
+\frac{1}{4\s^{2}}
\sum_{i=0}^{\infty}\frac{\xipara^{2i}}{(2i)!}(2i-1)(2i-3)
\;.
\nonumber
\\
\end{eqnarray}

Our last goal is to identify all the power series occurring in Eq.\ (\ref{Z:asympt:1}).
The series in the first term on the right-hand side is precisely that of the hyperbolic
cosine, $\cosh x=\sum_{i=0}^{\infty}x^{2i}/(2i)!$. The other two series can also be
identified after some redefinition of the summation indices ($k=i-1$):
\[
\sum_{i=0}^{\infty}\frac{x^{2i}}{(2i)!}(2i-1)
=
\sum_{k=0}^{\infty}\frac{x^{2k+2}}{(2k+1)!}
-\cosh x
=
x\sinh x-\cosh x
\;,
\]
while
\begin{eqnarray*}
\sum_{i=0}^{\infty}\frac{x^{2i}}{(2i)!}(2i-1)(2i-3)
&
=
&
\sum_{k=0}^{\infty}\frac{x^{2(k+1)}}{(2k)!}
-3(x\sinh x-\cosh x)
\\
&
=
&
(x^{2}+3)\cosh x-3x\sinh x
\;.
\end{eqnarray*}
Finally, we insert these results into Eq.\ (\ref{Z:asympt:1}), gather the
terms with the same power of $1/\s$, and extract a factor $\cosh\xipara$, obtaining
\begin{eqnarray}
\label{Z:asympt:2}
\Z\simeq
\frac{e^{\s}}{\s}\cosh\xipara
&
\bigg\{
&
1+\frac{1}{4\s}
\left[ (2+\xiperp^{2})-2\xipara\tanh\xipara
\right]
\nonumber
\\
&
& {}+\frac{1}{4\s^{2}}
\left[
\left(3+\xipara^{2}
+\xiperp^{2}
+\frac{1}{8}\xiperp^{4}\right)
-(3+\xiperp^{2})\xipara\tanh\xipara
\right]
\bigg\}
\;.
\nonumber
\\
\end{eqnarray}
This equation is the desired asymptotic expansion of the partition function. Note that,
as could be expected, the leading term in this equation is precisely what we called
partition function in the Ising regime  [Eq.\ (\ref{Z-F:ising})].

\subsection{Series expansions of the free energy}
\label{series:FE}

Once one has obtained an expansion of the partition function in a series of powers of a
given quantity, one needs to construct the corresponding expansion of $\ln\Z$ in order
to obtain the relevant thermal-equilibrium quantities (see Table
\ref{thermodyn:relations:table}). Here, we shall derive the expansions of the free
energy $\FE=-\T\ln\Z$ corresponding to those developed above for the partition
function.

\subsubsection{Expansion of the logarithm of a function}

The problem of constructing the series expansion of the logarithm of a function with a
given series representation appears in a number of physical and  mathematical
problems (e.g., in the construction of the {\em cumulants\/} of a probability
distribution in terms of the known {\em moments\/} of such distribution; see Risken,
1989). Thus, if one has derived an expansion of the partition function of the type
\begin{equation}
\label{Zexpansion:general}
\Z(y)
=
\Z(0)
\sum_{i=0}^{\infty}
\frac{A_{i}}{i!}y^{i}
\;,
\end{equation}
(note that $A_{0}=1$), the first few terms in the corresponding expansion of $\ln\Z$
are given by
\begin{eqnarray}
\label{lnZexpansion:general}
\ln\Z(y)
&
=
&
\ln\Z(0)
+
A_{1}y
+
\frac{1}{2}\left(A_{2}-A_{1}^{2}\right)y^{2}
+
\frac{1}{6}\left(A_{3}-3 A_{2}A_{1}+2 A_{1}^{3}\right)y^{3}
\nonumber
\\
&
& {}+
\frac{1}{24}
\left( A_{4}-4 A_{3}A_{1}-3 A_{2}^{2}+12A_{2}A_{1}^{2}-6A_{1}^{4}
\right) y^{4}
+\cdots
\;.
\end{eqnarray}
This formula, when multiplied by $-\T$, gives the first few terms of the
$y$-expansion of the free energy.

\subsubsection{Averages for anisotropy axes distributed at random}

In what follows we shall frequently consider the values of the relevant quantities for
an ensemble of magnetic moments whose anisotropy axes are distributed at random.
Note that averaging, in the sense of keep fixed some parameters and then {\em
sum\/} over the remainder ones (e.g., anisotropy-axis orientations), does not make
sense for the partition function since, {\em for independent entities}, $\Z$ is a 
multiplicative quantity. On the other hand, averaging makes sense for the customary
thermodynamical functions (free energy, entropy, energy, etc.) as they are additive
quantities.

When averaging the thermodynamical quantities over assemblies of equivalent 
magnetic moments (i.e., with the same characteristic parameters) whose anisotropy
axes are distributed at random, we shall need to calculate integrals of the general form
\[
\llangle f(\varphi_{\hat{n}},\alpha)\rrangle_{\ran}
=
\frac{1}{4\pi}
\int_{0}^{2\pi}\!\!\D{\varphi_{\hat{n}}}\,
\int_{0}^{\pi}\!\!\D{\alpha}\,
\senal f(\varphi_{\hat{n}},\alpha)
\;,
\]
where $\varphi_{\hat{n}}$ and $\alpha$ are, respectively, the azimuthal and polar
angles of the unit vector along the anisotropy axis $\hat{n}$. We shall be mainly
interested in the cases where
$f(\varphi_{\hat{n}},\alpha)=\cos^{2i}\!\alpha\sin^{2k}\!\alpha$, which does not
depend on the azimuthal angle. For these functions, one finds
\[
\llangle\cos^{2i}\!\alpha\sin^{2k}\!\alpha\rrangle_{\ran}
=
\frac{1}{2}
\int_{0}^{\pi}\!\!\D{\alpha}\,
\senal\cos^{2i}\!\alpha\sin^{2k}\!\alpha
\stackrel{x=\cosal}{=}
\int_{0}^{1}\!\!\D{x}\, x^{2i}(1-x^{2})^{k}
\;.
\]
Now, on comparing with the relation (\ref{averages-kummer}) between integrals of
$z^{2i}(1-z^{2})^{k}$ weighted by $\exp(\s z^{2})$, and Kummer functions, we get the
expression
\begin{equation}
\label{averages:general:0}
\llangle\cos^{2i}\!\alpha\sin^{2k}\!\alpha\rrangle_{\ran}
=
\frac{\Gamma(i+\half)k!}{2\Gamma(i+k+\threehalfs)}
\;,
\end{equation}
where we have employed $M(a,c\,;x=0)=1$ [see Eq.\ (\ref{kummer:series})] and
$\Gamma(k+1)=k!$. Alternatively, on using $\Gamma(z+1)=z\Gamma(z)$ to expand the
above quotient of gamma functions, we obtain
\[
\llangle\cos^{2i}\!\alpha\sin^{2k}\!\alpha\rrangle_{\ran}
=
\frac
{2^{k}\;k!}
{
\underbrace{
(2i+1)[(2i+1)+2]\cdots[(2i+1)+2k]
}_{k+1\;{\rm terms}}
}
\;.
\]

To conclude, we explicitly write down the particular cases of the above results that, in
what follows, will more frequently be used:
\begin{equation}
\label{averages:particular}
\begin{array}{rclrclrcl}
\avcosqal_{\ran}
&
=
&
1/3
\;,
&
\avsenqal_{\ran}
&
=
&
2/3
\;,
&
& {}
&
\\
\llangle\coscal\rrangle_{\ran}
&
=
&
1/5
\;,
&
\llangle\cosqal\senqal\rrangle_{\ran}
&
=
&
2/15
\;,
&
\llangle\sencal\rrangle_{\ran}
&
=
&
8/15
\;.
\end{array}
\end{equation}

\subsubsection{Field expansion of the free energy}

On considering the expansion (\ref{Zfinal}) of the partition function in powers of
$\xi=\mm B/\T$, one realizes that the function $2\F(\s)$, $\xi^{2}$, and $C_{i}$ play
the r\^{o}le, respectively, of $\Z(0)$, $y$, and $A_{i}$ in the generic $y$-expansion
(\ref{Zexpansion:general}). Consequently, the corresponding general series
(\ref{lnZexpansion:general}) for $\ln\Z$ yields in this case
\begin{equation}
\label{field:expansion:F}
\ln\Z
=
\ln[2\F(\s)]
+C_{1}(\s,\alpha)\xi^{2}
+\frac{1}{2}
\left[ C_{2}(\s,\alpha)-C_{1}(\s,\alpha)^{2}
\right]
\xi^{4}
+\cdots
\;.
\end{equation}
This result shows the convenience of the introduction of the factor $i!$ in the
definition (\ref{Ci}) of the coefficients $C_{i}$: the general expansion
(\ref{lnZexpansion:general}) can then be directly used by merely replacing the
coefficients $A_{i}$ by the $C_{i}$ ones.

Now, on introducing the first few angular terms $b_{i,k}(\alpha)$ [Eq.\ (\ref{bik})],
\[
\begin{array}{rclrclrcl} b_{0,0}
&
=
&
1
\;,
&
b_{1,0}
&
=
&
\half\cosqal
\;,
&
b_{0,1}
&
=
&
\frac{1}{4}\senqal
\;,
\\
b_{2,0}
&
=
&
\frac{1}{24}\coscal
\;,
&
b_{1,1}
&
=
&
\frac{1}{8}\cosqal\senqal
\;,
&
b_{0,2}
&
=
&
\frac{1}{64}\sencal
\;,
\end{array}
\]
into the definition (\ref{Ci}), one gets for the first coefficients $C_{i}$: $C_{0}=1$,
\begin{equation}
\label{C_1} C_{1}
=
\frac{1}{2}
\left(
\frac{\F'}{\F}\cosqal+\frac{\F-\F'}{2\F}\senqal
\right)
\;,
\end{equation}
and
\[
C_{2}
=
\frac{1}{4}
\left(
\frac{1}{3}\frac{\F''}{\F}\coscal
+\frac{\F'-\F''}{\F}\cosqal\senqal
+\frac{\F-2\F'+\F''}{8\F}\sencal
\right)
\;,
\]
where, instead of superscripts, we have used primes to indicate derivatives  of
$\F(\s)$ with respect to its argument. On using these formulae we get for the
coefficient of $\xi^{4}$ in the expansion (\ref{field:expansion:F}),
\begin{eqnarray}
\label{C_2:C_1}
\frac{1}{2}\left(C_{2}-C_{1}^{2}\right)
=
\frac{1}{8}
&
\Bigg\{
&
\left[
\frac{1}{3}\frac{\F''}{\F}-\bigg(\frac{\F'}{\F}\bigg)^{2}
\right]
\coscal
\nonumber
\\
&
& {}+
\left[
\bigg(\frac{\F'}{\F}\bigg)^{2}-\frac{\F''}{\F}
\right]
\cosqal\senqal
\nonumber
\\
&
& {}+\frac{1}{8}
\left[
-1+2\frac{\F'}{\F}
-2\bigg(\frac{\F'}{\F}\bigg)^{2}
+\frac{\F''}{\F}
\right]
\sencal
\Bigg\}
\;.
\end{eqnarray}
Equations (\ref{C_1}) and (\ref{C_2:C_1}), along with (\ref{field:expansion:F}), yield the
desired $\xi$-expansion of the free energy up to the fourth order.

In Section \ref{sect:quantities} we shall introduce the {\em reduced\/} linear and
non-linear susceptibilities. These quantities, which incorporate the {\em
anisotropy-induced\/} temperature dependence of the susceptibilities, are directly
related with $C_{1}$ and $(C_{2}-C_{1}^{2})$, respectively.

\paragraph*{Average for anisotropy axes distributed at random.}

On introducing the values of the averaged trigonometric coefficients
(\ref{averages:particular}) into Eq.\ (\ref{C_1}), we get $\llangle
C_{1}\rrangle_{\ran}=1/6$. Proceeding similarly with the expression (\ref{C_2:C_1})
for $(C_{2}-C_{1}^{2})/2$, one obtains
\begin{equation}
\label{C_2:C_1:ran}
\frac{1}{2}
\llangle C_{2}-C_{1}^{2}
\rrangle_{\ran}
=
\frac{1}{120}
\left[ 2\frac{\F'}{\F}-3\bigg(\frac{\F'}{\F}\bigg)^{2}-1
\right]
\;.
\end{equation}
If we introduce these results into the $\xi$-expansion of $\ln\Z$ [Eq.\
(\ref{field:expansion:F})], we finally get for the free energy of an ensemble of
equivalent dipoles with anisotropy axes distributed at random:
\begin{equation}
\label{field:expansion:F:ran}
\llangle\FE\rrangle_{\ran}
=
-\T
\left\{
\ln[2\F(\s)]
+\frac{1}{6}
\xi^{2}
+\frac{1}{120}
\left[ 2\frac{\F'}{\F}
-3\bigg(\frac{\F'}{\F}\bigg)^{2}-1
\right]
\xi^{4}
+\cdots
\right\}
\;.
\end{equation}
It is to be noted that the first correction, $-\T\xi^{2}/6$, to the unbiased free energy
$-\T\ln[2\F(\s)]$, does not depend on the magnetic anisotropy. This will take its
reflection in, for example, the independence of the linear  susceptibility on the
anisotropy energy for systems with axes distributed at random (see Subsec.\
\ref{subsect:X}).

\subsubsection{Expansion of the free energy in powers of the anisotropy parameter}

The expansion of the free energy in powers of $\s=Kv/\T$ can be obtained similarly.
Let us first rewrite the expansion (\ref{Zs2}) of the partition function in powers of $\s$
as
\[
\Z
=
\Z_{0}
\left( 1
+\frac{\Z_{1}}{\Z_{0}}\s
+\frac{1}{2}\frac{\Z_{2}}{\Z_{0}}\s^{2}
+\cdots
\right)
\;,
\]
where $\Z_{0}$ is a shorthand for $\Z_{\lan}=(2/\xi)\sinh\xi$. If one compares this
expansion with the general  one (\ref{Zexpansion:general}), one sees that $\Z_{0}$,
$\s$, and $\Z_{i}/\Z_{0}$ play the  r\^{o}le, respectively, of $\Z(0)$, $y$, and $A_{i}$
there. Accordingly, we can immediately write down an equation similar to that
obtained for the $\xi$-expansion of $\ln\Z$
\begin{equation}
\label{anisotropy:expansion:F}
\ln\Z
\simeq{\ln\Z}_{\lan}
+\frac{\Z_{1}}{\Z_{0}}\;\s
+\frac{1}{2}
\left[
\frac{\Z_{2}}{\Z_{0}}
-\left(\frac{\Z_{1}}{\Z_{0}}\right)^{2}
\right]
\s^{2}
\;.
\end{equation}
Concerning the coefficients in this expansion, $\Z_{1}/\Z_{0}$ was already written in
Eq.\ (\ref{Zsub1:final}), namely
\begin{equation}
\label{Z_1:over:Z_0}
\frac{\Z_{1}}{\Z_{0}}
=
\left(1-\frac{2}{\xi}L\right)\cosqal
+\frac{1}{\xi}L\senqal
\;,
\end{equation}
while, taking Eq.\ (\ref{Zsub2:final}) into account,  one obtains after some algebra
\begin{eqnarray}
\label{Z_2:Z_1:over:Z_0}
\frac{1}{2}
\left[
\frac{\Z_{2}}{\Z_{0}}
-\left(\frac{\Z_{1}}{\Z_{0}}\right)^{2}
\right]
=
\frac{2}{\xi^{2}}
&
\Bigg\{
&
\left[ 2\left(1-\frac{3}{\xi}L\right)-L^{2}
\right]
\coscal
\nonumber
\\
&
& {}-
\left[ 6\left(1-\frac{3}{\xi}L\right)-L^{2}-\xi L
\right]
\cosqal\senqal
\nonumber
\\
&
& {}+\frac{1}{4}
\left[ 3\left(1-\frac{3}{\xi}L\right)-L^{2}
\right]
\sencal
\Bigg\}
\;.
\end{eqnarray}
Equations (\ref{Z_1:over:Z_0}) and (\ref{Z_2:Z_1:over:Z_0}), together with Eq.\
(\ref{anisotropy:expansion:F}), yield the desired expansion of the free energy in
powers of the anisotropy parameter up to second order.

\paragraph*{Average for anisotropy axes distributed at random.}

On introducing now the averages (\ref{averages:particular}) of the trigonometric
coefficients into the expression for $\Z_{1}/\Z_{0}$, one gets
$\langle\Z_{1}/\Z_{0}\rangle_{\ran}=1/3$. Analogously, on averaging Eq.\
(\ref{Z_2:Z_1:over:Z_0}) one arrives at
\[
\frac{1}{2}
\llangle
\frac{\Z_{2}}{\Z_{0}}-\left(\frac{\Z_{1}}{\Z_{0}}\right)^{2}
\rrangle_{\ran}
=
\frac{2}{15}\left(2-\frac{3}{\xi}L\right)\frac{1}{\xi}L
\;.
\]
On introducing these results into the expansion (\ref{anisotropy:expansion:F}) of
$\ln\Z$, one gets for $\langle\FE\rangle_{\ran}$ the approximate result
\begin{equation}
\label{anisotropy:expansion:F:ran}
\llangle\FE\rrangle_{\ran}
=
-\T
\left\{
\ln\left(\frac{2}{\xi}\sinh\xi\right)+\frac{1}{3}\s
+\frac{2}{15}
\left[
\left(2-\frac{3}{\xi}L\right)\frac{1}{\xi}L
\right]
\s^{2}+\cdots
\right\}
\;.
\end{equation}
As $\T\s=Kv$ [see Eqs.\ (\ref{sigma-xi})] is a constant (neglecting the possible
temperature dependence of $K$), we get the important result that, for anisotropy axes
distributed at random, the corrections due to the magnetic anisotropy to the isotropic
free energy, begin at order $\s^{2}$. This will lead to, for example, a dramatic decrease
of the anisotropy effects on the magnetization curves for weakly anisotropic systems
($\s\lsim2$) with a random distribution of anisotropy axes (see Subsec.\
\ref{subsect:magnetization}).

\subsubsection{Asymptotic expansion of the free energy for strong anisotropy}

Finally, the $1/\s$-expansion of the free energy can be obtained similarly. If we
compare the asymptotic expansion (\ref{Z:asympt:2}) for the partition function with
the general one (\ref{Zexpansion:general}), we see that $(e^{\s}/\s)\cosh\xipara$ and
$1/\s$ play the r\^{o}le, respectively, of $\Z(0)$ and $y$ in that general formula.
Therefore, we can immediately write for $\ln\Z$
\begin{eqnarray*}
\ln\Z
&
\simeq
&
{\ln\left(\frac{e^{\s}}{\s}\cosh\xipara\right)}
+\frac{1}{\s}\times\frac{1}{4}
\left[ (2+\xiperp^{2})-2\xipara\tanh\xipara
\right]
\\
&
& {}+
\frac{1}{2\s^{2}}
\Bigg\{
\frac{1}{2}
\left[
\left(3+\xipara^{2}+\xiperp^{2}+\frac{1}{8}\xiperp^{4}\right)
-(3+\xiperp^{2})\xipara\tanh\xipara
\right]
\\
&
&
\hspace{3.45em}
-\frac{1}{16}
\left[ (2+\xiperp^{2})-2\xipara\tanh\xipara
\right]^{2}
\Bigg\}
\;,
\end{eqnarray*}
where, to get the coefficient of $1/\s^{2}$ [i.e., $(A_{2}-A_{1}^{2})/2$ in the general
expansion], we have subtracted from the corresponding coefficient in the expansion of
$\Z$ the square of the coefficient of $1/\s$ (i.e., $A_{1}$). Then, on explicitly squaring
such term, we finally get
\begin{eqnarray}
\label{lnZ:asympt}
\ln\Z
&
\simeq
&
{\ln\left(\frac{e^{\s}}{\s}\cosh\xipara\right)}
+
\frac{1}{4\s}
\left[ (2+\xiperp^{2})-2\xipara\tanh\xipara
\right]
\nonumber
\\
&
& {}+
\frac{1}{8\s^{2}}
\left[ 5+(2\xipara^{2}+\xiperp^{2})-(4+\xiperp^{2})\xipara\tanh\xipara
-\xipara^{2}\tanh^{2}\xipara
\right]
\;.
\qquad
\end{eqnarray}
Note that this expansion has as leading term the Ising-type free energy
(\ref{Z-F:ising}) (this corresponds to a potential with two deep minima), while the next
terms are corrections associated with the finite curvature of the potential at the
minima.

Note finally that, due to the presence of $\cosal$ (via $\xipara$) in
the arguments of the hyperbolic trigonometric functions, we cannot
write down an explicit analytical formula for the average of the above
expansion for anisotropy axes distributed at random.

%% file: garcms03.tex
\section
[Equilibrium properties: some important quantities]
{Equilibrium properties: some important\\quantities}
\label{sect:quantities}

\subsection{Introduction}

In this Section we shall use some of the general results of the
previous one, in order to calculate a number of thermodynamical
quantities for independent classical magnetic moments with axially
symmetric magnetic anisotropy. The results obtained would also apply
to systems approximately described as assemblies of classical dipole
moments with Hamiltonians like (\ref{U0}), i.e., Hamiltonians
comprising a coupling term to an external field plus an axially
symmetric orientational potential.

The organization of this Section is as follows. In Subsec.\
\ref{subsect:caloric} we shall study the thermal or caloric quantities
---energy, entropy, and specific heat--- in a number of particular
situations. Subsections \ref{subsect:magnetization},
\ref{subsect:X}, and \ref{subsect:Xnl} will be devoted, respectively,
to the study of the magnetization, the linear susceptibility, and the
non-linear susceptibilities. We shall mainly be interested in the
effects of the magnetic anisotropy on these quantities.

\subsection{Thermal (caloric) quantities}
\label{subsect:caloric}

We shall begin with a brief study of the thermal properties of non-interacting classical
magnetic moments. We shall merely consider the particular cases of zero anisotropy
and finite anisotropy in a zero field or in a constant longitudinal field.

\subsubsection{General definitions}

The thermodynamical energy, $\E$, is defined as the statistical-mechanical  average of
the Hamiltonian $\Hs$ [cf.\ Eq.\ (\ref{average})]
\begin{equation}
\label{U:def}
\E
=
\llangle\Hs\rrangle_{\eq}
=
\frac
{\int\!\D{\Omega}\,\Hs(\vartheta,\varphi)\exp[-\beta\Hs(\vartheta,\varphi)]}
{\int\!\D{\Omega}\,\exp[-\beta\Hs(\vartheta,\varphi)]}
\;,
\end{equation}
where $\int\!\D{\Omega}\,(\cdot)=(1/2\pi)
\int_{-1}^{1}\!\D{(\cos\vartheta)}\,
\int_{0}^{2\pi}\!\D{\varphi}\, (\cdot)$. From the above
definition one immediately gets the relation
\begin{equation}
\label{E-Z}
\E
=
-\frac{\partial {}}{\partial\beta} (\ln\Z)
\;,
\end{equation}
between $\E$ and the logarithm of the partition function
$\Z=\int\!\D{\Omega}\,\exp(-\beta\Hs)$ (or the free energy $\FE=-\beta^{-1}\ln\Z$).

The entropy, $\ent$, can formally be defined as minus the average of the logarithm of
the equilibrium probability distribution $\Weq=\exp(-\beta\Hs)/\Z$, i.e.,
\begin{equation}
\label{S:def}
\frac{\ent}{\kB}
=
-\llangle\ln\Weq\rrangle_{\eq}
=
\frac
{-\int\!\D{\Omega}\,\ln\Weq(\vartheta,\varphi)\exp[-\beta\Hs(\vartheta,\varphi)]}
{\int\!\D{\Omega}\,\exp[-\beta\Hs(\vartheta,\varphi)]}
\;.
\end{equation}
Note however that this quantity, in contrast to other thermodynamical
quantities, is not defined as the average of a physical quantity of
the system ---it is an intrinsic {\em thermal\/} quantity---. On the
other hand, by using $-\llangle\ln\Weq\rrangle_{\eq}=\beta\E+\ln\Z$,
which is essentially the celebrated thermodynamical relation
$\FE=\E-T\ent$, one gets from Eqs.\ (\ref{E-Z}) and (\ref{S:def}) the
entropy expressed in terms of the partition function as
\begin{equation}
\label{S-Z}
\frac{\ent}{\kB}
=
\ln\Z
-\beta
\frac{\partial {}}{\partial\beta} (\ln\Z)
\;.
\end{equation}

The last thermal quantity that we shall consider is the specific heat
at constant field, namely
\begin{equation}
\label{c:def}
\sh
=
\frac{\partial\E}{\partial T}\bigg|_{B}
\;.
\end{equation}
Taking into account the relation (\ref{E-Z}) between $\E$ and $\Z$,
one obtains from the above definition the well-known results
\begin{equation}
\label{c-Z}
\frac{\sh}{\kB}
=
-\beta^{2}\frac{\partial\E}{\partial\beta}
=
\beta^{2}\frac{\partial{}^2}{\partial\beta^{2}}(\ln\Z)
\;.
\end{equation}

Let us finally consider a quantity $A=A(\s,\xi)$ that is a function of
$\s=Kv/\T$ and $\xi=\mm B/\T$ [the dimensionless anisotropy and field
parameters (\ref{sigma-xi})].  Then, on using
$\beta\partial\s/\partial\beta=\s$ and
$\beta\partial\xi/\partial\beta=\xi$, one gets for the
$\beta$-derivatives of $A$
\[
\beta\frac{\partial A}{\partial\beta}
=
\frac{\partial A}{\partial\s}\s+\frac{\partial A}{\partial\xi}\xi
\;,
\qquad
\beta^{2}\frac{\partial {}^{2}A}{\partial\beta^{2}}
=
\frac{\partial {}^{2}A}{\partial\s^2}\s^{2}
+2\frac{\partial
{}^{2}A}{\partial\s\partial\xi}\s\xi
+\frac{\partial {}^{2}A}{\partial\xi^2}\xi^{2}
\;.
\]
Note that, when taking the $\beta$-derivatives, we have implicitly
assumed that the only dependence of $\s$ and $\xi$ on $T$ enters via
$\beta$, that is, we neglect the possible dependence on the
temperature of both $K$ and $\mm$, which otherwise might be relevant
in systems of magnetic nanoparticles at sufficiently high
temperatures. Next, if $A=\ln\Z$, on taking the relations (\ref{E-Z}),
(\ref{S-Z}), and (\ref{c-Z}) into account, we can express the thermal
quantities for a system described by $\s$ and $\xi$, as
\begin{eqnarray}
\label{E-Z:2}
\E
&
=
&
-\left(
\frac{\partial\Z/\partial\s}{\Z}Kv+\frac{\partial\Z/\partial\xi}{\Z}\mm B\right)
\;,
\\
\label{S-Z:2}
\frac{\ent}{\kB}
&
=
&
\ln\Z
-\left(
\frac{\partial\Z/\partial\s}{\Z}\s+\frac{\partial\Z/\partial\xi}{\Z}\xi\right)
\;,
\\
\label{c-Z:2}
\frac{\sh}{\kB}
&
=
&
\left[
\frac{\partial{}^{2}\Z/\partial\s^2}{\Z}
-\left(\frac{\partial\Z/\partial\s}{\Z}\right)^{2}
\right]
\s^{2}
\nonumber
\\
&
& {}+ 2
\Bigg[
\frac{\partial{}^{2}\Z/\partial\s\partial\xi}{\Z}
-\frac{(\partial\Z/\partial\s)(\partial\Z/\xi)}{\Z^{2}}
\Bigg]
\s\xi
\nonumber
\\
&
& {}+
\left[
\frac{\partial{}^{2}\Z/\partial\xi^2}{\Z}
-\left(\frac{\partial\Z/\partial\xi}{\Z}\right)^{2}
\right]
\xi^{2}
\;.
\end{eqnarray}
These formulae allow one to identify the contribution of the
anisotropy and Zeeman energies to the thermal quantities. However, one
does not need to use them in their general forms since, when both
types of energies are present, one can write $\xi=2\s h$ and
differentiate with respect to $\s$ keeping $h=B/\BK$, which is assumed
to be independent of the temperature, constant.

\subsubsection{Thermal quantities: particular cases}

\paragraph{Isotropic case.}

When the anisotropy energy is absent, the partition function reads
$\Z_{\lan}=(2/\xi)\sinh\xi$ [Eq.\ (\ref{Z-F:langevin})]. The
$\s$-derivatives of this partition function are identically zero,
while the required $\xi$-derivatives are given by Eqs.\
(\ref{Zsub0:derivative}) and (\ref{Zsub0:derivatives2}). Therefore, on
taking Eq.\ (\ref{E-Z:2}) into account, one obtains for the mean
energy
\begin{equation}
\label{U:langevin}
\E_{\lan}
=
-\mm\left(\coth\xi-\frac{1}{\xi}\right)B
=
-\mm L(\xi)B
\;,
\end{equation}
where $L(\xi)$ is the Langevin function. This is the natural result considering that in
this case $\Hs=-\mz B$ and that the Langevin result for the magnetization is
$\langle\mz\rangle_{\eq}=\mm L(\xi)$. Similarly, Eq.\ (\ref{S-Z:2}) yields the
following expression for the entropy
\begin{equation}
\label{S:langevin}
\frac{\ent_{\lan}}{\kB}
=
\ln
\left[ (2/\xi)\sinh\xi
\right]
-\xi L(\xi)
\;.
\end{equation}

Finally, on introducing Eqs.\ (\ref{Zsub0:derivative}) and
(\ref{Zsub0:derivatives2}) into Eq.\ (\ref{c-Z:2}), the isotropic specific heat can be
written as
\begin{equation}
\label{c:langevin}
\frac{c_{B,\lan}}{\kB}
=
1-\frac{\xi^{2}}{\sinh^{2}\xi}
=
\xi^{2}L'(\xi)
\;.
\end{equation}
At high temperatures, i.e., when $\xi\ll1$, we can approximate the square of the
hyperbolic sine in Eq.\ (\ref{c:langevin}) by
$\sinh^{2}\xi\simeq\xi^{2}+\xi^{4}/3$, while at low temperatures ($\xi\gg1$)
we have $\xi^{2}/\sinh^{2}\xi\simeq0$. Consequently, in these limiting ranges $\sh$
approximately reads
\begin{equation}
\label{c:langevin:appr} c_{B,\lan}\simeq
\left\{
\begin{array}{ll}
\kB\xi^{2}/3
&
\mbox{ for }\xi\ll1
\\
\kB
&
\mbox{ for }\xi\gg1 
\end{array}
\right.
\;.
\end{equation}
Thus, the specific heat obeys a customary $T^{-2}$ law in the high-temperature range,
whereas it tends to $\kB$ at low temperatures. This last limit does not obey Nerst's
theorem, which states that $\sh\to0$ as $T\to0$, and this is due to the classical
character of the magnetic moment (the energy levels of $\m$ are not discrete, which is
a proviso for the result mentioned, but they are continuously distributed).

\begin{figure}[t!]
\vspace{-3.ex}
\eps{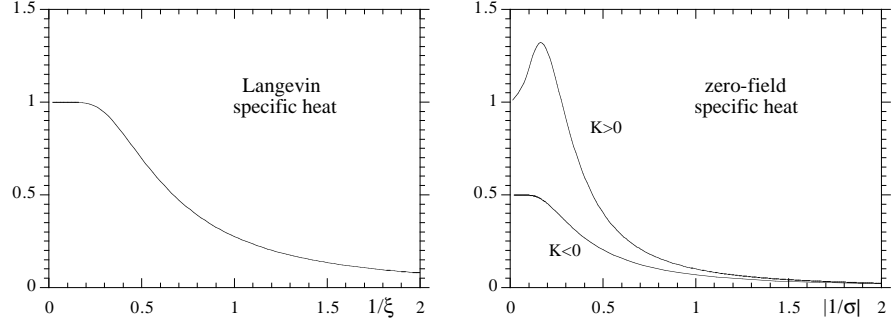}{1}
\vspace{-3.ex}
\caption[]
{
Temperature dependence of the specific heat, $\sh$, of a classical spin in the
isotropic and unbiased cases. $\sh$ is measured in units of $\kB$ and the
dimensionless temperatures are $1/\xi=\T/\mm B$ and $|1/\s|=\T/|K|v$, respectively.
\label{c:langevin-unbiased:plot}
}
\end{figure}

Figure \ref{c:langevin-unbiased:plot} shows the specific heat in the
isotropic case.  This increases monotonically from $0$ at high
temperatures to $\kB$ at low temperatures, where the curve exhibits a
plateau. This region corresponds to the high-field ($\xi\gg1$) range
where the average magnetic moment is close to saturation
$[1-L(\xi)]\propto\xi^{-1}$; the thermodynamical energy, which is
proportional to $L(\xi)$, then increases linearly with $T$, yielding a
constant $\sh$.

\paragraph{Zero-field case.}

In the absence of an external field (unbiased case), the partition function is
given by $\Z_{\unb}=2\F(\s)$ [Eq.\ (\ref{Z-F:unb})]. Owing to the fact that the
$\xi$-derivatives of $\Z_{\unb}$ are identically zero, the mean energy in the absence
of an external field obtained from Eq.\ (\ref{E-Z:2}) reads
\begin{equation}
\label{U:unb}
\E_{\unb}
=
-Kv\frac{\F'}{\F}
\;.
\end{equation}
This expression provides another simple physical interpretation for the familiar
combination $\F'/\F$
--- it is essentially minus the thermodynamical energy in the
absence of an external field---.

On the other hand, the zero-field entropy and specific heat, as derived  from Eqs.\
(\ref{S-Z:2}) and (\ref{c-Z:2}), read
\begin{equation}
\label{S:unb}
\frac{\ent_{\unb}}{\kB}
=
\ln(2\F)-\s\frac{\F'}{\F}
\;,
\end{equation}
and
\begin{equation}
\label{c:unb}
\frac{c_{B,\unb}}{\kB}
=
\left[
\frac{\F''}{\F}-
\left(\frac{\F'}{\F}\right)^{2}
\right]
\s^{2}
\;.
\end{equation}
In the high- ($|\s|\ll1$) and low-temperature ($|\s|\gg1$) ranges, we can use the
approximate Eq.\ (\ref{comb:F:leadingterm}) for $\F''/\F-(\F'/\F)^{2}$, to get the limit
behaviors of the zero-field specific heat:
\begin{equation}
\label{c:unb:appr} c_{B,\unb}\simeq
\left\{
\begin{array}{ll}
\kB/2
&
\mbox{ for } \s\ll-1
\\
(4/45)\kB\s^{2}
&
\mbox{ for }|\s|\ll1
\\
\kB
&
\mbox{ for }\s\gg1 
\end{array}
\right.
\;.
\end{equation}
As it should, the specific heat obeys a $T^{-2}$ law at high temperatures. At low
temperatures, owing to the classical nature of the spin (cf.\ Jacobs and Bean, 1963),
$\sh$ tends to $\kB$ and $\kB/2$, for easy-axis and easy-plane anisotropy,
respectively. The factor $1/2$ originates from the different geometry of the region of
the minima; for easy-axis anisotropy the minima are the poles of the unit sphere,
whereas for easy-plane anisotropy, the minima are continuously distributed on the
equatorial circle.

Figure \ref{c:langevin-unbiased:plot} also shows the specific heat in the unbiased case.
In contrast to the isotropic specific heat, in the easy-axis zero-field case, the specific
heat exhibits a maximum. This peak (located at $\s\sim5$) can be interpreted in
terms of the crossover from isotropic behavior at high temperatures to the two state
(Ising-type) behavior at low temperatures. This is supported by Fig.\
\ref{pdf:unbiased:plot}, where it was shown that, whereas at $\s\simeq2$,
$\W_{\eq,\unb}(\mz)$ is not far from uniform, for $\s\simeq5$, the probability
distribution is quite concentrated close to the poles. These features of the specific heat
resemble the Schottky effect, and, in this context, they could be attributed to the
``depopulation" of the high-energy ``equatorial levels." On the other hand, the specific
heat in the easy-plane unbiased case does not exhibit a peak but it also has a plateau
at low temperatures. The absence of maxima in $\sh(T)$ is to be attributed to the
geometrical structure of the Hamiltonian for easy-plane anisotropy.

\paragraph{Longitudinal-field case.}

We shall finally consider the caloric quantities when an external field  is applied
along the anisotropy axis. The corresponding partition function is given by Eq.\
(\ref{Zpara}), where $\spm=\s(1\pm h)^{2}$ and $h=\xi/2\s$. As was previously
remarked, in order to calculate the thermal quantities we do not need to make use of
Eqs.\ (\ref{E-Z:2}), (\ref{S-Z:2}), and (\ref{c-Z:2}) in their general forms; in this case we
only need to take $\s$-derivatives of $\Zp$ (denoted by primes) keeping $h=B/\BK$
constant.

On calculating $\Zp'/\Zp$, we get
\begin{equation}
\label{Zpara:derivative}
\frac{\Zp'}{\Zp}
=
-h^{2}+
\frac{(1+h)^{3}\F'(\spl)+(1-h)^{3}\F'(\smi)}{(1+h)\F(\spl)+(1-h)\F(\smi)}
\;,
\end{equation}
where we have used $\partial\spm/\partial\s=(1\pm h)^{2}$. Equation
(\ref{Zpara:derivative}) yields, essentially, minus the mean energy. However, before
writing down an equation for $\E$, we shall manipulate slightly the above expression
in order to eliminate $\F'(\spm)$. To this end, we can use $\F'=(e^{\s}-\F)/2\s$
[Eq.\ (\ref{F-Fp:1})], getting
\[
(1\pm h)^{3}\F'(\spm)
=
\frac{1\pm h}{2\s}
\left\{
\exp[\s(1+h^{2})\pm2\s h]-\F(\spm)
\right\}
\;.
\]
Then, on introducing the function
\[
J(\s,h)
=
2[\cosh(2\s h)+h\sinh(2\s h)]
\;,
\]
one can write the thermodynamical energy in a longitudinal field as
\begin{equation}
\label{U:para}
\E_{\|}
=
Kv
\left[ h^{2}+\frac{1}{2\s}
\left( 1-\frac{e^{\s}J}{\Zp}
\right)
\right]
\;.
\end{equation}
The entropy can then be derived by merely using $\FE=\E-T\ent$, to get
\begin{equation}
\label{S:para}
\frac{\ent_{\|}}{\kB}
=
\s h^{2}
+\ln(\Zp)
+\frac{1}{2}
\left( 1-\frac{e^{\s}J}{\Zp}
\right)
\;.
\end{equation}
Note that, since $(e^{\s}J/\Zp)\big|_{h=0}=e^{\s}/\F$ and $1-e^{\s}/\F=-2\s \F'/\F$,
Eqs.\ (\ref{U:para}) and (\ref{S:para}) duly reduce for $h=0$ to Eqs.\ (\ref{U:unb}) and
(\ref{S:unb}),  respectively.

Let us finally derive the specific heat in the longitudinal-field case. On taking the $\s$
derivative of Eq.\ (\ref{Zpara:derivative}) by using again
$\partial\spm/\partial\s=(1\pm h)^{2}$, we find
\begin{eqnarray}
\label{c:para:0}
\frac{c_{B,\|}}{\kB}
=
&
\Bigg\{
&
\frac
{(1+h)^{5}\F''(\spl)+(1-h)^{5}\F''(\smi)}
{(1+h)\F(\spl)+(1-h)\F(\smi)}
\nonumber\\
&
& {}-
\left[
\frac
{(1+h)^{3}\F'(\spl)+(1-h)^{3}\F'(\smi)}
{(1+h)\F(\spl)+(1-h)\F(\smi)}
\right]^{2}
\Bigg\}\s^{2}
\end{eqnarray}
which generalizes the zero-field expression (\ref{c:unb}). An alternative formula, more
suitable for computation, can be obtained by differentiating $\E$ in Eq.\
(\ref{U:para}), namely
\begin{equation}
\label{c:para}
\frac{c_{B,\|}}{\kB}
=
\frac{1}{2}
\left\{ 1
+\frac{e^{\s}J}{\Zp}
\left[
\s(1+h^{2})
-\frac{1}{2}
\left( 1+\frac{e^{\s}J}{\Zp}
\right)
+\s\frac{J'}{J}
\right]
\right\}
\;,
\end{equation}
where the prime in $J'$ stands for $\s$-derivative (keeping $h$ constant), i.e.,
\[
J'(\s,h)
=
4h[\sinh(2\s h)+h\cosh(2\s h)]
\;.
\]

In order to get the high-temperature behavior of $\sh$, we can expand Eq.\
(\ref{c:para:0}) in powers of $\s$ [to first order we evaluate $\F^{(\ell)}(\spm)$ at zero
with help from Eq.\ (\ref{F:zero})], getting
\begin{eqnarray*}
\left.\frac{c_{B,\|}}{\kB}\right|_{|\s|\ll1}
&
\simeq
&
\left\{
\frac{1}{5}\frac{(1+h)^{5}+(1-h)^{5}}{(1+h)+(1-h)}
-
\left[
\frac{1}{3}\frac{(1+h)^{3}+(1-h)^{3}}{(1+h)+(1-h)}
\right]^{2}
\right\}
\s^{2}
\;.
\end{eqnarray*}
The low temperature behavior (case $K<0$) can also be obtained by introducing the
asymptotic Eq.\ (\ref{Fderivatives:asympt2}) into Eq.\ (\ref{c:para:0}), whereas for
$K>0$ it is more easily obtained by differentiating twice the approximate partition
function (\ref{Zpara:approx:2}) with respect to $\s$ (keeping $h$ constant). Thus, one
arrives at the following limit behaviors of the specific heat
\begin{equation}
\label{c:para:appr}
c_{B,\|}
\simeq
\left\{
\begin{array}{ll}
\kB/2
&
\mbox{ for }\s\ll-1
\\
(4/45)\kB(1+15h^{2})\s^{2}
&
\mbox{ for }|\s|\ll1
\\
\kB
&
\mbox{ for }\s\gg1 
\end{array}
\right.
\;.
\end{equation}
Again, the specific heat obeys a $T^{-2}$ law at high temperatures while, due to the
classical character of the spin, $\sh$ tends to non-zero values at low temperatures.
\begin{figure}[t!]
\vspace{-3.ex}
\eps{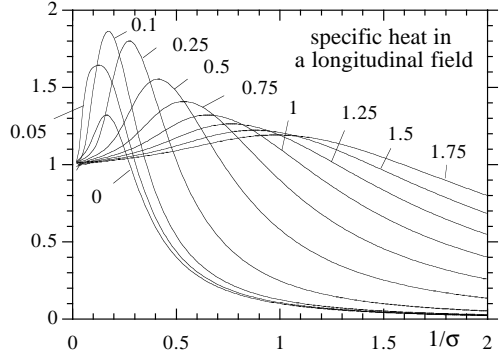}{0.56}
\vspace{-3.ex}
\caption[]
{
Temperature dependence of the specific heat, $\sh$, in various longitudinal fields
$h=B/\BK$ for easy-axis anisotropy. $\sh$ is measured in units of
$\kB$ and the dimensionless temperature is $1/\s=\T/Kv$.
\label{c:para:plot}
}
\end{figure}

Figure \ref{c:para:plot} displays the specific heat in the
longitudinal-field case. The $\sh$ curves exhibit a maximum, the
height and location of which depend on the magnitude of the applied
field. For $h\leq1$, these maxima can again be interpreted in terms of
the crossover from the isotropic regime at high temperatures to the
low-temperature regime in which the magnetic moments are concentrated
close to the potential minima. Besides, the height of the maximum
steeply increases for $h\le0.15$ and then decreases monotonically with
increasing $h$. At high fields, the maximum is actually rather smeared
and its height is small, approaching a plateau. This occurs because
the Zeeman energy dominates the magnetic-anisotropy energy for such
high fields, approaching the specific heat the zero-anisotropy
$c_{B,\lan}$, which, after exhibiting a plateau, decreases
monotonically (Fig.\ \ref{c:langevin-unbiased:plot}).

\subsection{Magnetization}
\label{subsect:magnetization}

We shall now study the magnetization of classical magnetic moments with axially
symmetric magnetic anisotropy. The magnetization along the external field direction,
$M_{B}\equiv\blangle\m\cdot\hat{b}\brangle_{\eq}$, where $\hat{b}=\B/B$, can in
the general case be derived from the partition function as follows. Consider that
$-\beta\Hs$ contains among others a Zeeman term
$\xi(\vec{e}\cdot\hat{b})$, where $\xi=\mm B/\T$ and $\vec{e}=\m/\mm$. Then,
because $\Z=\int\!\D{\Omega}\,\exp(-\beta\Hs)$, one has
\[
\blangle\m\cdot\hat{b}\brangle_{\eq}
\equiv
\Z^{-1}\int\!\!\D{\Omega}\,\mm(\vec{e}\cdot\hat{b})e^{-\beta\Hs}
=
\mm\Z^{-1}\frac{\partial {}}{\partial\xi}
\int\!\!\D{\Omega}\, e^{-\beta\Hs}
\;,
\]
whence one gets the known statistical-mechanical relation
\begin{equation}
\label{M-lnZ}
M_{B}
=
\mm\frac{\partial {}}{\partial\xi}\ln\Z
\;.
\end{equation}

The magnetization for an ensemble of non-interacting superparamagnetic particles
without magnetic anisotropy can be obtained by means of a simple translation of the
classical Langevin theory of  paramagnetism, and it is given by $M_{B,\lan}=\mm
L(\xi)$ where $L(\xi)$ is the Langevin function [Eq.\ (\ref{langevin:function})]. The
magnetization then depends on the field and temperature via $B/T$. A related salient
result is that in a liquid suspension of magnetic particles (usually called {\em
magnetic fluid\/} or {\em ferrofluid}) with a {\em general\/} single-particle magnetic
anisotropy, the magnetization is also given by the Langevin result (Krueger, 1979).
This holds essentially because the physical rotation of the particles in the liquid
decouples the anisotropy from the magnetization process. In fact, the same result
holds for a molecular beam of single-domain magnetic clusters, such as those deflected
in Stern-Gerlach experiments (Maiti and Falicov, 1993). However, the rotational
degrees of freedom are fastened in solid dispersions, giving rise to effects of the
magnetic anisotropy on the equilibrium quantities.

West (1961) studied the magnetization of an ensemble of non-interacting magnetic
nanoparticles with uniaxial anisotropy in a {\em longitudinal\/} constant field. He
derived an equation for the magnetization (see below) and studied the
anisotropy-induced non-$B/T$ superposition of the magnetization curves.
Unfortunately, his analytical calculation cannot be easily extended to situations where
the field and the anisotropy axis are not collinear, where only more or less complicated
expressions have been derived.

Lin (1961) and Chantrell (see, for example, Williams et~al., 1993), expressed the
magnetization for an arbitrary orientation of the magnetic field as quotients of two
infinite series. On the other hand, M{\o}rup (1983) derived an approximate expression
for the magnetization valid when $\T$ is much smaller than $\Hs$, which holds {\em irrespective\/} of the symmetry the Hamiltonian. However, inasmuch as is assumed
that the magnetic moment is effectively confined to {\em one\/} of the potential wells,
his formula does not hold for the full equilibrium (superparamagnetic) range.%
\footnote{
The mentioned approximation is different from what we are calling the
Ising regime, where the magnetic moment stays most of the time around
the potential minima, but it is still in complete equilibrium, and
performs a sufficiently large number of inter-potential-well rotations
during a typical observation time.  }

In what follows, we shall first consider the form of the magnetization in  various
simple cases. Then, we shall briefly analyze a general expression derived from the
field expansion (\ref{Zfinal}) of the partition function (this is our contribution to the
abovementioned class of ``more or less complicated expressions"). Finally, we shall
study the expressions for the magnetization derived from the weak- and
strong-anisotropy expansions of the free energy obtained in Subsec.\ \ref{series:FE}.

\subsubsection{Magnetization: particular cases}

We shall now study the expressions that emerge from Eq.\ (\ref{M-lnZ}) when one
introduces into it the particular cases of the partition function considered in Subsec.\
\ref{Z-F}.

\paragraph{Isotropic case.}

For $\s=0$ the partition function is given by $\Z_{\lan}=(2/\xi)\sinh\xi$ [Eq.\
(\ref{Z-F:langevin})], so that the magnetization reads
\begin{equation}
\label{m:langevin} M_{B,\lan}
=
\mm\left(\coth\xi-\frac{1}{\xi}\right)
=
\mm L(\xi)
\;,
\end{equation}
where $L(\xi)$ is the Langevin function (\ref{langevin:function}).

\paragraph{Ising regime.}

For $\s\to\infty$, the partition function is $\Z_{\ising}\simeq(e^{\s}/\s)\cosh\xipara$
[Eq.\ (\ref{Z-F:ising})]. Since $\xipara=\xi\cosal$, the magnetization derived from
Eq.\ (\ref{M-lnZ}) reads
\begin{equation}
\label{m:ising} M_{B,\ising}
=
\mm\cosal\tanh(\xipara)
\;,
\end{equation}
which naturally vanishes when $\B$ is perpendicular to the ``Ising axis" $\hat{n}$.

\paragraph{Plane-rotator regime.}

The $\s\to-\infty$ partition function is
$\Z_{\rotator}\simeq(-\pi/\s)^{1/2}I_{0}(\xiperp)$ [Eq.\ (\ref{Z-F:planerotator})], so
that the plane-rotator magnetization is given by
\begin{equation}
\label{m:rotator} M_{B,\rotator}
=
\mm\senal I_{1}(\xiperp)/I_{0}(\xiperp)
\;,
\end{equation}
where we have used $I'_{0}(y)=I_{1}(y)$ [see the integral representation
(\ref{mod:bessel:1st}) for $I_{n}(y)$]. In this case, $M_{B}$ is zero when $\B$ is
perpendicular to the easy plane.

Note that, when the magnitude of the magnetic moment is independent of the
temperature, $M_{B}$ depends on $B$ and $T$ via $\xi$ ($\propto B/T$) in all three
considered cases. This is called the {\em $B/T$ superposition of $M_{B}$}; the
magnetization vs.\ field curves corresponding to different temperatures, when plotted
against $B/T$, collapse onto a single master curve. However, outside those limit
ranges, $T$ does not enter in $M_{B}(B,T)$ via $B/T$ only, but $M_{B}$ depends on
$\xi$ as well as on $\s$. This will be illustrated now with the magnetization in a
longitudinal field.

\paragraph{Longitudinal-field case.}

When $\B\parallel\hat{n}$, the partition function is given by Eq.\ (\ref{Zpara}). In
order to derive the associated magnetization, we need to take the derivatives
($h=\xi/2\s$)
\[
\frac{\partial {}}{\partial\xi}
\left[ (1\pm h)\F(\spm)
\right]
=
\pm
\left[
\frac{1}{2\s}\F(\spm)+(1\pm h)^{2}\F'(\spm)
\right]
=
\pm\frac{e^{\spm}}{2\s}
\;,
\]
where we have used $\partial\spm/\partial\xi=\pm(1\pm h)$ and the
terms $\F'(\spm)$ have been eliminated by dint of Eq.\ (\ref{F-Fp:1}). Then, with
help from $\exp(\spm)
=\exp[\s(1+h^{2})]\exp(\pm\xi)$, we get from Eq.\
(\ref{M-lnZ}) the magnetization in a longitudinal field as
\begin{equation}
\label{m:para}
\frac{M_{B,\|}}{\mm}
=
\frac
{e^{\s(1+h^{2})}}{\s}\,\frac{\sinh\xi}
{(1+h)\F(\spl)+(1-h)\F(\smi)}
-h
\;,
\end{equation}
which, by using Eq.\ (\ref{Zpara}), can more compactly be written as
\begin{equation}
\label{m:para:Z}
\frac{M_{B,\|}}{\mm}
=
\frac{e^{\s}}{\s}\,\frac{\sinh\xi}{\Zp}
-\frac{\xi}{2\s}
\;.
\end{equation}

\begin{figure}[t!]
\vspace{-3.ex}
\eps{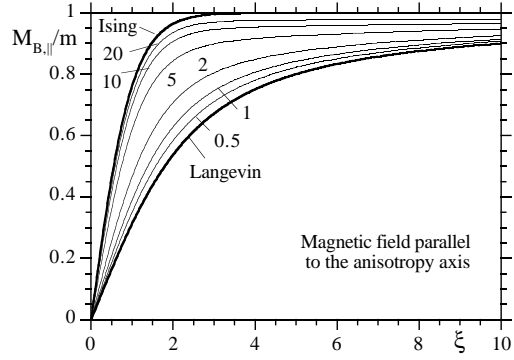}{0.7}
\vspace{-3.ex}
\caption[]
{
Magnetization vs.\ longitudinal field $\xi=\mm B/\T$ [Eq.\ (\ref{m:para:Z})] for various
values of the dimensionless anisotropy parameter $\s=Kv/\T$, showing the
anisotropy-induced non-$B/T$ superposition of the magnetization curves.
\label{magnetization:west:plot}
}
\end{figure}

Figure \ref{magnetization:west:plot} displays the magnetization vs.\
the longitudinal field, showing that $M_{B,\|}$ does not depend on $B$
and $T$ via $\xi$ only. As $T$ decreases one finds the crossover,
induced by the uniaxial magnetic anisotropy, from the high-temperature
($|\s|\ll1$) isotropic regime, to the low-temperature ($\s\gg1$) Ising
regime. Note that, even for $\s\sim20$, the typical measurement times
for the magnetization ($\sim1$--$100$\,s) would be much longer than
the relaxation times of the magnetic moment. Therefore, all the
displayed curves could be observed experimentally without leaving the
equilibrium (superparamagnetic) range.

Finally, we shall compare the above results with other expressions derived for the
magnetization. For $\s>0$, Eq.\ (\ref{m:para:Z}) reduces to the expression obtained by
West (1961). Indeed, if we use the alternative expression (\ref{Zpara:dawson}) for
$\Zp$ in terms of the Dawson integral $D$, we get
\begin{equation}
\label{m:para:west}
\frac{M_{B,\|}}{\mm}
=
\frac{1}{\sqrt{\s}}\,
\frac{\sinh\xi}{e^{\xi}D(\sqrt{\spl})+e^{-\xi}D(\sqrt{\smi})}-\frac{\xi}{2\s}
\;,
\end{equation}
which is the result of West almost in its original form. Another formula was derived
by Coffey, Cregg and Kalmykov (1993) when calculating relaxation times for magnetic
nanoparticles by the effective eigenvalue method, namely
\[
\frac{M_{B,\|}}{\mm}
=
\frac
{\xi}
{
\sqrt{\s}
\left[ (\xi L(\xi)+1+\xi)D(\sqrt{\s}+\frac{\xi}{2\sqrt{\s}})
+(\xi L(\xi)+1-\xi)D(\sqrt{\s}-\frac{\xi}{2\sqrt{\s}})
\right]
}
-\frac{\xi}{2\s}
\;,
\]
where $L(\xi)$ is the Langevin function. However, on merely noting that
$\sqrt{\s}\pm\xi/2\sqrt{\s}=\sqrt{\spm}$, and using
\begin{eqnarray*}
\xi L(\xi)+1\pm\xi
=
\xi\coth\xi\pm\xi
=
\frac{\xi}{\sinh\xi}(\cosh\xi\pm\sinh\xi)
=
\frac{\xi}{\sinh\xi}e^{\pm\xi}
\;,
\end{eqnarray*}
their formula can be cast into the form (\ref{m:para:west}) of West. Likewise the
latter, the above alternative expression for $M_{B,\|}$ is written by assuming
easy-axis anisotropy implicitly [recall the discussion as regards the validity of the
expression (\ref{Zpara:dawson}) for $\Zp$].

\subsubsection{General formula for the magnetization}

On inserting the field expansion of the partition function (\ref{Zfinal}) into the
statistical-mechanical relation (\ref{M-lnZ}), the magnetization emerges in the form
\begin{equation}
\label{M:gral} M_{B}
=
\mm
\sum_{i=1}^{\infty}\frac{2C_{i}}{(i-1)!}\xi^{2i-1}
\bigg/
\sum_{i=0}^{\infty}\frac{C_{i}}{i!}\xi^{2i}
\;.
\end{equation}
This formula gives a general expression for $M_{B}$ as a quotient of two series of
powers of $\xi$ whose coefficients are expressible in terms of Kummer functions [Eq.\
(\ref{Ci:alt})]. Such a mathematical object is clearly not easy to deal with. Nevertheless,
one can check by an explicit identification of the corresponding series, that when the
limit cases of the coefficients $C_{i}$ (see Table \ref{Ci:limits:table}) are introduced
into Eq.\ (\ref{M:gral}), one gets the isotropic, Ising, and plane-rotator results for the
magnetization. Indeed, for the series in the numerator and the denominator (the
magnetic-field dependent factor in the partition function) we obtain
\begin{center}
\begin{tabular}{c||c|c|c}
&
$\s=0$
&
$\s\to\infty$
&
$\s\to-\infty$
\cr
\hline
\hline
$\sum_{i=1}^{\infty}\frac{2C_{i}}{(i-1)!}\xi^{2i-1}$
&
$\frac{1}{\xi}\left(\cosh\xi-\frac{1}{\xi}\sinh\xi\right)$
&
$\cosal\sinh(\xipara)$
&
$\senal I_{1}(\xiperp)$
\cr
\hline $\sum_{i=0}^{\infty}\frac{C_{i}}{i!}\xi^{2i}$
&
$\frac{1}{\xi}\sinh\xi$
&
$\cosh(\xipara)$
&
$I_{0}(\xiperp)$
\end{tabular}
\end{center} Therefore, Eq.\ (\ref{M:gral}) contains, as particular cases, the limit
formulae for the magnetization discussed above.

\subsubsection{Series expansions of the magnetization}

\paragraph{Expansion of the magnetization in powers of the anisotropy parameter.}

Here we shall derive the magnetization from the weak-anisotropy expansion of the
free energy obtained in Subsec.\ \ref{series:FE}. In this way, we shall arrive at an
approximate analytical expression for $M_{B}$ that comprises the first corrections to
the Langevin magnetization due to non-zero magnetic anisotropy.

To this end, we must differentiate the expansion of $\FE$ in powers of $\s=Kv/\T$
[Eq.\ (\ref{anisotropy:expansion:F})] with respect to the field. Prior to taking the
$\xi$-derivatives of the first two coefficients of that expansion, we shall rewrite them
in alternative forms. Equation (\ref{Z_1:over:Z_0}) for $\Z_{1}/\Z_{0}$ can be written
as
\[
\frac{\Z_{1}}{\Z_{0}}
=
\cosqal-(3\cosqal-1)\frac{1}{\xi}L
\;,
\]
while Eq.\ (\ref{Z_2:Z_1:over:Z_0}) for the coefficient in $\s^{2}$ can be cast into the
form
\begin{eqnarray*}
\frac{1}{2}
\left[
\frac{\Z_{2}}{\Z_{0}}-\left(\frac{\Z_{1}}{\Z_{0}}\right)^{2}
\right]
&
=
&
(35\coscal-30\cosqal+3)\frac{1}{2\xi^{2}}\left(1-\frac{3}{\xi}L\right)
\\
&
& {}-(9\coscal-6\cosqal+1)\frac{1}{2\xi^{2}}L^{2}
-(\coscal-\cosqal)\frac{2}{\xi}L
\;.
\end{eqnarray*}
Now, on taking the derivatives of the above coefficients with help form Eq.\
(\ref{fracxiL:derivative}) for $(L/\xi)'$, we get
\begin{eqnarray}
\label{Z_1:over:Z_0:derivative}
\left(\frac{\Z_{1}}{\Z_{0}}\right)'
&
=
&
(3\cosqal-1)\frac{1}{\xi}
\left[ L^{2}-\left(1-\frac{3}{\xi}L\right)
\right]
\;,
\\
\label{Z_2:Z_1:over:Z_0:derivative}
\frac{1}{2}
\left[
\frac{\Z_{2}}{\Z_{0}}-\left(\frac{\Z_{1}}{\Z_{0}}\right)^{2}
\right]'
&
=
&
(35\coscal-30\cosqal+3)
\frac{1}{2\xi^{3}}
\left[ 3L^{2}-5\left(1-\frac{3}{\xi}L\right)
\right]
\nonumber
\\
&
& {}+ (9\coscal-6\cosqal+1)
\frac{1}{\xi^{2}}L
\left[ L^{2}-\left(1-\frac{3}{\xi}L\right)
\right]
\nonumber
\\
&
& {}+ (\coscal-\cosqal)\frac{2}{\xi}
\left[ L^{2}-\left(1-\frac{3}{\xi}L\right)
\right]
\;.
\end{eqnarray}
These expressions, when introduced into
\begin{equation}
\label{anisotropy:expansion:m}
\frac{M_{B}}{\mm}
\simeq L(\xi)+\left(\frac{\Z_{1}}{\Z_{0}}\right)'\s
+\frac{1}{2}
\left[
\frac{\Z_{2}}{\Z_{0}}-\left(\frac{\Z_{1}}{\Z_{0}}\right)^{2}
\right]'
\s^{2}
\;,
\end{equation}
yield the first terms of the desired weak-anisotropy expansion of the magnetization.

Some relevant particular cases are those where the field points along the  anisotropy
axis, perpendicular to it, and when the anisotropy axes  are distributed at random. In
the first two cases we find
\begin{eqnarray}
\label{anisotropy:expansion:m:para}
\frac{M_{B,\|}}{\mm}
&
\simeq
&
L(\xi)
+\frac{2}{\xi}
\left[ L^{2}-\left(1-\frac{3}{\xi}L\right)
\right]
\s
\nonumber
\\
&
& {}+
\frac{4}{\xi^{3}}
\left\{
\left[ 3L^{2}-5\left(1-\frac{3}{\xi}L\right)
\right]
+\xi L
\left[ L^{2}-\left(1-\frac{3}{\xi}L\right)
\right]
\right\}
\s^{2}
\;,
\qquad
\quad
\\
\label{anisotropy:expansion:m:perp}
\frac{M_{B,\perp}}{\mm}
&
\simeq
&
L(\xi)
-\frac{1}{\xi}
\left[ L^{2}-\left(1-\frac{3}{\xi}L\right)
\right]
\s
\nonumber
\\
&
& {}+
\frac{1}{\xi^{3}}
\left\{
\frac{3}{2}
\left[ 3L^{2}-5\left(1-\frac{3}{\xi}L\right)
\right]
+\xi L
\left[ L^{2}-\left(1-\frac{3}{\xi}L\right)
\right]
\right\}
\s^{2}
\;,
\qquad
\quad
\end{eqnarray}
while $\langle M_{B}\rangle_{\ran}$ is obtained by introducing the averages
(\ref{averages:particular}) into  Eqs.\ (\ref{Z_1:over:Z_0:derivative}) and
(\ref{Z_2:Z_1:over:Z_0:derivative}), getting%
\footnote{
Note that
$\langle 3\cosqal-1\rangle_{\ran}
=\langle 35\coscal-30\cosqal+3\rangle_{\ran}
=0$, the terms into the brackets being
proportional to the second and fourth Legendre polynomials, respectively [see Eq.\
(\ref{legendre}) below].
}
\begin{equation}
\label{anisotropy:expansion:m:ran}
\frac{\llangle M_{B}\rrangle_{\ran}}{\mm}
\simeq L(\xi)
-\frac{4}{15}
\left(1-\frac{3}{\xi}L\right)
\frac{1}{\xi}
\left[ L^{2}-\left(1-\frac{3}{\xi}L\right)
\right]
\s^{2}
\;.
\end{equation}
Naturally, one can also obtain this result by taking the $\xi$-derivative of the
$\s$-expansion of $\langle\FE\rangle_{\ran}$ [Eq.\ (\ref{anisotropy:expansion:F:ran})].
As was anticipated there, {\em for anisotropy axes distributed at random, the
corrections to the Langevin magnetization due to the magnetic anisotropy begin at
second order}.

\begin{figure}[t!]
\vspace{-3.ex}
\eps{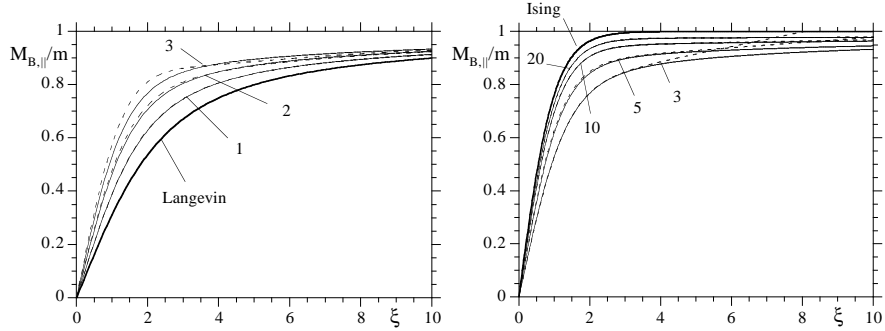}{1}
\vspace{-3.ex}
\caption[]
{
Magnetization vs.\ longitudinal field $\xi=\mm B/\T$  [Eq.\ (\ref{m:para:Z}); solid
lines], along with the weak-anisotropy formula (\ref{anisotropy:expansion:m:para})
(left panel, dashes) and the asymptotic formula (\ref{m:asympt:para}) (right
panel, dashes), for various values of the dimensionless anisotropy parameter
$\s=Kv/\T$.
\label{m:appr-asympt-west:plot}
}
\end{figure}

In order to estimate the range of validity of the weak-anisotropy
expansion of the magnetization, this has been compared with the exact
analytical formula (\ref{m:para:Z}) for the longitudinal
magnetization. It is shown in Fig.\
\ref{m:appr-asympt-west:plot} that the approximate
(\ref{anisotropy:expansion:m:para}) works reasonably well up to
$\s\sim2$.  Considering that the expansion has been performed by
assuming $\s$ as the small parameter, the range of validity obtained
is quite wide.

\begin{figure}[t!]
\vspace{-3.ex}
\eps{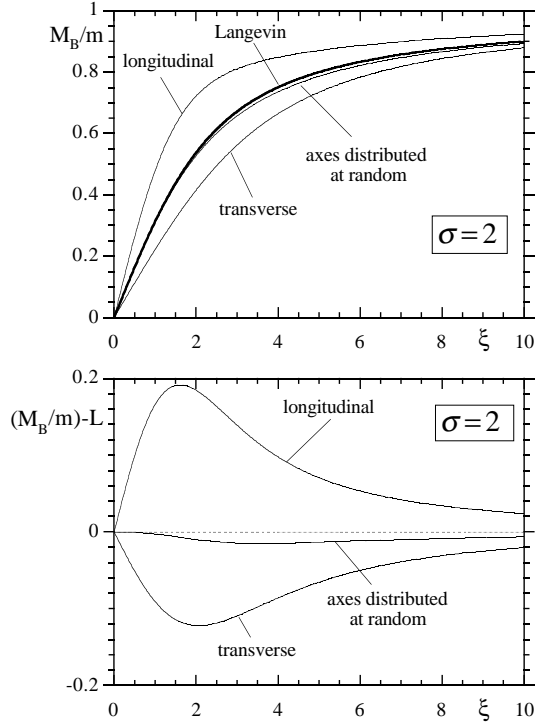}{0.7}
\vspace{-3.ex}
\caption[]
{
Upper panel: Magnetization vs.\ longitudinal field [Eq.\
(\ref{anisotropy:expansion:m:para})] and transverse field [Eq.\
(\ref{anisotropy:expansion:m:perp})], and for anisotropy axes distributed at random
[Eq.\ (\ref{anisotropy:expansion:m:ran})]. Lower panel: Anisotropy induced
contribution to the magnetization.
\label{magnetization:appr_alpha:plot}
}
\end{figure}

The effect of the orientation of the field with respect to the
anisotropy axis is shown in Fig.\
\ref{magnetization:appr_alpha:plot}. In contrast to the
longitudinal-field case, where the anisotropy energy favors the
alignment of the magnetic moment in the field direction, in the
transverse case the anisotropy hinders the magnetization process, and
the magnetization curve goes below the Langevin curve. In addition,
for an ensemble of spins with anisotropy axes distributed at random,
this phenomenon slightly dominates the favored alignment of the
longitudinal-field case, so that the corresponding magnetization is
slightly lower the Langevin magnetization.

The anisotropy-induced contribution to the magnetization,
$M_{B}(\xi)-\mm L(\xi)$, has been isolated in the lower panel of Fig.\
\ref{magnetization:appr_alpha:plot}. This representation neatly shows
that the random orientation of the anisotropy axes significantly
reduces the anisotropy-induced contribution to the magnetization
process.  In the range of low fields, moreover, that significant
reduction becomes an exact cancellation. This is due to the fact that
the {\em linear\/} susceptibility is independent on the anisotropy
energy when the anisotropy axes are distributed at random. This
result, which was advanced when considering such an average of the
field expansion of the free energy [Eq.\
(\ref{field:expansion:F:ran})], is not restricted to the
weak-anisotropy range (see Subsec.\ \ref{subsect:X}).

We finally remark that for easy-plane anisotropy ($\s<0$), the results described are
only slightly modified. Here, the longitudinal- and transverse-field cases, interchange
in some sense their r\^{o}les. For $\B\parallel\hat{n}$ and $\s<0$, the magnetic
anisotropy hinders the magnetization process, whereas this is naturally favored in the
transverse field case. However, for anisotropy axes distributed at random, the net
magnetization curve again goes slightly below the Langevin curve.

\paragraph{Asymptotic expansion of the magnetization for strong anisotropy.}

We shall now derive the magnetization from the asymptotic expansion of the free
energy for large $\s=Kv/\T$. In this way, we shall obtain an analytical formula that
contains the first corrections to the Ising-type magnetization due to non-infinite
magnetic anisotropy.

We proceed by differentiating the $1/\s$-expansion of $\ln\Z$ [Eq.\
(\ref{lnZ:asympt})] with respect to the field. The $\xi$-derivative of the coefficient of
$1/\s$, reads
\[
\left[ (2+\xiperp^{2})-2\xipara\tanh\xipara
\right]'
=
-2
\bigg[
-\senal\xiperp
+\cosal\tanh\xipara
+\frac{\cosal\xipara}{\cosh^{2}\xipara}
\bigg]
\;,
\]
where $\D\xipara/\D\xi=\cosal$ and $\D\xiperp/\D\xi=\senal$ have been used. The
$\xi$-derivative of the coefficient of $1/\s^{2}$ is taken similarly, yielding
\begin{eqnarray*}
\lefteqn{
\left[ 5+(2\xipara^{2}+\xiperp^{2})
-(4+\xiperp^{2})\xipara\tanh\xipara
-\xipara^{2}\tanh^{2}\xipara
\right]' }
\qquad
\qquad
&
&
\\
&
=
&
-\cosal\tanh\xipara
\bigg\{ 4+3\xiperp^{2}
-2\xipara
\bigg[
\tanh\xipara-\frac{\xipara}{\cosh^{2}\xipara}
\bigg]
\bigg\}
\\
&
& {}+ 2\senal\xiperp-\cosal\xiperp^{2}\frac{\xipara}{\cosh^{2}\xipara}
\;.
\end{eqnarray*}
On collecting these results and using $M_{B}=\mm(\partial\ln\Z/\partial\xi)$, the
approximate magnetization can finally be written as
\begin{eqnarray}
\label{m:asympt}
\frac{M_{B}}{\mm}
&
\simeq
&
\cosal\tanh\xipara
\Bigg\{ 1-\frac{1}{2\s}
\left[ 1+\frac{2\xipara}{\sinh(2\xipara)}
\right]
-\frac{1}{8\s^{2}}
\left[ 4-\xipara\frac{\sinh(2\xipara)-2\xipara}{\cosh^{2}\xipara}
\right]
\Bigg\}
\nonumber\\
&
& {}+\senal\xiperp\left(\frac{1}{2\s}+\frac{1}{4\s^{2}}\right)
-\cosal\xiperp^{2}
\frac{3\sinh(2\xipara)+2\xipara}{\cosh^{2}\xipara}
\frac{1}{16\s^{2}}
\;.
\end{eqnarray}
This formula extends the asymptotic result of Garanin (1996, Eq.\ (3.13)) in the
longitudinal-field case ($\xipara=\xi$, $\xiperp=0$) to an arbitrary orientation of the
field.

Let us explicitly write down the above approximate expression when the field points
along the anisotropy axis and perpendicular to it, namely
\begin{eqnarray}
\label{m:asympt:para}
\frac{M_{B,\|}}{\mm}
&
\simeq
&
\tanh\xi
\left\{
1-\frac{1}{2\s}
\left[ 1+\frac{2\xi}{\sinh(2\xi)}
\right]
-\frac{1}{8\s^{2}}
\left[ 4-\xi\frac{\sinh(2\xi)-2\xi}{\cosh^{2}\xi}
\right]
\right\}
\;,
\nonumber
\\
&
&
\\
\label{m:asympt:perp}
\frac{M_{B,\perp}}{\mm}
&
\simeq
&
\xi\left(\frac{1}{2\s}+\frac{1}{4\s^{2}}\right)
\;.
\end{eqnarray}
Note that in the transverse-field case the leading (Ising) result is identically zero and
one gets a linear increase of the magnetization with $\xi$. On the other hand, the
occurrence of $\alpha$ in the arguments of the hyperbolic functions in Eq.\
(\ref{m:asympt}), precludes the obtainment of a simple formula for $M_{B}$ when the
anisotropy axes are distributed at random.

As we did when studying the magnetization for weak anisotropy, we may
estimate the range of validity of the asymptotic expansion of $M_{B}$,
by comparing it with the exact analytical formula for
$M_{B,\|}$. Figure \ref{m:appr-asympt-west:plot} also displays such a
comparison showing that, for the field range considered, the
approximation derived works reasonably well down to quite small values
of $\s$.  There is however an important difference with the
weak-anisotropy formula for $M_{B}$, the accuracy of which was not
significantly sensitive to the magnitude of the field. Here, all the
approximate curves depart from the exact results at a certain value of
the field, which decreases as the anisotropy does. The breaking down
of the asymptotic expansion at high fields is apparent in the
transverse-field case (\ref{m:asympt:perp}), which yields a linear
dependence of $M_{B}$ on $\xi$, whereas at high fields the
magnetization must saturate.

These limitations occur because of the $\s\gg1$ expansions have as
leading terms Ising-type results (i.e., they correspond to a potential
with two deep minima), and the next-order terms are corrections
associated with the finite curvature of the potential at the bottom of
the wells. However, at sufficiently high fields the two-minima
structure of the potential disappears (for example, for $B=\BK$ in a
longitudinal field), and the expansion breaks down. In fact, already
for $B\sim\BK/2$, which corresponds to $\xi\sim\s$ [see Eq.\
(\ref{BK-h})], the upper potential well is quite shallow (see Fig.\
\ref{barrier:plot}) and the inverse of the potential curvature at the
minimum is large, so the expansion must already fail. This is
consistent with the asymptotic results shown in Fig.\
\ref{m:appr-asympt-west:plot}: the approximate $M_{B}$ departs from
the exact one at $\xi\sim3$ for $\s=3$, at $\xi\sim4$ for $\s=5$, at
$\xi\sim8$ for $\s=10$, and so on.

We finally mention that, as Fig.\ \ref{m:appr-asympt-west:plot}
suggests, the use of the weak-anisotropy formula, swapped at some
point between $\s=2$ and $\s=5$ by the asymptotic expression, yields a
reasonable approximation of the exact magnetization, except for the
discrepancies discussed of the asymptotic $\xi\gsim\s$ results. In
this connection, as the $\s=3$ curve suggests, one can replace the
asymptotic expansion for $\xi\gsim\s$ by the weak-anisotropy formula
in order to improve the overall approximation.

\paragraph{Field expansion of the magnetization.}

Let us finally discuss the low-field expansion of the magnetization
($H=B/\mu_{0}$),
\begin{equation}
\label{Mexpansion} M_{B}
=
\chi_{1}H+\chi_{3} H^{3}+\chi_{5}H^{5}+\cdots
\;,
\end{equation}
which defines the linear, $\chi_{1}$ (or simply $\chi$), and
non-linear, $\chi_{2n+1}$, $n=1,2,3,\ldots$, susceptibilities.

In order to derive general expressions for the susceptibilities, we can take the
$\xi$-derivative of the low-$\xi$ expansion of $\ln\Z$ [Eq.\ (\ref{field:expansion:F})],
getting
\begin{eqnarray}
\label{M2} M_{B}
=
\mm
&
\Big[
&
2C_{1}
\xi
+2(C_{2}-C_{1}^{2})
\xi^{3}
+(C_{3}-3C_{2}C_{1}+2C_{1}^{3})
\xi^{5}
\nonumber
\\
&
& {}+\third (C_{4}-4C_{3}C_{1}-3C_{2}^{2}+12C_{2}C_{1}^{2}-6C_{1}^{4})
\xi^{7}
+\cdots 
\Big]
\;,
\end{eqnarray}
where the coefficients $C_{i}$ are given by Eqs.\ (\ref{Ci}) or
(\ref{Ci:alt}). One also arrives at Eq.\ (\ref{M2}) by expanding in
powers of $\xi$ the inverse of the denominator of the general formula
(\ref{M:gral}), and multiplying this expansion by the first terms of
the series in the numerator.
\begin{table}
\vspace{-3.ex}
\caption[]
{
Combinations of the coefficients $C_{i}$ occurring in the first terms of the
expansion (\ref{M2}) of the magnetization in powers of the magnetic field, in the
isotropic, Ising, plane-rotator, and longitudinal-field cases.
\label{Ci:combinations:limits:table}
}
\begin{center}
\begin{tabular}{|c||r|r|r|c|}
&
$\s=0$
&
$\s\to\infty$
&
$\s\to-\infty$
&
$\B\parallel\hat{n}$
\cr
\hline
\hline
${\scriptstyle 2C_{1}}$
&
$\frac{1}{3}$
&
$\cosqal$
&
$\half\senqal$
&
$\frac{\F'}{\F}$
\cr
\hline
${\scriptstyle 2(C_{2}-C_{1}^{2})}$
&
$-\frac{1}{45}$
&
$-\frac{1}{3}\coscal$
&
$-\frac{1}{16}\sencal$
&
$\frac{1}{2}
\left[
\frac{1}{3}\frac{\F''}{\F}-\big(\frac{\F'}{\F}\big)^{2}
\right]$
\cr
\hline
${\scriptstyle C_{3}-3C_{2}C_{1}+2C_{1}^{3}}$
&
$\frac{2}{945}$
&
$\frac{2}{15}\cos^{6}\!\alpha$
&
$\frac{1}{96}\sin^{6}\!\alpha$
&
$\frac{1}{4}
\left[
\frac{1}{30}\frac{\F'''}{\F} -\frac{1}{2}\frac{\F''}{\F}\frac{\F'}{\F}
+\big(\frac{\F'}{\F}\big)^{3}
\right]$
\end{tabular}
\end{center}
\end{table}

The expansion (\ref{M2}) embodies $\chi$, $\chi_{3}$, $\chi_{5}$, and
$\chi_{7}$; in general, $\chi_{2n+1}$ can be obtained by inserting the
appropriate $C_{i}$ into the expression for the $n$th-order
cumulant. The coefficients of the first three terms at
$\s\to0,\pm\infty$, and for $\B\parallel\hat{n}$, are given in Table
\ref{Ci:combinations:limits:table} (they can be obtained from the
expressions of Table \ref{C1C2C3:limits}). On inserting those
coefficients into the above expansion of $M_{B}$, one gets the
approximate formulae
\begin{eqnarray}
\label{m:expansions:langevin}
M_{B,\lan}
&
=
&
\mm
\left[
\frac{1}{3}\xi-\frac{1}{45}\xi^{3}
+\frac{2}{945}\xi^{5}+\cdots
\right]
\;,
\\*
\label{m:expansions:ising}
M_{B,\ising}
&
=
&
\mm\cosal
\left[
\xipara-\frac{1}{3}\xipara^{3}
+\frac{2}{15}\xipara^{5}+\cdots
\right]
\;,
\\*
\label{m:expansions:rotator}
M_{B,\rotator}
&
=
&
\mm\senal
\left[
\half\xiperp-\frac{1}{16}\xiperp^{3}
+\frac{1}{96}\xiperp^{5}+\cdots
\right]
\;,
\\*
\label{m:expansions:para} M_{B,\|}
&
=
&
\mm\Bigg\{
\frac{\F'}{\F}\xi
+\frac{1}{2}
\left[
\frac{1}{3}\frac{\F''}{\F}
-\bigg(\frac{\F'}{\F}\bigg)^{2}
\right]
\xi^{3}
\nonumber
\\*
&
&
\hspace{1.75em}
{}+\frac{1}{4}
\left[
\frac{1}{30}\frac{\F'''}{\F}
-\frac{1}{2}\frac{\F''}{\F}\frac{\F'}{\F}
+\bigg(\frac{\F'}{\F}\bigg)^{3}
\right]
\xi^{5}
+\cdots
\Bigg\}
\;.
\qquad
\end{eqnarray}
Note that, in the first three cases $\chi_{2n+1}$ depends on $T$ with a $T^{-(2n+1)}$
law. This is the translation to linear and non-linear susceptibilities of the $B/T$
superposition of the corresponding magnetization curves. Outside these limit ranges,
however, the temperature dependence of $C_{i}(\s,\alpha)$ through $\s$,
provokes that $\chi_{2n+1}(T)$ no longer satisfies such a simple $T^{-(2n+1)}$ law.
This is already illustrated by the above expansion of $M_{B,\|}$, in which it can be
recognized the extra dependence of the susceptibilities on $T$, provided by the
functions $\F^{(\ell)}(\s)/\F(\s)$ via $\s$.

These points will be further investigated in the following two subsections devoted to
the linear and non-linear susceptibilities, respectively.

\subsection{Linear susceptibility}
\label{subsect:X}

We shall now study the linear susceptibility of classical spins with axially
symmetric magnetic anisotropy. The linear susceptibility, $\chi$, can be defined as the
coefficient of the linear term in the expansion of the magnetization in powers of the
external field. On comparing the $H$-expansion of $M_{B}$ (\ref{Mexpansion}) with
the $\xi$-expansion (\ref{M2}), and using $\xi=\mu_{0}\mm H/\T$, one gets the
following expression for $\chi$
\begin{equation}
\label{X:1}
\chi
=
\Xo 2C_{1}(\s,\alpha)
\;,
\end{equation}
which involves the first coefficient in the expansion of the partition function in powers
of $\xi$. Recall that $\alpha$ is the angle between the anisotropy axis $\hat{n}$ and
the field, while $\s=Kv/\T$.

\subsubsection{Linear susceptibility: particular cases}

Let us first consider the expressions that emerge from Eq.\ (\ref{X:1}) when one
inserts the particular cases of $2C_{1}$ into it (Table \ref{Ci:combinations:limits:table}).

\paragraph{Isotropic case.}

For $\s\to0$, $2C_{1}=1/3$, whence one gets the Curie law for the susceptibility
\begin{equation}
\label{X:langevin}
\chi_{\lan}
=
\frac{\mu_{0}\mm^{2}}{3\T}
\;.
\end{equation}
For classical spins, this result naturally follows from the absence of anisotropy.

\paragraph{Ising regime.}

For $\s\to\infty$, $2C_{1}=\cosqal$, so that
\begin{equation}
\label{X:ising}
\chi_{\ising}
=
\Xo\cosqal
\;,
\end{equation}
which is analogous to the susceptibility of an Ising spin. Thus, when the field points
along a direction perpendicular to the ``Ising axis" ($\cosal=0$), $\chi$ vanishes.

\paragraph{Plane-rotator regime.}

For $\s\to-\infty$, $2C_{1}=\senqal/2$, so that the plane-rotator linear susceptibility
is given by
\begin{equation}
\label{X:rotator}
\chi_{\rotator}
=
\frac{\mu_{0}\mm^{2}}{2\T}\senqal
\;.
\end{equation}
In this case the response is identically zero when the field points perpendicular to the
easy plane.

\paragraph{Longitudinal-field case.}

On introducing $2C_{1}|_{\alpha=0}=\F'/\F$ in Eq.\ (\ref{X:1}) one gets the longitudinal
susceptibility
\begin{equation}
\label{X:para}
\chi_{\|}
=
\Xo\frac{\F'}{\F}
\;,
\end{equation}
where the factor $\F'/\F$ induces an extra dependence on $T$ via $\s$,
``interpolating" between the isotropic ($\F'/\F|_{\s=0}=1/3$) and Ising
($\F'/\F|_{\s\to\infty}=1$) results.

\subsubsection{Formulae for the linear susceptibility}

When the general expression (\ref{C_1}) for $C_{1}$ is introduced into Eq.\ (\ref{X:1}),
the linear susceptibility emerges in the form
\begin{equation}
\label{X:2}
\chi
=
\Xo
\left(
\frac{\F'}{\F}\cosqal+\frac{\F-\F'}{2\F}\senqal
\right)
\;.
\end{equation}
It is convenient to introduce the longitudinal and transverse components of $\chi$
(which are related with the diagonal elements of the susceptibility tensor; see below)
\begin{equation}
\label{X:para:perp}
\chi_{\|}
=
\Xo\frac{\F'}{\F}
\;,
\qquad
\chi_{\perp}
=
\Xo\frac{\F-\F'}{2\F}
\;,
\end{equation}
so that $\chi$ can be written as
\begin{equation}
\label{X:4}
\chi
=
\chi_{\|}\cosqal+\chi_{\perp}\senqal
\;.
\end{equation}
The quantities $\chi_{\|}$ and $\chi_{\perp}$ characterize, respectively, the
equilibrium response to a longitudinal (parallel to $\hat{n}$) and transverse
(perpendicular to $\hat{n}$) probing field. Due to the linearity of the response, when
the probing field points along an arbitrary direction, the projection of the response
along the probing-field direction in given by the weighted sum (\ref{X:4}) of the
longitudinal and transverse responses.

Other derivations of the equilibrium linear susceptibility of a dipole moment in the
simplest axially symmetric anisotropy potential were carried out by Lin (1961),
Ra{\u{\i}}kher and Shliomis (1975) (see also Shliomis and Stepanov, 1993),
Shcherbakova (1978), and Chantrell et~al.\ (1985).

\paragraph{Average of the linear susceptibility for anisotropy axes distributed at
random.}

For an ensemble of equivalent magnetic moments (i.e., with the same characteristic
parameters) whose anisotropy axes are distributed at random, one finds
\begin{equation}
\label{X:ran}
\avX
=
\Xo
\left(
\frac{\F'}{\F}\frac{1}{3}+\frac{\F-\F'}{2\F}\frac{2}{3}
\right)
=
\frac{\mu_{0}\mm^{2}}{3\T}
\;,
\end{equation}
which is merely the Curie law for the linear susceptibility. This equation entails that,
irrespective of the magnitude of the magnetic anisotropy as compared with the
thermal energy, the linear susceptibility of the randomly oriented ensemble is equal
to the susceptibility of isotropic magnetic moments. This also holds in the extreme
anisotropy cases: for an ensemble of Ising spins, with Ising axes distributed at
random, $\langle\chi_{\ising}\rangle_{\ran} =\mu_{0}\mm^{2}/3\T$; likewise, for
an ensemble of plane rotators, with axes of rotation distributed at random,
$\langle\chi_{\rotator}\rangle_{\ran}$ is given by the Curie law (\ref{X:ran}).

We shall see below that Eq.\ (\ref{X:ran}) is in fact rather general; it holds
whenever the Hamiltonian of the spin (in the absence of the probing field) has
inversion symmetry ($\m\leftrightarrow-\m$).

\paragraph{Reduced linear susceptibility.}

\begin{figure}[t!]
\vspace{-3.ex}
\eps{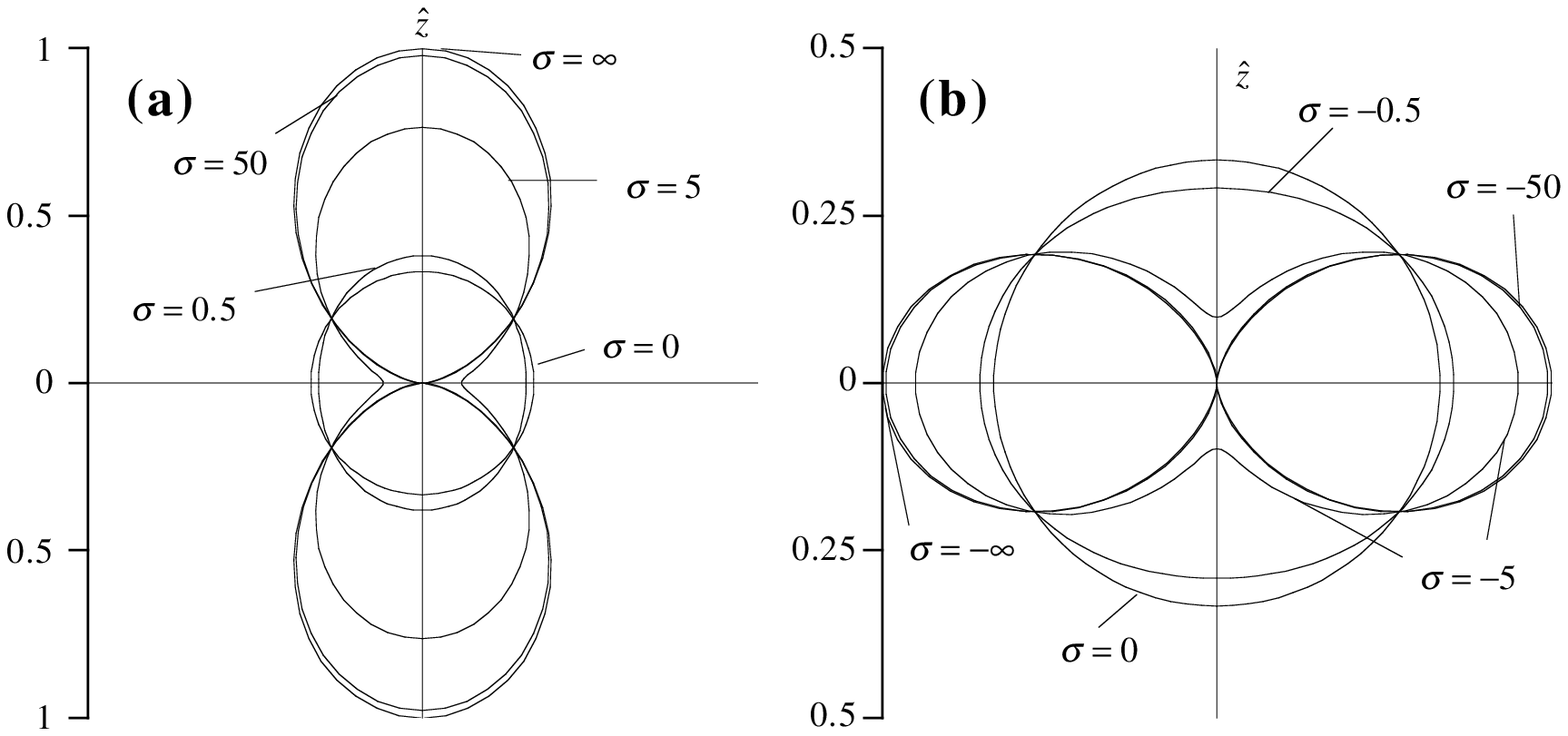}{1.15}
\vspace{-3.ex}
\caption[]
{
Polar plots showing the angular dependence of the reduced linear susceptibility
$\chi^{\red}$ [Eq.\ (\ref{X:red})] for various values of the dimensionless anisotropy
parameter $\s=Kv/\T$. (a) Easy-axis anisotropy. (b) Easy-plane anisotropy.
\label{X:polar:plot}
}
\end{figure}

An informative quantity is the {\em reduced\/} linear susceptibility defined as
$\chi^{\red} =\chi(\T/\mu_{0}\mm^{2})
=2C_{1}$, whence
\begin{equation}
\label{X:red}
\chi^{\red}(\s,\alpha)
=
\frac{\F'}{\F}\cosqal+\frac{\F-\F'}{2\F}\senqal
\;.
\end{equation}
This quantity has the property that isolates the
temperature-dependence of $\chi$ induced by the magnetic
anisotropy. Besides, it embodies the angular dependence of
$\chi$. Figure \ref{X:polar:plot} shows $\chi^{\red}$ as a function of
the angle between the anisotropy axis and the probing field (cf.\ Lin,
1961). As expected, the larger the $|\s|$, the more anisotropic the
$\chi^{\red}$ curves, becoming rather different from circles already
for $|\s|\simeq5$.

\begin{figure}[t!]
\vspace{-3.ex}
\eps{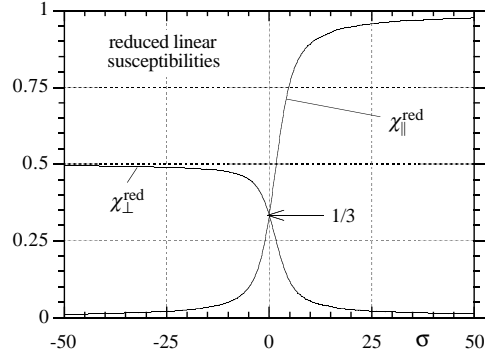}{0.7}
\vspace{-3.ex}
\caption[]
{
Reduced linear susceptibility (\ref{X:red}) in the longitudinal,
$\chi_{\|}^{\red}$, and transverse, $\chi_{\perp}^{\red}$, field cases vs.\ the
dimensionless anisotropy parameter $\s=Kv/\T$ (cf.\ Fig.\ \ref{F-der:plot}).
\label{X:red:plot}
}
\end{figure}

Figure \ref{X:red:plot} shows $\chi^{\red}$ for the longitudinal and
transverse components of the linear susceptibility (in this
representation $\langle\chi^{\red}\rangle_{\ran}$ would take the
constant value $1/3$). Both curves coincide for $\s=0$, where the
orientation of the field plays no r\^{o}le, taking the Langevin value
$1/3$. It can also be seen that the maximum variation of $\chi^{\red}$
with $\s$, occurs when the probing field is parallel to the anisotropy
axis. Note also that, {\em qualitatively}, the longitudinal- and the
transverse-field cases interchange their r\^{o}les when the sign of
the anisotropy is reversed. This statement, which is supported by
Fig.\ \ref{X:polar:plot}, is associated with the qualitatively
``equivalent" magnetization behavior in the easy-axis and easy-plane
anisotropy cases when the probing field points in the
``easy-magnetization region" or in the ``hard-magnetization region,"
regions that interchange themselves when the sign of the anisotropy is
changed.

\subsubsection{Generalizations}

\paragraph{Probing-field derivative of the magnetization.}

The definition of the linear susceptibility as the coefficient of the linear term in the
expansion of the magnetization in powers of the external field, of course agrees with
that in terms of the field derivative of the magnetization at zero field, i.e.,
$\chi=\mu_{0}[\partial\blangle\m\cdot\hat{b}\brangle_{\eq} /\partial B]\big|_{B=0}$.
This definition suggests the immediate generalization
\begin{equation}
\label{X:0}
\chi
=
\mu_{0}
\left.
\frac
{\partial\blangle\m\cdot\bp\brangle}
{\partial (\dBo)}
\right|_{\dBo=0}
\;,
\end{equation}
where $\dB=\dBo\,\bp$ is an external probing field ($\bp$ stands now for the
unit vector in the direction of the probing field). The absence of the subscript ``$\eq$"
in the thermal-equilibrium averages is used to indicate that they are taken with
respect to the total energy (system plus perturbation). The unperturbed system can
already be subjected to a constant (bias) field, $\B$, not necessarily collinear with
$\dB$.

Indeed, the calculation of the linear susceptibility can be carried out by
starting from a total Hamiltonian $\Hs_{\rm T}=\Hs-\m\cdot\dB$, where the
knowledge of the actual form of $\Hs$ is not required. Let us calculate first
\begin{eqnarray*}
\lefteqn{
\frac
{\partial\blangle(\m\cdot\bp)^{n}\brangle}
{\partial (\dBo)}
=
\frac{\partial\;\;}{\partial (\dBo)}
\frac
{\int\!\D{\Omega}\, (\m\cdot\bp)^{n} e^{-\beta\Hs_{\rm T}}}
{\int\!\D{\Omega}\,
e^{-\beta\Hs_{\rm T}}} }
\\
&
=
&
\beta\,
\frac
{
\Z\int\!\D{\Omega}\, (\m\cdot\bp)^{n+1} e^{-\beta\Hs_{\rm T}}
-\int\!\D{\Omega}\,
(\m\cdot\bp)^{n} e^{-\beta\Hs_{\rm T}}
\int\!\D{\Omega}\, (\m\cdot\bp) e^{-\beta\Hs_{\rm T}}
}
{\Z^{2}}
\;,
\end{eqnarray*}
where as usual
$\int\!\D{\Omega}\,(\cdot)
=(1/2\pi)
\int_{-1}^{1}\!\D{(\cos\vartheta)}\,
\int_{0}^{2\pi}\!\D{\varphi}\, (\cdot)$.
From the above result we get the general relation
\begin{equation}
\label{FDR:equilibrium:n}
\frac
{\partial\blangle(\m\cdot\bp)^{n}\brangle}
{\partial (\dBo)}
=
\beta
\left[
\blangle(\m\cdot\bp)^{n+1}\brangle
-\blangle(\m\cdot\bp)^{n}\brangle
\blangle\m\cdot\bp\brangle
\right]
\;,
\end{equation}
the $n=1$ particular case of which merely reads
\begin{equation}
\label{FDR:equilibrium}
\frac
{\partial\blangle\m\cdot\bp\brangle}
{\partial (\dBo)}
=
\beta
\left[
\blangle(\m\cdot\bp)^{2}\brangle
-\blangle\m\cdot\bp\brangle^{2}
\right]
\;,
\end{equation}
and holds irrespective of the magnitude of $\dB$. When this equation is
evaluated at $\dBo=0$ and inserted in Eq.\ (\ref{X:0}), one gets the celebrated
expression for the linear susceptibility in terms of the statistics of the
thermal-equilibrium fluctuations of the magnetic moment in the absence of the
probing field, namely
\begin{equation}
\label{X:FDR}
\chi
=
\frac{\mu_{0}}{\T}
\left[
\blangle(\m\cdot\bp)^{2}\brangle_{\eq}
-\blangle\m\cdot\bp\brangle_{\eq}^{2}
\right]
\;,
\end{equation}
where $\langle\,\rangle_{\eq}$ denotes the equilibrium average in the absence of the
perturbation.

The relation (\ref{X:FDR}) is valid for {\em any\/} form of the Hamiltonian. When $\Hs$
is given by $\Hs=-(Kv/\mm^{2})(\m\cdot\hat{n})^{2}$ [cf.\ Eq.\ (\ref{U0})], the above
averages in the absence of the probing field are in fact zero-field averages, which are
directly related with the coefficients $C_{i}$ by Eq.\ (\ref{Ci:averages}). Thus, by
inserting
\[
\blangle
\m\cdot\bp
\brangle_{\eq}\big|_{B=0}
=
0
\;,
\qquad
\blangle (\m\cdot\bp)^{2}
\brangle_{\eq}
\big|_{B=0}
=
\mm^{2}2C_{1}
\;,
\]
into Eq.\ (\ref{X:FDR}), one recovers the expression (\ref{X:1}) for $\chi$.

\paragraph{Tensor structure.}

The linear susceptibility is in fact a tensor defined by
\begin{equation}
\label{X:0:tensor}
\chi_{ij}
=
\mu_{0}
\left.
\frac{\partial\llangle\mi\rrangle}{\partial (\Delta B_{j})}
\right|_{\dBo=0}
\;.
\end{equation}
Note that the diagonal elements are given by Eq.\ (\ref{X:0}) when $\bp$ points along
$\hat{x}$, $\hat{y}$, and $\hat{z}$. By a derivation analogous to that leading to Eq.\
(\ref{X:FDR}), one arrives at the result
\begin{equation}
\label{X:FDR:tensor}
\chi_{ij}
=
\frac{\mu_{0}}{\T}
\left[
\llangle\mi\mj\rrangle_{\eq}
-\llangle\mi\rrangle_{\eq}\llangle\mj\rrangle_{\eq}
\right]
\;.
\end{equation}

Owing to the fact that $\chi_{ij}$ is a symmetrical second-rank tensor, it can be
diagonalized by a suitable change of coordinates. Let us assume that this
diagonalization has already been carried out. Then, if a probing field
$\dB=\dBo\,\bp$ is applied, the projection of the average magnetic moment
onto $\bp$ is given in the linear response range by (we use
$\mu_{0}\langle\mi\rangle
\simeq
\mu_{0}\langle\mi\rangle|_{\dBo=0}
+\sum_{j}\chi_{ij}\Delta B_{j}
+\cdots$)
\begin{equation}
\label{X:def}
\mu_{0}
\Delta\blangle\m\brangle
\cdot
\bp
\simeq
\left(\chi_{xx}\cos^{2}\!\alpha
+\chi_{yy}\cos^{2}\!\beta
+\chi_{zz}\cos^{2}\!\gamma\right)
\dBo
\;,
\end{equation}
where $(\alpha,\beta,\gamma)$ are the direction cosines of $\bp$ (in the coordinate
system that diagonalizes $\chi_{ij}$). The quantity into the brackets defines an {\em
effective\/} linear susceptibility $\chi$, which is in fact what we have been calling
linear susceptibility throughout.

\paragraph{The average of the linear susceptibility for anisotropy axes distributed at
random revisited.}

On the basis of the above expressions, we can derive the result mentioned for the
linear susceptibility of an ensemble of equivalent dipole moments whose Hamiltonian
has inversion symmetry and their intrinsic axes are distributed at random.

For an ensemble of independent dipole moments, the contribution of each dipole to
$\chi$ is analogous to that occurring in Eq.\ (\ref{X:def}), with (in principle) different
direction cosines and diagonal elements $\chi_{ii}$ for each dipole. However, if these
elements are equal we can write the total effective susceptibility as
\begin{equation}
\label{X:ensemble}
\chi
=
\chi_{xx}\llangle\cos^{2}\!\alpha\rrangle
+\chi_{yy}\llangle\cos^{2}\!\beta\rrangle
+\chi_{zz}\llangle\cos^{2}\!\gamma\rrangle
\;,
\end{equation}
where $\langle\,\rangle$ denotes average over the ensemble of dipoles. Note that for
the assumption about the equality of the tensor elements to hold, the dipole moments
must be equivalent (in the sense of having the same characteristic parameters) and the
orientation of the intrinsic axes (which diagonalize the linear susceptibility tensor for
each $\m$) with respect to the main reference frame, must be irrelevant in
determining the $\chi_{ii}$; this excludes, for instance, the occurrence of an external
(bias) field. Then, if those intrinsic axes are distributed at random, the effective linear
susceptibility (\ref{X:ensemble}) reads
\[
\avX
=
\frac{1}{3}
\left(\chi_{xx}+\chi_{yy}+\chi_{zz}\right)
=
\frac{\mu_{0}}{3\T}
\left\{
\mm^{2}
-
\left[
\llangle\mx\rrangle_{\eq}^{2}
+\llangle\my\rrangle_{\eq}^{2}
+\llangle\mz\rrangle_{\eq}^{2}
\right]
\right\}
\;,
\]
where Eq.\ (\ref{X:FDR:tensor}) has been used to express the $\chi_{ii}$. Finally, if the
Hamiltonian of each dipole has inversion symmetry
($\langle\mi^{2n+1}\rangle_{\eq}=0$), one has $\langle\mi\rangle_{\eq}=0$, $i=x,y,z$,
so that the above formula reduces to
\begin{equation}
\label{X:ran:general}
\avX
=
\frac{\mu_{0}\mm^{2}}{3\T}
\;,
\end{equation}
(note that presence of a bias field could as well be excluded on the basis of the
inversion-symmetry condition). Equation (\ref{X:ran:general}) is the announced result:
{\em for an ensemble of equivalent dipole moments whose Hamiltonian has inversion
symmetry, the effective linear susceptibility is given by the Curie law when their
intrinsic axes are distributed at random}.

\paragraph{General formula for any axially symmetric Hamiltonian.}

We shall now calculate the linear susceptibility of a magnetic moment with an
arbitrary axially symmetric Hamiltonian. The corresponding equilibrium probability
distribution of $z=\mz/\mm$ is given by [cf.\ Eq.\ (\ref{distribution:para})]
\begin{equation}
\label{distribution:para:gral}
\W_{\eq,\|}(z)
=
\Zp^{-1}\exp[-\beta\Hs(z)]
\;,
\qquad
\Zp
=
\int_{-1}^{1}\!\!\D{z}\,\exp[-\beta\Hs(z)]
\;,
\end{equation}
where we have assumed that the symmetry axis points along $\hat{z}$. In such a
reference frame, the susceptibility tensor is diagonal and the diagonal elements are
given by
\begin{equation}
\label{X:components}
\chi_{ii}
=
\frac{\mu_{0}}{\T}
\left[
\llangle\mi^{2}\rrangle_{\eq}-\llangle\mi\rrangle_{\eq}^{2}
\right]
\;,
\quad i
=
x,y,\mbox{~and~} z
\;.
\end{equation}
Besides, due to the axial symmetry of the Hamiltonian, the susceptibility tensor has
only two independent elements $\chi_{\|}=\chi_{zz}$ and
$\chi_{\perp}=\chi_{xx}=\chi_{yy}$.

Let us introduce the averages of the Legendre polynomials $p_{n}(z)$,
\begin{equation}
\label{legendre}
\begin{array}{rclrcl} p_{1}(z)
&
=
&
z
\;,
&
p_{2}(z)
&
=
&
\half(3z^{2}-1)
\;,
\\
p_{3}(z)
&
=
&
\half(5z^{3}-3z)
\;,
&
p_{4}(z)
&
=
&
\frac{1}{8}(35z^{4}-30z^{2}+3)
\;,
\ldots
\end{array}
\end{equation}
with respect to the equilibrium probability distribution $\W_{\eq,\|}(z)$, namely
\begin{equation}
\label{Sn}
S_{n}
\stackrel{{\rm def}}{=}
\llangle p_{n}(z)\rrangle_{\eq}
=
\int_{-1}^{1}\!\!\D{z}\,p_{n}(z)\W_{\eq,\|}(z)
\;.
\end{equation}
In terms of these quantities, we can write $\chi_{\|}$ and $\chi_{\perp}$ as
\begin{equation}
\label{X:para:perp:bias}
\chi_{\|}
=
\Xo\left(\frac{1+2S_{2}}{3}-S_{1}^{2}\right)
\;,
\qquad
\chi_{\perp}
=
\Xo\frac{1-S_{2}}{3}
\;,
\end{equation}
for the writing of which we have employed
\[
\begin{array}{rclrcl}
\llangle\mz\rrangle_{\eq}
&
=
&
\mm S_{1}
\;,
&
\llangle\mz^{2}\rrangle_{\eq}
&
=
&
\mm^{2}(1+2S_{2})/3
\;,
\\
\llangle\mm_{x,y}\rrangle_{\eq}
&
=
&
0
\;,
\hspace{2.5em}
&
\llangle\mm_{x,y}^{2}\rrangle_{\eq}
&
=
&
\left(\mm^{2}-\llangle\mz^{2}\rrangle_{\eq}\right)/2
\;.
\end{array}
\]

The above expressions for $\chi$ are valid, for example, for {\em any\/} axially
symmetric anisotropy potential in a longitudinal bias field. For the simplest uniaxial
anisotropy in a longitudinal bias field
\begin{equation}
\label{U_z}
-\beta\Hs
=
\s z^{2}+\xi z
\;,
\end{equation}
one can derive the following explicit expressions for $S_{1}$ and $S_{2}$
\begin{eqnarray}
\label{S1}
	S_{1}
&
=
&
\frac{e^{\s}}{\s\Zp}\sinh\xi-h
\;,
\\
\label{S2}
	S_{2}
&
=
&
\frac{3}{2}
\left[
\frac{e^{\s}}{\s\Zp} (\cosh\xi-h\sinh\xi)+h^{2}-\frac{1}{2\s}
\right]
-\frac{1}{2}
\;,
\end{eqnarray}
where $h=\Bred=\xi/2\s$ and $\Zp$ is given by Eq.\ (\ref{Zpara}).%
\footnote{
The formula for $S_{1}=\langle z\rangle_{\eq}$, is essentially Eq.\
(\ref{m:para:Z}) for the longitudinal magnetization. In order to derive the formula for
$S_{2}$, we can take advantage of some previous results. Note first that the
thermodynamical energy in the longitudinal-field case can be written as
\begin{eqnarray*}
\E_{\|}
=
\langle
-Kvz^{2}-\mm Bz
\rangle_{\eq}
=
-Kv\left(
\langle z^{2}\rangle_{\eq}
+2h\langle z\rangle_{\eq}
\right)
=
-Kv
\left[
(1+2S_{2})/3
+2hS_{1}
\right]
\;.
\end{eqnarray*}
Then, on using Eq.\ (\ref{U:para}) for $\E_{\|}$, taking Eq.\ (\ref{S1}) into account,  and
recalling that $J=2(\cosh\xi+h\sinh\xi)$, one gets
\begin{eqnarray*} (1+2S_{2})/3
=
-(\E_{\|}/Kv)-2hS_{1}
=
(e^{\s}/\s\Zp)(\cosh\xi-h\sinh\xi)+h^{2}-1/2\s
\;,
\end{eqnarray*}
from which Eq.\ (\ref{S2}) follows.\qed }

In the $K=0$ case, the linear susceptibility is more easily obtained directly from the
definition (\ref{Sn}) of the $S_{n}$ with help from Eqs.\
(\ref{dZ0})--(\ref{Zsub0:derivatives2}). On doing so, one obtains
\begin{equation}
\label{X:para:perp:langevin}
\chi_{\|}
=
\Xo
L'
\;,
\qquad
\chi_{\perp}
=
\Xo
\frac{1}{\xi}L
\;,
\end{equation}
where $L(\xi)$ is the Langevin function. Note that, since $L(\xi)=\xi/3+\cdots$ for
low fields [Eq.\ (\ref{m:expansions:langevin})], both components of the above formula
merge on the Curie law $\chi=\mu_{0}\mm^{2}/3\T$ as the bias field goes to zero.

For $B=0$, the linear susceptibility is sometimes found written in a number of
alternative forms. Note first that in this case one has $S_{1}=0$. Therefore, on
introducing the notation $\legunb=S_{2}(\s,\xi)|_{\xi=0}$, one gets from Eq.\
(\ref{X:para:perp:bias}) the following formulae
\begin{equation}
\label{X:para:perp:shs}
\chi_{\|}
=
\Xo
\frac{1+2\legunb}{3}
\;,
\qquad
\chi_{\perp}
=
\Xo
\frac{1-\legunb}{3}
\;.
\end{equation}
(The quantity $\legunb$ is sometimes written as $S$ or merely $S_{2}$.) In order to
directly check Eqs.\ (\ref{X:para:perp:shs}) against Eqs.\ (\ref{X:para:perp}) one only
needs to use
\begin{equation}
\label{F-S}
\frac{\F'}{\F}
=
\frac
{\int_{-1}^{1}\!\D{z}\,z^{2}\exp(\s z^{2})}
{\int_{-1}^{1}\!\D{z}\,\exp(\s z^{2})}
=
\left.
\llangle z^{2}\rrangle_{\eq}
\right|_{B=0}
=
\frac{1}{3}
\left. (1+2S_{2})
\right|_{B=0}
=
\frac{1}{3}(1+2\legunb)
\;.
\end{equation}
Alternative expressions for $\chi$ at $B=0$ can also be written in terms of Kummer
functions. Thus, on introducing $C_{1}$ from Eq.\ (\ref{C_1:alt}) into Eq.\ (\ref{X:1}),
one directly gets (cf.\ Coffey, Crothers, Kalmykov and Waldron, 1995{\em b})
\begin{equation}
\label{X:para:perp:coffey}
\chi_{\|}
=
\frac{\mu_{0}\mm^{2}}{3\T}
\frac{M(\threehalfs,{\textstyle \frac{5}{2}};\s)}{M(\half,\threehalfs;\s)}
\;,
\qquad
\chi_{\perp}
=
\frac{\mu_{0}\mm^{2}}{3\T}
\frac{M(\half,{\textstyle \frac{5}{2}};\s)}{M(\half,\threehalfs;\s)}
\;.
\end{equation}

\subsubsection{Approximate formulae for the linear susceptibility}

We shall now derive approximate formulae for $\chi$ with the aim of by-pass, when
possible, the use of expressions involving non-elementary functions. The formulae
obtained, based on weak- and strong-anisotropy expansions, reasonably compare
with the exact results in whole temperature range.

We find it convenient to rewrite first the exact expression (\ref{X:2}) for $\chi$ as
follows
\begin{equation}
\label{X:5}
\chi
=
\Xo\frac{1}{3}
\left[ 1+\frac{1}{2}\left(3\frac{\F'}{\F}-1\right) (3\cosqal-1)
\right]
\;,
\end{equation}
where the factor multiplying $(3\cosqal-1)$ is precisely $\legunb$ [see Eq.\
(\ref{F-S})]. In order to derive approximate formulae for $\chi$ in the unbiased case
we shall use the approximate results for $\F'/\F$ derived in Appendix \ref{app:F}. We
can also get most of the following results (up to second order) if we start from the
expansions of $M_{B}$ in powers of $\s$ [Eq.\ (\ref{anisotropy:expansion:m})] and the
asymptotic expansion (\ref{m:asympt}).

\paragraph{Weak-anisotropy range.}

In order to obtain an approximate formula for $\chi$ valid in the $|\s|\ll1$ range,
we insert the approximate $\F'/\F$ from Eq.\ (\ref{F-der:approx1}) into Eq.\
(\ref{X:5}), getting
\begin{equation}
\label{X:approx1}
\left.
\chi
\right|_{|\s|\ll1}
\simeq
\frac{\mu_{0}\mm^{2}}{3\T}
\left[
   1+\left(\frac{2}{15}\s+\frac{4}{315}\s^{2}
-\frac{8}{4725}\s^{3}\right) (3\cosqal-1)
\right]
\;.
\end{equation}
This equation yields a good approximation of the exact $\chi$ for $|\s|\leq 2$. Note
that, as it should, when the anisotropy axes are distributed at random, the corrections
to the leading (isotropic) result vanish at {\em all\/} orders.

\paragraph{Strong-anisotropy ranges.}

Similarly, to obtain approximate formulae for $\chi$ valid in the $|\s|\gg1$ ranges, we
shall use the corresponding approximate expressions for $\F'/\F$ derived in Appendix
\ref{app:F}.

For $\s\ll-1$, we insert $\F'/\F$ from Eq.\ (\ref{F-der:approx2a}) into Eq.\ (\ref{X:5}),
getting
\begin{equation}
\label{X:approx2a}
\left.
\chi
\right|_{\s\ll-1}
\simeq
\Xo
\left[
\frac{1}{2}\senqal
-\frac{1}{4\s}(3\cosqal-1)
\right]
\;.
\end{equation}
An approximate formula for the extreme easy-axis case can be derived in a similar
way. On substituting the $\s\gg1$ result (\ref{F-der:approx2b}) for $\F'/\F$ in
Eq.\ (\ref{X:5}), we obtain
\begin{equation}
\label{X:approx2b}
\left.
\chi
\right|_{\s\gg1}
\simeq\Xo
\left[
\cosqal
-\left(
\frac{1}{2\s}
+\frac{1}{4\s^{2}}
+\frac{5}{8\s^{3}}
\right) (3\cosqal-1)
\right]
\;.
\end{equation}

Again, when the anisotropy axes are distributed at random, all the
corrections to the leading plane-rotator and Ising results vanish
identically. These approximate formulae compare well with the
corresponding exact results for $|\s|\geq5$, so that, on complementing
Eqs.\ (\ref{X:approx1}), (\ref{X:approx2a}), and (\ref{X:approx2b})
one can cover the entire $\s$-range reasonably. This merely follows
from the patching (shown in Fig.\ \ref{Fp-Fpp_approx:plot} of Appendix
\ref{app:F}) of the exact $\F'/\F$ provided by the approximate
formulae with which the above approximate results for $\chi$ have been
constructed.

For future reference, we finally write down the longitudinal and transverse
components of $\chi$ for strong anisotropy to order $1/|\s|$, namely
\begin{equation}
\label{X:para:perp:approx}
\chi_{\|}
\simeq
\Xo-\frac{\mu_{0}\mm^{2}}{Kv}
\;,
\qquad
\chi_{\perp}
\simeq
\frac{\mu_{0}\mm^{2}}{2Kv}
\;,
\qquad
(K>0)
\;,
\end{equation}
and
\begin{equation}
\label{X:para:perp:approx:easyplane}
\chi_{\|}
\simeq
\frac{\mu_{0}\mm^{2}}{2|K|v}
\;,
\qquad
\chi_{\perp}
\simeq
\frac{\mu_{0}\mm^{2}}{2\T}-\frac{\mu_{0}\mm^{2}}{4|K|v}
\;,
\qquad
(K<0)
\;.
\end{equation}
Note the qualitative interchange of the r\^{o}les of $\chi_{\|}$ and $\chi_{\perp}$
with the transformation $K\to-K$.

\paragraph{Formulae in the presence of a longitudinal bias field.}

We can also obtain high-barrier approximations of the exact equilibrium
susceptibilities in the presence of a longitudinal bias field. Those equations, which will
be valid for $h\ll1$, can be obtained by starting from the approximate expression
(\ref{Zpara:approx}) for the partition function. Thus, on applying the relations [readily
obtainable from Eqs.\ (\ref{distribution:para:gral}) and (\ref{Sn})]
\begin{equation}
\label{S-Zderivatives} S_{1}
=
\frac{1}{\Zp}\frac{\partial\Zp}{\partial\xi}
\;,
\qquad
\frac{1}{3}(1+2S_{2})
=
\frac{1}{\Zp}\frac{\partial\Zp}{\partial\s}
\;,
\end{equation}
to the approximate $\Zp$ mentioned, one gets from Eqs.\ (\ref{X:para:perp:bias})
\begin{eqnarray}
\label{X:para:bias:approx}
\chi_{\|}
&
\simeq
&
\Xo
\frac{1}{(\cosh\xi-h\sinh\xi)^{2}}
\nonumber
\\
&
& {}\times
\bigg\{ (1-h^{2})-\frac{1}{\s}
+\frac{1}{8\s^{2}}
\bigg[
          1-\frac{1+6h^{2}+h^{4}}{(1-h^{2})^{2}}\cosh(2\xi)
\nonumber
\\
&
&
\hspace{12em}
{}+\frac{4h(1+h^{2})}{(1-h^{2})^{2}}\sinh(2\xi)
\bigg]
\bigg\}
\;,
\\
\label{X:perp:bias:approx}
\chi_{\perp}
&
\simeq
&
\Xo
\frac{1}{2\s}
\frac
{(1+h^{2})\cosh\xi-2h\sinh\xi}
{(1-h^{2})(\cosh\xi-h\sinh\xi)}
\;.
\end{eqnarray}
For $B=0$, these formulae duly reduce to Eqs.\ (\ref{X:para:perp:approx}). Finally, on
taking formally the $K\to\infty$ limit in these formulae (i.e., $\s\to\infty$ and
$h=\xi/2\s\to0$), one gets the ``Ising-type" equilibrium susceptibilities in a
longitudinal bias field [cf. Eq.\ (\ref{X:ising})]
\begin{equation}
\label{X:para:perp:bias:ising}
\chi_{\|}
\simeq
\Xo\frac{1}{\cosh^{2}\xi}
\;,
\qquad
\chi_{\perp}\simeq0
\;.
\end{equation}
Equations (\ref{X:para:bias:approx}), (\ref{X:perp:bias:approx}), and
(\ref{X:para:perp:bias:ising}) will be used in Section \ref{sect:stochastic}.

\subsubsection{Temperature dependence of the linear susceptibility}

\begin{figure}[t!]
\vspace{-3.ex}
\eps{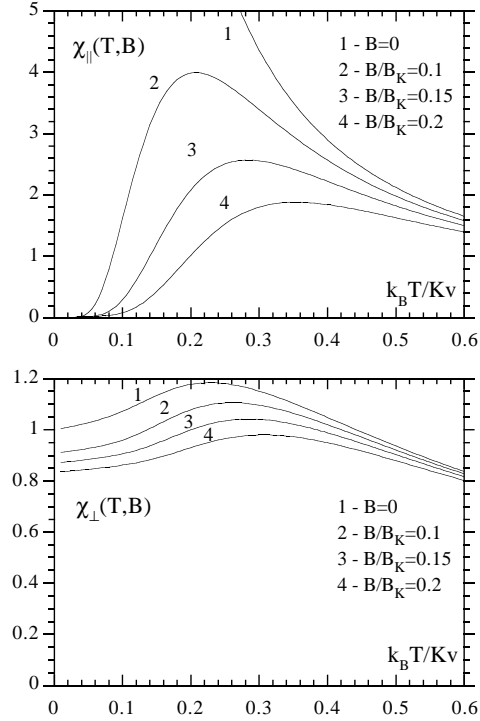}{0.7}
\vspace{-3.ex}
\caption[]
{
Longitudinal and transverse components of the linear susceptibility
vs.\ $T$ in the unbiased case and in the presence of longitudinal bias
fields [Eqs.\ (\ref{X:para:perp:bias})]. The anisotropy is assumed to
be of easy-axis type ($K>0$) and the susceptibilities are measured in
units of $\mu_{0}\mm/\BK=\mu_{0}m^{2}/2Kv$ [the transverse equilibrium
susceptibility at $T=0$ in the unbiased case; see Eq.\
(\ref{X:para:perp:approx})].
\label{X:bias:plot}
}
\end{figure}
Figure \ref{X:bias:plot} displays the linear susceptibility in a
longitudinal bias field.  Concerning the longitudinal component, this
decreases with increasing $B$ for a given $T$, since $\chi_{\|}(T,B)$
is the slope of the longitudinal magnetization curve at $B$ (see Fig.\
\ref{magnetization:west:plot}). As regards the temperature dependence
of $\chi_{\|}$, because $\chi_{\|}(T,B=0)$ is the {\em initial\/}
slope of $M_{B,\|}$, it always increases as the thermal agitation is
reduced. In contrast, $\chi_{\|}(T,B\neq0)$ has a maximum as a
function of the temperature and tends to zero at low
temperatures. This is also a result of $\chi_{\|}(T,B\neq0)$ being the
slope of $M_{B,\|}$ at $B\neq0$. Indeed, at high temperatures
($\xi\ll1$), $\chi_{\|}$ also increases with decreasing thermal
agitation.  However, at low temperatures ($\xi\gg1$), the slope of
$M_{B,\|}$ vs.\ $B$ decreases as $T$ is lowered---``high-field"
magnetization approaching a straight line due to the saturation of
$\langle\m\rangle_{\eq}$ (cf.\ Fig.\ \ref{magnetization:west:plot}).
Therefore, in the intermediate temperature range $\chi_{\|}(T,B\neq0)$
exhibits a maximum at the temperature where the ``shoulder" of the
magnetization curve passes through $B$. Note finally that, for this
maximum to exist the anisotropy is secondary, whereas a non-zero bias
field is essential. Indeed, the longitudinal component of Eq.\
(\ref{X:para:perp:langevin}) for an {\em isotropic\/} spin also
exhibits a maximum in $\chi_{\|}$ vs.\ $T$ if $B\neq0$.

Concerning the transverse susceptibility, it exhibits a maximum as a
function of $T$ even for $B=0$, so it cannot be attributed to the
presence of the bias field. This maximum is to be interpreted in terms
of the anisotropy-induced crossover from the free-rotator (isotropic)
regime at high $T$ to the discrete-orientation regime as $T$ is
lowered. Indeed, at low temperatures the transverse probing field
competes with the anisotropy energy in aligning the magnetic moments,
which are concentrated close to the potential minima. Then, the
increase of the thermal agitation permits $\m$ to (statistically)
separate from the poles and the (transverse) response increases.
However, if the temperature is further increased $\m$ becomes
progressively unfastened from the anisotropy and the transverse field
competes mainly with the thermal agitation in aligning $\m$; the
response then exhibits a maximum and decreases as $T$ is increased. In
this transverse probing-field case, is the anisotropy, not the bias
field, the essential element for the appearance of the maximum in the
response. Indeed, the transverse component of Eq.\
(\ref{X:para:perp:langevin}), i.e.,
$\chi_{\perp}=(\mu_{0}\mm^{2}/\T)L/\xi$, starting from the non-zero
value $\mu_{0}\mm/B$ at $T=0$, decreases {\em monotonically\/} with
$T$ in the whole temperature range [as
$\chi_{\perp}\simeq(\mu_{0}\mm/B)(1-\T/\mm B)$ for $\xi\gg1$ the
decreasing is linear at low $T$].

\subsection{Non-linear susceptibilities}
\label{subsect:Xnl}

We shall now consider the non-linear susceptibilities of classical spins with axially
symmetric magnetic anisotropy. Part of the motivation to study the non-linear
susceptibilities is the suitability of these quantities in the study of collective
phenomena in glassy systems, together with the glassy-like features exhibited by
interacting magnetic nanoparticles (see, for example, Jonsson et~al., 1995). Most of the
following results were obtained by Garc{\'{\i}}a-Palacios and L{\'{a}}zaro (1997), while
the extension of the theory to the dynamical case was done by Ra{\u{\i}}kher and
Stepanov (1997).

The non-linear susceptibilities are defined as the coefficients of the non-linear terms
in the expansion of the magnetization in powers of the external field. To our
knowledge, these quantities had never been derived from the available expressions
for the magnetization that take the magnetic anisotropy into account. In fact, these
formulae are either not expressly suitable to extract the non-linear susceptibilities,
because they are not expressed as series of powers of the field (see Chantrell's
formula in Williams et~al., 1993), or would yield the non-linear susceptibilities as
series of powers of the anisotropy parameter (Lin, 1961).

Here, some of the parallel properties of the non-linear susceptibilities of
non-interacting classical spins will be illustrated with the first one of the series,
$\chi_{3}$. The basic expression for this quantity can be obtained by comparing the
$H$-expansion of $M_{B}$ (\ref{Mexpansion}) with its $\xi$-expansion (\ref{M2}), to
get
\begin{equation}
\label{Xnl:1}
\chi_{3}
=
\frac{\mu_{0}^{3}\mm^{4}}{(\T)^{3}}2(C_{2}-C_{1}^{2})
\;,
\end{equation}
which involves the first two coefficients of the field-expansion (\ref{Zfinal}) of the
partition function.

\subsubsection{Non-linear susceptibilities: particular cases}

Let us first write down the expressions that emerge from Eq.\ (\ref{Xnl:1}) when
one considers various particular cases of the combination $2(C_{2}-C_{1}^{2})$ (see
Table \ref{Ci:combinations:limits:table}).

\paragraph{Isotropic case.}

For $\s\to0$, one has $2(C_{2}-C_{1}^{2})=-1/45$, so that the Langevin $\chi_{3}$
reads
\begin{equation}
\label{Xnl:langevin}
\chi_{3,\lan}
=
-\frac{\mu_{0}^{3}\mm^{4}}{45(\T)^{3}}
\;.
\end{equation}

\paragraph{Ising regime.}

For $\s\to\infty$, the combination of the $C_{i}$ required reads
$2(C_{2}-C_{1}^{2})=-\coscal/3$; accordingly, the Ising $\chi_{3}$ is given by
\begin{equation}
\label{Xnl:ising}
\chi_{3,\ising}
=
-\frac{\mu_{0}^{3}\mm^{4}\coscal}{3(\T)^{3}}
\;,
\end{equation}
which vanishes when the field points along a direction perpendicular to the anisotropy
axis.

\paragraph{Plane-rotator regime.}

For $\s\to-\infty$, we have $2(C_{2}-C_{1}^{2})=-\sencal/16$, whence
\begin{equation}
\label{Xnl:rotator}
\chi_{3,\rotator}
=
-\frac{\mu_{0}^{3}\mm^{4}\sencal}{16(\T)^{3}}
\;.
\end{equation}
Here, the non-linear susceptibility vanishes when the field points along the direction
perpendicular to the plane of the rotator.

\paragraph{Longitudinal-field case.}

Finally, when the field is parallel to the anisotropy axis, one has $2(C_{2}-C_{1}^{2})
=[\F''/3\F-(\F'/\F)^{2}]/2$, so that the corresponding non-linear susceptibility reads
\begin{equation}
\label{Xnl:para}
\chi_{3,\|}
=
\frac{\mu_{0}^{3}\mm^{4}}{(\T)^{3}}\frac{1}{2}
\left[
\frac{1}{3}\frac{\F''}{\F}
-\left(\frac{\F'}{\F}\right)^{2}
\right]
\;.
\end{equation}
As occurs with the linear susceptibility, the magnetic anisotropy induces an additional
dependence of $\chi_{3}$ on $T$ via the functions $\F^{(\ell)}/\F$, with the
consequent departure from the $T^{-3}$ dependences of the above limit cases.

\subsubsection{Formulae for the non-linear susceptibility}

On introducing the complete expression for $2(C_{2}-C_{1}^{2})$ obtained from Eq.\
(\ref{C_2:C_1}) into Eq.\ (\ref{Xnl:1}), we get the following general formula
for $\chi_{3}$
\begin{eqnarray}
\label{Xnl:2}
\chi_{3}
=
\frac{\mu_{0}^{3}\mm^{4}}{(\T)^{3}}
&
\Bigg\{
&
\frac{1}{2}
\bigg[
\frac{1}{3}\frac{\F''}{\F}-\bigg(\frac{\F'}{\F}\bigg)^{2}
\bigg]
\coscal
\nonumber
\\
&
& {}+\frac{1}{2}
\bigg[
\bigg(\frac{\F'}{\F}\bigg)^{2}-\frac{\F''}{\F}
\bigg]
\cosqal\senqal
\nonumber
\\
&
& {}+\frac{1}{16}
\bigg[
-1+2\frac{\F'}{\F}-2\bigg(\frac{\F'}{\F}\bigg)^{2}+\frac{\F''}{\F}
\bigg]
\sencal
\Bigg\}
\;.
\qquad
\end{eqnarray}
This expression can alternatively be written in terms of the averages
of the Legendre polynomials (\ref{Sn}) evaluated at zero field
(Ra{\u{\i}}kher and Stepanov, 1997)
\begin{eqnarray}
\label{Xnl:raiste}
\chi_{3}
=
\frac{\mu_{0}^{3}\mm^{4}}{(\T)^{3}}
\frac{1}{315}
&
\bigg[
&
(12\legunbfour-70\legunb^{2}-40\legunb-7)
\coscal
\nonumber
\\
&
&
{}-2(18\legunbfour-35\legunb^{2}+10\legunb+7)
\cosqal\senqal
\nonumber
\\
&
&
{}+\frac{1}{2}
(9\legunbfour-35\legunb^{2}+40\legunb-14)
\sencal
\bigg]
\;,
\end{eqnarray}
where $\tilde{S}_{n}=S_{n}(\s,\xi)|_{\xi=0}$. These formulae simplify notably when
averaged over an ensemble of equivalent dipole moments with a random distribution
of anisotropy axes.

\paragraph{Average of the non-linear susceptibility for anisotropy axes distributed at
random.}

When the expressions (\ref{averages:particular}) for the averages of the angular
terms are introduced into Eq.\ (\ref{Xnl:2}), one gets the following formula for
$\langle\chi_{3}\rangle_{\ran}$ [cf.\ Eq.\ (\ref{C_2:C_1:ran})]
\begin{equation}
\label{Xnl:ran}
\llangle\chi_{3}\rrangle_{\ran}
=
\frac{\mu_{0}^{3}\mm^{4}}{(\T)^{3}}
\frac{1}{30}
\bigg[ 2\frac{\F'}{\F}-3\bigg(\frac{\F'}{\F}\bigg)^{2}-1
\bigg]
\;,
\end{equation}
or, by using the relation $\F'/\F=(1+2\legunb)/3$, the more compact form
\begin{equation}
\label{Xnl:raiste:ran}
\llangle\chi_{3}\rrangle_{\ran}
=
-\frac{\mu_{0}^{3}\mm^{4}}{(\T)^{3}}
\frac{1+2\legunb^{2}}{45}
\;.
\end{equation}
Note that, unlike $\avX$, which is given by the Curie law, $\chi_{3}$
depends on the anisotropy energy even for anisotropy axes distributed
at random. Indeed, we had already seen in Fig.\
\ref{magnetization:appr_alpha:plot} that, while for low fields one has
$\langle M_{B}\rangle_{\ran}\simeq\mm L(\xi)$, as the field is
increased $\langle M_{B}\rangle_{\ran}$ bends downwards more rapidly
than the Langevin magnetization. Thus, not only
$\langle\chi_{3}\rangle_{\ran}\neq\chi_{3,\lan}$, but
$|\langle\chi_{3}\rangle_{\ran}|>|\chi_{3,\lan}|$ (up to factors of
$3$ and $1.5$ at low $T$ for $K>0$ and $K<0$, respectively).

\paragraph{Reduced non-linear susceptibility.}

\begin{figure}[t!]
\vspace{-3.ex}
\eps{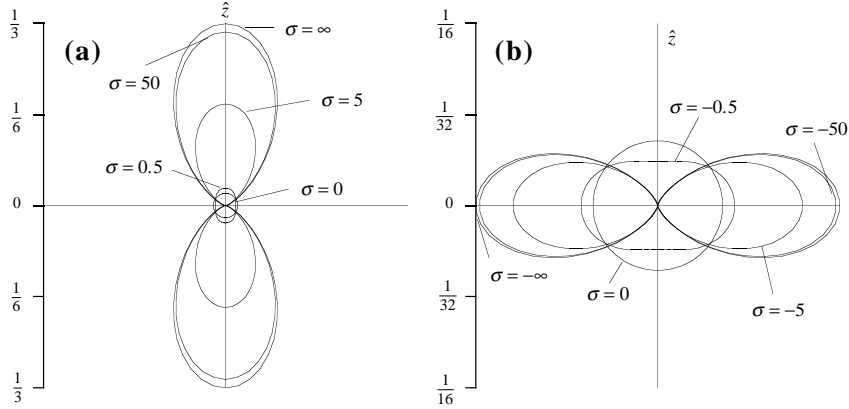}{1.15}
\vspace{-3.ex}
\caption[]
{
Polar plots showing the angular dependence of the reduced
non-linear susceptibility $-\chi_{3}^{\red}$ [Eq.\ (\ref{Xnl:red})]
for various values of the dimensionless anisotropy parameter
$\s=Kv/\T$. (a) Easy-axis anisotropy. (b) Easy-plane anisotropy.
\label{X3:polar:plot}
}
\end{figure}

In analogy with the reduced linear susceptibility (\ref{X:red}), we
can define a reduced non-linear susceptibility isolating the
anisotropy-induced temperature dependence of $\chi_{3}$ as follows
\begin{equation}
\label{Xnl:red}
\chi_{3}^{\red}(\s,\alpha)
=
\chi_{3}(\s,\alpha)\frac{(\T)^{3}}{\mu_{0}^{3}\mm^{4}}
=
2(C_{2}-C_{1}^{2})
\;.
\end{equation}
Figure \ref{X3:polar:plot} displays $-\chi_{3}^{\red}$ as a function
of the angle between the anisotropy axis and the external field. It is
shown that the $\chi_{3}^{\red}$ curves become increasingly
anisotropic as $|\s|$ increases, being quite different from circles
already for $|\s|\simeq1$. (The circles for the isotropic
$-\chi_{3}^{\red}|_{\s=0}$ correspond to the same radius ($1/45$), but
they have different sizes in the plots since the maximum value of
$-\chi_{3}^{\red}$ is $1/3$ for $K>0$ and $1/16$ for $K<0$.)

\begin{figure}[t!]
\vspace{-3.ex}
\eps{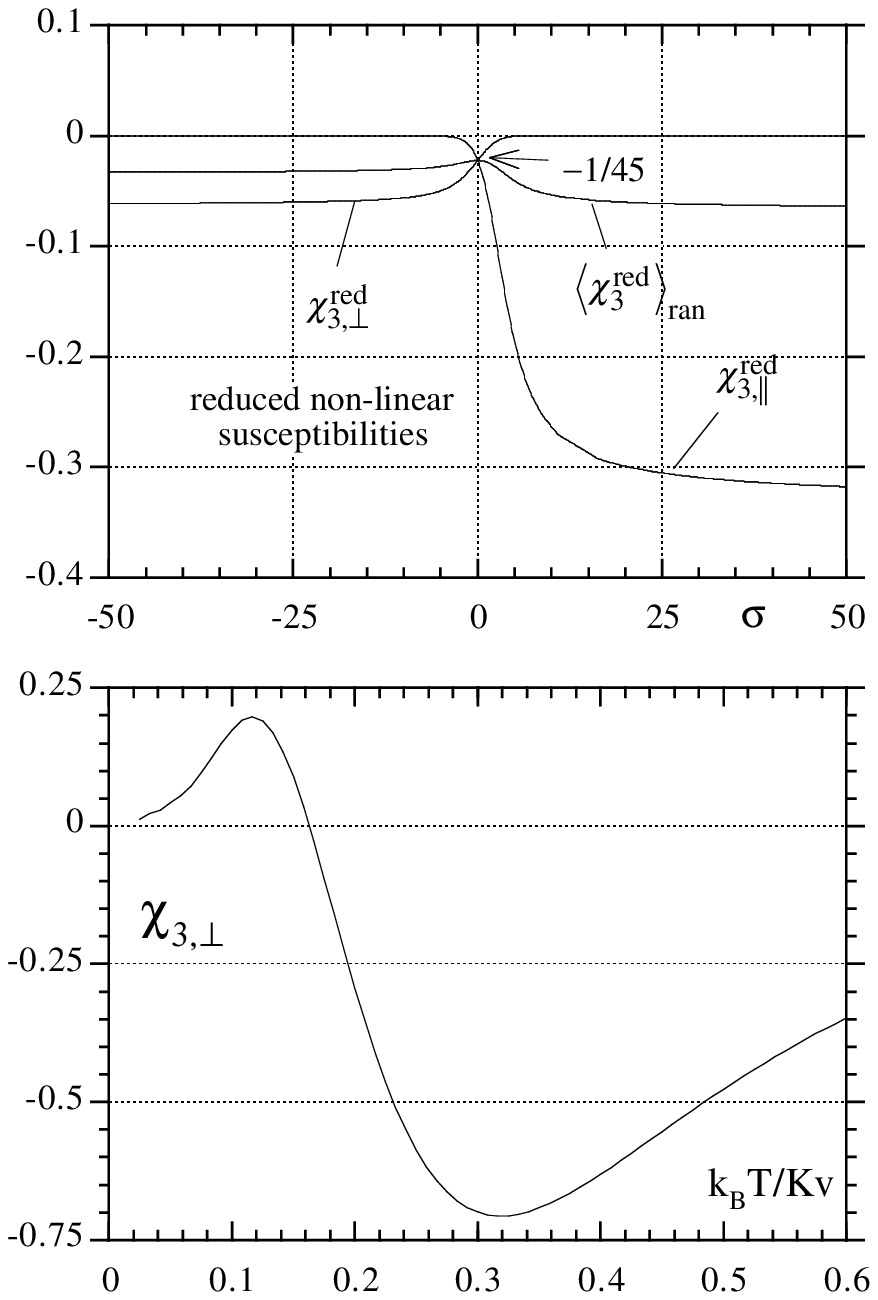}{0.56}
\vspace{-3.ex}
\caption[]
{
Upper panel: Reduced non-linear susceptibility (\ref{Xnl:red}) in the
longitudinal, $\chi_{3,\|}^{\red}$, and transverse,
$\chi_{3,\perp}^{\red}$, field cases, and for anisotropy axes
distributed at random, $\langle\chi_{3}^{\red}\rangle_{\ran}$, vs.\
the dimensionless anisotropy parameter $\s=Kv/\T$.  Lower panel:
Temperature dependence of the transverse component of the non-linear
susceptibility [from Eq.\ (\ref{Xnl:2})]. $\chi_{3,\perp}$ is measured
in units of $\mm(\mu_{0}/\BK)^{3}$
\label{X3:red:plot}
}
\end{figure}

The upper panel of Fig.\ \ref{X3:red:plot} shows $\chi_{3}^{\red}$
vs.\ $\s$ in the longitudinal and transverse field cases, as well as
for anisotropy axes distributed at random.
The three curves coincide at $\s=0$, where the orientation of the
magnetic field plays no r\^{o}le, taking the Langevin value
$-1/45$. It is noticeable the large variation of $\chi_{3}^{\red}$
with respect to $\s$ for anisotropy axes parallel to the field. Note
also that, although dramatically reduced, the anisotropy-induced
temperature dependence of $\chi_{3}$ is kept for anisotropy axes
distributed at random. On the other hand, we can again remark that,
{\em qualitatively}, the longitudinal and the transverse field cases
interchange their r\^{o}les when the sign of the anisotropy is
reversed (see also Fig.\ \ref{X3:polar:plot}). For instance, for
easy-plane anisotropy $\chi_{3,\|}$ rapidly vanishes as $|\s|$ departs
from zero. The analogous result for easy-axis anisotropy occurs in the
presence of a transverse field; then $\chi_{3,\perp}$ rapidly
decreases as $\s$ departs from zero. However, in this case $\chi_{3}$
does not exactly vanish, but it goes to a finite non-zero value for
large $\s$ (which is not resolved with the scale used in Fig.\
\ref{X3:red:plot}). This will be discussed below.

\paragraph{The sign of the non-linear susceptibility.}

As the non-linear susceptibility is a measure of the departure of the magnetization
from the linear regime, and this departure usually consists of a bending downwards,
one is tempted to conclude that $\chi_{3}$ is a negative quantity. Indeed, the above
formula for anisotropy axes distributed at random [Eq.\ (\ref{Xnl:raiste:ran})] clearly
shows that this is indeed the case for $\langle\chi_{3}\rangle_{\ran}$ (in accordance
with the downward bending of the corresponding magnetization in Fig.\
\ref{magnetization:appr_alpha:plot}). However, this result is not general as will be
illustrated now with $\chi_{3,\perp}$. Let us compute the low temperature ($\s\gg1$)
expression for $\chi_{3,\perp}$ by using the asymptotic methods of Appendix
\ref{app:F}:%
\footnote{
As the first non-vanishing term in $\chi_{3,\perp}$ is of fourth order [see Eq.\
(\ref{F-der:approx2b:comb})], we need to compute one more coefficient
$b_{i}$ in the $\s\gg1$ expansion of Appendix \ref{app:F}. On doing this we get
$b_{4}=-37/8$, from which we obtain the fourth order term of $\F'/\F$, and from this
we can calculate the corresponding terms in $(\F'/\F)^{2}$ and $\F''/\F$.
}
\[
\chi_{3,\perp}
=
\frac{1}{16}
\frac{\mu_{0}^{3}\mm^{4}}{(\T)^{3}}
\bigg[
-1+2\frac{\F'}{\F}-2\bigg(\frac{\F'}{\F}\bigg)^{2}+\frac{\F''}{\F}
\bigg]
\simeq
\frac{1}{16}
\frac{\mu_{0}^{3}\mm^{4}}{(\T)^{3}}
\frac{1}{\s^{4}}
\;,
\]
so that
\begin{equation}
\label{Xnl:perp}
\chi_{3,\perp}
\simeq
\frac{1}{2}m
\Big(\frac{\mu_{0}}{\BK}\Big)^{3}
\,
\frac{\T}{Kv}
\;.
\end{equation}
Therefore, we see that, not only is $\chi_{3,\perp}$ positive at low
temperatures, but it indeed increases linearly with $T$. At higher
temperatures the above expansion must break down and the corresponding
corrections bring $\chi_{3,\perp}$ to the negative values that it must
take at sufficiently high temperatures
($\chi_{3,\perp}|_{\s\ll1}\simeq\chi_{3,\lan}
=-[\mu_{0}^{3}\mm^{4}/45(\T)^{3}]$).  Thus, from the knowledge of the
limit temperature dependences ($\chi_{3,\perp}\propto T$ and
$-1/T^{3}$) one concludes that $\chi_{3,\perp}$ must have two peaks
and cross the temperature axis at a certain intermediate temperature.
This is precisely what it can be seen in the lower panel of Fig.\
\ref{X3:red:plot}, showing that $\chi_{3}\leq0$ is not a general
result. As $T$ decreases, $\chi_{3,\perp}$ has a negative minimum,
increases, crosses zero, exhibits a secondary positive maximum, and
eventually tends to zero at low temperatures. These are the typical
features exhibited by the {\em dynamical\/} non-linear susceptibility
$\chi_{3}(\w,T)$ (Ra{\u{\i}}kher and Stepanov, 1997), but their
occurrence in the equilibrium susceptibility is somewhat
unexpected. This is another good example of the effects of the
magnetic anisotropy on the properties of superparamagnetic systems.

\subsubsection{Generalizations}

One can also derive the non-linear susceptibility by means of the relation between the
thermal-equilibrium fluctuations of $\m$, in the absence of a probing field, and the
actual magnetic response of the system, by-passing the explicit expansion of the
magnetization in a series of powers of the field.

On inspecting the definition (\ref{Mexpansion}), one realizes that $\chi_{3}$ can be
obtained by differentiating the magnetization as $\chi_{3}=\frac{1}{6}\mu_{0}^{3}
\partial^{3}\blangle\m\cdot\hat{b}\brangle_{\eq} /\partial B^{3}|_{B=0}$. This is
directly generalized to
\begin{equation}
\label{Xnl:0}
\chi_{3}
=
\frac{1}{6}\mu_{0}^{3}
\left.
\frac
{\partial^{3}\blangle\m\cdot\bp\brangle}
{\partial (\dBo)^{3}}
\right|_{\dBo=0}
\;,
\end{equation}
where $\dB=\dBo\,\bp$ is an external probing field and the averages are now
taken with respect to the total energy of the system in the presence of $\dB$.

On calculating the above third-order derivative by making repeated use of the Eq.\
(\ref{FDR:equilibrium:n}), one arrives at the general result [cf.\ Eq.\ (\ref{X:FDR})]
\begin{eqnarray*}
\chi_{3}
=
\frac{\mu_{0}^{3}}{(\T)^{3}}
\frac{1}{6}
&
\Big[
&
\blangle(\m\cdot\bp)^{4}\brangle_{\eq}
-4\blangle(\m\cdot\bp)^{3}\brangle_{\eq}
\blangle\m\cdot\bp\brangle_{\eq}
-3\blangle(\m\cdot\bp)^{2}\brangle_{\eq}^{2}
\\
&
& {}+12\blangle(\m\cdot\bp)^{2}\brangle_{\eq}
\blangle\m\cdot\bp\brangle_{\eq}^{2}
-6\blangle\m\cdot\bp\brangle_{\eq}^{4}
\Big]
\;,
\end{eqnarray*}
where the averages are finally taken in the absence of the probing field. Note however
that if a bias field is applied, there is also a non-zero term in $(\dBo)^{2}$, which
defines the corresponding susceptibility $\chi_{2}$ (see, for example, Ra{\u{\i}}kher
et~al., 1997). Nevertheless, on assuming that no constant field is applied and noting
that, consequently, the above averages at zero probing field are then zero-field
averages, we can use $\langle(\m\cdot\bp)^{2n+1}\rangle_{\eq}|_{B=0}=0$, to get
\begin{equation}
\label{Xnl:FDR}
\chi_{3}
=
\frac{\mu_{0}^{3}}{(\T)^{3}}
\frac{1}{6}
\left.
\left[
\blangle(\m\cdot\bp)^{4}\brangle_{\eq}
-3\blangle(\m\cdot\bp)^{2}\brangle_{\eq}^{2}
\right]
\right|_{B=0}
\;.
\end{equation}
This relation between the non-linear susceptibility and the thermal-equilibrium
fluctuations of the magnetic moment in zero field, is valid for any form of the
magnetic-anisotropy energy provided that this has inversion symmetry
$\langle(\m\cdot\bp)^{2n+1}\rangle_{\eq}=0$.

Finally, on returning to the simplest uniaxial-anisotropy case and recalling that the
zero-field averages of $(\m\cdot\bp)^{2i}$ are directly related with the coefficients
$C_{i}$ by Eq.\ (\ref{Ci:averages}), specifically
\[
\blangle (\m\cdot\bp)^{2}
\brangle_{\eq}
\big|_{B=0}
=
\mm^{2}2C_{1}
\;,
\qquad
\blangle(\m\cdot\bp)^{4}\brangle_{\eq}\big|_{B=0}
=
\mm^{4}12C_{2}
\;,
\]
one gets
\[
\frac{1}{6}
\left.
\left[
\blangle(\m\cdot\bp)^{4}\brangle_{\eq}
-3\blangle(\m\cdot\bp)^{2}\brangle_{\eq}^{2}
\right]
\right|_{B=0}
=
\mm^{4}2\left(C_{2}-C_{1}^{2}\right)
\;,
\]
so that the expression (\ref{Xnl:1}) for $\chi_{3}$ is reobtained.

\subsubsection{Approximate formulae for the non-linear susceptibility}

We shall now derive approximate expressions for $\chi_{3}$, with the aim of establish
simple approximate expressions valid in wide temperature ranges. Again, in order to
obtain the approximate formulae we shall use the corresponding expressions for
$\F'/\F$ and $\F''/\F$ derived in Appendix \ref{app:F}. (We could also proceed from
the weak- and strong-anisotropy formulae for $M_{B}$.) The approximate expressions
for the combinations of the functions $\F^{(\ell)}/\F$ entering in the general formula
(\ref{Xnl:2}) are given by Eqs.\ (\ref{F-der:approx1:comb}),
(\ref{F-der:approx2a:comb}), and (\ref{F-der:approx2b:comb}).

\paragraph{Weak-anisotropy range.}

To obtain an approximate formula for $\chi_{3}$ valid for weak anisotropy, we insert
Eqs.\ (\ref{F-der:approx1:comb}) into Eq.\ (\ref{Xnl:2}), gather the terms with the
same power of $\s$, and express the trigonometric factors in terms of
$\cosqal$ and $\coscal$ only, obtaining
\begin{eqnarray}
\label{Xnl:approx1}
\left.
\chi_{3}
\right|_{|\s|\ll1}
\simeq-\frac{\mu_{0}^{3}\mm^{4}}{45(\T)^{3}}
&
\bigg[
&
1+\frac{8}{21}(3\cosqal-1)\s
\nonumber
\\
&
& {}+\frac{8}{105}(4\coscal-\cosqal)\s^{2}
\\
&
& {}+\frac{32}{10395}(21\coscal-18\cosqal+4)\s^{3}
\bigg]
\;.
\nonumber
\end{eqnarray}
This equation is a good approximation of the exact $\chi_{3}$ for $|\s|\leq 2$. Note
that, in contrast to $\chi$, only the {\em first\/} correction to the leading (isotropic)
result vanishes when the anisotropy axes are distributed at random [recall Eq.\
(\ref{anisotropy:expansion:m:ran})].

\paragraph{Strong-anisotropy ranges.}

Let us first consider the $\s\ll-1$ range. If we insert Eqs.\ (\ref{F-der:approx2a:comb})
into Eq.\ (\ref{Xnl:2}) and gather the terms with the same power of $1/\s$, we obtain
\begin{equation}
\label{Xnl:approx2a}
\left.
\chi_{3}
\right|_{\s\ll-1}
\simeq
-\frac{\mu_{0}^{3}\mm^{4}\sencal}{16(\T)^{3}}
\left[ 1+\frac{1}{\s}
+(16\cot^{2}\alpha-1)\frac{1}{4\s^{2}}
\right]
\;.
\end{equation}
This the desired approximate formula for $\chi_{3}$ valid in extreme easy-plane 
range. An approximate expression for $\s\gg1$ can be obtained in a similar way. On
inserting Eqs.\ (\ref{F-der:approx2b:comb}) into Eq.\ (\ref{Xnl:2}) and gathering the
terms with the same power of $1/\s$, we arrive at
\begin{equation}
\label{Xnl:approx2b}
\left.
\chi_{3}
\right|_{\s\gg1}
\simeq
-\frac{\mu_{0}^{3}\mm^{4}\coscal}{3(\T)^{3}}
\left[ 1-\frac{2}{\s}+(3\tan^{2}\alpha-1)\frac{1}{2\s^{2}}
+(3\tan^{2}\alpha-4)\frac{1}{2\s^{3}}
\right]
\;.
\end{equation}

These strong-anisotropy equations match the corresponding exact results for $|\s|\geq
5$. In fact, with the combined use of Eqs.\ (\ref{Xnl:approx1}), (\ref{Xnl:approx2a}),
and (\ref{Xnl:approx2b}), one can almost cover the exact $\chi_{3}$ in the whole
temperature range. Again this arises directly from the reasonable patching shown in
Appendix \ref{app:F} of the exact $\F'/\F$ and $\F''/\F$ curves yielded by the
approximate formulae employed.

\subsubsection{Temperature dependence of the non-linear susceptibility}

\paragraph{Theoretical results.}

We shall now study in more detail the temperature dependence of $\chi_{3}$. Facing
the subsequent particularization of the results to a number of systems of magnetic
nanoparticles, we shall consider the occurrence of a distribution of particle volumes.
We shall however take the anisotropy constant $K$ and the spontaneous
magnetization $M_{s}=\mm/v$ as fixed, i.e., neither distribution in particle shape, nor
size effects on $M_{s}$ or $K$ will be considered. Then, if the anisotropy axes of the
particles with the same volume are distributed at random, one can write
\[
\chi_{3}
=
\int_{0}^{\infty}\!\!\D{v}\, v^{-1}\llangle\chi_{3}\rrangle_{\ran}f(v)
\;,
\]
where the factor $v^{-1}$ occurs since $f(v)\D v$ is taken as the fraction of the total
volume occupied by particles with volumes in the interval $(v,v+\D v)$.

In order to isolate the effect of the magnetic anisotropy on $\chi_{3}(T)$, we shall
assume that $M_{s}$ is independent of $T$. This condition, which is obeyed at
temperatures well below the ordering temperature of the magnetic material
constituting the particles, yields also temperature independent anisotropy constants
[this is apparent when the anisotropy is due to the magnetostatic self-energy, see Eq.\
(\ref{Kdem})]. The computed quantity will be the dimensionless
$\tilde{\chi}_{3}=\chi_{3}[K^{3}/(\mu_{0}^{3}M_{s}^{4})]$ and we shall employ a
logarithmic-normal distribution for $f(v)$, namely
\[
f(v)
=
\frac{1}{\sqrt{2\pi}\,\rho_{v} v}
\exp
\bigg\{
-\frac{\left[\ln(v/v_{{\rm m}})\right]^{2}}{2\rho_{v}^{2}}
\bigg\}
\;,
\]
where $v_{{\rm m}}$ is the {\em median\/} of the distribution and $\rho_{v}$ is the
standard deviation of $\ln(v)$.

\begin{figure}[t!]
\vspace{-3.ex}
\eps{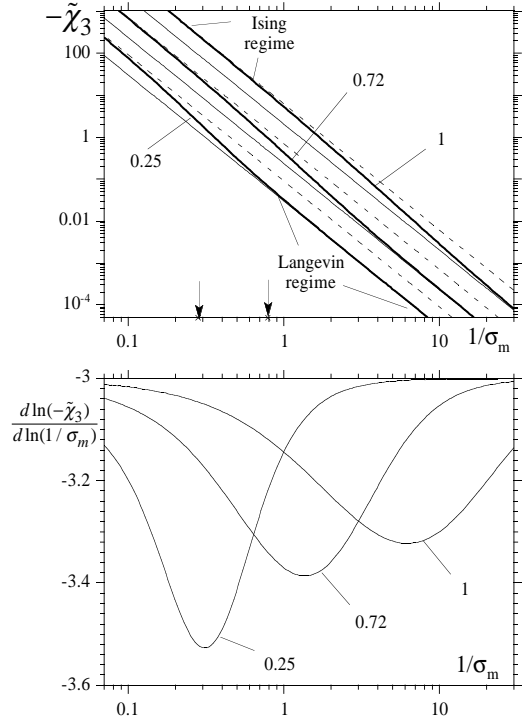}{0.7}
\vspace{-3.ex}
\caption[]
{
Upper panel: Log-log plot of $-\tilde{\chi}_{3}$ vs.\ $1/\s_{{\rm m}}$
($=\T/Kv_{{\rm m}}$) for a system with randomly distributed anisotropy
axes. The straight lines correspond to the isotropic (thin solid) and
Ising (dashed) non-linear susceptibilities. The numbers mark the width
$\rho_{v}$ of the volume distribution. The mean slope of the
$\rho_{v}=0.72$ curve between the arrows is compared with the
experiment of Bitoh et~al.\ (1993) in the text. Lower panel:
Logarithmic slopes.
\label{X3:plot}
}
\end{figure}

Figure \ref{X3:plot} displays $\chi_{3}$ and the corresponding Ising
and isotropic results vs.\ the temperature.
As the influence of the anisotropy decreases with increasing $T$,
$\chi_{3}$ undergoes a smooth crossover from the low-temperature Ising
regime to the high-temperature isotropic regime. For $\s_{{\rm
m}}\gg1$ ($\s_{{\rm m}}=Kv_{{\rm m}}/\T$) and $|\s_{{\rm m}}|\ll1$,
the logarithmic slope $\D\ln(-\chi_{3})/\D\ln(1/\s_{{\rm m}})$ tends
to $-3$, indicating the limit $T^{-3}$ dependences. However,
logarithmic slopes lesser than $-3$ emerge in the transitional regime,
where the departure of $\chi_{3}(T)$ from an inverse-temperature-cubed
law is sizable. As the width of the volume distribution increases, the
crossover region widens and shifts to higher temperatures. This is due
to the fact that the function $v^{3}f(v)$, which determines the
particles making the most substantial contribution to $\chi_{3}$,
broadens and moves to larger volumes, the $\chi_{3}$ of which is of
Ising type over a wider interval of the displayed temperature range.

\begin{figure}[t!]
\vspace{-3.ex}
\eps{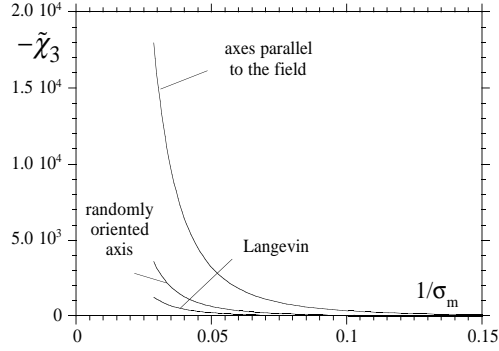}{0.7}
\vspace{-3.ex}
\caption[]
{
Effect of the alignment of the anisotropy axes towards $\B$ on the
temperature dependence of the non-linear susceptibility. The width of
the volume distribution is $\rho_{v}=0.25$.
\label{X3_orientation:plot}
}
\end{figure}

The rate of change of $\chi_{3}(T)$, moreover, increases as the
anisotropy axes are aligned towards $\B$ (see Fig.\
\ref{X3_orientation:plot}).
To illustrate, for a volume distribution with $\rho_{v}=0.25$, the
maximum logarithmic slope changes from $-3.53$ for anisotropy axes
distributed at random (see the lower panel of Fig.\ \ref{X3:plot}) to
$-3.98$ for axes collinear with the field. On the other hand, although
less dramatic, the discussed effects also occur for easy-plane
anisotropy ($K<0$), being then magnified as $\B$ points towards the
easy plane. Considering these significant deviations of $\chi_{3}(T)$
from a $T^{-3}$ dependence, arguments discarding superparamagnetism
based on this type of departure, such as those employed by Schiffer
et~al.\ (1995), should be carefully scrutinized.

On the other hand, when observed over limited temperature windows
(e.g., those imposed by the unavoidably finite measurement time), an
increase of the equilibrium $\chi_{3}(T)$ steeper than $T^{-3}$ could
resemble the high-temperature range of a quantity with a
low-temperature divergence. This might misleadingly suggest the
presence of appreciable inter-particle interactions in the ensemble
and, consequently, one could try a fit of the non-linear
susceptibility to, for example $\chi_{3}(T)\propto(T-T_{c})^{-3}$,
obtaining false ``critical" temperatures. If we do so with the
$\chi_{3}(T)$ {\em theoretically\/} computed for the most diluted
sample of Jonsson et~al.\ (1995) over $100$\,K $\leq T\leq 180$\,K, we
get the sizable value $T_{c}\simeq17.3$\,K (regression of the fit
$0.99992$). Note that, for $\omega/2\pi\sim1$--$10^{3}$\,Hz, effects
associated with the finite measurement time appear at
$T\lsim40$--$100$\,K, below of which one cannot measure the
equilibrium $\chi_{3}$.

\paragraph{Comparison with experimental data.}

Bitoh et~al.\ (1993; 1995) measured the non-linear dynamical susceptibility,
$\chi_{3}(\w,T)$, for  cobalt particles precipitated in a Cu$_{97}$Co$_{3}$ alloy. From
the equilibrium (high-temperature) part of the $\chi_{3}$ vs.\ $T$ curve they
obtained a mean logarithmic slope $-3.17$, whose departure from $-3$ was not
considered.

Their sample appears suitable to check the studied deviation of $\chi_{3}$ from a
$T^{-3}$ law, since:
\begin{enumerate}
\item
The high Curie temperature of the particles ($\simeq1400$\,K) yields $M_{s}$
feebly dependent on $T$ in the range of the experiment ($\leq280$\,K). 
\item
The equilibrium linear susceptibility can be fitted to a Curie law with a mean
logarithmic slope $\langle\D\ln\chi/\D\ln T\rangle=-1.01$, compatible with the
absence of dipole-dipole interaction effects and anisotropy axes distributed at random.
\end{enumerate}
On the other hand, one can still argue that, due to finite size effects,
the temperature dependence of the spontaneous magnetization of the Co particles
could be larger than that of the bulk material, so that the measured temperature
dependence of $\chi_{3}$ could be attributed to such phenomenon. However, the
ascription of the extra $T^{-0.17}$ factor in $\chi_{3}(T)$ to $M_{s}(T)^{4}$, entails the
occurrence of its square root in the Curie law  [$\chi\propto M_{s}(T)^{2}$], yielding a
total exponent $-(1+0.17/2)=-1.085$ for $\chi$, which is not consistent with the
measured one ($-1.01$).

Unfortunately, the high amplitude of the oscillating field employed in their experiment
($v_{{\rm m}}M_{s}\dBo/\kB\simeq17$\,K) might have induced non-linear
``saturation" effects on the measured susceptibilities at low temperatures, moving the
volume distribution $f(v)$ that they derived from the $\chi(\w,T)$ data, from the
actual one. Even so, we have specialized the above calculation of $\chi_{3}(T)$ to the
so-derived logarithmic-normal $f(v)$. The temperature range of their experiment, in
the dimensionless units $\T/Kv_{{\rm m}}$, is delimited in Fig.\ \ref{X3:plot} by the
arrows. Our calculation yields a mean logarithmic slope $-3.25$ that is within $2.5$\%
of the experimentally determined value $-3.17$. One must anyway conclude that the
sizable departure of the {\em theoretical\/} exponent from $-3$, makes mandatory the
inclusion of anisotropy effects on the temperature dependence of $\chi_{3}$ to
achieve a complete understanding of this kind of experiments.

\paragraph{Proposed experiments.}

In addition to search for deviations of $\chi_{3}(T)$ from a $T^{-3}$ law, the
dependence of $\chi_{3}$ on the angle between the anisotropy axis and the applied
field could be measured in systems with oriented anisotropy axes. (Molecular
magnetic clusters and textured frozen magnetic fluids are examples of systems with
parallel axes where such experiments could be performed.) In a polar plot (see Fig.\
\ref{X3:polar:plot}), $\chi_{3}(\alpha)$ will undergo an increasing deformation from a
circle at high temperatures (isotropic $\chi_{3}$) towards the characteristic two-looped
shape of the Ising regime ($\chi_{3}|_{\ising}\propto\coscal$) as $T$ decreases.

Other possible experiment could be to measure $\chi_{3}(T)$ in a magnetic fluid
through the freezing point of the solvent, $T_{{\rm f}}$. Recall that, due to the physical
rotation of the particles in the fluid state, the magnetization is given by the Langevin
law for each particle, irrespective of the anisotropy energy (Krueger, 1979). On the
other hand, at temperatures below the freezing point, the anisotropy axes become
immobilized; the magnetic anisotropy then takes reflection in the equilibrium
quantities and $\chi_{3}(T)$ would undergo a discontinuous change at $T_{{\rm f}}$.%
\footnote{
This jump could be smeared out around $T_{{\rm f}}$ due to effects related with the
immediacy of the critical point of the carrier. }
In contrast, if at $T_{{\rm f}}$
the anisotropy axes become immobilized in a random pattern, the linear equilibrium
susceptibility would be continuous there (recall that $\avX$ does not depend on the
anisotropy energy in a solid dispersion).

The relative size of the discontinuity in the non-linear susceptibility
$\Delta\chi_{3}/\chi_{3}$ at the freezing temperature is determined by the value of
$T_{{\rm f}}$ in magnetic-anisotropy units, so that the size of the jump also depends
on the anisotropy constants and the actual volume distribution. We have computed
$\Delta\chi_{3}/\chi_{3}$ with the parameters of two magnetic fluids in the literature.
First, for most diluted sample of Luo et~al.\ (1991), $\Delta\chi_{3}/\chi_{3}$ would
be small, because the freezing point of the carrier liquid is close to the isotropic
regime. On the other hand, for the most diluted sample of Jonsson et~al.\ (1995),
$\Delta\chi_{3}/\chi_{3}$ would be about $90$\%. Once more, if the anisotropy axes
are frozen collinear with $\B$, this effect will be even more dramatic.%
\footnote{
However, for oriented anisotropy axes, $\chi(T)$ would also exhibit a
discontinuity at the freezing point of the magnetic fluid.
}

%% file: garcms04.tex
\section{Dynamical properties: heuristic approach}
\label{sect:heuristic}

\subsection{Introduction}

In this Section we shall briefly consider a heuristic approach to the
dynamics of classical magnetic moments in anisotropy potentials. We
shall focus on the linear dynamical response, i.e., the response of
the system to a small-amplitude, oscillating or constant, magnetic
field. The responses to both types of stimulus are related in a simple
way, so that we shall merely employ the language of the linear
dynamical response in the frequency domain---the linear dynamical
susceptibility $\chi(\w)$.  This quantity, in addition to supplying
valuable information about the intrinsic dynamics of the spins, is of
relevance for general studies on magnetic nanoparticle systems. For
instance, under certain conditions $\chi(\w)$ can be used to
approximately determine the distribution of energy barriers
(essentially particle volumes), occurring in assemblies of
non-interacting magnetic nanoparticles (Shliomis and Stepanov,
1994). Besides, a rough estimate of the pre-exponential factor of the
longitudinal relaxation time in the Arrhenius regime can also be
derived from the $\chi(\w)$ data.

The organization of this Section is as follows. In Subsec.\
\ref{subsect:models} various heuristic expressions that have been
proposed to describe the linear dynamical response are discussed (they
will be compared with exact numerical results in Section
\ref{sect:stochastic}). In Subsec.\ \ref{analysis:ShS}, the most
general of those expressions will be analyzed in detail, illustrating
how it can be used to get the energy-barrier distribution of magnetic
nanoparticle ensembles. Finally, in Subsec.\ \ref{experiments} some of
the previous results will be illustrated with experiments performed on
a frozen magnetic fluid of maghemite ($\gamma$--Fe$_{2}$O$_{3}$)
nanoparticles.  Part of the results of this Section were presented by
Svedlindh, Jonsson and Garc{\'{\i}}a-Palacios (1997).

\subsection{Heuristic treatment of the linear dynamical response}
\label{subsect:models}

Let us commence by considering the expression (\ref{X:4}) for the linear {\em
equilibrium\/} susceptibility in terms of its longitudinal and transverse contributions,
namely
\begin{equation}
\label{X:4:ensemble}
\chi
=
\chi_{\|}\cosqal+\chi_{\perp}\senqal
\;,
\end{equation}
where $\alpha$ is the angle between the anisotropy axis and the probing field. The
term $\chi_{\|}\cosqal$ is proportional to the projection along the probing field
direction of the response of the magnetic moment to the longitudinal component (with
respect to the anisotropy axis) of the field. Likewise, $\chi_{\perp}\senqal$ is
proportional to the projection onto the probing field of the response of the spin to the
transverse component of the field. As we know from Subsec.\
\ref{subsect:X}, averaging this equation with $\chi_{\|}$ and $\chi_{\perp}$ from Eq.\
(\ref{X:para:perp}), one gets $\langle\chi\rangle_{\ran}=\mu_{0}\mm^{2}/3\T$.
Consequently, in a non-interacting magnetic nanoparticle ensemble with anisotropy
axes distributed at random, the linear {\em equilibrium\/} susceptibility {\em in the
absence of an external bias field\/} is independent of the magnetic anisotropy ($\chi$
is then identical with that derived in a na\"{\i}ve superparamagnetic model where the
anisotropy is neglected). The main effect of the anisotropy is to introduce energy
barriers that the spins need to overcome before equilibrium is reached, implying that
the ensemble could, depending on the measurement time, display {\em magnetic
relaxation}.

The relaxational mechanism consists of an orientational redistribution of the magnetic
moments according to the conditions set by the magnetic anisotropy, temperature, and
external field. The relaxation can be envisaged as a two-stage process: first, the dipoles
redistribute inside the potential wells, with a characteristic time $\ttr$ related with
the inverse of the precession frequency of the magnetic moments in the anisotropy
field ($\sim10^{-10}$--$10^{-12}$\,s); then, the equilibration between the potential
wells, which is a thermally activated process, proceeds. This second mechanism can
result in exceedingly slow magnetic relaxation since its characteristic time $\tlo$,
which essentially follows an Arrhenius law [see Eq.\ (\ref{arrhenius:tau:0})], ranges
from picoseconds to geological time scales depending on the magnetic anisotropy,
temperature, and external field.

A rigorous theoretical derivation of the linear {\it dynamical\/} susceptibility of
classical magnetic moments in anisotropy potentials, as well as other dynamical
quantities, is hindered by a number of mathematical difficulties (see Section
\ref{sect:stochastic}). Thus, in order to describe the linear dynamical response of
non-interacting magnetic nanoparticles, various simple expressions  have been
proposed in the literature. We shall mainly consider the expression suggested, on the
basis of the two-stage relaxation process mentioned, by Shliomis and Stepanov (1993)
to describe $\chi(\w)$ at frequencies below the ferromagnetic-resonance frequency
range. Besides, we shall show that this model contains as particular cases some models
previously proposed.

\subsubsection*{Shliomis and Stepanov model}

In a study of magnetic fluids these authors suggested that $\chi(\w)$ could be
described as a sum of two independent Debye-type relaxation mechanisms: one for the
response to the longitudinal component of the probing field and the other for the
response to the transverse component (see also Ra{\u{\i}}kher and Stepanov, 1997).
The expression proposed can be generalized in order to describe the effect of a
longitudinal {\em bias\/} field by merely writing
\begin{equation}
\label{shschi:bias}
\chi_{{\rm ShS}}
=
\frac{\chi_{\|}(T,B)}{1+i\w\tlo}\cosqal
+\frac{\chi_{\perp}(T,B)}{1+i\w\ttr}\senqal
\;,
\end{equation}
where $\chi_{\|}$ and $\chi_{\perp}$ are the exact equilibrium susceptibilities
(\ref{X:para:perp:bias}).

Various expressions can be used for the characteristic times appearing in the above
formula (see Subsec.\ \ref{subsect:brown}). However, for the purposes of this Section
it is sufficient to consider that in the high-barrier range $\tlo$ can be written in the
Arrhenius form $\tlo=\tau_{0}\exp(\dU/\T)$, where $\tau_{0}$ is assumed to be a
constant $\sim10^{-10}$--$10^{-12}$\,s (that is, we disregard the dependences of the
pre-exponential factor on the temperature, external field, and the parameters of the
particles in comparison with the dependences of the exponential term). Concerning the
transverse relaxation time, for not very high frequencies (say,
$\omega\lsim10^{6}$\,Hz), the condition $\omega\ttr\ll1$ holds (Subsec.\
\ref{subsect:brown}). One can then approximate $1/(1+i\omega\ttr)$ by unity in Eq.\
(\ref{shschi:bias}), to get the {\em low-frequency\/} equation
\begin{equation}
\label{shschi:lowfrec}
\left.
\chi_{\rm ShS}
\right|_{\w\ttr\ll1}
\simeq\frac{\chi_{\|}}{1+i\wtlo}\cosqal
+\chi_{\perp}\senqal
\;.
\end{equation}
The approximation used is equivalent to assume from the outset that the response to
the transverse components of the probing field is instantaneous. In fact, very short
measurement times, such as those obtained in neutron scattering or ferromagnetic
resonance experiments, are required to probe the intra-potential-well dynamics (see
Table \ref{measurement_times:table}).

From now on Eq.\ (\ref{shschi:bias}) with the {\em exact\/} equilibrium susceptibilities,
will be referred to as the Shliomis and Stepanov equation. Further, the formula
obtained when in the low-frequency Eq.\ (\ref{shschi:lowfrec}) one uses the {\em high-barrier approximations\/} (\ref{X:para:bias:approx}) and
(\ref{X:perp:bias:approx}) of the equilibrium susceptibilities, will be called the
Gittleman, Abeles, and Bozowski (1974) equation, since it properly generalizes their
formula to $B\neq0$ and an arbitrary anisotropy-axis orientation. Indeed, on
introducing Eqs.\ (\ref{X:para:bias:approx}) and (\ref{X:perp:bias:approx}) evaluated at
$B=0$ [that is, Eqs.\ (\ref{X:para:perp:approx})] into Eq.\ (\ref{shschi:lowfrec}), one
first gets
\begin{equation}
\label{gabchi}
\chi_{{\rm GAB}}
\simeq
\bigg[
\Xo\cosqal
+\frac{\mu_{0}\mm^{2}}{Kv}\big(\threehalfs\senqal-1\big)
+i\w\tlo\frac{\mu_{0}\mm^{2}}{2Kv}\senqal
\bigg]
\frac{1}{1+i\w\tlo}
\;,
\end{equation}
which, when averaged over an ensemble with randomly distributed anisotropy axes
(the second term in the square brackets then vanishes), reduces to the equation
proposed in by the authors mentioned. Finally, the expression obtained when one
introduces the Ising-type Eqs.\ (\ref{X:para:perp:bias:ising}) into Eq.\
(\ref{shschi:bias}) [or Eq.\ (\ref{shschi:lowfrec})], namely
\begin{equation}
\label{isingchi:bias}
\chi_{{\rm Ising}}
=
\Xo\frac{1}{\cosh^{2}\xi}
\frac{\cosqal}{1+i\w\tlo}
\;,
\end{equation}
is called the discrete-orientation or Ising dynamical susceptibility.

\subsection{Analysis of the low-frequency Shliomis and Stepanov model}
\label{analysis:ShS}

We shall now analyze the low-frequency Eq.\ (\ref{shschi:lowfrec}) for an ensemble of
non-interacting magnetic nanoparticles where there exists a distribution in particle
parameters.

If the distribution occurs mainly in one of the parameters, say, the volumes of the
particles, and one assumes that the contribution of each particle to the linear
susceptibility is given by an expression like the low-frequency (\ref{shschi:lowfrec}),
one can write the linear susceptibility of the ensemble as
\begin{equation}
\label{shschi:int}
\chi(\w,T)
=
\frac{\Mss}{\T}
\frac{1}{K}
\int_{0}^{\infty}\!\!\D{E}\, f(E)E
\left[
\frac{\F'}{\F}\frac{\avcosqal}{1+i\wtlo}
+\frac{\F-\F'}{2\F}\avsenqal
\right]
\;.
\end{equation}
In this equation the functions $\F^{(\ell)}$ are evaluated at $\s=E/\T$, $E=Kv$ (with
$K$ assumed equal for all particles) and $f(E)\D E$ is the fraction of the total
``magnetic" volume occupied by those particles with energy barriers in the interval
$(E,E+\D E)$. Note that the square of the magnetic moment has been written in terms
of the spontaneous magnetization $M_{s}$ as $\mm^{2}=M_{s}^{2}v^{2}$ and, since we
are using the ``occupied volume" representation of the distribution, one $v$ is already
incorporated into $f(E)$.

In the above formula, the orientational averages are taken with respect to the
particles in $(E,E+\D E)$ and could, in principle, depend on $E$. We shall not study this
situation but merely consider that $\avcosqal$ and $\avsenqal$ are the same for each
energy interval. One could also consider the cases where, due to finite size effects,
$M_{s}$ and $K$ depend on $v$. Although this could be incorporated in the following
considerations, we shall not take those dependences into account explicitly.

\subsubsection{The out-of-phase linear dynamical susceptibility and the
energy-barrier distribution}

The out-of-phase component (imaginary part) of Eq.\ (\ref{shschi:int}) reads
\begin{equation}
\label{X:out}
\chi''(\w)
=
\frac{\Mss\avcosqal}{\T}
\frac{1}{K}
\int_{0}^{\infty}\!\!\D{E}\, f(E)E\,\frac{\F'}{\F}\frac{\wtlo}{1+(\wtlo)^{2}}
\;,
\end{equation}
to which the response to the transverse components of the probing field (with respect
to the different anisotropy axes) does not contribute due to the low-frequency
assumption ($\omega\lsim10^{6}$\,Hz).

\begin{figure}[t!]
\vspace{-3.ex}
\eps{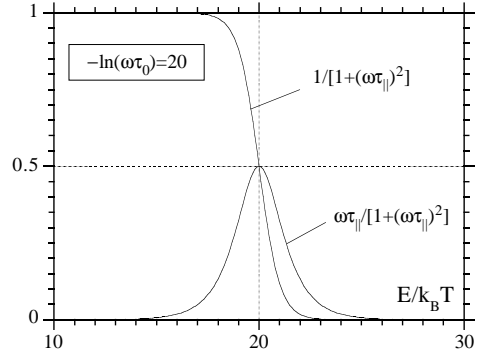}{0.7}
\vspace{-3.ex}
\caption[]
{
Real and imaginary parts of the Debye factor $1/(1+i\w\tlo)$ vs.\ the energy-barrier
height $E/\T$. The relaxation time is given by the Arrhenius law
$\tlo=\tau_{0}\exp(E/\T)$.
\label{debye-factor:plot}
}
\end{figure}

The term $\wtlo/[1+(\wtlo)^{2}]$ in the integrand of Eq.\
(\ref{X:out}), has a maximum at the energy barrier, $\Eb$, for which
$\wtlo=1$ (see Fig.\ \ref{debye-factor:plot}).
On assuming a simple Arrhenius form for the relaxation time,
$\tlo=\tau_{0}\exp(E/\T)$, one finds $\Eb=-\T\lnwto$, which explicitly
depends on the temperature and the frequency. Besides, due to the {\em
exponential\/} dependence assumed for $\tlo$, it follows from the
definition of $\Eb$ that: (i) $\tlo(E)\ll\tlo(\Eb)=1/\w$, if $E<\Eb$,
whereas (ii) $\tlo(E)\gg\tlo(\Eb)=1/\w$, if $E>\Eb$. In virtue of
these properties, and considering that $1/\omega$ is the {\em
measurement time\/} in a dynamical experiment, $\Eb$ is called the
{\em blocking barrier\/} (recall the considerations in Section
\ref{sect:introduction}). Similarly, one can define the corresponding {\em
dimensionless blocking barrier\/} $\sb=\Eb/\T$, whence
\begin{equation}
\label{blocking}
\Eb
=
-\T\lnwto
\;,
\qquad
\sb
=
-\lnwto
\;.
\end{equation}
For these two quantities one has, by definition, $\w\tlo=1$.

We shall not consider the finite height and width of the function
$\wtlo/[1+(\wtlo)^{2}]$, but we shall take this function as a (unnormalized) Dirac delta
centered at $\sb$. This replacement works when the remainder terms in the
integrand of the formula for $\chi''(\w)$ (e.g., the energy-barrier distribution) change
slowly enough in the interval about $\sb$ where $\wtlo/[1+(\wtlo)^{2}]$ differs
appreciably from zero. Concerning the term $\F'/\F$, when
$\sb\gsim15$--$25$, its changes are not very large, because
\[
\frac{\D {}}{\D\s}\bigg(\frac{\F'}{\F}\bigg)
=
\frac{\F''}{\F}-\bigg(\frac{\F'}{\F}\bigg)^{2}
\stackrel{{\rm Eq.\ (\ref{comb:F:leadingterm})}}\simeq\frac{1}{\s^{2}}
\sim\frac{1}{400}
\;,
\quad
\mbox{for}\quad
\s\sim20
\;.
\]
Values of $\sb\gsim15$--$25$ are typical for probing fields with $\w\lsim10^{3}$\,Hz.

Under the conditions mentioned, $\wtlo/[1+(\wtlo)^{2}]$ plays the r\^{o}le of a function
proportional to a Dirac delta. In order to calculate the proportionality factor, one
integrates that function over the entire energy range by means of the substitution
$\D\tlo=(\tlo/\T)\D E$
\[
\int_{0}^{\infty}\!\!\D{E}\,
\frac{\wtlo}{1+(\wtlo)^{2}}
=
\T
\int_{\tau_{0}}^{\infty}\!\!\D{\tlo}\,
\frac{\omega}{1+(\wtlo)^{2}}
\simeq
\hfpi\T
\;,
\]
where, on considering the low-frequency assumption ($\omega\lsim10^{6}$\,Hz) and
taking the tiny value of $\tau_{0}$ ($\sim10^{-10}$--$10^{-12}$\,s) into account, we
have used the approximation $\arctan(\wto)\lsim\arctan(10^{-4}$--$10^{-6})\simeq0$.
Therefore, when integrating functions slowly varying about $\Eb=-\T\lnwto$, one can
make use of the approximation
\begin{equation}
\label{delta:approx}
\frac{\wtlo}{1+(\wtlo)^{2}}\simeq\hfpi\T\,\delta(E-\Eb)
\;.
\end{equation}

Thus, on calculating the integral in Eq.\ (\ref{X:out}) by means of Eq.\
(\ref{delta:approx}), one obtains (cf.\ Eq.\ (41) by Shliomis and Stepanov, 1994)
\begin{equation}
\label{X:out:approx}
\chi''(\w,T)
=
\hfpi\frac{\Mss}{K}\avcosqal
\frac{\F'(\sb)}{\F(\sb)}f(\Eb)\Eb
\;,
\end{equation}
which directly relates the energy-barrier distribution and the out-of-phase linear
dynamical susceptibility. Note that, even if we consider the weak temperature
dependence of $\tau_{0}$, this is further weakened when occurring inside the
logarithm $\lnwto$, so that $\sb=-\lnwto$ and thus the factor $\F'(\sb)/\F(\sb)$, are
almost independent of $T$. Then, because $\F'(\sb)/\F(\sb)$ is also weakly dependent
on $\w$, Eq.\ (\ref{X:out:approx}) shows that, approximately, all the dependence of
$\chi''$ on $T$ and $\w$ enters via the combination $\Eb=-\T\lnwto$.%
\footnote{
We are also implicitly assuming that $\D(M_{s}^{2}/K)/\D T\simeq0$. For
instance, for the ``shape" anisotropy of ellipsoids of revolution, $M_{s}^{2}/K$ is in fact
a geometric term [see Eq.\ (\ref{Kdem})].
}
Therefore, if we plot $\chi''$ vs.\
$-\T\lnwto$, all the $\chi''(T)$ curves corresponding to different frequencies collapse
onto a single ``master" curve [proportional to $f(E)E$ and with maximum at $\EM$].
Conversely, by fitting the frequency-dependent temperature of the maximum of
$\chi''(T)$, denoted by $\TM(\w)$, to the ``Arrhenius law" $\EM=-\kB\TM(\w)\lnwto$,
one can get $\EM$ and $\tau_{0}$.

Note however that the parameter $\EM$, which is sometimes called ``average  energy
barrier", is merely the maximum of the function $f(E)E$. Therefore, it is not
necessarily related with a characteristic parameter of the energy-barrier distribution
(incidentally, for the gamma and logarithmic-normal distributions $\EM$ is equal to
the {\em mean\/} and the {\em median\/} of the distribution, respectively).

\subsubsection{The in-phase linear dynamical susceptibility}

The in-phase component (real part) of Eq.\ (\ref{shschi:int}) is given by
\begin{equation}
\label{X:in}
\chi'
=
\frac{\Mss}{\T}
\frac{1}{K}
\int_{0}^{\infty}\!\!\D{E}\, f(E)E
\left[
\frac{\F'}{\F}\frac{\avcosqal}{1+(\wtlo)^{2}}
+\frac{\F-\F'}{2\F}\avsenqal
\right]
\;,
\end{equation}
where, because of the low-frequency assumption ($\omega\lsim10^{6}$\,Hz), the
response to the transverse components of the probing field contribute to $\chi'(\w)$
with its thermal-equilibrium value.

The term $1/[1+(\wtlo)^{2}]$ as a function of $\s=E/\T$ has the form of a smooth step
about $\sb$, whose width is of the order of the width of the peak of
$\wtlo/[1+(\wtlo)^{2}]$ (see Fig.\ \ref{debye-factor:plot}). However, when that term is
under the integral sign and multiplied by functions that vary slowly around $\sb$, we
can approximate
$1/[1+(\wtlo)^{2}]$ by a step function, namely
\begin{equation}
\label{step:approx}
\frac{1}{1+(\wtlo)^{2}}\simeq
\left\{
\begin{array}{ll} 1
&
\mbox{~for~} E<\Eb
\\
0
&
\mbox{~for~} E>\Eb
\end{array}
\right.
\;.
\end{equation}

Thus, on introducing Eq.\ (\ref{step:approx}) into Eq.\ (\ref{X:in}) and rearranging the
integration limits, one gets the approximate result
\begin{eqnarray}
\label{X:in:approx}
\chi'(\w)
&
=
&
\frac{\Mss}{\T}\frac{1}{K}
\int_{0}^{\Eb}\!\!\D{E}\, f(E)E
\left[
\frac{\F'}{\F}\avcosqal
+\frac{\F-\F'}{2\F}\avsenqal
\right]
\nonumber
\\
&
& {}+\frac{\Mss}{\T}\frac{1}{K}
\int_{\Eb}^{\infty}\!\!\D{E}\, f(E)E\,
\frac{\F-\F'}{2\F}\avsenqal
\;,
\end{eqnarray}
which can be interpreted as follows. Note first that only the particles with $E<\Eb$, i.e.,
those obeying $\tlo(E)\ll\tlo(\Eb)=1/\w$, contribute to the first term. However,
$1/\omega$ is the measurement time in a dynamical experiment, so that those
particles are the superparamagnetic particles ($\tlo\ll1/\omega$), and the first term
is indeed their contribution to the linear {\em equilibrium\/} susceptibility. On the
other hand, the particles with $E>\Eb$, which are those contributing to the second
term, satisfy $\tlo(E)\gg\tlo(\Eb)=1/\w$, so that the over-barrier rotation process is
not effective for them. These are the {\em blocked\/} particles, and contribute to
$\chi'(\w)$ via  the fast rotations of their magnetic moments {\em inside\/} the
potential wells towards the transverse components of the field. In fact, the second
term in Eq.\ (\ref{X:in:approx}) is $\avsenqal$ times the equilibrium transverse
susceptibility of the blocked particles.

We finally note that, since $\Eb=-\T\lnwto$, the second term in Eq.\
(\ref{X:in:approx}) is small in comparison with the first one at sufficiently high
temperatures, so that $\chi'$ is then approximately equal to the equilibrium
susceptibility. In addition, for anisotropy axes distributed at random we can write
\begin{equation}
\label{X:curie}
\left.
\avXr
\right|_{{\rm high~} T}
=
\frac{\Mss}{3\T}\frac{1}{K}
\int_{0}^{\infty}\!\!\D{E}\, f(E)E
\equiv
\frac{C}{T}
\;,
\end{equation}
where $C$ is the Curie constant.

\subsubsection{The $\pi/2$-law}

We shall now explicitly derive, starting from the low-frequency Shliomis and Stepanov
equation (\ref{shschi:int}), a celebrated relation between $\partial\chi'/\partial\ln\w$
and $\chi''$ known as the {\em $\pi/2$-law}.

First, on rearranging the integration limits in Eq.\ (\ref{X:in:approx}),  we can write
$\chi'(\w)$ as
\begin{eqnarray}
\label{X:in:approx:2}
\chi'(\w)
&
=
&
\frac{\Mss\avcosqal}{\T}\frac{1}{K}
\int_{0}^{\Eb}\!\!\D{E}\, f(E)E\frac{\F'}{\F}
\nonumber
\\
&
& {}+
\frac{\Mss\avsenqal}{\T}\frac{1}{K}
\int_{0}^{\infty}\!\!\D{E}\, f(E)E\,\frac{\F-\F'}{2\F}
\;,
\end{eqnarray}
where the last term, which is $\avsenqal$ times the transverse equilibrium
susceptibility of the {\em whole\/} ensemble, does not depend on $\w$. Then, on using
$\partial\Eb/\partial\ln\w=-\T$ and the {\em Leibniz formula}
\begin{equation}
\label{leibnitz_formula}
\frac{\D {}}{\D x}
\int_{g(x)}^{h(x)}\!\!\D{t}\, F(x,t)
=
\left\{ F[x,h(x)]h'(x)-F[x,g(x)]g'(x)
\right\}
+\int_{g(x)}^{h(x)}\!\!\D{t}\,
\frac{\partial {}}{\partial x} F(x,t)
\;.
\end{equation}
one gets
\begin{equation}
\label{dXdlnw}
\frac{\partial\chi'}{\partial\ln\w}
=
-\frac{\Mss}{K}\avcosqal\frac{\F'(\sb)}{\F(\sb)}f(\Eb)\Eb
\;.
\end{equation}
Finally, on comparing this equation with Eq.\ (\ref{X:out:approx}), we get the desired
relation between $\partial\chi'/\partial\ln\w$ and $\chi''$, namely
\begin{equation}
\label{pihalf:law}
\chi''
=
-\hfpi\frac{\partial\chi'}{\partial\ln\w}
\;.
\end{equation}

For systems with a sufficiently wide distribution of relaxation times, the $\pi/2$-law
is in fact a quite general result and independent of the dynamical model used, since it
can then be derived from the Kramers--Kronig relations. These relations are merely
based on general principles as the {\em linearity of the response}, and {\em
causality\/} (i.e., the response at time $t$ only depends on the values of the stimulus
at times $t'<t$). For the sake of completeness, we shall repeat here one such derivation
of the $\pi/2$-law by B{\"{o}}ttcher and Bordewijk (1978, p.~58).

On writing one of the Kramers--Kronig relations in the form
\[
\chi'(\w)
=
\chi_{S}
+\frac{2}{\pi}
\int_{0}^{\infty}\!\!\D{\tilde{\w}}\,
\frac{\tilde{\w}\chi''(\tilde{\w})}{\tilde{\w}^{2}-\w^{2}}
=
\chi_{S}
+\frac{2}{\pi}
\int_{-\infty}^{\infty}\!\!\D{(\ln\tilde{\w})}\,
\frac{\tilde{\w}^{2}\chi''(\tilde{\w})}{\tilde{\w}^{2}-\w^{2}}
\;,
\]
where $\chi_{S}$ is the adiabatic ($\w\to\infty$) susceptibility ($\chi_{\perp}$ in our
case), and approximating in the last integral the factor
$\tilde{\w}^{2}/(\tilde{\w}^{2}-\w^{2})$ by a unit step function (with step at
$\w$), one obtains
\[
\chi'(\w)\simeq
\chi_{S}
+\frac{2}{\pi}
\int_{\ln\w}^{\infty}\!\!\D{(\ln\tilde{\w})}\,
\chi''(\tilde{\w})
\;.
\]
Then, on differentiating this equation with respect to $\ln\w$ by means of the
Leibniz formula (\ref{leibnitz_formula}) one finally gets the $\pi/2$-law.

The assumption of broad relaxation-time spectrum enters implicitly when
approximating the factor $\tilde{\w}^{2}/(\tilde{\w}^{2}-\w^{2})$ by a step function:
the broad spectrum entails flat curves for $\chi''(\w)$, so that the replacement
mentioned  does not introduce a significant error. This approximation is equivalent to
the assumptions made above concerning the change of the functions appearing in the
integrand of the equations for $\chi(\w)$, in the range where the Debye factor has its
maximum variation.

\subsubsection{$\partial(T\chi')/\partial T$ and its relation with $\chi''$ and the
energy-barrier distribution}

Wohlfarth (1979), when studying spin glasses in the context of the superparamagnetic
cluster model, proposed a method to obtain the energy-barrier distribution from the
derivative $\partial(T\chi')/\partial T$. He considered a distribution of ``blocking
temperatures," which in our notation are $\Tb=\Eb/\kB$ (``blocking energies" in
temperature units), and disregarded the contribution of the blocked clusters, and
wrote
\begin{equation}
\label{Wohlfarth:chi}
\chi(T)
\simeq
\frac{C}{T}
\int_{0}^{T}\!\!\D{\Tb}\, f(\Tb)
\;.
\end{equation}
Here $C$ is the Curie constant, and the susceptibility is the non-equilibrium
susceptibility obtained in a dc experiment with a typical measurement time
$\sim100$\,s. Then, by means of the inversion procedure [see Eq.\
(\ref{leibnitz_formula})]
\begin{equation}
\label{Wohlfarth:inversion} f(T)
=
\frac{1}{C}\frac{\partial (T\chi)}{\partial T}
\;,
\end{equation}
he expressed the distribution of blocking temperatures in terms of the linear
susceptibility.

Note that Eq.\ (\ref{Wohlfarth:chi}) can be considered as the particular case of Eq.\
(\ref{X:in:approx}) where the anisotropy axes are distributed at random (the term in
the square brackets in the first integral then equals $1/3$) and the second integral
(the $\chi_{\perp}$ of the blocked clusters) is neglected (Ising-type case). Besides, in
order to establish this correspondence we must assume that his $f(\Tb)$ incorporates
the extra energy factor, i.e., that $f(\Tb)\propto Ef(E)$.

Lundgren, Svedlindh and Beckman (1981) derived a relation between $\chi''$ and
$\partial(T\chi')/\partial T$ for the following model
\begin{equation}
\label{chi:LSB}
\chi(\w)
=
\int_{\ln\tau_{\Min}}^{\ln\tau_{\Max}}\!\!\D{(\ln\tau)}\,
g(\tau)\chi(\tau)\frac{1}{1+i\w\tau}
\;,
\end{equation}
where $\chi(\tau)$ is the equilibrium susceptibility and $g(\tau)$ the distribution of
relaxation times. They assumed $\chi(\tau)\propto1/T$ and an Arrhenius dependence
for $\tau$, getting
\begin{equation}
\label{LSB:relation}
\chi''
=
-\hfpi\frac{1}{\lnwto}\frac{\partial (T\chi')}{\partial T}
\;.
\end{equation}
Because in the model (\ref{chi:LSB}), $\chi''$ is also directly related with the
distribution of relaxation times, the above relation yields an inversion procedure
analogous to that of Wohlfarth.

We shall now calculate $\partial(T\chi')/\partial T$ for the low-frequency
(\ref{shschi:int}). In this way, we shall take into account the effect of the finite width
and depth of the anisotropy potential wells.

Let us begin by taking the $T$-derivative of $T\chi'$, with $\chi'$ given by Eq.\
(\ref{X:in:approx}) [or Eq.\ (\ref{X:in:approx:2})]. Since the integrals in those equations
also depend on $T$ via the integration limits, the required $T$-derivative can be
taken by dint of the Leibniz formula (\ref{leibnitz_formula}). On doing so, we get after
the rearrangement of the integration limits,
\begin{eqnarray}
\label{X:in:der:approx:2}
\frac{\partial (T\chi')}{\partial T}
&
=
&
-\lnwto
\frac{\Mss}{K}\avcosqal
\frac{\F'(\sb)}{\F(\sb)}f(\Eb)\Eb
\nonumber
\\
&
& {}+
\frac{\Mss}{2K}\avsenqal
\int_{\Eb}^{\infty}\!\!\D{E}\, f(E)
\bigg[
\frac{\F''}{\F}
-\bigg(\frac{\F'}{\F}\bigg)^{2}
\bigg]
\s^{2}
\nonumber
\\
&
& {}+
\frac{\Mss}{2K}
\left[ 3\avcosqal-1
\right]
\int_{0}^{\Eb}\!\!\D{E}\, f(E)
\bigg[
\frac{\F''}{\F}
-\bigg(\frac{\F'}{\F}\bigg)^{2}
\bigg]
\s^{2}
\;,
\nonumber\\
\end{eqnarray}
where we have assumed that neither $M_{s}$ nor $K$ depend on $T$, and used
$\partial\Eb/\partial T=-\kB\lnwto$ as well as $(E/\kB)\partial\s/\partial T=-\s^{2}$.

Note that the first line on the right-hand side of Eq.\ (\ref{X:in:der:approx:2}) is
directly related with the energy-barrier distribution. If the remainder terms were
absent, this equation would give the inversion procedure of Wohlfarth
(\ref{Wohlfarth:inversion}). However, since the last two lines contain information
about $f(E)$ in integral form, we see that the quantity $\partial(T\chi')/\partial T$
does not directly scan the energy-barrier distribution. Note in this connection that,
unlike $\chi''$ (or $\partial\chi'/\partial\ln\w$) the quantity
$[1/\lnwto]\partial(T\chi')/\partial T$ does not properly scale when represented
against $-\T\lnwto$ due to the presence of the mentioned integral terms.

Next, on taking Eq.\ (\ref{X:out:approx}) into account we get the following relation
between $\chi''$ and $\partial(T\chi')/\partial T$
\begin{eqnarray*}
\chi''
=
-\hfpi\frac{1}{\lnwto}
&
\Bigg\{
&
\frac{\partial (T\chi')}{\partial T}
-\frac{\Mss}{2K}\avsenqal
\int_{\Eb}^{\infty}\!\!\D{E}\, f(E)
\bigg[
\frac{\F''}{\F}
-\bigg(\frac{\F'}{\F}\bigg)^{2}
\bigg]
\s^{2}
\\
&
& {}-\frac{\Mss}{2K}
\left[ 3\avcosqal-1
\right]
\int_{0}^{\Eb}\!\!\D{E}\, f(E)
\bigg[
\frac{\F''}{\F}
-\bigg(\frac{\F'}{\F}\bigg)^{2}
\bigg]
\s^{2}
\Bigg\}
\;,
\end{eqnarray*}
which is the counterpart of Eq.\ (\ref{LSB:relation}) in the low-frequency Shliomis and
Stepanov model. Furthermore, since the angular factor in the last term on the
right-hand side vanishes for anisotropy axes distributed at random, the above relation
simplifies in that case to
\begin{equation}
\label{X:in:der:ran:1}
\avXi
=
-\hfpi\frac{1}{\lnwto}
\Bigg\{
\frac{\partial(T\avXr)}{\partial T}
-\frac{\Mss}{3K}
\int_{\Eb}^{\infty}\!\!\D{E}\, f(E)
\bigg[\frac{\F''}{\F}
-\bigg(\frac{\F'}{\F}\bigg)^{2}
\bigg]
\s^{2}
\Bigg\}
\;,
\end{equation}
Finally, since $\s>\sb\sim20$--$25$, if $E>\Eb$, one can replace $\F''/\F-(\F'/\F)^{2}$
in the above integral by its high-barrier approximation (\ref{comb:F:leadingterm}), to
get
\begin{equation}
\label{X:in:der:ran:2}
\avXi
=
-\hfpi\frac{1}{\lnwto}
\bigg\{
\frac{\partial{}}{\partial T}(T\avXr)
-\frac{\Mss}{3K}
\int_{\Eb}^{\infty}\!\!\D{E}\, f(E)
\bigg\}
\;.
\end{equation}
This is an interesting result: in spite of the differences between $\chi''$ and
$\partial(T\chi')/\partial T$ being reduced upon averaging for anisotropy axes
distributed at random, some of them remain. These differences, and accordingly those
of $\partial(T\chi')/\partial T$ with respect to the energy-barrier distribution, are
again due to the presence of the second term on the right-hand side, which contains
information about $f(E)$ in integral form. In addition, the lower the temperature, the
larger the differences mentioned, because the lower integration limit in Eq.\
(\ref{X:in:der:ran:2}) decreases with $T$ (recall that $\Eb\propto\T$).

Note finally that, by using the high-barrier formula
$\chi_{\perp}\simeq\mu_{0}\mm^{2}/2Kv$ per particle [Eqs.\
(\ref{X:para:perp:approx})], the integral in Eq.\ (\ref{X:in:der:ran:2}) can alternatively
be written in terms of the approximate transverse susceptibility of the blocked
particles (at the temperature and frequency considered), namely
\begin{equation}
\label{X:in:der:ran:3}
\avXi
=
-\hfpi\frac{1}{\lnwto}
\left\{
\frac{\partial{}}{\partial T}(T\avXr)
-\frac{2}{3}\chi_{\perp,{\rm blo}}
\right\}
\;.
\end{equation}
Therefore, we find the $T$- and $\w$-dependent criterion
$(2/3)\chi_{\perp,{\rm blo}}
\ll\partial(T\chi')/\partial T$,
for the quantity $\partial(T\chi')/\partial T$ scanning the energy-barrier distribution
as properly as $\chi''$ (for anisotropy axes distributed at random only). Recall that no
restriction of this type exists for the obtainment of the energy-barrier distribution
from $\chi''$ (or $\partial\chi'/\partial\ln\w$). Note also that, not only
$\chi_{\perp,{\rm blo}}$ is the transverse susceptibility of the blocked particles but,
when multiplied by $\avsenqal_{\ran}=2/3$, is their total contribution to the
susceptibility, because the over-barrier relaxation mechanism is indeed
``blocked" for those particles.

\subsection{Comparison with experiment}
\label{experiments}

To conclude, we shall briefly illustrate some of the results of the
previous subsection with experiments performed on a {\em frozen\/}
magnetic fluid containing nanometric maghemite
($\gamma$--Fe$_{2}$O$_{3}$) particles.
\begin{figure}[b!]
\vspace{-3.ex}
\eps{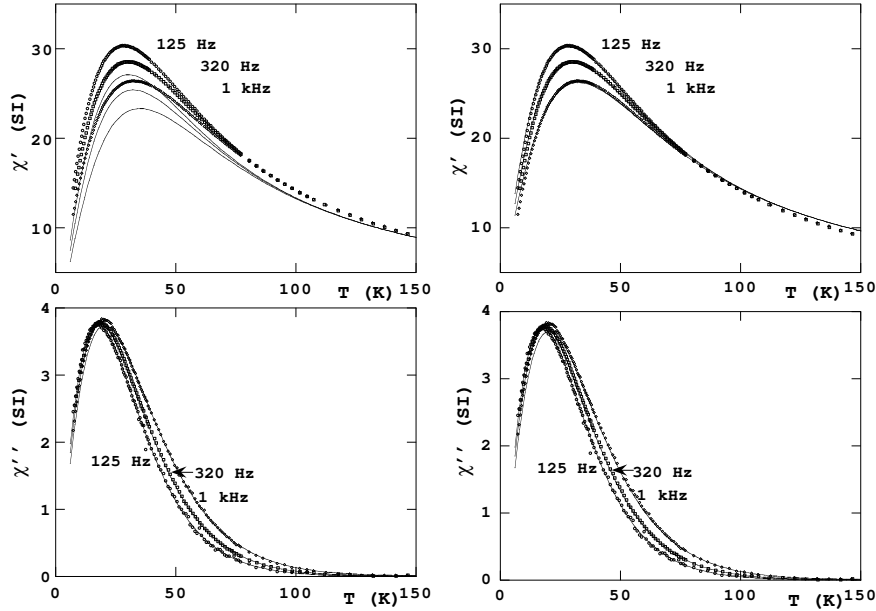}{1}
\vspace{-3.ex}
\caption[]
{
Temperature dependence of the in-phase (upper panels) and
out-of-phase (lower panels) components of the dynamical susceptibility
of a frozen magnetic fluid of maghemite particles. Left panels: solid
lines computed with the Ising-type model where
$\chi_{\|}=\mu_{0}\mm^{2}/\T$ and $\chi_{\perp}=0$ (per
particle). Right panels: solid lines computed with the low-frequency
Shliomis and Stepanov equation (\ref{shschi:int}).
\label{ferrchi_doble:plot}
}
\end{figure}

The degree of dilution of the sample studied was $\sim0.03$\% by
volume, in order to avoid dipole-dipole interaction effects. This
illustrates one of the advantages of the use of frozen magnetic fluids
for fundamental studies on systems of magnetic nanoparticles: by
simple dilution and subsequent freezing of the magnetic fluid, one can
get a series of {\em solid\/} dispersions of nanoparticles where the
strength of the interactions is tuned almost as desired. (This method
also guarantees that all the samples have the same distribution in
particle parameters.) Another advantage of these systems is that by
means of the application of magnetic fields when freezing the samples,
one can produce systems with different anisotropy-axis
distributions. The sample considered here (Svedlindh et~al., 1997) was
frozen in zero field, so that a random distribution of the anisotropy
axes is to be expected.

\subsubsection{Comparison with the Ising-type and Shliomis and Stepanov models}

Figure \ref{ferrchi_doble:plot} displays the measured dynamical
susceptibility and the Ising-type theoretical curves computed with the
energy-barrier distribution derived from $\chi''$. While the
calculated and experimental out-of-phase susceptibilities compare to a
high degree of precision (by construction), the matching of the
in-phase curves is comparatively poor. One may guess that the reason
for this poor matching is the absence of the transverse response in
the model employed.%
\footnote{
We use the terms ``take the transverse response into account" to
abbreviate ``take the finite width and depth of the anisotropy
potential wells into account", since the lack of response to the
transverse components of the probing field is perhaps the most
characteristic feature of the Ising-type response.} In order to check
this hypothesis, Fig.\ \ref{ferrchi_doble:plot} also displays the same
experimental results together with the curves computed with the
low-frequency Shliomis and Stepanov equation. One can see that the
description of the experimental curves provided by this model has
improved significantly.
\begin{figure}[t!]
\vspace{-3.ex}
\eps{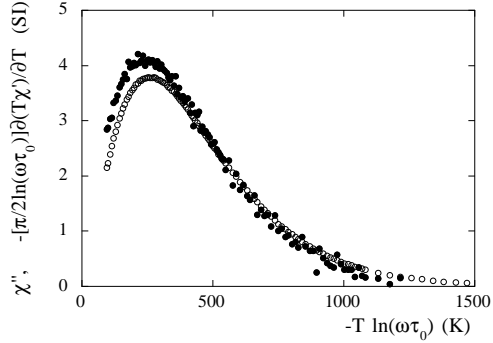}{0.56}
\vspace{-3.ex}
\caption[]
{
$\chi''(T)$ (open symbols) and
$-[\pi/2\lnwto]\partial(T\chi')/\partial T$ (filled symbols) vs.\
$-T\lnwto$ of a frozen magnetic fluid of maghemite particles at the
frequency $\w/2\pi=320$\,Hz.
\label{ferrchi3:plot}
}
\end{figure}

\subsubsection{Comparison of $\chi''$ with $\partial(T\chi')/\partial T$}

Equation (\ref{X:in:der:ran:1}) suggests that a joint plot of $\chi''$
and $\partial(T\chi')/\partial T$ could be an alternative means to
show the necessity of including the transverse response of the
nanoparticles. When this contribution to the total response is
negligible, those two curves should trace out the same energy-barrier
dependence, whereas one would expect
$(\pi/2\sb)\partial(T\chi')/\partial T$ being larger than $\chi''$
otherwise. Moreover, one would also expect that, the lower the
temperature, the larger the differences between the two curves,
because the lower limit in the integral of Eq.\ (\ref{X:in:der:ran:1})
decreases with $T$. This is what is indeed observed in Fig.\
\ref{ferrchi3:plot}, giving further evidence of the necessity of
including the transverse contribution to the total response of the
studied magnetic nanoparticle system. The figure also confirms the
point that $\partial(T\chi')/\partial T$ does not determine the
energy-barrier distribution as accurately as $\chi''$ does.

%% file: garcms05.tex
\section{Dynamical properties: stochastic approach}
\label{sect:stochastic}

\subsection{Introduction}

In this Section we shall study the dynamics of classical spins in the
context of the {\em theory of stochastic processes}.

In order to study the properties of classical magnetic moments,
numerical simulation techniques can also be used, with most of the
studies that have been performed being based on the Monte Carlo
method. Although this method is a rigorous and efficient tool to
compute thermal-equilibrium quantities, the interpretation of the
dynamical properties derived by means of Monte Carlo techniques,
especially for non-Ising spins, is not free from criticism (Ettelaie
and Moore, 1984; Binder and Stauffer, 1984). On the contrary, when
using stochastic methods based on Fokker--Planck or Langevin
equations, time does not merely label the sequential order of
generated states when sampling the phase space, but is related with
physical time.

For classical spins, the basic Langevin equation is the stochastic
Landau--Lifshitz (--Gilbert) equation introduced by Brown (1963) (see also, Kubo and
Hashitsume, 1970). The {\em multiplicative\/} fluctuating terms occurring in this
Langevin equation were treated in Brown's work, as well as in the subsequent
theoretical developments, by means of the {\em Stratonovich stochastic calculus}. In
this context, Brown constructed the celebrated Fokker--Planck (diffusion) equation for
the time evolution of the {\it non-equilibrium\/} probability distribution of magnetic
moment orientations.

In order to solve Brown's Fokker--Planck equation (a partial differential equation of
parabolic type) a number of techniques have been used, such as direct solution
techniques (Rod{\'{e}}, Bertram and Fredkin, 1987) or more elaborate approaches
involving continued-fractions techniques or the numerical calculation of the
eigenvalues and amplitudes of the relevant dynamical modes (Aharoni, 1964; Bessais,
Ben~Jaffel and Dormann, 1992; Coffey, Crothers, Kalmykov, Massawe and Waldron,
1994; Ra{\u{\i}}kher and Stepanov, 1995{\em b}; Coffey, Crothers, Kalmykov and
Waldron, 1995{\em a}).

An approach equivalent to solving a Fokker--Planck equation is to construct solutions
of the underlying stochastic equation of motion of the system. This {\em
Langevin-dynamics\/} approach by-passes the Fokker--Planck equation as it directly
generates the stochastic trajectories of the variables of the system, from which
averages can be computed. This is a relevant point since the solution of the
Fokker--Planck equation for multivariate systems, either numerically or analytically,
is usually a formidable task.

In this Section we shall integrate the stochastic Landau--Lifshitz--Gilbert equation
numerically in the context of the Stratonovich stochastic calculus. This is undertaken
taking account of the underlying subtleties of the stochastic calculus as compared with
the deterministic calculus. As the Langevin-dynamics method employed generates the
selfsame stochastic trajectories of each individual magnetic moment, it provides much
insight into the dynamics of the system. In addition, the theoretical study of
single-particle phenomena is of special interest because dynamical measurements of
{\em individual\/} magnetic {\em nano}particles have recently been performed
(Wernsdorfer et~al., 1997).

Concerning the response of an ensemble of classical magnetic moments (averaged
quantities), the Langevin-dynamics method allows one to compute any desired
quantity, e.g.: hysteresis loops, field-cooled and zero-field-cooled magnetization
curves, relaxation times, linear and non-linear susceptibilities, thermal quantities, and,
with appropriate relationships between line-shapes and correlation functions of the
system, even spectroscopic quantities. We shall restrict our study to the linear
dynamical response, which is chosen since it is a probe that enables one to examine
the intrinsic dynamics of the system. In addition, because some relevant parameters of
nanoparticle ensembles can be extracted from the analysis of the dynamical response
data (see Section \ref{sect:heuristic}), an assessment of the accuracy of the heuristic
equations employed in such analyses is necessary.

We finally note that, when studying {\it averaged\/} quantities, the
Langevin-dynamics method requires an extensive computational effort and is then
less efficient than numerical methods especially suitable for non-interacting magnetic
moments, such as those based on the Fokker--Planck equation mentioned above.
However, with a significant increase of the computational effort, the
Langevin-dynamics technique can also be used to study assemblies of interacting
spins.

The organization of this Section, which is an extended version of the results presented
by Garc{\'{\i}}a-Palacios and L{\'{a}}zaro (1998), is as follows. In order to provide the
necessary background to undertake the study of the stochastic dynamics of classical
spins, we begin in Subsec.\ \ref{subsect:detdyn} with the study of the {\em
deterministic\/} Landau--Lifshitz equation. Then, the Brown--Kubo--Hashitsume
model for the stochastic dynamics of classical magnetic moments is discussed in
Subsec.\ \ref{subsect:brown}. The numerical method used to solve the stochastic
Landau--Lifshitz (--Gilbert) equation is discussed in Subsec.\
\ref{subsect:nummethod}. Finally, the results of the numerical integration of this
Langevin equation are presented in subsections \ref{subsect:simulationsI} and
\ref{subsect:simulationsII}. Specifically, Subsec.\ \ref{subsect:simulationsI} is
devoted to the study of the trajectories of individual magnetic moments, while the
dynamical response of the spin ensemble is studied in Subsec.\
\ref{subsect:simulationsII}.

\subsection{Deterministic dynamics of classical spins}
\label{subsect:detdyn}

To begin with, we shall study some aspects of the deterministic dynamics of classical
magnetic moments.

\subsubsection{The  Gilbert and Landau-Lifshitz equations}

Let us start by considering the Gilbert equation of motion for a classical magnetic
moment $\m$ (unpublished work, mentioned in Gilbert, 1955)
\begin{equation}
\label{detgileq}
\frac{\D\m}{\D t}
=
\gmr\m
\vecpro\left[\Beff-(\gmr\mm)^{-1}\la\frac{\D\m}{\D t}\right]
\;,
\end{equation}
where $\gmr$ is the gyromagnetic ratio and $\la$ is a dimensionless damping
coefficient (the coefficient appearing when one writes the equation for the
magnetization $\vec{M}=\m/v$ is equal to the one used here multiplied by $v$). The
{\em effective\/} field in Eq.\ (\ref{detgileq}) is given by
\begin{equation}
\label{beff:def}
\Beff
=
-\frac{\partial\Hs}{\partial\m}
\;,
\end{equation}
where $\Hs$ is the Hamiltonian of $\m$ and $\partial {}/\partial\m$ stands for the
gradient operator [$\partial f/\partial\m
=(\partial f/\partial\mx)\hat{x}
+(\partial
f/\partial\my)\hat{y}
+(\partial f/\partial\mz)\hat{z}$]. For the justification of the
occurrence of the expression (\ref{beff:def}) in the dynamical equations the reader is
referred to Subsec.\ \ref{subsect:free_dynamics}. Anyway, note that for
$\Hs=-\m\cdot\B$ one indeed has $\Beff=\B$, while in a more general situation
$\Beff$ incorporates the (deterministic) effects of the magnetic-anisotropy energy, the
interaction with other spins, etc., on the dynamics of $\m$.

To illustrate, if the magnetic anisotropy is assumed to have the simplest axial
symmetry (with symmetry axis $\hat{n}$) and $\m$ is subjected to an external
constant field, $\B$, and a low probing field, $\dB(t)$, the Hamiltonian reads [cf. Eq.\
(\ref{U0})]
\begin{equation}
\label{U}
\Hs(\m,t)
=
-\m\cdot
\big[
\B+\dB(t)
\big]
-\frac{Kv}{\mm^{2}}(\m\cdot\hat{n})^{2}
\;.
\end{equation}
In terms of $\BK=2Kv/\mm$ [Eq.\ (\ref{BK-h})], the effective field associated with this
Hamiltonian can be written as
\begin{equation}
\label{Beff}
\Beff
=
\B+\dB(t)
+(\BK/\mm)(\m\cdot\hat{n})\hat{n}
\;.
\end{equation}
Note that the quantity $|\BK|$ is the magnitude of the maximum {\em anisotropy
field\/}
\[
\B_{{\rm a}}
=
(\BK/\mm)(\m\cdot\hat{n})\hat{n}
\;,
\]
which occurs when $\m=\pm\mm\,\hat{n}$. The anisotropy field decreases as $\m$
approaches the equatorial region ($\m\perp\hat{n}$), where it vanishes. Recall finally
that for easy-axis anisotropy in a longitudinal bias field ($\B\parallel\hat{n}$), the
Hamiltonian has two minima at $\m=\pm\mm\,\hat{n}$ for $|B|<|\BK|$, with a
potential barrier between, whereas the upper (shallower) potential minimum
disappears for $|B|\geq|\BK|$ (see Subsec.\ \ref{hamiltonian}).

An equation of Gilbert type can be cast into the archetypal Landau--Lifshitz form
(1935) as follows. Take the vector product of $\m$ with both sides of Eq.\
(\ref{detgileq})
\[
\m\vecpro\frac{\D\m}{\D t}
=
\gmr\m\vecpro\left(\m\vecpro\Beff\right)
-\frac{\la}{\mm}
\bigg[
\m
\bigg(
\underbrace{
\m\cdot\frac{\D\m}{\D t}
}_{0}
\bigg)
-\mm^{2}\frac{\D\m}{\D t}
\bigg]
\;,
\]
where the triple vector product $\m\vecpro[\m\vecpro(\D\m/\D t)]$ has been
expanded by using the rule
\begin{equation}
\label{BAC-CAB}
\vec{A}\vecpro\left(\vec{B}\vecpro\vec{C}\right)
=
\vec{B}\left(\vec{A}\cdot\vec{C}\right)
-\vec{C}\left(\vec{A}\cdot\vec{B}\right)
\;,
\end{equation}
and $\m\cdot(\D\m/\D t)=0$ (conservation of the magnitude of $\m$) follows from
the starting equation (\ref{detgileq}). On introducing the above result for
$\m\vecpro(\D\m/\D t)$ in the right-hand side of Eq.\ (\ref{detgileq}), passing
$-\la^{2}\D\m/\D t$ to the left-hand side, and introducing the ``renormalized"
gyromagnetic ratio $\tilde{\gmr}=\gmr/(1+\la^{2})$, one finally gets the desired
Landau--Lifshitz form of the Gilbert equation
\begin{equation}
\label{detlleq:0}
\frac{\D\m}{\D t}
=
\tilde{\gmr}\m\vecpro\Beff
-\tilde{\gmr}\frac{\la}{\mm}\m\vecpro\left(\m\vecpro\Beff\right)
\;.
\end{equation}
The celebrated Landau--Lifshitz relaxation (damping) term proportional to
$-\m\vecpro(\m\vecpro\Beff)$ drives $\m$ to the direction of $\Beff$, while $\la$
measures the magnitude of the relaxation term relative to the gyromagnetic term in
the dynamical equation.

Conversely, one can start from Eq.\ (\ref{detlleq:0}) with $\tilde{\gmr}$ replaced
by $\gmr$ and then write down its Gilbert equivalent equation. This is like Eq.\
(\ref{detgileq}) with $\gmr$ being replaced by a different ``renormalized"
gyromagnetic ratio: $\tilde{\gmr}'=\gmr\times(1+\la^{2})$.

There exist some controversy concerning which equation (Gilbert or Landau--Lifshitz)
is more basic, or, equivalently, when one must use a renormalized $\gmr$. However,
on recalling that both equations are anyway phenomenological ones, we can consider
$\tilde{\gmr}$ (or $\tilde{\gmr}'$) to be a given constant for each magnetic moment.
In addition, when $\la^{2}\ll1$ (weak damping), which is the common situation at
least for bulk magnets, one has $\tilde{\gmr}'\simeq\tilde{\gmr}\simeq\gmr$, so that
one does not need to worry about whether the gyromagnetic ratio occurring in a given
formula is a bare or renormalized one.

Henceforth, we shall merely use the symbol $\gmr$ in the dynamical quantities (as if
we would have started from the Landau--Lifshitz equation). If one wishes to consider
the Gilbert form as the commencing equation, one just needs to substitute
$\gmr/(1+\la^{2})$ for $\gmr$ in the corresponding formulae.

\subsubsection{General solution for axially symmetric Hamiltonians}

We shall now investigate solutions of the deterministic Landau--Lifshitz equation
\begin{equation}
\label{detlleq}
\frac{\D\m}{\D t}
=
\gmr\m\vecpro\Beff
-\gmr\frac{\la}{\mm}\m
\vecpro\big(\m\vecpro\Beff\big)
\;,
\end{equation}
restricting our attention to the case in which $\Hs(\m)$ is axially symmetric. In this
case, the effective field $\Beff(\m)=-\partial\Hs/\partial\m$ is parallel to the
symmetry axis, which can be chosen as the $z$ axis, $\Beff=B_{{\rm eff}}(\m)\hat{z}$.
Then, on introducing the $\m$-dependent ``frequency" $\weff(\m)=\gmr B_{{\rm
eff}}(\m)$, we can explicitly write the deterministic Landau--Lifshitz equation
(\ref{detlleq}) as a system of coupled ordinary differential equations:
\begin{eqnarray*}
\frac{\D\mx}{\D t}
&
=
&
\weff\Big(\my-\frac{\la}{\mm}\mx\mz\Big)
\;,
\\
\frac{\D\my}{\D t}
&
=
&
\weff\Big(-\mx-\frac{\la}{\mm}\my\mz\Big)
\;,
\\
\frac{\D\mz}{\D t}
&
=
&
\weff\frac{\la}{\mm}(\mm^{2}-\mz^{2})
\;.
\end{eqnarray*}
Next, on introducing spherical coordinates
$\mz=\mm\cos\vartheta$
and
$\mx+i\my=\mm\sin\vartheta\exp(-i\varphi)$
(we measure here the azimuthal angle clock-wise), the above system of differential
equations can equivalently be written as
\begin{eqnarray}
\label{detlleq:component_theta}
\frac{\D\vartheta}{\D t}
&
=
&
-\la\weff\sin\vartheta
\;,
\\
\label{detlleq:component_phi}
\frac{\D\varphi}{\D t}
&
=
&
-\frac{1}{\la\sin\vartheta}\frac{\D\vartheta}{\D t}
\;,
\quad (\mbox{or~}
\D\varphi/\D t=\weff)
\;.
\end{eqnarray}
Equation (\ref{detlleq:component_phi}) can be solved by separation of variables, to get
\begin{equation}
\label{phi_theta:general}
\varphi(\vartheta)
-\varphi(\vartheta_{0})
=
-\frac{1}{\la}
\ln\left[\q\big/\qo\right]
\;,
\end{equation}
where $\int\!\D{x}\,/\sin x=\ln[\tan(x/2)]$ has been used and
$\vartheta_{0}=\vartheta(t_{0})$, $t_{0}$ being the initial time. Concerning the
equation (\ref{detlleq:component_theta}) for $\vartheta$, since
$\weff=\weff(\vartheta)$, we can also separate the variables to obtain the following
implicit expression for $\vartheta(t)$
\begin{equation}
\label{t_theta:general}
-\la\left(t-t_{0}\right)
=
\int_{\vartheta_{0}}^{\vartheta(t)}
\!\!
\frac{\D\vartheta'}{\weff(\vartheta')\sin\vartheta'}
\;.
\end{equation}
Equations (\ref{phi_theta:general}) and (\ref{t_theta:general}) are the solution of the
deterministic Landau--Lifshitz equation (\ref{detlleq}) for {\em any\/} axially
symmetric Hamiltonian $\Hs(\vartheta)$.

\paragraph*{Weak damping case.}

An important case is that in which $\la\ll1$. Note first that Eq.\
(\ref{detlleq:component_phi}) can also be written as $\D\vartheta=-\la\sin\vartheta
\D\varphi$, which, for weak damping yields $|\D\vartheta|\ll|\sin\vartheta
\D\varphi|$. Then, the ``displacement" of the tip of $\m$ along the polar direction
($\Delta\vartheta$) in a time interval $\Delt$ is much smaller than the displacement
along the tangential direction ($\sin\vartheta\Delta\varphi$). It makes then sense to
introduce a ``position-dependent" frequency of rotation about $\hat{z}$, which is
precisely given by $\weff=\gmr B_{{\rm eff}}$ [see the alternative form of Eq.\
(\ref{detlleq:component_phi})].

\subsubsection{The simplest axially symmetric Hamiltonian}

Let us now specialize the above general solutions to the Hamiltonian obtained by the
sum of the simplest axially symmetric anisotropy potential plus a longitudinal Zeeman
term. Then [cf.\ Eq.\ (\ref{Beff})]
\begin{equation}
\label{Beff:axial}
\Beff
=
B\hat{z}
+(\BK/\mm)\mz\hat{z}
\;,
\end{equation}
and $\weff=\gmr B_{{\rm eff}}$ can be written as
\begin{equation}
\label{weff:axial}
\weff(\vartheta)
=
\wB+\wK\cos\vartheta
\;,
\qquad
\wB
=
\gmr B
\;,
\quad
\wK
=
\gmr\BK
\;.
\end{equation}
On the other hand, the integral in the solution (\ref{t_theta:general}) is now given by
\begin{eqnarray*}
\int
\frac{\D{\vartheta}}{\left(\wB+\wK\cos\vartheta\right)\sin\vartheta}
&
=
&
\frac{1}{\wB+\wK}\ln\left[\q\right]
\nonumber
\\
&
& {}+
\frac{\wK}{\wB^{2}-\wK^{2}}
\ln
\left[ 1+\left(\frac{\wB-\wK}{\wB+\wK}\right)\qs
\right]
\nonumber
\\
&
& {}+
\frac{\wK}{\wB^{2}-\wK^{2}}
\ln\left(\wB+\wK\right)
\;,
\end{eqnarray*}
as can be checked by differentiation of the right-hand side. Therefore, from the
general result (\ref{t_theta:general}) we get the still implicit solution
\begin{equation}
\label{t_theta:2}
Ce^{-\la(\wB+\wK)t}
=
\q
\left[ 1+\left(\frac{\wB-\wK}{\wB+\wK}\right)\qs
\right]^{\frac{\wK}{\wB-\wK}}
\;,
\end{equation}
where the constant of integration $C$ involves the terms evaluated at $t=t_{0}$.

\subsubsection{Particular cases}

The above implicit solution for $\vartheta(t)$ turns into an explicit solution in various
particular cases.

\paragraph{Dynamics in the isotropic case.}

Here $\wK=0$, so that Eqs.\ (\ref{phi_theta:general}) and (\ref{t_theta:2}) reduce to
the celebrated results (see, for example, Chikazumi, 1978, Ch.\ 16)
\[
\q
=
\qo e^{-\la\wB(t-t_{0})}
\;,
\quad
\varphi(t)-\varphi_{0}
=
\wB(t-t_{0})
\;.
\]
Thus, the motion of $\m$ consist of a precession with frequency $\wB=\gmr B$ about
$\hat{z}$ and a spiralling towards this axis with a characteristic time constant
\begin{equation}
\label{tauB}
\tB
=
\frac{1}{\la\wB}
=
\frac{1}{\la\gmr B}
\;.
\end{equation}
Note that this is the characteristic decay time of $\q$; for $\mz=\mm\cos\vartheta$ in
the vicinity of the minimum [$\q\simeq\vartheta/2$ and
$\cos\vartheta\simeq1-\vartheta^{2}/2$], the characteristic time constant is $\tB/2$.
Note also that, for $B<0$, one has $\wB<0$ and therefore $\lim_{t\to\infty}\q=\infty$,
that is, $\vartheta\to\pi$ as $t\to\infty$, as it should.

\paragraph{Dynamics in the zero-field case.}

Here $\wB=0$, so that, by using $\tan\vartheta=2\q/[1-\qs]$ in Eq.\ (\ref{t_theta:2}),
one gets
\begin{equation}
\label{theta_t:wK}
\tan\vartheta
=
\tan\vartheta_{0} e^{-\la\wK(t-t_{0})}
\;.
\end{equation}
Thus, the spiralling towards the minima has for $K>0$ a characteristic time constant
\begin{equation}
\label{tauK}
\tK
=
\frac{1}{\la\wK}
=
\frac{1}{\la\gmr\BK}
\;,
\end{equation}
or its absolute value if $K<0$. In this easy-plane case one has $\BK,\wK<0$, so that
$\lim_{t\to\infty}\tan\vartheta=\infty$, that is, $\vartheta\to\pi/2$ as $t\to\infty$,
and the magnetic moment eventually rests in the equatorial plane. This behavior upon
the change $\BK\to-\BK$ is different from the behavior upon the transformation
$B\to-B$ in the isotropic case (where $\m$ then falls into the $-\hat{z}$ minimum),
and it is mathematically reflected by the occurrence of $\tan\vartheta$ in the solution
of the unbiased case, whereas $\q$ appears in the solution of the isotropic case.

Note that for both signs of $K$, Eq.\ (\ref{theta_t:wK}) yields
$\vartheta\in[0,\pi/2]$ if $\vartheta_{0}\in[0,\pi/2]$ and $\vartheta\in[\pi/2,\pi]$
when $\vartheta_{0}\in[\pi/2,\pi]$. This expresses that, during the time evolution,
$\vartheta(t)$ remains in the same hemisphere in which it was initially. For instance,
$\m$ does not surmount the anisotropy-potential barrier when $K>0$, as it should in a
deterministic damped dynamics, while for $K<0$, $\m$ does not oscillate about (cross)
the equatorial circle when spiralling towards the easy plane.

Concerning the azimuthal angle, by expressing $\q$ in terms of $\tan\vartheta$, one
gets from Eq.\ (\ref{phi_theta:general})
\[
\varphi(t)-\varphi_{0}
=
\wK(t-t_{0})
-\frac{1}{\la}
\ln
\Bigg[
\frac
{1+\sec\vartheta_{0}}
{1\pm\sqrt{1+\tan^{2}\vartheta_{0}e^{-2\la\wK(t-t_{0})}}}
\Bigg]
\;,
\]
where the plus sign corresponds to $\vartheta\in[0,\pi/2]$ and the minus sign to
$\vartheta\in[\pi/2,\pi]$. From this equation it follows that the asymptotic
$\la\wK(t-t_{0})\gg1$ behavior of the azimuthal angle for $K>0$ is
\[
\Delta\varphi(t)
\simeq
\pm
\wK(t-t_{0})
\;,
\]
which corresponds to a precession close to the bottom of the corresponding potential
well with an angular velocity $\wK\hat{z}$ in the $z>0$ well and $-\wK\hat{z}$ in the
$z<0$ well. For easy-plane anisotropy, one has $\q\stackrel{t\to\infty}{\rightarrow}1$,
so that we find from Eq.\ (\ref{phi_theta:general}) that the magnetic moment finally
rests in the equatorial plane at
$\varphi=\varphi(\vartheta_{0})+\la^{-1}\ln[\tan(\vartheta_{0}/2)]$ (unless it starts
at $\vartheta_{0}=0,\pi$ which are unstable equilibrium points).

\paragraph{Dynamics close to the potential minima.}

The implicit solution (\ref{t_theta:2}) for $\vartheta(t)$ can also be explicitly written
in the general case (both $\wK$ and $\wB$ different from zero) for the dynamics close
to the potential minima (we only consider the case $\BK>0$). Let us initially assume
$\vartheta\simeq0$ [i.e., $\q\ll1$]. Then, on retaining terms of order $\q$ in Eq.\
(\ref{t_theta:2}), we get $\q\simeq\qo\exp[-\la(\wB+\wK)(t-t_{0})]$ and
$\varphi(t)-\varphi_{0}\simeq(\wB+\wK)(t-t_{0})$ by Eq.\ (\ref{phi_theta:general}).
However, within the same approximation ($\vartheta\ll1$) we can replace the
tangents by their arguments, getting
\[
\vartheta(t)
\simeq\vartheta_{0} e^{-\la(\wB+\wK)(t-t_{0})}
\;,
\quad
\varphi(t)-\varphi_{0}
\simeq(\wB+\wK)(t-t_{0})
\;.
\]
Thus, $\m$ precesses with frequency $\wB+\wK$ when spiralling towards the
$\vartheta=0$ potential minimum and the time constant of the decay of $\vartheta$ is
$1/[\la(\wB+\wK)]=\tB\tK/(\tB+\tK)$. Note that the characteristic decay time of
$\mz\propto\cos\vartheta\simeq1-\vartheta^{2}/2$, is a half of this result.

From the above equations we see that the approximation used ($\vartheta\ll1$) is
self-consistent if $\wB+\wK>0$, that is, for any positive $B$ and also for negative
external fields of magnitude less than the anisotropy field $|B|<\BK$ (i.e., inasmuch as
the $\vartheta=0$ potential minimum exists; recall the discussion in Subsec.\
\ref{hamiltonian}).

On the other hand, in the $\vartheta\simeq\pi$ case one has
$\vartheta/2\simeq\pi/2$ and, hence, $\q\gg1$. Then, we can use
$[(\wB-\wK)/(\wB+\wK)]\qs\gg1$ in Eq.\ (\ref{t_theta:2}) to get
$\q\simeq\qo\exp[\la(\wK-\wB)(t-t_{0})]$, whence
$\varphi(t)-\varphi_{0}\simeq-(\wK-\wB)(t-t_{0})$ by Eq.\ (\ref{phi_theta:general}).
However, when $\q\gg1$, we can use the approximation $\tan\vartheta\simeq-2/\q$,
so that on expanding $\tan\vartheta$ about $\vartheta=\pi$, we finally get
\[
\vartheta(t)-\pi
\simeq(\vartheta_{0}-\pi)e^{-\la(\wK-\wB)(t-t_{0})}
\;,
\quad
\varphi(t)-\varphi_{0}
\simeq-(\wK-\wB)(t-t_{0})\;.
\]
Therefore, $\m$ precesses with frequency $\wK-\wB$ (about $-\hat{z}$) when
spiralling towards the $\vartheta=\pi$ minimum, while $\vartheta$ decays with a
characteristic time constant $1/[\la(\wK-\wB)]=\tB\tK/(\tB-\tK)$ (and $\mz$ with a
half of this value).

Note finally that the approximation used ($\pi-\vartheta\ll1$) is self-consistent if
$\wK-\wB>0$, that is, for any negative $B$ and also for positive $B$ of magnitude less
than the anisotropy field ($B<\BK$). Thus, in this case, and exhibiting a natural
symmetry with the $\vartheta\simeq0$ case, the motion is stable inasmuch as the
$\vartheta=\pi$ minimum exists.

\subsection[Stochastic dynamics of classical spins
(Brown--Kubo--Hashitsume model)]
{Stochastic dynamics of classical spins\\(Brown--Kubo--Hashitsume model)}
\label{subsect:brown}

Due to the interaction of a spin with the surrounding medium (phonons, conducting
electrons, nuclear spins, etc.) its $T\neq0$ dynamics is quite complicated. The
complexity itself, however, permits an idealization of the phenomenon, by replacing
the effect of the environment by a magnetic field randomly varying in time.
Nevertheless, in order to describe the environmental effects properly and to attain a
thermodynamically consistent description, the fluctuating terms must be
supplemented with the analogue of a {\em relaxation\/} (damping or dissipative)
term, to which must be linked by {\em fluctuation-dissipation\/} relations.

We shall begin with a survey of how this general programme is specialized to the
study of the stochastic dynamics of classical magnetic moments. This was done by
Brown (1963), in the context of the small-particle magnetism, and by Kubo and
Hashitsume (1970), who studied generic classical spins. The subsequent developments
based on each of these works have taken place separately in the literature.
Nevertheless, both approaches are essentially equivalent and we shall present here a
unified discussion of them.%
\footnote{
Notice that Kubo and Hashitsume say in their article that the main part of
their work was done in the summer of 1963, so that both approaches are in addition
contemporary.
}

\subsubsection{Stochastic dynamical (Langevin) equations}

In the Brown--Kubo--Hashitsume model the starting dynamical equation is the Gilbert
equation (\ref{detgileq}) where the total field acting on $\m$ is obtained by
augmenting the deterministic effective field $\Beff$ by a fluctuating or stochastic field
$\bfl(t)$, namely
\begin{equation}
\label{stogileq}
\frac{\D\m}{\D t}
=
\gmr\m
\vecpro
\left[
\Beff+\bfl(t)-(\gmr\mm)^{-1}\la\frac{\D\m}{\D t}
\right]
\;.
\end{equation}
This equation, which is technically a non-linear {\em stochastic differential (Langevin)
equation}, is called the {\em stochastic Gilbert equation}. It suggest a heuristic analogy
with the Langevin equation for ordinary Brownian motion since the ``friction field" is
proportional to minus the ``velocity," $-(\D\m/\D t)$. However, the analogy ends here;
in the dynamical equation for a Brownian particle [see, for example, Eq.\
(\ref{brownian_particle})], a friction term proportional to minus the velocity enters in
the Newton equation (i.e., in the equation for the acceleration), whereas
$-(\D\m/\D t)$ enters in the equation for the ``velocity" itself. Besides, the
fluctuating terms enter in Eq.\ (\ref{stogileq}) in a multiplicative way (see below).

As has been mentioned,  the fluctuating  field $\bfl(t)$ accounts for the effects of the
interaction of $\m$ with the microscopic degrees of freedom (phonons, conducting
electrons, nuclear spins, etc.), which cause fluctuations of the magnetic moment
orientation. Those environmental degrees of freedom are {\em also\/} responsible for
the damped precession of $\m$, since fluctuations and dissipation are related
manifestations of one and the same interaction of the magnetic moment with its
environment (see Section \ref{sect:gle}).

The customary assumptions about $\bfl(t)$ are that it is a Gaussian ``stochastic
process" with  the following statistical properties
\begin{equation}
\label{bcorr}
\llangle b_{\fl,k}(t)\rrangle
=
0
\;,
\qquad
\llangle b_{\fl,k}(t)b_{\fl,\ell}(\tp)\rrangle
=
2D\delta_{k\ell}\delta(t-\tp)
\end{equation}
(the first two {\em moments\/} determine a Gaussian process), where $k$ and $\ell$ are
Cartesian indices, the constant $D$ measures the strength of the thermal fluctuations
(assumed isotropic), and $\langle\,\rangle$ denotes an average taken over different
{\em realizations\/} of the fluctuating field. (The constant $D$ is determined on the
grounds of statistical-mechanical considerations; see below.) The Gaussian property of
the fluctuations arises because they emerge from the interaction of $\m$ with a large
number of microscopic degrees of freedom with equivalent statistical properties
(Central Limit Theorem). On the other hand, the Dirac delta in the second Eq.\
(\ref{bcorr}) expresses that above certain temperature the auto-correlation time of
$\bfl(t)$ (of microscopic scale) is much shorter than the rotational-response time of
the system (``white" noise), while the Kronecker delta expresses that the different
components of $\bfl(t)$ are assumed to be uncorrelated. Finally, it is also customarily
assumed that the fluctuating fields acting on different magnetic moments are
independent.

On starting from the stochastic Gilbert equation (\ref{stogileq}), the discussed
transformation to the equivalent Landau--Lifshitz form yields (recall our convention
for the gyromagnetic ratio)
\begin{equation}
\label{stollgeq}
\frac{\D\m}{\D t}
=
\gmr\m\vecpro
\left[\Beff+\bfl(t)\right]
-\gmr\frac{\la}{\mm}\m
\vecpro
\left\{
\m\vecpro
\left[\Beff+\bfl(t)\right]
\right\}
\;,
\end{equation}
which will be called the {\em stochastic Landau--Lifshitz--Gilbert equation}.
As will be shown below, the thermodynamical consistency of the approach entails that
$|\bfl|\sim\la^{1/2}$. Therefore, for weak damping ($\la\ll1$) we can drop the
fluctuating field from the relaxation term of Eq.\ (\ref{stollgeq}), to arrive at
\begin{equation}
\label{stolleq}
\frac{\D\m}{\D t}
=
\gmr\m\vecpro
\left[\Beff+\bfl(t)\right]
-\gmr\frac{\la}{\mm}\m
\vecpro
\left(\m\vecpro\Beff\right)
\;.
\end{equation}
This equation, which was in fact the equation studied by Kubo and Hashitsume (1970),
will be called the {\em stochastic Landau--Lifshitz equation}, since in accordance with
the spirit of its original deterministic counterpart, it describes weakly damped
precession. Equation (\ref{stolleq}) is besides a Langevin equation more archetypal
than Eq.\ (\ref{stollgeq}), because the fluctuating and relaxation terms are not
entangled.

On the other hand, one can by-pass the reasoning employed to obtain Eq.\
(\ref{stolleq}) from Eq.\ (\ref{stollgeq}), and consider the former as an alternative
stochastic model. It will be shown below that, when the condition of thermodynamical
consistency is applied, the {\em average\/} properties derived both from Eqs.\
(\ref{stollgeq}) and from (\ref{stolleq}) are completely equivalent.

\paragraph*{The multiplicative noise terms.}

Apparently, for a given $D$, Eqs.\ (\ref{stollgeq}) or (\ref{stolleq}), supplemented by
Eqs.\ (\ref{bcorr}), fully determine the dynamical problem under consideration.
Nevertheless, due to the vector {\em products\/} of $\m$ and $\bfl(t)$ occurring in
those equations, the fluctuating field $\bfl(t)$ enters in a {\em multiplicative\/} way.
This fact gives rise to some formal problems because, for white multiplicative noise,
any Langevin equation must be supplemented by an interpretation rule to properly
define it (see, for example, van Kampen, 1981, p.~246).

Two dominant interpretations, which lead to either the It\^{o} or the Stratonovich {\em
stochastic calculus}, are usually considered, yielding different dynamical properties for
the system. For instance, depending on the stochastic calculus used, disparate
Fokker--Planck equations for the time evolution of the non-equilibrium probability
distribution are obtained. The It\^{o} calculus is commonly chosen on certain
mathematical grounds, since rather general results of probability theory can then be
employed. On the other hand, since the white noise is an idealization of physical noise
with short auto-correlation time, the Stratonovich calculus is usually preferred in
physical applications, since the associated results coincide with those obtained in the
formal zero-correlation-time limit of fluctuations with finite auto-correlation time
(see, for example, Risken, 1989).

Both the seminal works of Brown (1963) and, Kubo and Hashitsume (1970), as well as
all the subsequent theoretical developments, are based, implicitly or explicitly, on the
Stratonovich stochastic calculus.

\subsubsection{Fokker--Planck equations}

We shall now consider the Fokker--Planck equations governing the time evolution of
the non-equilibrium probability distribution of magnetic moment orientations. Brown
(1963) derived the Fokker--Planck equation associated with the stochastic
Landau--Lifshitz--Gilbert equation (\ref{stollgeq}). By a different method and starting
from the stochastic Landau--Lifshitz equation (\ref{stolleq}), Kubo and Hashitsume
(1970) arrived at an equation for the probability distribution, which, when the
auto-correlation times of $\bfl(t)$ are much shorter than the precession period of
$\m$, coincides with the Fokker--Planck equation of Brown in the absence of the
anisotropy potential (they studied the case $\Beff=\B$) (for an alternative derivation
starting from Eq.\ (\ref{stolleq}) see, for example, Garanin, 1997). We shall begin by
giving a {\em unified\/} derivation of the Fokker--Planck equations associated with
Eqs.\ (\ref{stollgeq}) and (\ref{stolleq}).

\paragraph{Derivation of the Fokker--Planck equations.}

Let us consider the general system of Langevin equations
\begin{equation}
\label{langevinequation:n-dim}
\frac{\D y_{i}}{\D t}
=
\drift_{i}(\ym,t)
+\sum_{k}\diff_{ik}(\ym,t)\Lan_{k}(t)
\;,
\end{equation}
where $\ym=(y_{1},\ldots,y_{n})$ (the variables of the system), $k$ runs over a given
set of indices, and the ``Langevin" sources $\Lan_{k}(t)$ are independent Gaussian
stochastic processes satisfying
\begin{equation}
\label{langevin:n-dim}
\llangle\Lan_{k}(t)\rrangle
=
0
\;,
\qquad
\llangle\Lan_{k}(t)\Lan_{\ell}(\tp)\rrangle
=
2D\delta_{k\ell}\delta(t-\tp)
\;.
\end{equation}
When the functions $\diff_{ik}(\ym,t)$ depend on $\ym$, the noise in the above
equations is termed {\em multiplicative\/}, whereas for $\partial\diff_{ik}/\partial
y_{j}\equiv0$ the noise is called {\em additive\/} (here the It\^{o} and Stratonovich
stochastic calculi coincide).

The time evolution of $\W(\ym,t)$, the non-equilibrium probability distribution of
$\ym$ at time $t$, is given by the Fokker--Planck equation (see, for example, Risken,
1989)
\begin{equation}
\label{fokkerplanck:langevin:n-dim}
\frac{\partial\W}{\partial t}
=
-\sum_{i}\frac{\partial {}}{\partial y_{i}}
\bigg[
\bigg(
\drift_{i}
+D\sum_{jk}\,\diff_{jk}\frac{\partial\diff_{ik}}{\partial y_{j}}
\bigg)\W
\bigg]
+\sum_{ij}
\frac{\partial^{2}}{\partial y_{i}\partial y_{j}}
\bigg[
\bigg(
D\sum_{k}\,\diff_{ik}\diff_{jk}
\bigg)
\W
\bigg]
\;,
\end{equation}
where the Stratonovich calculus has been used to treat the (in general) multiplicative
fluctuating terms in the Langevin equations (\ref{langevinequation:n-dim}) [when
using the It\^{o} calculus the {\em noise-induced\/} drift coefficient
$D\sum_{jk}\,\diff_{jk}(\partial\diff_{ik}/\partial y_{j})$ is simply omitted]. On taking
the $y_{j}$-derivatives of the second term on the right-hand side (the diffusion term),
one alternatively gets the Fokker--Planck equation in the form of a {\em continuity
equation\/} for the probability distribution, namely
\begin{equation}
\label{fokkerplanck:langevin:n-dim:cont}
\frac{\partial\W}{\partial t}
=
-\sum_{i}\frac{\partial {}}{\partial y_{i}}
\bigg\{\bigg[
\drift_{i}
-D\sum_{k}\diff_{ik}
\bigg(\sum_{j}\frac{\partial\diff_{jk}}{\partial y_{j}}\bigg)
-D\sum_{jk}\,\diff_{ik}\diff_{jk}
\frac{\partial {}}{\partial y_{j}}\,
\bigg]\W
\bigg\}
\;,
\end{equation}
where term within the curly brackets defines the $i$th component of the current of
probability $J_{i}(\ym,t)$.

Next, on considering the {\em stochastic Landau--Lifshitz (--Gilbert) equation},
supplemented by the statistical properties (\ref{bcorr}), the following substitutions
cast them into the form of the general system of Langevin equations
(\ref{langevinequation:n-dim}): $(y_{1},y_{2},y_{3})=(\mx,\my,\mz)$,
$\Lan_{k}(t)=b_{{\rm fl},k}(t)$, and
\begin{eqnarray}
\label{F:ll-llg}
\drift_{i}
&
=
&
\gmr
\left[
\m\vecpro\Beff
-\frac{\la}{\mm}
\m\vecpro\left(\m\vecpro\Beff\right)
\right]_{i}
\;,
\\
\label{G:ll-llg}
\diff_{ik}
&
=
&
\gmr
\bigg[
\sum_{j}\epsilon_{ijk}\mj
+\llg
\frac{\la}{\mm} (\mm^{2}\delta_{ik}-\mi\mk)
\bigg]
\;,
\end{eqnarray}
where $\epsilon_{ijk}$ is the antisymmetrical unit tensor of rank three (Levi-Civita
symbol)%
\footnote{ This tensor is defined as the tensor antisymmetrical in all
three indices with $\epsilon_{xyz}=1$. Therefore, one can write the
vector product of $\vec{A}$ and $\vec{B}$ as
$\left(\vec{A}\vecpro\vec{B}\right)_{i}
=\sum_{jk}\epsilon_{ijk}A_{j}B_{k}$.  In addition, one has the useful
contraction property $\sum_{k}\epsilon_{ijk}\epsilon_{i'j'k}
=\delta_{ii'}\delta_{jj'}-\delta_{ij'}\delta_{ji'}$.  } and we have
expanded the triple vector product $-\m\vecpro(\m\vecpro\bfl)$ by
using the rule (\ref{BAC-CAB}). The parameter $\llg$ enables us to
deal with both equations simultaneously: to obtain the stochastic
Landau--Lifshitz--Gilbert equation (\ref{stollgeq}) we put $\llg=1$,
whereas the stochastic Landau--Lifshitz equation (\ref{stolleq}) is
recovered if $\llg=0$, since in this case $\bfl(t)$ only enters in the
precession term. Note that the $\diff_{ik}$ depend on $\m$ in both
cases, i.e., {\em the noise terms in the stochastic Landau--Lifshitz
(--Gilbert) equation are multiplicative}.

Next, on using $\partial\mi/\partial\mj=\delta_{ij}$, one first gets
\begin{equation}
\label{Gij:derivative:ll-llg}
\frac{\partial\diff_{ik}}{\partial\mj}
=
\gmr
\left[\epsilon_{ijk}
-\llg
\frac{\la}{\mm}
\left(
\delta_{ij}\mk
+\delta_{kj}\mi
\right)
\right]
\;,
\end{equation}
where the terms dependent on $\mm=(\sum_{i}\mi^{2})^{1/2}$ have not been
differentiated due to the conservation of the magnitude of $\m$. (One can indeed check
that differentiating those terms by using $\partial m/\partial \mj=\mj/\mm$ and
repeating the following calculations we arrive at the same final results.) Then,  on
taking
$\epsilon_{jjk}=0$ into account one finds
$\sum_{j}\partial\diff_{jk}/\partial\mj=-4\llg\gmr(\la/\mm)\mk$. From this result
and Eq.\ (\ref{G:ll-llg}) we get
$\sum_{k}\diff_{ik}
(\sum_{j}\partial\diff_{jk}/\partial\mj)=0$
by using
$\sum_{jk}\epsilon_{ijk}\mj\mk=0$
(due to the contraction of a symmetrical tensor with an antisymmetrical tensor)
and
$\sum_{k}(\mm^{2}\delta_{ik}-\mi\mk)\mk=0$.
Therefore, the second term on the right-hand side of the general Fokker--Planck
equation (\ref{fokkerplanck:langevin:n-dim:cont}) vanishes identically in this case. In
order to obtain the third term we need to calculate first
\begin{eqnarray*}
\lefteqn{
\frac{1}{\gmr^{2}}
\sum_{k}\diff_{ik}\diff_{jk}
}
&
&
\\
&
=
&
\sum_{k}
\bigg[
\sum_{r}\epsilon_{irk}\mr
+\llg
\frac{\la}{\mm} (\mm^{2}\delta_{ik}-\mi\mk)
\bigg]
\bigg[
\sum_{s}\epsilon_{jsk}\ms
+\llg
\frac{\la}{\mm}(\mm^{2}\delta_{jk}-\mj\mk)
\bigg]
\\
&
=
&
\sum_{rs} (\delta_{ij}\delta_{rs}-\delta_{is}\delta_{rj})
\mr\ms
\\
&
&
{}+
\llg
\frac{\la}{\mm}
\bigg(
\mm^{2}\sum_{r}
\underbrace{ (\epsilon_{irj}\mr
+
\overbrace{
\epsilon_{jri}
}^{-\epsilon_{irj}}
\mr)
}_{0}
-\mj\underbrace{
\sum_{kr}
\epsilon_{irk}\mr\mk
}_{0}
-\mi
\underbrace{
\sum_{ks}
\epsilon_{jsk}\ms\mk
}_{0}
\bigg)
\\
&
&
{}+
\llg
\left(\frac{\la}{\mm}\right)^{2}
\bigg[
\mm^{4}\delta_{ij}
-\mm^{2}(\mi\mj+\mj\mi)
+\mi\mj\sum_{k}\mk^{2}
\bigg]
\\
&
=
&
(1+\llg\la^{2})(\mm^{2}\delta_{ij}-\mi\mj)
\;,
\end{eqnarray*}
where we have taken into account that $\llg^{2}=\llg$ and employed the mentioned
contraction rule of $\epsilon_{ijk}$. Then, on introducing the {\em N\'{e}el time},
\begin{equation}
\label{neeltime:D}
\frac{1}{\tN}
=
2D\gmr^{2}(1+\llg\la^{2})
\;,
\end{equation}
which is the characteristic time of diffusion in the absence of potential (free-diffusion
time; see below), we get for the third term in
Eq.\ (\ref{fokkerplanck:langevin:n-dim:cont})
\begin{equation}
\label{fpe:3rd_term}
-D\sum_{jk}\,
\diff_{ik}\diff_{jk}
\frac{\partial\W}{\partial\mj}
=
\frac{1}{2\tN}
\bigg[
\m\vecpro
\bigg(
\m\vecpro\frac{\partial\W}{\partial\m}
\bigg)
\bigg]_{i}
\;.
\end{equation}

On introducing these results into Eq.\ (\ref{fokkerplanck:langevin:n-dim:cont}) one
finally arrives at the Fokker--Planck equation
\begin{equation}
\label{brownfpe}
\frac{\partial\W}{\partial t}
=
-\frac{\partial {}}{\partial\m}\cdot
\left[
\gmr\m\vecpro\Beff
-\gmr\frac{\la}{\mm}
\m\vecpro\left(\m\vecpro\Beff\right)
+\frac{1}{2\tN}\m\vecpro
\left(\m\vecpro\frac{\partial {}}{\partial\m}\right)
\right]
\W
\;,
\end{equation}
where $(\partial/\partial\m)\cdot$ stands for the divergence operator
[$(\partial/\partial\m)\cdot\vec{J}
=\sum_{i} (\partial J_{i}/\partial\mi)$].
Thus, the Fokker--Planck equations associated with the stochastic
Landau--Lifshitz--Gilbert equation (\ref{stollgeq}) and the stochastic Landau--Lifshitz
equation (\ref{stolleq}) are {\it both\/} given by Eq.\ (\ref{brownfpe}), the only
difference being the relation between the N\'{e}el time and the amplitude of the
fluctuating field:
\[
\frac{1}{\tN}
=
2D\gmr^{2}(1+\la^{2})
\;\; {\rm (LLG)}
\;,
\qquad
\frac{1}{\tNalt}
=
2D\gmr^{2}
\;\; {\rm (LL)}
\;.
\]
Equation (\ref{brownfpe}) is equivalent to the Fokker--Planck equation derived by
Brown (1963) (see below).

\paragraph{Stationary solution of the Fokker--Planck equation and comparison
between the stochastic models.}

In order to ensure that the stationary properties of the system, derived from the
Langevin equations (\ref{stollgeq}) or (\ref{stolleq}), coincide with the correct
thermal-equilibrium properties, the Fokker--Planck equation associated with these
Langevin equations is forced to have the Boltzmann distribution
\[
\Weq(\m)\propto\exp[-\beta\Hs(\m)]
\;,
\]
as stationary solution.

To do so, note first that, by means of $\Beff=-\partial\Hs/\partial\m$, one can write
$\partial\Weq/\partial\m$ as
\begin{equation}
\label{BeffWeq}
\frac{\partial\Weq}{\partial\m}
=
\beta\Beff\,\Weq
\;.
\end{equation}
From this result one can easily show that $\m\vecpro\Beff\,\Weq$ is divergenceless
(solenoidal).%
\footnote{
This result follows from the general one
\[
\frac{\partial {}}{\partial\m}
\cdot
\left(\m\vecpro\vec{A}\right)
=
\sum_{i}
\bigg(
\sum_{jk}\epsilon_{ijk}\mj\frac{\partial A_{k}}{\partial\mi}
\bigg)
=
-\m\cdot\left(\frac{\partial {}}{\partial\m}\vecpro\vec{A}\right)
\;,
\]
when applied to $\vec{A}=\Beff\,\Weq$, since $\Beff\,\Weq$ can be written as the
gradient of a scalar by Eq.\ (\ref{BeffWeq}) and, thus, its rotational is zero.
\qed } Therefore, on taking these results into account when introducing the Boltzmann
distribution in the Fokker--Planck equation (\ref{brownfpe}), one gets
\[
0
=
\frac{\partial\Weq}{\partial t}
=
-\frac{\partial {}}{\partial\m}\cdot
\left[
-\gmr\frac{\la}{\mm}
\m\vecpro\left(\m\vecpro\Beff\right)\Weq
+\frac{\beta}{2\tN}\m\vecpro
\left(\m\vecpro\Beff\right)
\Weq
\right]
\;.
\]
One then sees by inspection that, in order to have the Boltzmann distribution as
stationary solution of the Fokker--Planck equation (\ref{brownfpe}), it is sufficient to
put
\begin{equation}
\label{stationary_condition}
\gmr\frac{\la}{\mm}
=
\frac{\beta}{2\tN}
\;,
\end{equation}
from which one gets the following expression for the N\'{e}el time
\begin{equation}
\label{neeltime}
\tN
=
\frac{1}{\la}\frac{\mm}{2\gmr\T}
\;.
\end{equation}
Note that, since this result does not depend on the actual form of the Hamiltonian
$\Hs$, it also holds for assemblies of interacting magnetic moments.

Therefore, as the thermodynamical consistency of the approach determines $\tN$
completely, we arrive at the important result that, once that the consistency condition
is applied, {\em the Fokker--Planck equations associated with the stochastic
Landau--Lifshitz--Gilbert and stochastic Landau--Lifshitz equations result to be
identical}.%
\footnote{
Since the stochastic Gilbert equation (\ref{stogileq}) is equivalent to the stochastic
Landau--Lifshitz--Gilbert equation (\ref{stollgeq}) with $\gmr\to\gmr/(1+\la^{2})$,
the Fokker--Planck equation associated with the former is also given by Eq.\
(\ref{brownfpe}) with $\tN$ from (\ref{neeltime}) after substituting
$\gmr/(1+\la^{2})$ for $\gmr$. As $\tN^{-1}\propto\gmr$ this gives a global
time-scale factor.
}

As $\tN$ is related with the amplitude $D$ of the fluctuating field by different
expressions [Eq.\ (\ref{neeltime:D})], the only difference between the two stochastic
models lies in the relation among $D$, $\la$, and $T$, namely
\begin{equation}
\label{coeffdif:ll-llg}
D
=
\frac{\la}{1+\llg\la^{2}}\frac{\T}{\gmr\mm}
\;.
\end{equation}
Let us also write this result explicitly
\[
D_{{\rm LLG}}
=
\frac{\la}{1+\la^{2}}\frac{\T}{\gmr\mm}
\;,
\qquad
D_{{\rm LL}}
=
\la\frac{\T}{\gmr\mm}
\;,
\]
so that we can compare with Brown's (1963) result. He wrote the right-hand side of
the first of these equations as $(\eta/v)\T$, since he began with the Gilbert equation
[$\gmr\to\gmr/(1+\la^{2})$] and our $\la/\gmr\mm$ is equivalent to his $\eta/v$.

The above Einstein-type relations between the amplitude of the thermal-agitation field
and the temperature, via the damping coefficient, ensure that the proper
thermal-equilibrium properties are obtained from the stochastic Landau--Lifshitz
(--Gilbert) equation. They also ensure that the average dynamical properties
associated with each one of these stochastic models are identical with each other
(those properties are determined by the same Fokker--Planck equation), even though
the stochastic trajectories for a given realization of the fluctuating field
$\bfl(t)$ are in principle different.

Later on we shall integrate the stochastic Landau--Lifshitz--Gilbert equation
(\ref{stollgeq}) numerically. Nevertheless, the above considerations ensure that, if we
integrate the stochastic Landau--Lifshitz equation (\ref{stolleq}) instead, we shall
obtain the same results for the {\em averaged\/} quantities.

\paragraph{It\^{o} case.}

It is to be noted that the relations (\ref{coeffdif:ll-llg}) between the temperature and
the amplitude of the fluctuating field [or equivalently Eq.\ (\ref{neeltime})], being
derived from Brown's Fokker--Planck equation (\ref{brownfpe}), {\em pertain to the
Stratonovich stochastic calculus}. Indeed, after constructing the corresponding
Fokker--Planck equation by using the It\^{o} calculus, one finds that Eq.\
(\ref{neeltime}) does not ensure that the Boltzmann distribution is a solution of such
an equation. Let us prove this.

Let us first calculate the so-called {\em noise-induced\/} drift coefficient of the
Fokker--Planck equation, namely $D\sum_{jk}\,\diff_{jk}(\partial\diff_{ik}/\partial
y_{j})$, which is the extra term accompanying $\drift_{i}$ in Eq.\
(\ref{fokkerplanck:langevin:n-dim}). On introducing Eq.\ (\ref{G:ll-llg}) for $\diff_{ik}$
and the partial derivative (\ref{Gij:derivative:ll-llg}) in the definition of the
noise-induced drift, one finds
\begin{eqnarray*}
\frac{1}{\gmr^{2}}
\sum_{jk}\diff_{jk}\frac{\partial\diff_{ik}}{\partial\mj}
&
=
&
\sum_{\ell j}
\overbrace{
\Big(
\sum_{k}\epsilon_{j\ell k}\epsilon_{ijk}
\Big)
}^{\delta_{ji}\delta_{\ell j}-\delta_{jj}\delta_{\ell i}}
\ml
\\
&
&
-\llg
\left(\frac{\la}{\mm}\right)^{2}
\sum_{jk}(\mm^{2}\delta_{jk}-\mj\mk)
\left(\delta_{ij}\mk+\delta_{kj}\mi\right)
\\
&
=
&
\sum_{\ell}(\delta_{i\ell}-3\delta_{i\ell})\ml
-\llg
\left(\frac{\la}{\mm}\right)^{2}
\mi
\sum_{k}(\mm^{2}\delta_{kk}-\mk\mk)
\\
&
=
&
-2(1+\llg\la^{2})\mi
\;,
\end{eqnarray*}
where all the terms linear in $\la$ have cancelled out due to the contraction of
symmetrical tensors with antisymmetrical ones. Therefore, on using the unified
expression (\ref{neeltime:D}) for the N\'{e}el time, we can write the noise-induced drift
coefficient as
\begin{equation}
\label{noiseinduceddrift:ll-llg}
D\sum_{jk}\diff_{jk}\frac{\partial\diff_{ik}}{\partial\mj}
=
-\frac{1}{\tN}\mi
\;.
\end{equation}

The It\^{o} case of the Fokker--Planck equation is readily constructed by omitting the
noise-induced drift coefficient in Eq.\ (\ref{fokkerplanck:langevin:n-dim}). As Eq.\
(\ref{noiseinduceddrift:ll-llg}) shows, this term yields a contribution
$-\tN^{-1}\mi\W$ to the $i$th component of the current of probability $J_{i}$.
However, in the Stratonovich case, that contribution is cancelled by a term
$\tN^{-1}\mi\W$ originating from the second-order derivatives in the Fokker--Planck
equation [this is a restatement of the vanishing of the second term on the right-hand
side of the general Fokker--Planck equation (\ref{fokkerplanck:langevin:n-dim:cont})
for the stochastic Landau--Lifshitz (--Gilbert) equation]. Thus, the absence of the
noise-induced contribution in the It\^{o} equation yields a term $\tN^{-1}\mi\W$
added to the Stratonovich $J_{i}$. Therefore, the Fokker--Planck equation associated
with the stochastic Landau--Lifshitz (--Gilbert) equation when this is interpreted in
the It\^{o} sense, can be written as [cf.\ Eq.\ (\ref{brownfpe})]
\begin{eqnarray}
\label{brownfpe:ito}
\frac{\partial\W}{\partial t}
=
-\frac{\partial {}}{\partial\m}\cdot
&
\bigg[
&
\gmr\m\vecpro\Beff
-\gmr\frac{\la}{\mm}
\m\vecpro\left(\m\vecpro\Beff\right)
\nonumber
\\
&
& {}+
\frac{1}{\tN}\m
+\frac{1}{2\tN}\m\vecpro
\left(\m\vecpro\frac{\partial {}}{\partial\m}\right)
\bigg]
\W
\;.
\end{eqnarray}

Again, for the equilibrium distribution $\m\vecpro\Beff\,\Weq$ is divergenceless and,
if $\tN$ is given by Eq.\ (\ref{neeltime}), the second and fourth terms in the square
brackets of Eq.\ (\ref{brownfpe:ito}) cancel each other (by construction). Therefore,
the It\^{o} Fokker--Planck equation yields for $\W=\Weq$
\[
0
=
\frac{\partial {}}{\partial\m}
\cdot\big(\m\Weq\big)
=
\big(3+\beta\m\cdot\Beff\big)\Weq
\;,
\qquad
\mbox{(It\^{o} case)}
\]
which is not necessarily satisfied by a general form of the Boltzmann distribution
$\Weq(\m)$ (that is, by a general form of the Hamiltonian). The simplest example is
that of the dynamics in a constant potential. Then $\Beff=0$ and the equilibrium
distribution ---$\Weq(\m)$ uniform--- is not a solution of the It\^{o} case of the
Fokker--Planck equation. {\em Therefore, the stochastic Landau--Lifshitz (--Gilbert)
equation, when interpreted in the It\^{o} sense, does not yield the correct
thermal-equilibrium properties}.

We can give an even stronger argument against the interpretation of Eqs.\
(\ref{stollgeq}) and (\ref{stolleq}) as It\^{o} stochastic differential equations, based in
the non-conservation of the magnitude of the magnetic moment. The deterministic
counterpart of those equations [Eq.\ (\ref{detlleq:0})] yields $0=\m\cdot(\D\m/\D
t)=\half\D(\m^{2})/\D t$, so that the magnitude of $\m$ is preserved during the time
evolution.  Nevertheless, when passing from ordinary to stochastic differential
equations, specific rules of calculus (integration and differentiation) are required. In
the context of the Stratonovich calculus, such rules are {\em formally\/} identical with
the rules of the ordinary calculus. Therefore, $0=\m\cdot(\D\m/\D t)$, which always
follows from Eqs.\ (\ref{stollgeq}) and (\ref{stolleq}), also entails $\D(\m^{2})/\D t=0$.
However, when using the specific rules of differentiation of the It\^{o} calculus, one
finds that $\m\cdot(\D\m/\D t)\neq\half\D(\m^{2})/\D t$ for those equations, which
therefore {\em do not\/} conserve the magnitude of $\m$.%
\footnote{
This can be demonstrated by using the Stratonovich {\em equivalents\/} of
Eqs.\ (\ref{stollgeq}) and (\ref{stolleq}) {\em when they are interpreted as It\^{o}
equations}. Those Stratonovich equivalent equations are obtained by augmenting the
(now It\^{o}) Eqs.\ (\ref{stollgeq}) and (\ref{stolleq}) by $\tN^{-1}\m\W$, so that
the stated result directly follows from the application the ordinary rules of
differentiation to the resulting equations.
}

\paragraph{Fokker--Planck equation in spherical coordinates.}

For future use, let us write the Fokker--Planck equation (\ref{brownfpe}) in a
spherical coordinate system, as was originally presented by Brown (1963).

First, on using $\gmr\la/\mm=\beta/2\tN$ [Eq.\ (\ref{stationary_condition})], the
Fokker--Planck equation (\ref{brownfpe}) can be written as
\begin{equation}
\label{brownfpe:2}
\frac{\partial\W}{\partial t}
=
-\frac{\partial {}}{\partial\m}\cdot
\left\{
\gmr\m\vecpro\Beff
-\frac{1}{2\tN}
\m
\vecpro
\left[
\m\vecpro
\left(\beta\Beff-\frac{\partial {}}{\partial\m}\right)
\right]
\right\}
\W
\;.
\end{equation}
Then, on introducing the dimensionless effective field
$\vec{\xi}_{\eff}=\beta\mm\Beff$ and using again the expression
(\ref{stationary_condition}) for $\tN$, we can rewrite Eq.\ (\ref{brownfpe:2}) in the
form
\begin{equation}
\label{brownfpe:3}
2\tN\frac{\partial\W}{\partial t}
=
-\frac{\partial
{}}{\partial\m}\cdot
\left\{
\frac{1}{\la}\m\vecpro\vec{\xi}_{\eff}
-\frac{1}{\mm}\m\vecpro
\left[
\m\vecpro\left(\vec{\xi}_{\eff}
-\mm\frac{\partial {}}{\partial\m}\right)
\right]
\right\}
\W
\;.
\end{equation}
On using now the formulae for the gradient and divergence operators in spherical
coordinates ($\vec{r}=\m$)
\begin{eqnarray}
\label{gradient}
\frac{\partial u}{\partial\vec{r}}
&
=
&
\hat{r}\frac{\partial u}{\partial r}
+\hat{\vartheta}\frac{1}{r}\frac{\partial u}{\partial\vartheta}
+\hat{\varphi}\frac{1}{r\sin\vartheta}\frac{\partial u}{\partial\varphi}
\;,
\\
\label{divergence}
\frac{\partial {}}{\partial\vec{r}}\cdot\vec{A}
&
=
&
\frac{1}{r^{2}}\frac{\partial {}}{\partial r}(r^{2}A_{r})
+\frac{1}{r\sin\vartheta}
\frac{\partial {}}{\partial\vartheta}(\sin\vartheta A_{\vartheta})
+\frac{1}{r\sin\vartheta}
\frac{\partial A_{\varphi}}{\partial\varphi}
\;,
\end{eqnarray}
along with the result $\m\vecpro\vec{A}
=\mm\big(-A_{\varphi}\hat{\vartheta}+A_{\vartheta}\hat{\varphi}\big)$, one can
write Eq.\ (\ref{brownfpe:3}) in a spherical coordinate system as
\begin{equation}
\label{brownfpe:spherical}
2\tN\frac{\partial\W}{\partial t}
=
-\frac{1}{\sin\vartheta}
\left[
\frac{\partial {}}{\partial\vartheta} (\sin\vartheta\widetilde{J}_{\vartheta})
+\frac{\partial {}}{\partial\varphi}(\widetilde{J}_{\varphi})
\right]
\;,
\end{equation}
where the spherical components of the {\em reduced\/} current of probability
[$\widetilde{J}_{i}=(2\tN/\mm)J_{i}$] are given by
\begin{eqnarray}
\label{probability_current:ll-llg:theta:2}
\widetilde{J}_{\vartheta}
&
=
&
-
\left[
\frac{1}{\T}
\left(
\frac{\partial\Hs}{\partial\vartheta}
-\frac{1}{\la}
\frac{1}{\sin\vartheta}\frac{\partial\Hs}{\partial\varphi}
\right)
\W
+\frac{\partial\W}{\partial\vartheta}
\right]
\;,
\\
\label{probability_current:ll-llg:phi:2}
\widetilde{J}_{\varphi}
&
=
&
-
\left[
\frac{1}{\T}
\left(
\frac{1}{\la}\frac{\partial\Hs}{\partial\vartheta}
+\frac{1}{\sin\vartheta}\frac{\partial\Hs}{\partial\varphi}
\right)
\W
+\frac{1}{\sin\vartheta}\frac{\partial\W}{\partial\varphi}
\right]
\;.
\end{eqnarray}
To get these expressions we have also taken into account the definition (\ref{beff:def})
of the effective field in terms of the Hamiltonian $\Hs(\m)$, which, together with Eq.\
(\ref{gradient}), has allowed us to write the components of $\vec{\xi}_{\eff}$ as
\begin{equation}
\label{Beff:dimless:components}
\xi_{\eff,\vartheta}
=
-\frac{1}{\T}\frac{\partial\Hs}{\partial\vartheta}
\;,
\qquad
\xi_{\eff,\varphi}
=
-\frac{1}{\T}\frac{1}{\sin\vartheta}\frac{\partial\Hs}{\partial\varphi}
\;.
\end{equation}
Finally, when Eqs.\ (\ref{probability_current:ll-llg:theta:2}) and
(\ref{probability_current:ll-llg:phi:2}) are introduced in (\ref{brownfpe:spherical}),
Brown's Fokker--Planck equation emerges in its original form (1963).

\paragraph{The axially symmetric Fokker--Planck equation as a Sturm--Liouville
problem.}

In an axially symmetric situation, that is, for $B_{\eff,\varphi}=0$ and
$B_{\eff,\vartheta}=B_{\eff,\vartheta}(\vartheta)$, and restricting ourselves to
solutions with axial symmetry $\partial\W/\partial\varphi\equiv0$ (the ones of
interest when, for example, determining the steady-state solution in the presence of a
longitudinal probing field), the Fokker--Planck equation (\ref{brownfpe:spherical})
reduces to
\begin{equation}
\label{brownfpe:axi}
2\tN\frac{\partial\W}{\partial t}
=
-\frac{1}{\sin\vartheta}
\frac{\partial {}}{\partial\vartheta}
\left[
\sin\vartheta
\left(
-\beta\frac{\partial\Hs}{\partial\vartheta}\W
-\frac{\partial\W}{\partial\vartheta}
\right)
\right]
\;.
\end{equation}
Then, if we introduce the substitution $z=\cos\vartheta$ and use the
relation $(\partial f/\partial\vartheta) =-\sin\vartheta(\partial
f/\partial z)$, the axially symmetric Fokker--Planck equation
(\ref{brownfpe:axi}) can be written as
\begin{equation}
\label{brownfpe:axi:2}
2\tN\frac{\partial\W}{\partial t}
=
\frac{\partial {}}{\partial z}
\left[
\Omega(z)
\left(\frac{\partial\W}{\partial z}+\beta\Hs'\W\right)
\right]
\;,
\end{equation}
where we have used the shorthand $\Omega(z)=1-z^{2}$ and the prime denotes
differentiation with respect to $z$. Note that, in this axially symmetric case, the
gyromagnetic terms [those multiplied by $\la^{-1}$ in Eq.\ (\ref{brownfpe:3}) or in
Eqs.\ (\ref{probability_current:ll-llg:theta:2}) and
(\ref{probability_current:ll-llg:phi:2})] are absent from the Fokker--Planck equation.
This entails that the effect of the damping parameter $\la$ on the averaged quantities
enters via the N\'{e}el time (\ref{neeltime}) only. Note that this no longer holds in
non-axially symmetric situations (for example, in the presence of a transverse field).

The current of probability is defined by writing the Fokker--Planck equation
(\ref{brownfpe:axi:2}) as a continuity equation for the probability distribution, namely
$2\tN(\partial\W/\partial t)=-(\partial\widetilde{J}_{z}/\partial z)$, whence
\begin{equation}
\label{jz}
\widetilde{J}_{z}
=
-\Omega(z)
\left(\frac{\partial\W}{\partial z}+\beta\Hs'\W
\right)
\;.
\end{equation}
Note that this expression can also be obtained from Eq.\
(\ref{probability_current:ll-llg:theta:2}) for $\widetilde{J}_{\vartheta}$, by using
$\widetilde{J}_{z}=-\widetilde{J}_{\vartheta}\sin\vartheta$.

On the other hand, by assuming a solution of Eq.\ (\ref{brownfpe:axi:2}) of the form
$\W(z,t)=T(t)F(z)$ (separation of variables), one gets $T(t)\propto\exp(-\Lambda
t)$, while $F(z)$ then satisfies
\begin{equation}
\label{brownfpe:sturm-liouville:1}
\frac{\D {}}{\D z}
\left\{
\Omega(z)e^{-\beta\Hs(z)}
\frac{\D {}}{\D z}
\left[ e^{\beta\Hs(z)}F(z)
\right]
\right\}
=
-(2\tN\Lambda) F(z)
\;,
\end{equation}
for the writing of which we have used the identity
\begin{equation}
\label{identity:SL} e^{-\beta\Hs(z)}
\frac{\D {}}{\D z}
\left[ e^{\beta\Hs(z)}F(z)
\right]
=
\frac{\D F}{\D z}+\beta\Hs'F
\;.
\end{equation}
Further, on introducing the function $\phi(z)=e^{\beta\Hs(z)}F(z)$, Eq.\
(\ref{brownfpe:sturm-liouville:1}) can equivalently be written as
\begin{equation}
\label{brownfpe:sturm-liouville:2}
\frac{\D {}}{\D z}
\left[
\Omega(z)e^{-\beta\Hs(z)}
\frac{\D\phi}{\D z}
\right]
=
-(2\tN\Lambda)e^{-\beta\Hs(z)}\phi(z)
\;.
\end{equation}
Therefore, to solve the Fokker--Planck equation in the axially symmetric case is
transformed into the Sturm--Liou\-ville problem of finding the eigenvalues
$\Lambda_{k}$, and eigenfunctions $\phi_{k}(z)$ of Eq.\
(\ref{brownfpe:sturm-liouville:2}).%
\footnote{
One can also define the more customary dimensionless eigenvalues by
$\lambda_{k}=2\tN\Lambda_{k}$.
}

In order to prove that the problem defined by Eq.\ (\ref{brownfpe:sturm-liouville:2})
is in fact a Sturm--Liou\-ville problem, note first that
$\Omega(z)e^{-\beta\Hs(z)}\neq0$ inside the relevant interval $(-1,1)$. The same
holds for the function $e^{-\beta\Hs(z)}$ multiplying $\phi(z)$ on the right-hand side
of Eq.\ (\ref{brownfpe:sturm-liouville:2}). In addition, the differential operator on the
left-hand side is {\em self-adjoint}, since, when expanding it, the coefficient of
$\D\phi/\D z$ is equal to the derivative of the coefficient of $\D^{2}\phi/\D z^{2}$.
This completes the proof of that Eq.\ (\ref{brownfpe:sturm-liouville:2}) defines a
Sturm--Liou\-ville problem.

On the other hand, we must also check that the common boundary condition in
Sturm--Liou\-ville problems (see, for example, Arfken, 1985, p.~503)
\begin{equation}
\label{sturm-liouville:boundary}
\left. p(z)\phi_{1}^{\ast}(z)\frac{\D\phi_{2}}{\D z}
\right|_{z=-1}
=
\left. p(z)\phi_{1}^{\ast}(z)\frac{\D\phi_{2}}{\D z}
\right|_{z=1}
\;,
\end{equation}
is satisfied. Here, $\phi_{1}(z)$ and $\phi_{2}(z)$ are two solutions of the differential
equation being considered, $(\,)^{\ast}$ stands for complex conjugation, and
$p(z)=\Omega(z)e^{-\beta\Hs(z)}$ for the Sturm--Liou\-ville problem
(\ref{brownfpe:sturm-liouville:2}). The proof is based on the fact that
$\Omega(z)e^{-\beta\Hs(z)}\D\phi/\D z$ is proportional to $J_{z}$, as can be
checked by using the definition (\ref{jz}) and the identity (\ref{identity:SL}). However,
the current of probability is tangent to the unit sphere, so that $J_{z}$
must vanish at the poles. Therefore $J_{z}|_{z=\pm1}=0$, from which Eq.\
(\ref{sturm-liouville:boundary}) follows, the two sides of the equation being equal to
zero.

The property (\ref{sturm-liouville:boundary}) is very important since from
self-adjointness plus that boundary condition it follows the {\em Hermitian\/}
character of the differential operator in the Sturm--Liou\-ville problem. Hermitian
operators have the following three important properties:
\begin{enumerate}
\item\label{hermitian:real} The eigenvalues $\Lambda_{k}$ are real.
\item\label{hermitian:orthogonal} The eigenfunctions $\phi_{k}(z)$ are orthogonal
with respect to a suitably chosen scalar product. \item\label{hermitian:completeness}
The eigenfunctions $\phi_{k}(z)$ [and therefore the $F_{k}(z)$] form a complete set.
\end{enumerate}
Therefore, by using the completeness property
\ref{hermitian:completeness}, the general solution of the Fokker--Planck equation,
$\W(z,t)$, can be expanded in terms of the (basis) functions
$F_{k}(z)=e^{-\beta\Hs(z)}\phi_{k}(z)$ as
\begin{equation}
\label{Wneq:gral}
\W(z,t)
=
\Z^{-1}\exp[-\beta\Hs(z)]+
\sum_{k\geq1}c_{k}F_{k}(z)\exp(-\Lambda_{k}t)
\;,
\end{equation}
where $\Z^{-1}\exp[-\beta\Hs(z)]$ is the equilibrium ($t\to\infty$) solution (associated
with the eigenvalue $\Lambda_{0}=0$) and the coefficients of the expansion $c_{k}$
depend on the ``initial conditions" (initial probability distribution).

In general, the eigenvalues and eigenfunctions of the Sturm--Liou\-ville problem
discussed above must be computed by means of numerical techniques. However,
analytical results can be obtained for certain quantities without solving the full
Sturm--Liou\-ville problem (see below).

\subsubsection{Equations for the averages of the magnetic moment}

Let us now consider the dynamical equations for the averages of the magnetic moment
with respect to the non-equilibrium probability distribution $\W(\m,t)$. (As these
equations involve averaged quantities, they will be identical for the stochastic
Landau--Lifshitz--Gilbert and stochastic Landau--Lifshitz models.)

The dynamical equations for the first two moments of a stochastic variable
$\ym=(y_{1},\ldots,y_{n})$ that obeys the Fokker--Planck equation
\[
\frac{\partial \W}{\partial t}
=
-\sum_{i}\frac{\partial {}}{\partial y_{i}}
\left[
a_{i}^{(1)}(\ym,t)\W
\right]
+\frac{1}{2}
\sum_{ij}
\frac{\partial^{2}}{\partial y_{i}\partial y_{j}}
\left[
a_{ij}^{(2)}(\ym,t)\W
\right]
\;,
\]
are given by (see van Kampen, 1981, p.~130)
\begin{eqnarray}
\label{avdyneq:n-dim}
\frac{\D {}}{\D t}
\llangle y_{i}\rrangle
&
=
&
\blangle a_{i}^{(1)}(\ym,t)
\brangle
\\
\label{avdyneq:2nd:n-dim}
\frac{\D {}}{\D t}
\llangle y_{i}y_{j}\rrangle
&
=
&
\blangle a_{ij}^{(2)}(\ym,t)
\brangle
+\blangle y_{i}a_{j}^{(1)}(\ym,t)
\brangle
+\blangle y_{j}a_{i}^{(1)}(\ym,t)
\brangle
\;.
\end{eqnarray}
On comparing with the Fokker--Planck equation (\ref{fokkerplanck:langevin:n-dim}),
taking Eqs.\ (\ref{F:ll-llg}) and (\ref{G:ll-llg}) into account, and using Eq.\
(\ref{noiseinduceddrift:ll-llg}) for the noise-induced drift coefficient, we get for the
functions $a_{i}^{(1)}$ and $a_{ij}^{(2)}$ associated with the stochastic Landau--Lifshitz
(--Gilbert) equation
\begin{eqnarray*} a_{i}^{(1)}(\m,t)
&
=
&
\gmr
\left[
\m\vecpro\Beff
-\frac{\la}{\mm}
\m\vecpro\left(\m\vecpro\Beff\right)
\right]_{i}
-\frac{1}{\tN}\mi
\;,
\\
a_{ij}^{(2)}(\m,t)
&
=
&
\frac{1}{\tN}
\left(\mm^{2}\delta_{ij}-\mi\mj\right)
\;.
\end{eqnarray*}
Thus, the dynamical equation for the first moment
$\langle\mi\rangle(t)=\int_{|\m|=\mm}\!\D^{3}{\m}\,\W(\m,t)\mi$ reads
\begin{equation}
\label{avdyneq:1st:ll-llg}
\frac{\D {}}{\D t}
\blangle\m\brangle
=
\gmr\llangle\m\vecpro\Beff\rrangle
-\gmr\frac{\la}{\mm}
\llangle\m\vecpro\left(\m\vecpro\Beff\right)\rrangle
-\frac{1}{\tN}\blangle\m\brangle
\;,
\end{equation}
where the results for the Cartesian components have been gathered in vector form.
Note that the term $-\llangle\m\rrangle/\tN$ is analogous to the relaxation term in a
Bloch-type equation (Garanin, Ishchenko and Panina, 1990). Analogously, for the
second-order moments $\langle\mi\mj\rangle(t)$ one finds
\begin{eqnarray}
\label{avdyneq:2nd:ll-llg}
\frac{\D {}}{\D t}
\llangle\mi\mj\rrangle
&
=
&
-\frac{3}{2\tN}
\left(\llangle\mi\mj\rrangle-\frac{1}{3}\mm^{2}\delta_{ij}\right)
\nonumber
\\
&
& {}+
\gmr
\Big\langle{
\mi\left(\m\vecpro\Beff\right)_{j}
\Big\rangle}
-\gmr\frac{\la}{\mm}
\Big\langle{
\mi
\left[
\m\vecpro\left(\m\vecpro\Beff\right)
\right]_{j}
\Big\rangle}
\qquad
\nonumber
\\
&
& {}
+i\leftrightarrow j
\;,
\end{eqnarray}
where $i\leftrightarrow j$ stands for the interchange in the entire previous expression
of the subscripts $i$ and $j$.

Equations (\ref{avdyneq:1st:ll-llg}) and (\ref{avdyneq:2nd:ll-llg}) show that, for a
general form of the Hamiltonian, no closed equation exists for the time evolution of the
averages of the magnetic moment. For instance, even if $\Beff$ does not depend on
$\m$, for example, for $\Beff=\B$, the Landau--Lifshitz-type relaxation term
introduces $\langle\mi\mj\rangle(t)$ in the equation (\ref{avdyneq:1st:ll-llg}) for
$\langle\mi\rangle(t)$. Therefore, an additional differential equation for
$\langle\mi\mj\rangle(t)$ is required [i.e., Eq.\ (\ref{avdyneq:2nd:ll-llg})], however
that equation involves $\langle\mi\mj\mk\rangle(t)$, and so on. {\em The absence of
such a closed dynamical equation is a major source of mathematical difficulties in the
theoretical study of the dynamical properties of classical spins}.

\paragraph*{Free-diffusion case.}

A situation in which the equations for the averages are not coupled and can in addition
be explicitly solved, is that where the Hamiltonian is constant (independent of $\m$).
Then, one has $\Beff=0$ so that the equations for the first two moments reduce to
\begin{equation}
\label{avdyneq:ll-llg:free}
\frac{\D {}}{\D t}
\llangle\mi\rrangle
=
-\frac{1}{\tN}\llangle\mi\rrangle
\;,
\qquad
\frac{\D {}}{\D t}
\llangle\mi\mj\rrangle
=
-\frac{3}{\tN}
\left(\llangle\mi\mj\rrangle-\frac{1}{3}\mm^{2}\delta_{ij}\right)
\;.
\end{equation}
Note that, because $\tN^{-1}\propto\T$ [Eq.\ (\ref{neeltime})], this apparently
academic case can be a reasonable approximation for sufficiently high temperatures,
where the terms multiplied by $\tN^{-1}$ in Eqs.\ (\ref{avdyneq:1st:ll-llg}) (the
mentioned Bloch-type term) and (\ref{avdyneq:2nd:ll-llg}) dominate the remaining
terms.

The solution for the first moment is
\begin{equation}
\label{avdyneq:1st:ll-llg:free:solution}
\llangle\mi\rrangle\!(t)
=
\llangle\mi\rrangle\!(t_{0})e^{-(t-t_{0})/\tN}
\;,
\end{equation}
which justifies to call the characteristic time constant $\tN$ the {\em free-diffusion\/}
time. Similarly, on using $\D\llangle\mi\mj\rrangle/\D t
=\D(\llangle\mi\mj\rrangle-\frac{1}{3}\mm^{2}\delta_{ij})/\D t$, one gets for the
second-order moments
\begin{equation}
\label{avdyneq:2nd:ll-llg:free:solution}
\llangle\mi\mj\rrangle\!(t)
=
\frac{1}{3}\mm^{2}\delta_{ij}
+
\left[
\llangle\mi\mj\rrangle\!(t_{0})-\frac{1}{3}\mm^{2}\delta_{ij}
\right] e^{-3(t-t_{0})/\tN}
\;.
\end{equation}
For $(t-t_{0})\gg\tN$, one therefore finds $\langle\mi\rangle(t)\to0$ and
$\langle\mi\mj\rangle(t)\to\frac{1}{3}\mm^{2}\delta_{ij}$. Thus, the initial
correlations between different components of the magnetic moment are lost for long
times, while $\langle\mi^{2}\rangle(t)\to\frac{1}{3}\mm^{2}$, $\forall i$ (random
distribution of $\m$) as it should for the diffusion in a constant orientational potential
or at very high temperatures. Note finally that these natural results are not obtained
when one interprets the stochastic Landau--Lifshitz (--Gilbert) equation {\em \`{a}
la\/} It\^{o}.

\subsubsection{Relaxation times}

We shall finally review various expressions for the relaxation times
of independent classical magnetic moments in the context of the
Brown--Kubo--Hashitsume stochastic model.

\paragraph{Longitudinal relaxation time.}

The longitudinal relaxation time is associated with the response to a
field applied along the anisotropy axis. Therefore, the very
definition of this quantity requires a previous discussion of the
relaxation under such conditions.

Let us assume that the Hamiltonian $\Hs$ has uniaxial symmetry, so
that the transformation discussed above of the Fokker--Planck equation
into a Sturm--Liouville problem holds. Let us also assume that $\Hs$
contains, among other terms, a (longitudinal) Zeeman term
$-\beta\Hs_{{\rm Zeeman}}=\beta(\mz B)=z\xi$, where $\xi=\beta\mm B$
is the customary dimensionless magnetic field parameter. By averaging
$\mz(t)$ with respect to the non-equilibrium probability distribution
(\ref{Wneq:gral}), the relaxation curve, after a sudden infinitesimal
change on the applied field $B$ by $\dBo$ at $t=0$, reads
\begin{equation}
\label{mrelax}
\llangle\mz(\infty)\rrangle-\llangle\mz(t)\rrangle
=
\mu_{0}^{-1}\dBo\chi_{\|}\sum_{k\geq1}a_{k}\exp(-\Lambda_{k}t)
\;.
\end{equation}
Here $\chi_{\|}=\mu_{0}\partial\llangle\mz\rrangle_{\eq}/\partial B$
is the longitudinal equilibrium susceptibility
[$\langle\cdot\rangle_{\eq}$ denotes the thermal average in the
absence of the perturbing field $\dBo$, i.e., with respect to the
initial distribution $\Weq=\Z_{0}^{-1}\exp(-\beta\Hs_{0})$].%
\footnote{ We omit the subscript $\|$ in the equilibrium distribution
function and in the corresponding partition function.  } In Eq.\
(\ref{mrelax}) the $\Lambda_{k}$ are the eigenvalues of the associated
Sturm--Liouville problem and the $a_{k}$ are the corresponding
amplitudes, which are related with the constants $c_{k}$ of Eq.\
(\ref{Wneq:gral}) and also involve integrals of the form
$\int_{-1}^{1}\!\D{z}\,F_{k}(z)z$. Those amplitudes, by construction,
obey the sum rule ${\sum_{k\geq1}a_{k}}=1$, as can be seen by
considering that at $t=0$ the system was in thermal equilibrium, so
that $\mu_{0}\llangle\mz(\infty)-\mz(0)\rrangle=\dBo\chi_{\|}$.

The eigenvalues are usually sorted in increasing order
$0=\Lambda_{0}<\Lambda_{1}\leq\Lambda_{2}\cdots$. The first
non-vanishing eigenvalue, $\Lamone$, is commonly associated with the
inter-potential-well dynamics, while the information about the
intra-pot\-ential-well relaxation appears in the higher-order
eigenvalues $\Lambda_{k}$, $k\geq2$. In some cases, however, $\Lamone$
corresponds to a ``long-lived" mode and characterizes reasonably well
the relaxation (except for its earliest stages).

On defining the longitudinal relaxation time as $\tlo=\Lamone^{-1}$,
Brown (1963) derived the approximate results
\begin{equation}
\label{browntau}
\tlo\simeq\left\{
\begin{array}{lr}
\tN
\left[
  1-\frac{2}{5}\s+\frac{48}{875}\s^{2}\left(1+\frac{175}{24}h^{2}\right)
\right]^{-1}
\;,
&
\s\ll1
\\
\tN
\frac{\sqrt{\pi}}{2}
\s^{-3/2}
{\displaystyle
\frac
{\exp\left[\s(1+h^{2})\right]}
{(1-h^{2})(\cosh\xi-h\sinh\xi)}}
\;,
&
\s\gg1
\end{array}
\right.
\;,
\end{equation}
where $\s=Kv/\T$ is the dimensionless barrier-height parameter,
$h=\xi/2\s$, and $\tN$ is the N\'{e}el time (\ref{neeltime}). To get
these formulae Brown solved the Fokker--Planck equation perturbatively
in the low potential-barrier case and with the use of Kramers
transition-state method in the high-barrier limit.

Cregg, Crothers and Wickstead (1994) {\em proposed\/} a simple
expression for $\Lamone$ that is remarkably close to the exact
$\Lamone$ in the entire $\s$ range.  It is essentially a formula that
interpolates between the above limiting results of Brown and reads
($\tlo^{-1}=\Lamone$)
\begin{eqnarray}
\label{creggtau}
\tlo^{-1}
&
\simeq
&
\tN^{-1}
\half(1-h^{2})
\left(\frac{2}{\sqrt{\pi}}\frac{\s^{3/2}}{1+1/\s}+\s 2^{-\s}\right)
\nonumber
\\
&
& {}\times
\left\{
\frac{1-h}{\exp\left[\s(1-h)^{2}\right]-1}
+\frac{1+h}{\exp\left[\s(1-h)^{2}\right]-1}
\right\}
\;.
\end{eqnarray}

Nevertheless, when the relaxation comprises different decay modes, a
more useful characterization of the thermo-activation rate is provided
by the {\em integral relaxation time}, $\tint$, which is in general
defined as the area under the relaxation curve (normalized at $t=0$)
after a sudden infinitesimal change at $t=0$ of the external control
parameter. A general expression for the integral relaxation time
associated with any one-dimensional Fokker--Planck equation was
obtained by Jung and Risken (1985). Moro and Nordio (1985), in the
context of the thermo-activation phenomena in chemical-physics
problems, also derived a similar expression.

In the context of the Brown--Kubo--Hashitsume model for classical
spins, $\tint$ was calculated for systems with uniaxial anisotropy in
a longitudinal magnetic field by Garanin, Ishchenko, and Panina
(1990). Here, the relaxing quantity is the average magnetic moment
$\langle\mz(t)\rangle$, and the external control parameter is the
magnetic field. Thus, the above general definition reduces in this
case to
\begin{equation}
\label{tauint:def}
\tint
=
\int_{0}^{\infty}\!\!\D{t}\,
\frac
{\llangle\mz(\infty)\rrangle-\llangle\mz(t)\rrangle}
{\llangle\mz(\infty)\rrangle-\llangle\mz(0)\rrangle}
\;.
\end{equation}
For a single exponential decay, i.e.,
$[\llangle\mz(\infty)\rrangle-\llangle\mz(t)\rrangle]
\propto\exp(-t/\tau)$, the above definition indeed yields
$\tint=\tau$, whereas for a multi-exponential decay, $\tint$ is given
by the average of the corresponding relaxation times weighted by the
associated amplitudes. Indeed, when
$\langle\mz(\infty)\rangle-\langle\mz(t)\rangle$ from Eq.\
(\ref{mrelax}) is substituted into the above definition, $\tint$
emerges in the form
\begin{equation}
\label{tauint-lambda}
\tint
=
\frac{1}{\sum_{k\geq1}a_{k}}
\sum_{k\geq1}a_{k}\int_{0}^{\infty}\!\!\D{t}\,
\exp(-\Lambda_{k}t)
=
\sum_{k\geq1}a_{k}\Lambda_{k}^{-1}
\;,
\end{equation}
where we have used the sum rule $\sum_{k\geq1}a_{k}=1$.

In order to calculate $\tint$ without finding the eigenvalues and
amplitudes of the Sturm--Liou\-ville problem, Garanin, Ishchenko, and
Panina (1990) used the relation between $\tint$ and the low-frequency
longitudinal susceptibility to get (see also Garanin, 1996, and
Appendix \ref{app:taus})
\begin{equation}
\label{tauint}
\tint
=
\frac{2\tN}{\partial\langle z\rangle_{\eq}/\partial\xi}
\int_{-1}^{1}\frac{\D z}{1-z^{2}}\Phi(z)^{2}/\Weq(z)
\;,
\end{equation}
where the function $\Phi(z)$ is given by
\begin{equation}
\label{Phi}
\Phi(z)
=
\int_{-1}^{z}\!\!\D{z_{1}}\,
\Weq(z_{1})
\left[
\llangle z\rrangle_{\eq}-z_{1}
\right]
\;.
\end{equation}
Equation (\ref{tauint}), which is valid for {\em any\/} axially
symmetric Hamiltonian, can readily be computed by means of the
numerical integration of a double definite integral. Moreover,
explicit expressions for $\Phi(z)$ can be derived for particular forms
of the Hamiltonian (see Appendix \ref{app:taus}).

In the absence of a constant magnetic field (unbiased case), the
integral relaxation time yields the results for $\Lamone^{-1}$ of
Brown in the appropriate limiting cases (Garanin, Ishchenko and
Panina, 1990; Garanin, 1996). However, $\tint$ depends on the whole
set of eigenvalues $\Lambda_{k}$, and is therefore more informative
than $\Lamone$. Indeed, in the presence of a bias field, the
higher-order modes can make a substantial contribution to the
relaxation, and $\Lamone^{-1}$ can largely (exponentially) deviate
from $\tint$ in the low-temperature region (Coffey, Crothers, Kalmykov
and Waldron, 1995{\em a}; Garanin, 1996). Besides, in contrast to
$\Lamone$, the integral relaxation time is, by its very definition, a
directly mensurable quantity (for a comprehensive review including the
comparison of different definitions and methods for the calculation of
relaxation times, see Coffey, 1998).

\paragraph{Transverse relaxation time.}

The formula usually employed for the transverse relaxation time is that yielded by
the effective eigenvalue method (see, for example, Coffey, Kalmykov and Massawe,
1993)
\begin{equation}
\label{effeigentau:bias}
\ttr^{{\rm od}}
=
2\tN\frac{1-S_{2}(\s,\xi)}{2+S_{2}(\s,\xi)}
\;,
\end{equation}
where $S_{2}$ is the average of the second Legendre polynomial with respect to the
thermal-equilibrium distribution [Eq.\ (\ref{Sn})]. This equation, although valid for
any axially symmetric Hamiltonian, does not take gyromagnetic effects into account
[Eq.\ (\ref{effeigentau:bias}) only holds in the {\it overdamped\/} case,
$\la\gg1$].

On noting that from Eq.\ (\ref{F-S}) one gets the relation
\[
\legunb
=
\frac{1}{2}\left(3\frac{\F'}{\F}-1\right)
\;,
\]
between $\legunb\equiv S_{2}(\s,\xi)|_{\xi=0}$ and $\F'/\F$, one sees that Eq.\
(\ref{effeigentau:bias}) reduces in the unbiased case to
\begin{equation}
\label{effeigentau}
\ttr^{{\rm od}}|_{\xi=0}
=
2\tN\frac{1-\F'/\F}{1+\F'/\F}
\;.
\end{equation}
Now, if we employ the small and large $\s$ approximations for $\F'/\F$ (see Appendix
\ref{app:F}), we find
\[
\frac{1-\F'/\F}{1+\F'/\F}
\simeq
\frac{1/\s}{2-1/\s}
\quad (\s\gg1)
\;,
\qquad
\frac{1-\F'/\F}{1+\F'/\F}
\simeq
\frac{2-4\s/15}{4+4\s/15}
\quad(\s\ll1)
\;,
\]
whence one gets the limit behaviors of $\ttr^{{\rm od}}|_{\xi=0}$
\begin{equation}
\label{ttr:limits}
\ttr^{{\rm od}}|_{\xi=0}
\longrightarrow
\left\{
\begin{array}{cl}
1/(\la\gBK)
&
\mbox{as~} T\to0
\\
0
&
\mbox{as~} T\to\infty
\end{array}
\right.
\;.
\end{equation}
Thus, as it should, $\ttr^{{\rm od}}|_{\xi=0}$ goes to zero at high temperatures,
whereas it tends to the constant deterministic result $\tK$ for $T\to0$ [Eq.
(\ref{tauK})].

\begin{figure}[t!]
\vspace{-3.ex}
\eps{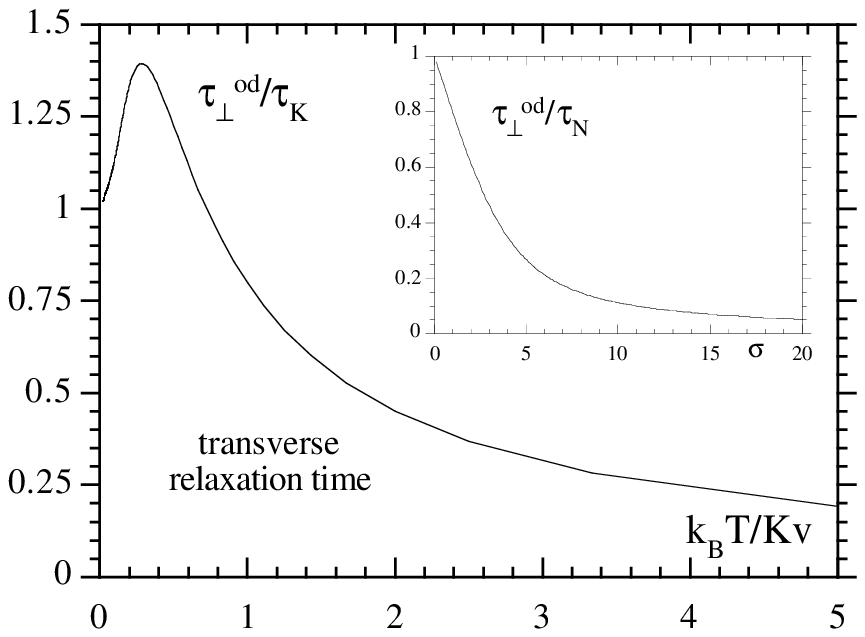}{0.7}
\vspace{-3.ex}
\caption[]
{
Transverse relaxation time, $\ttr^{{\rm od}}|_{\xi=0}$ [in units of $\tK=1/(\la\gBK)$],
vs.\ the temperature. Inset: $\ttr^{{\rm od}}|_{\xi=0}/\tN$
vs.\ $\s=Kv/\T$.
\label{tperp:plot}
}
\end{figure}

Finally, it is shown in Fig.\ \ref{tperp:plot} that, in contrast to
the longitudinal relaxation time, which may increase exponentially at
low temperatures, $\ttr^{{\rm od}}|_{\xi=0}$ is bounded from
above.
Indeed, from the graph one concludes that this transverse
relaxation time is, at most, of the order of $\tK$ (specifically,
$\ttr^{{\rm od}}|_{\xi=0}<1.5\tK$).

On the other hand, we have mentioned that the expression (\ref{effeigentau:bias}) for
the transverse relaxation time does not take the effects of the gyromagnetic terms into
account. In order to investigate the effects of these terms on the transverse response,
Ra{\u{\i}}kher and Shliomis (1975; 1994) studied the transverse
dynamical susceptibility for $B=0$, by a decoupling ansatz for the infinite system of
differential equations for the averages of the magnetic moment [recall the discussion
after Eqs.\ (\ref{avdyneq:1st:ll-llg}) and (\ref{avdyneq:2nd:ll-llg})]. They derived an
expression for $\chi_{\perp}(\w)$ and studied mainly the ferromagnetic-resonance
frequency range. If one is interested in the low-frequency range, their
$\chi_{\perp}(\w)$ can be expanded in powers of $\w$, and then cast into the
Debye-type form
\[
\chi(\w,T)\simeq\chi(T)(1-i\w\tau)\simeq\frac{\chi(T)}{1+i\w\tau}
\qquad (\w\tau\ll1)
\;,
\]
where the last step is done with help from the binomial expansion
$(1+x)^{\epsilon}=1+\epsilon x+\cdots$. Then, the quantity multiplying $i\w$ defines a
effective relaxation time useful in the {\em low\/}-frequency range, which is given by
(see Appendix \ref{app:taus})
\begin{equation}
\label{effeigentaumod}
\ttr|_{\xi=0}
=
2\tN\frac{1-\legunb}{2+\legunb}
\frac{1}{1+p(\s)/\la^{2}}
\;,
\end{equation}
where
\begin{equation}
\label{p} p(\s)
=
\frac
{(3\legunb)^{2}}
{(2+\legunb)\big[2+\legunb(1-6/\s)\big]}
\;.
\end{equation}
Note that, in the absence of gyromagnetic effects ($\la\to\infty$), Eq.\
(\ref{effeigentaumod}) reduces to the unbiased case (\ref{effeigentau}) of the
overdamped result (\ref{effeigentau:bias}).

Finally, on considering that
\[
\ttr|_{\xi=0}
=
2\tN\frac{1-\legunb}{2+\legunb}
\frac{1}{1+p(\s)/\la^{2}}
\leq 2\tN\frac{1-\legunb}{2+\legunb}
=
\ttr^{{\rm od}}|_{\xi=0}
\;,
\]
we also find that $\ttr|_{\xi=0}$ is, at most, of the order of $\tK=(\la\gBK)^{-1}$. For
typical values of the quantities occurring in $\tK$ one has
\begin{equation}
\label{tauK:estimation}
\left.
\begin{array}{rcl}
\gmr
&
=
&
1.76\times 10^{11}\,{\rm T}^{-1}{\rm s}^{-1}
\\
\BK
&
\sim
&
50\,{\rm mT}
\\
\la
&
\sim
&
0.01\!\!-\!\!1
\end{array}
\right\}
\quad
\longrightarrow
\quad
\tK^{-1}\sim10^{8}\!\!-\!\!10^{10}\,{\rm s}^{-1}
\;.
\end{equation}
Thus, for frequencies not very high (say, $\omega<10^{6}$\,Hz) the condition
$\omega\ttr|_{\xi=0}\ll1$ is obeyed, justifying (admittedly, non-rigorously) the step
leading to the low-frequency Shliomis and Stepanov equation (\ref{shschi:lowfrec}).

\subsection{Numerical method}
\label{subsect:nummethod}

In the remainder of this Section we shall study the $T\neq0$ dynamics by solving
stochastic dynamical equations for classical spins numerically. To this end, we shall
now discuss some topics related with the numerical integration of those Langevin
equations.

\subsubsection{Dimensionless quantities}

Let us first introduce a number of dimensionless quantities. The maximum anisotropy
field $\BK=2Kv/\mm$ provides a suitable reference magnetic-field scale that yields
the dimensionless fields (in what follows we shall only consider easy-axis anisotropy
$K>0$)
\begin{equation}
\label{reduced_fields}
\vec{h}
=
\frac{\B}{\BK}
\;,
\qquad
\vec{h}_{\eff}
=
\frac{\Beff}{\BK}
\;,
\qquad
\vec{h}_{\fl}(t)
=
\frac{\bfl(t)}{\BK}
\;.
\end{equation}
A suitable time scale is provided by $\tK$, the deterministic relaxation time at $\B=0$
[Eq.\ (\ref{tauK})], which yields the dimensionless time
\begin{equation}
\label{tauwell}
\bar{t}
=
t/\tK
\;,
\qquad
\tK^{-1}
=
\la\gBK
\;.
\end{equation}
Note that in terms of $\tK$ and $\s=Kv/\T$, the N\'{e}el time (\ref{neeltime}) merely
reads
\begin{equation}
\label{neeltime-tauwell}
\tN
=
\s\tK
\;.
\end{equation}

\subsubsection{Dimensionless stochastic Landau--Lifshitz (--Gilbert) equation}

On using the dimensionless quantities introduced, the stochastic Landau--Lifshitz
(--Gilbert) equation can be rewritten in a dimensionless form suitable for computation,
namely
\begin{equation}
\label{dimstoeq:ll-llg}
\frac{\D\vec{e}}{\D\bar{t}}
=
\frac{1}{\la}\vec{e}\vecpro
\left[
\vec{h}_{\eff}+\vec{h}_{\fl}(\bar{t})
\right]
-\vec{e}\vecpro
\left\{
\vec{e}\vecpro
\left[
\vec{h}_{\eff}+\llg\,\vec{h}_{\fl}(\bar{t})
\right]
\right\}
\;,
\end{equation}
where $\vec{e}=\m/\mm$ is a unit vector in the direction of $\m$ and $\llg=1$ for
Eq.\ (\ref{stollgeq}) while $\llg=0$ for Eq.\ (\ref{stolleq}). The statistical properties of
the dimensionless fluctuating field $\vec{h}_{\fl}(\bar{t})$, which arise directly from
those of $\bfl(t)$ [Eqs.\ (\ref{bcorr})], are given by
\begin{equation}
\llangle h_{\fl,k}(\bar{t})\rrangle
=
0
\;,
\qquad
\llangle h_{\fl,k}(\bar{t}) h_{\fl,\ell}(\bar{\tp})\rrangle
=
2\bar{D}\delta_{k\ell}\delta(\bar{t}-\bar{\tp})
\;,
\end{equation}
where, by using Eq.\ (\ref{coeffdif:ll-llg}) for $D$ and
$\delta(t)=\delta(\bar{t})\D\bar{t}/\D t=\delta(\bar{t})/\tK$, we find for the
dimensionless coefficient $\bar{D}$:
\begin{equation}
\label{coeffdif:dimless}
\bar{D}
=
\frac{D}{\tK\BK^{2}}
=
\frac{\la^{2}}{1+\llg\la^{2}}\frac{\T}{\mm\BK}
=
\frac{\la^{2}}{1+\llg\la^{2}}\frac{\T}{2Kv}
\;.
\end{equation}

Let us finally cast the dimensionless Eq.\ (\ref{dimstoeq:ll-llg}) into the form of the
general system of Langevin equations (\ref{langevinequation:n-dim}):
\begin{equation}
\label{dimstoeq:ll-llg:components}
\frac{\D e_{i}}{\D\bar{t}}
=
\bar{\drift}_{i}
+\sum_{k}\bar{\diff}_{ik}h_{\fl,k}(\bar{t})
\;,
\end{equation}
where $k$ runs over $x,y,z$, and [cf.\ Eqs.\ (\ref{F:ll-llg}) and (\ref{G:ll-llg})]
\begin{eqnarray}
\label{F:ll-llg:dimless}
\bar{\drift}_{i}
&
=
&
\sum_{k}
\bigg[
\frac{1}{\la}
\sum_{j}
\epsilon_{ijk}e_{j}
+(\delta_{ik}-e_{i}e_{k})
\bigg] h_{\eff,k}
\;,
\\
\label{G:ll-llg:dimless}
\bar{\diff}_{ik}
&
=
&
\frac{1}{\la}
\sum_{j}\epsilon_{ijk}e_{j}
+\llg (\delta_{ik}-e_{i}e_{k})
\;.
\end{eqnarray}

\subsubsection{The choice of the numerical scheme}

As has been mentioned, the stochastic Landau--Lifshitz (--Gilbert) equation
contains multiplicative white-noise terms [Eq.\ (\ref{G:ll-llg}), or its dimensionless
counterpart (\ref{G:ll-llg:dimless}) clearly depend on $\m$ both for $\llg=0$ and
$\llg=1$]. Together with difficulties at the level of definition, the occurrence of
multiplicative white noise in a Langevin equation entails some technical problems as
well. For instance, serious difficulties arise in developing high-order numerical
integration schemes for this case (Kloeden and Platen, 1995). In  general, the simple
translation of a numerical scheme valid for deterministic differential equations does
not necessarily yield a proper scheme in the stochastic case:
\begin{enumerate}
\item
Depending on the original deterministic scheme chosen, its na\"{\i}ve stochastic
translation might converge to an It\^{o} solution, to a Stratonovich solution, or to none
of them. \item
Even if there exists proper convergence of the scheme chosen in the
context of the stochastic calculus used, the {\em order of convergence\/} obtained is
usually lower than that of the original deterministic scheme.
\end{enumerate}

Let us consider the stochastic generalization of the deterministic Heun scheme, namely
\begin{eqnarray}
\label{heun:scheme} y_{i}(t+\Delt)
&
=
&
y_{i}(t)
+\half
\big[\drift_{i}(\tilde{\ym},t+\Delt)
+\drift_{i}(\ym,t)\big]\Delt
\nonumber\\
&
& {}+\half
\sum_{k}
\big[\diff_{ik}(\tilde{\ym},t+\Delt)
+\diff_{ik}(\ym,t)\big]\DelW_{k}
\;,
\end{eqnarray}
where $\Delt$ is the discretization time interval, $\ym=\ym(t)$, the $\tilde{y}_{i}$ are
Euler-type supporting values,
\begin{equation}
\label{euler:support}
\tilde{y}_{i}
=
y_{i}(t)+\drift_{i}(\ym,t)\Delt
+\sum_{k}\diff_{ik}(\ym,t)\DelW_{k}
\;,
\end{equation}
and the $\DelW_{k}=\int_{t}^{t+\Delt}\!\D{\tp}\,\Lan_{k}(\tp)$ are Gaussian random
numbers whose first two moments are
\begin{equation}
\label{Wcorr}
\llangle\DelW_{k}\rrangle
=
0
\;,
\quad
\llangle\DelW_{k}\DelW_{\ell}\rrangle
=
(2D\Delt)\delta_{k\ell}
\;.
\end{equation}
The Heun scheme converges {\em in quadratic mean\/} to the solution of the general
system of stochastic differential equations (\ref{langevinequation:n-dim})
supplemented by Eqs.\ (\ref{langevin:n-dim}), {\em when interpreted in the sense of
Stratonovich\/} (see, for example, R{\"{u}}melin, 1982).

On the other hand, if one uses the Euler-type Eq.\ (\ref{euler:support}) as the
numerical integration scheme [by identifying $y_{i}(t+\Delt)=\tilde{y}_{i}$], the
constructed trajectory {\em converges to the It\^{o} solution\/} of the same system of
equations (\ref{langevinequation:n-dim}). A proper Euler-type scheme in the context
of the Stratonovich stochastic calculus is obtained when the deterministic drift in Eq.\
(\ref{euler:support}), $\drift_{i}$, is augmented  by the noise-induced drift, namely
\begin{equation}
\label{euler:scheme} y_{i}(t+\Delt)
=
y_{i}(t)
+
\bigg[
\drift_{i}
+D\sum_{jk}\diff_{jk}\frac{\partial\diff_{ik}}{\partial y_{j}}
\bigg]_{(\ym,t)}
\Delt
+\sum_{k}\diff_{ik}(\ym,t)\DelW_{k}
\;,
\end{equation}
(for an alternative Euler-type algorithm for multiplicative noise see
Ram{\'{\i}}rez-Piscina, Sancho and Hern{\'{a}}ndez-Machado, 1993). In order to use the
scheme (\ref{euler:scheme}), one needs to calculate the corresponding noise-induced
drift. This was already done yielding Eq.\ (\ref{noiseinduceddrift:ll-llg}), which can
readily be adapted to the dimensionless Eq.\ (\ref{dimstoeq:ll-llg:components}):
\[
\bar{D}\sum_{jk}\bar{\diff}_{jk}\frac{\partial\bar{\diff}_{ik}}{\partial e_{j}}
=
-\frac{1}{\tN/\tK}e_{i}
=
-\frac{\T}{Kv}e_{i}
\;,
\]
where Eq.\ (\ref{neeltime-tauwell}) has been used to write down the last equality.%
\footnote{
On recalling that in Eq.\ (\ref{dimstoeq:ll-llg}) the time is measured in units
of $\tK$, one realizes that the term $-(\tK/\tN)e_{i}$ corresponds to
$-\llangle\m\rrangle/\tN$ in the averaged dynamical equation
(\ref{avdyneq:1st:ll-llg}). Indeed,  by using $\langle\DelW_{k}\rangle=0$ for
averaging Eq.\ (\ref{euler:scheme}) when particularized to the stochastic
Landau--Lifshitz (--Gilbert) equation, one gets the discretized version of Eq.\
(\ref{avdyneq:1st:ll-llg}).
}

However, in order to choose the numerical scheme to undertake the integration of Eq.\
(\ref{dimstoeq:ll-llg}), it is convenient to determine first the {\em character\/} of the
multiplicative noise in that equation. When the $\diff_{ik}$ fulfill the relation
\begin{equation}
\label{commutativenoise:def}
\sum_{j}\diff_{jk}\frac{\partial\diff_{i\ell}}{\partial y_{j}}
=
\sum_{j}\diff_{j\ell}\frac{\partial\diff_{ik}}{\partial y_{j}}
\;,
\quad
\forall i
\end{equation}
(i.e., {\em symmetry\/} with respect to the subscripts $k$ and $\ell$), the noise in the
Langevin equations is said to be {\em commutative}. The condition of commutative
noise is rather general and includes additive noise, $\partial\diff_{ik}/\partial
y_{j}\equiv0$, diagonal multiplicative noise,
$\diff_{ik}(\ym,t)=\delta_{ik}\diff_{ii}(y_{i})$, and linear multiplicative noise,
$\diff_{ik}(\ym,t)=\diff_{ik}(t)y_{i}$ (see, for example, Kloeden and Platen, 1995,
p.~348). In addition, when Eq.\ (\ref{commutativenoise:def}) is satisfied, the stochastic
Heun scheme (\ref{heun:scheme}) has an order of convergence higher than the order
of convergence of the Euler scheme (\ref{euler:scheme}) (see, for example,
R{\"{u}}melin, 1982).

Unfortunately, not only the noise in the stochastic Landau--Lifshitz (--Gilbert)
equation is multiplicative, but is {\em non-commutative\/} as well. Indeed, on
calculating the right-hand side of Eq.\ (\ref{commutativenoise:def}) with $\diff_{ik}$
from Eq.\ (\ref{G:ll-llg}), we find
\begin{eqnarray*}
\lefteqn{
\frac{1}{\gmr^{2}}
\sum_{j}\diff_{j\ell}\frac{\partial\diff_{ik}}{\partial\mj}
}\qquad\quad
\\
&
=
&
-\mi
\underbrace{
\delta_{k\ell}
}_{{\rm S}}
+\underbrace{
\mk\delta_{i\ell}
}_{{\rm ND}}
\\
&
& {}+
\llg
\frac{\la}{\mm}
\bigg(
\underbrace{
\mm^{2}\epsilon_{i\ell k}
}_{{\rm A}}
\underbrace{
-\ml\sum_{j}\epsilon_{ijk}\mj
-\mk\sum_{r}\epsilon_{ir\ell}\mr
}_{{\rm S}}
-\mi
\underbrace{
\sum_{r}\epsilon_{kr\ell}\mr
}_{{\rm A}}
\bigg)
\\
&
& {}-\llg
\left(\frac{\la}{\mm}\right)^{2}
\big(
\underbrace{
\mm^{2}\mk\delta_{i\ell}
}_{{\rm ND}}
+\mm^{2}\mi
\underbrace{
\delta_{k\ell}
}_{{\rm S}}
-2\mi
\underbrace{
\mk\ml
}_{{\rm S}}
\big)
\;,
\end{eqnarray*}
where ${\rm S}$, ${\rm A}$, and ${\rm ND}$, indicate, respectively, symmetry,
anti-symmetry, and not defined symmetry with respect to the subscripts $k$ and
$\ell$. Therefore, owing to the presence of these last two types of terms, the
commutative noise condition {\em is not\/} obeyed by either the stochastic
Landau--Lifshitz--Gilbert or the stochastic Landau--Lifshitz equation.

For non-commutative noise the best order of convergence is attained (R{\"{u}}melin,
1982) with the Heun scheme (\ref{heun:scheme}) but also with the simpler Euler
algorithm (\ref{euler:scheme}) or with the scheme of Ram{\'{\i}}rez-Piscina, Sancho
and Hern{\'{a}}ndez-Machado (1993). Although the Heun scheme requires the
evaluation of $\drift_{i}$ and $\diff_{ik}$ at two points per time step (at the initial
and support ones), we have chosen it to integrate the stochastic Landau--Lifshitz
(--Gilbert) equation. This is done because:
\begin{enumerate}
\item
The Heun scheme yields Stratonovich solutions of the stochastic differential
equations naturally, without alterations to the drift term.
\item
The deterministic part of the differential equations is treated with a
second-order accuracy in $\Delt$, which renders the Heun scheme numerically more
stable than the Euler-type schemes.
\end{enumerate}

We finally emphasize that, in order to integrate the stochastic Landau--Lifshitz
(--Gilbert) equation numerically one cannot merely employ a bare Euler-like scheme
like (\ref{euler:support}), since this scheme yields It\^{o} solutions of the differential
equations. Even the stationary properties derived by means of such an approach
would not coincide with the correct thermal-equilibrium properties [recall the
discussion after Eq.\ (\ref{brownfpe:ito})].

\subsubsection{Implementation}

The integration of the stochastic Landau--Lifshitz (--Gilbert) equation is performed by
starting from a given initial configuration, and updating recursively the state of the
system, $\m(t)\to\m(t+\Delt)$, by means of the set of finite-difference equations
(\ref{heun:scheme}). This generates {\em stochastic trajectories\/} from which, when
required, averages are directly computed. If one extrapolates the results obtained to
zero discretization time interval $\Delt$, the only error in the {\em averaged\/}
quantities is a statistical error bar that can, in principle, be made arbitrarily small by
averaging over a sufficiently large number of trajectories. We usually not carry out
such $\Delt\to0$ limiting procedure, but we employ a discretization time interval small
enough. Unless otherwise stated, the choice $\Delt=0.01\tK$ is made.

When computing average quantities, in order to minimize effects that are not caused
by the application of the probing field $\dB(t)$, the following {\em subtraction\/}
method is used. Starting from the same initial configuration, the equations of motion
are solved for two identical ensembles, one in the presence of $\dB(t)$ and the other
subjected to $-\dB(t)$, and the time evolution analyzed is that of
\[
M_{\rm sub}(t)
=
\half
\left\{
\sum\m\big[\dB(t)\big]
-
\sum\m\big[-\dB(t)\big]
\right\}
\;.
\]
Moreover, we have found that this technique significantly diminishes the number of
stochastic trajectories required to achieve convergence in the averaged results. On the
other hand, the subtraction technique automatically eliminates the non-linear terms
{\em quadratic\/} in the probing field that could emerge.

Finally, the {\em Gaussian\/} random numbers required to simulate the $\DelW_{k}$
entering in the above schemes, are constructed from {\em uniformly\/} distributed
random numbers by means of the Box--Muller algorithm. Thus, if $r_{1}$ and $r_{2}$
are random numbers uniformly distributed in the interval $(0,1)$ (as those
pseudo-random numbers provided by a computer), the transformation
\begin{eqnarray*}
w_{1}
&
=
&
\sqrt{-2\ln(r_{1})}\cos(2\pi r_{2})
\\
w_{2}
&
=
&
\sqrt{-2\ln(r_{1})}\sin(2\pi r_{2})
\;,
\end{eqnarray*}
outputs $w_{1}$ and $w_{2}$, which are Gaussian-distributed random numbers of
zero mean and variance unity (if one needs Gaussian numbers with variance $\s^{2}$,
these are immediately obtained by multiplying the above $w_{i}$ by $\s$). Owing to
the fact that the generation of the random numbers is the slowest step in the recursive
scheme, when computing an averaged quantity at various temperatures we generate
all the trajectories at once, by using the same sequence of random numbers for the
different temperatures.

\subsection{Stochastic trajectories of individual spins}
\label{subsect:simulationsI}

We shall now study the $T\neq0$ dynamics of {\em individual\/} magnetic moments.
To this end, we shall integrate the {\em stochastic\/} Landau--Lifshitz--Gilbert
equation (\ref{stollgeq}) numerically in the context of the Stratonovich calculus, by
means of the stochastic generalization (\ref{heun:scheme}) of the Heun scheme. If one
wishes to have a reference of the time scales involved, one can assume values like
those of Eq.\ (\ref{tauK:estimation}), so that $\tK\sim10^{-10}$--$10^{-8}$\,s.

\subsubsection{The over-barrier rotation process}

\begin{figure}[t!]
\vspace{-6.5ex}
\eps{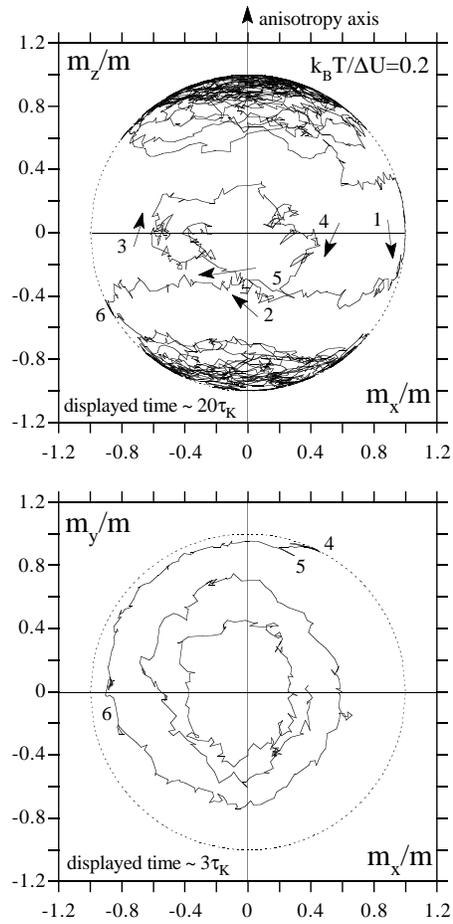}{0.65}
\vspace{-6.5ex}
\caption[]
{ 2D projections of $\m(t)$, as determined by integration of the
stochastic Landau--Lifshitz Eq.\ (\ref{stollgeq}). The anisotropy
energy is $-\dU(\mz/\mm)^{2}$, no field is applied, and the damping
coefficient is $\la=0.1$. Upper panel: Projection onto a plane
containing the anisotropy axis. Lower panel: Projection onto the plane
perpendicular to the anisotropy axis of the first stages of the damped
precession down to the $\m=-\mm\hat{z}$ potential minimum, after the
last barrier crossing.
\label{randomwalks:plot}
}
\end{figure}

Figure \ref{randomwalks:plot} displays the projection of the
trajectory of a magnetic moment with the simplest axially symmetric
anisotropy onto selected planes. No magnetic field has been applied,
so the graphs show the (in this sense) ``intrinsic" dynamics.

The projection of $\m(t)$ onto a plane containing the anisotropy axis
$\hat{n}$ (defining the $\hat{z}$ direction in Fig.\
\ref{randomwalks:plot}), corresponds to a typical stochastic
trajectory that starts close to one of the potential minima
($\m=\mm\hat{z}$) and, after some irregular rotations about it,
reaches the potential-barrier (equatorial) region, where it wanders
for a while, and eventually descends to the other potential
minimum. Concerning the projection of this motion onto a plane
perpendicular to the anisotropy axis, we have only shown the first
stages, after the last potential-barrier crossing, of the damped
precession of $\m$ about the anisotropy field when spiralling down to
the bottom of the $\mz<0$ potential well.

These graphs reveal the important r\^{o}le of the gyromagnetic terms in the stochastic
dynamics of the magnetic moment. Thus, the projection of $\m(t)$ onto the
equatorial plane shows some of the typical irregular features of ordinary Brownian
motion, although the rotary character is clearly exhibited. Concerning the projection of
$\m(t)$ onto a plane containing the anisotropy axis, it can clearly be seen that crossing
the potential barrier does not entail an immediate descent to the other potential
minimum, but the gyromagnetic terms together with an appropriate sequence of
fluctuating fields can produce a rapid crossing back to the initial potential well.

For an ordinary, non-gyromagnetic system, i.e., a mechanical system with inertia, the
inertia guarantees that, unless the system reaches the potential barrier with zero
velocity, it will descend to the other potential well with a large probability. Moreover,
the forces, after the potential-barrier crossing, accelerate the system downward.
In contrast, in the gyromagnetic case the dynamics is ``non-inertial" (the equations of
motion are of first order in the time). Besides, the anisotropy field $\B_{{\rm
a}}=(\BK/\mm)\mz\hat{z}$ indeed drives $\m$ down to the bottom of the potential
well, but this is effected via a damped precession about the anisotropy axis.
Moreover, the effective precession ``frequency" of this motion $\weff\propto\mz$ is
initially rather low because the anisotropy field is low in the potential-barrier region
($\mz\simeq0$). Consequently, in the beginning of the spiraling down after a
potential-barrier crossing, the magnetic moment rotates (say, along a parallel of
latitude) quite slowly not far from the potential-barrier, so that an appropriate
sequence of fluctuations can drive it back to the initial potential well.

What is shown in Fig.\ \ref{randomwalks:plot} is precisely a multiple
occurrence of this phenomenon; more than 10 potential-barrier
crossings can be identified in the overall excursion between the two
potential minima. Besides, the magnetic moment might also have
eventually fallen into the original potential well. As will be shown
below, none of these processes is infrequent. The physical acumen of
Brown (1959) is noteworthy since, on considering the gyromagnetic
nature of the dynamics, he posed the possible occurrence of this kind
of phenomena in his criticism of the calculation of N\'{e}el (1949) of
the relaxation time as the inverse of the rate of equatorial crossings
of the magnetic moment.

\subsubsection{The effect of the temperature}

\begin{figure}[t!]
\vspace{-3.ex}
\eps{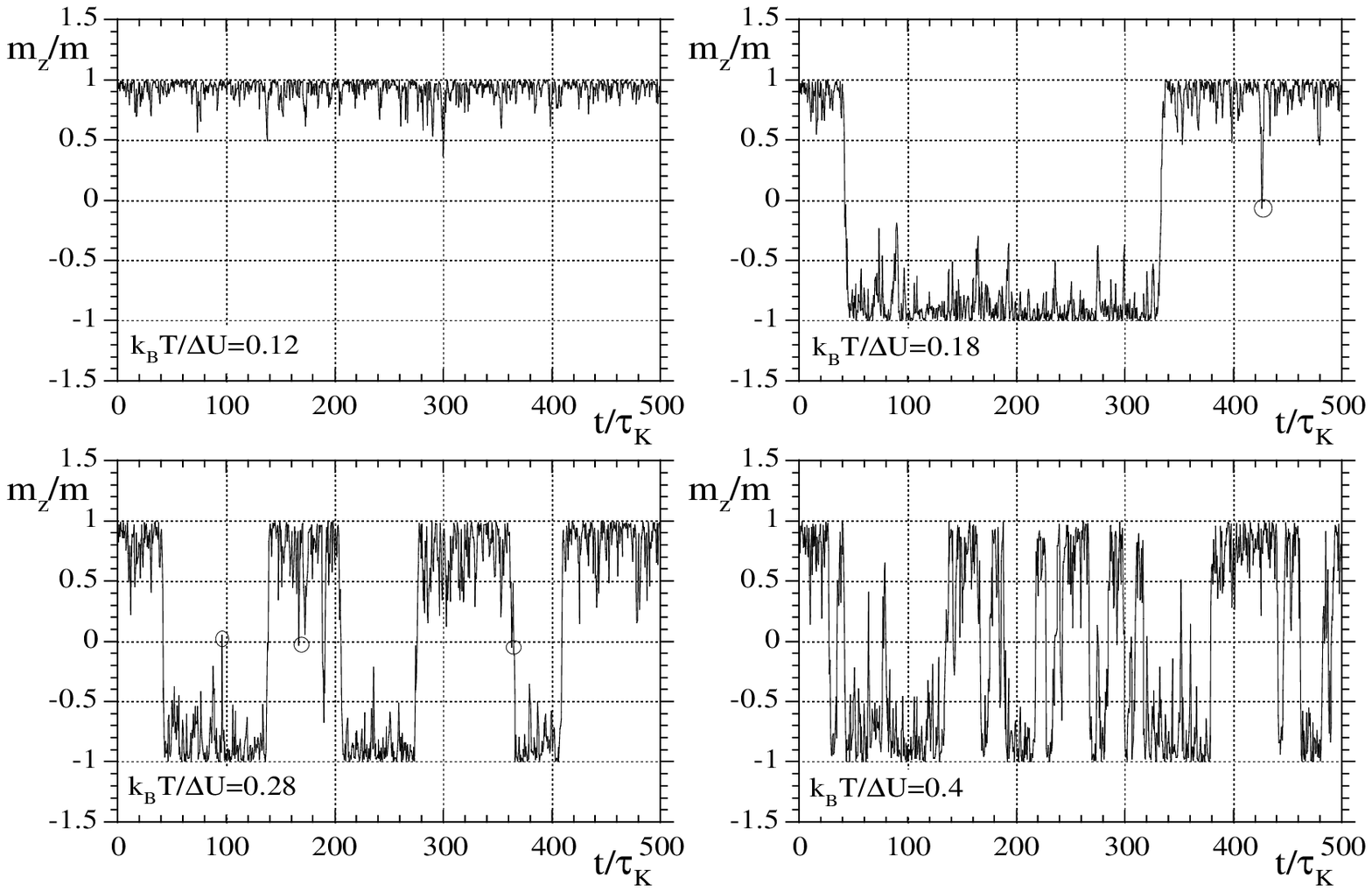}{1.15}
\vspace{-3.ex}
\caption[]
{ Projection onto the anisotropy axis of $\m(t)$, as determined by
numerical integration of the stochastic Landau--Lifshitz--Gilbert
equation (\ref{stollgeq}), for various temperatures. The
magnetic-anisotropy energy is $-\dU(\mz/\mm)^{2}$, $\B=0$, and
$\la=0.1$. The small circles mark potential-barrier crossings followed
by a back rotation to the initial potential well.
\label{jumps:plot}
}
\vspace{-3.ex}
\vspace{-3.ex}
\eps{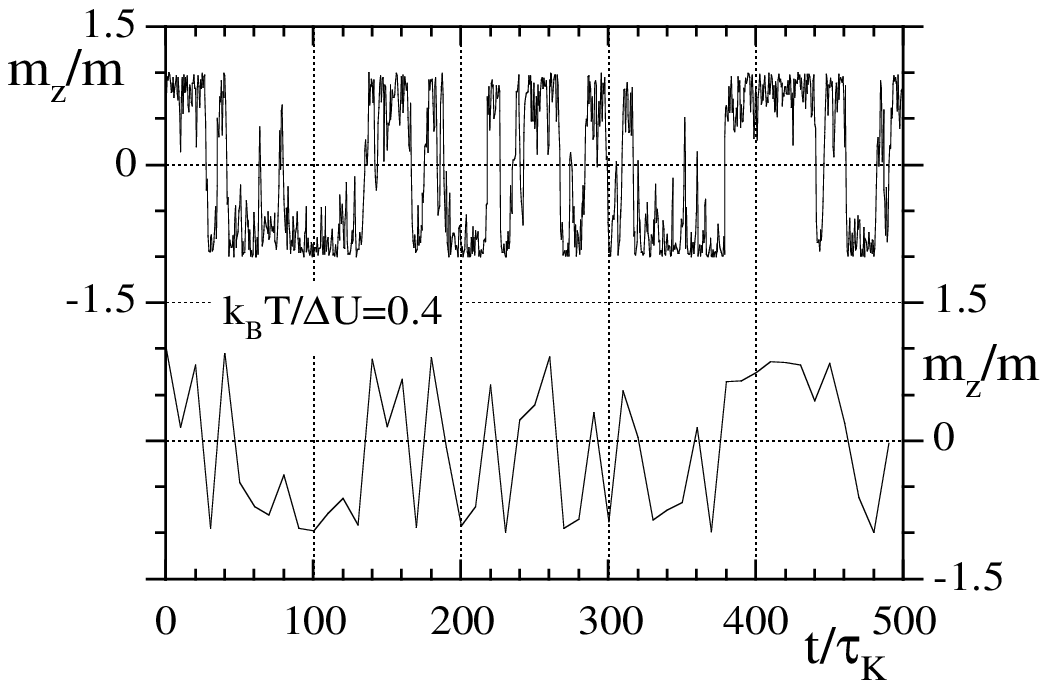}{0.7}
\vspace{-3.ex}
\caption[]
{
The same as in panel $\T/\dU=0.4$ of Fig.\ \ref{jumps:plot}, but the
trajectory has also been plotted with a larger sampling time interval.
\label{jumps_double:plot}
}
\end{figure}

In order to assess the effect of the temperature on the dynamics of
the magnetic moment, we have displayed in Fig.\ \ref{jumps:plot} some
typical time evolutions of the projection of $\m$ onto the anisotropy
axis.

As can be seen, at low temperatures (panel $\T/\dU=0.12$), the dynamics merely
consist of the rotations of the magnetic moment close to the bottom of the potential
wells (intra-potential-well relaxation modes), with the over-barrier relaxation
mechanism being ``blocked." As the temperature is increased, the magnetic moment
can effect over-barrier rotations at the expense of the energy gained from the heat
bath, and a number of them do occur during the displayed time interval (panels
$\T/\dU=0.18$ and $0.28$). Finally, at higher temperatures (panel $\T/\dU=0.4$), the
magnetic moment effects a considerable number of over-barrier rotations during the
observation time interval, exhibiting almost the thermal-equilibrium distribution of
orientations.

The curves of Fig.\ \ref{jumps:plot} resemble those of the experiments
of Wernsdorfer et~al.\ (1997) on individual ferromagnetic
nanoparticles (see Fig.\ 6 in that reference).  Furthermore, if the
same trajectory is plotted with a larger sampling time interval, in
order to mimic the finite resolution time of a measuring device, the
resemblance is more apparent, since the curves then have less and
wider angles (Fig.\
\ref{jumps_double:plot}). (Recall that the strong dependence of the appearance of the
time evolution curves on the sampling period is a typical feature of the stochastic
dynamics.)

Note finally that in Fig.\ \ref{jumps:plot} a number of
potential-barrier crossings followed by a rotation back to the
original potential well can be identified (marked with small circles):
one for $\T/\dU=0.18$; three for $\T/\dU=0.28$, the one occurring at
$\sim360\,t/\tK$ being a double crossing-back; and about seven for
$\T/\dU=0.4$ (not marked for the sake of clarity). It is also to be
noted that an apparent single (or double) crossing-back can be
multiple instead. Indeed, when the about 10 potential-barrier
crossings of Fig.\ \ref{randomwalks:plot} are represented as $\mz$
vs.\ $t$, they seem to be a mere double crossing-back of the potential
barrier.

\subsubsection{Projection of the magnetic moment onto the direction of a probing
field}

\begin{figure}[t!]
\vspace{-3.ex}
\eps{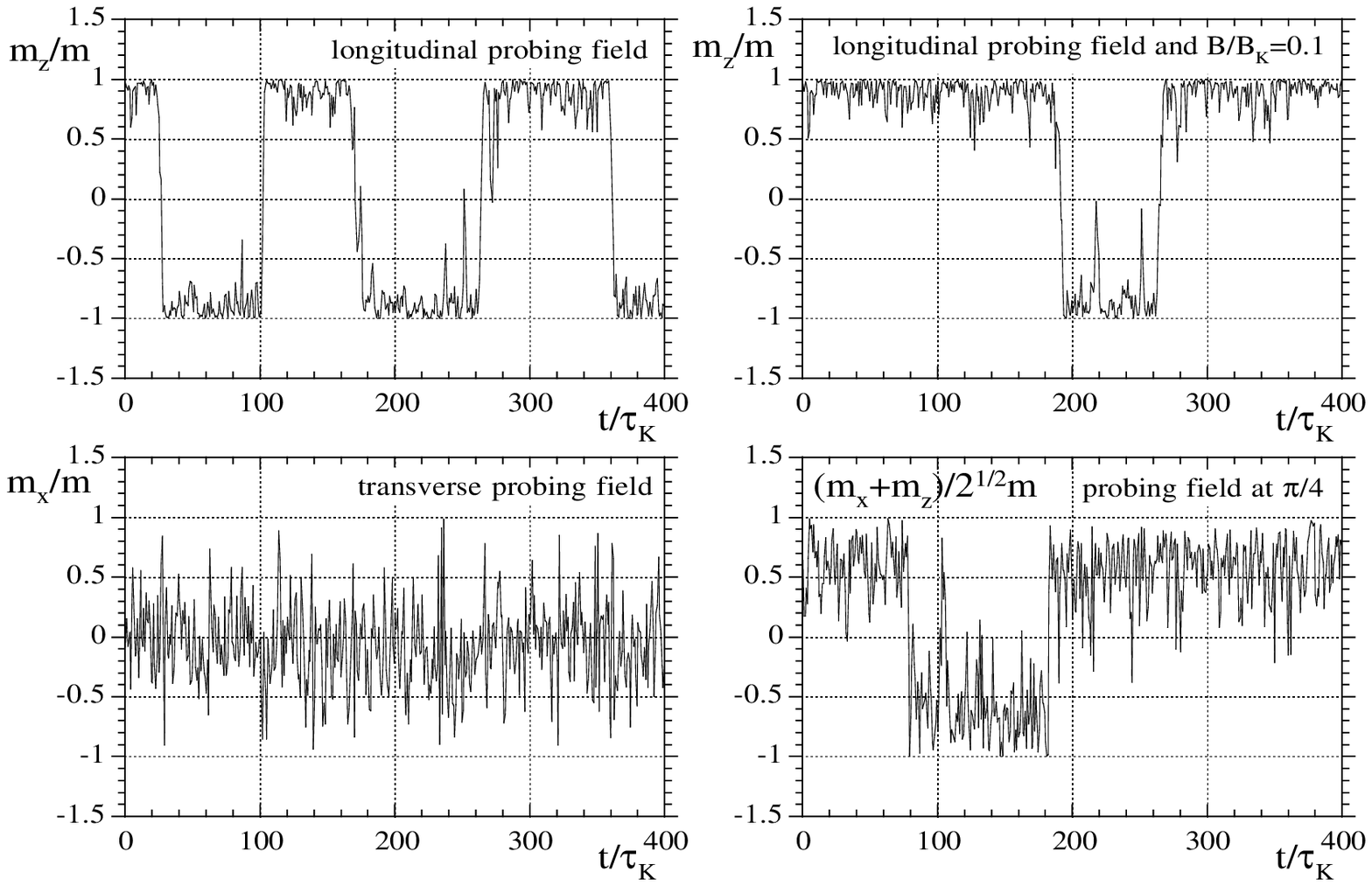}{1.15}
\vspace{-3.ex}
\caption[]
{
Projection onto the direction of a probing field
$\dB(t)=\dB\cos(\w t)$ of $\m(t)$, as determined by numerical integration of
the stochastic Landau--Lifshitz--Gilbert equation (\ref{stollgeq}). The
magnetic-anisotropy energy is $-\dU(\mz/\mm)^{2}$ and all the results are for
$\T/\dU=0.2$ and $\la=0.1$. The displayed time interval corresponds to a complete
cycle of the oscillating field ($\w\tK/2\pi=0.0025$). In the longitudinal probing field
case, results in the presence of a longitudinal bias field are also shown.
\label{jumps:ac:plot}
}
\end{figure}

It is also illuminating to show the projection of the trajectories of
individual spins onto the direction of a probing field
$\dB(t)=\dB\cos(\w t)$. Figure \ref{jumps:ac:plot} shows this kind of
trajectories in the intermediate temperature range.

The projection onto the anisotropy axis direction
($\dB\parallel\hat{z}$) exhibits, as in the corresponding case of
Fig.\ \ref{jumps:plot}, a well resolved bistability, and $\m$ ``jumps"
from one well to the other a number of times during a cycle of the
probing field. Similar features are encountered when a longitudinal
bias field is also applied, the main difference being that the lower
potential well is less frequented by the magnetic moment. In contrast,
the features of the stochastic trajectory obtained by projecting
$\m(t)$ onto a direction perpendicular to the anisotropy axis
($\dB\perp\hat{z}$) are markedly different (for example, this
projection corresponds to plotting the trajectory of the upper panel
of Fig.\ \ref{randomwalks:plot} as $\mx$ vs.\ $t$). Here, the response
is dominated by the fast ($\sim\tK$) intra-potential-well relaxation
modes, and the transverse projection is a highly irregular sequence of
sharp peaks. Finally, the projection of $\m(t)$ onto $\dB$ making an
intermediate angle with the anisotropy axis ($\pi/4$ for the displayed
curve), shows the magnetic bistability of the longitudinal projection,
but the fast intra-potential-well motions are superimposed on it. This
leads to a less well-resolved magnetic bistability.

Note finally that curves like those of Fig.\ \ref{jumps:ac:plot} are
the ones ``analyzed" by the probing field in a dynamical
``measurement." Recall also that the application of the oscillating
field hardly changes the overall features of the curves from the free
evolution ones. This is naturally so, since one applies a low enough
field in order to probe the intrinsic dynamics of the system.

\subsection{Dynamical response of the ensemble of spins}
\label{subsect:simulationsII}

Keeping Figs.\ \ref{jumps:plot}--\ref{jumps:ac:plot} in mind, we shall
undertake the study of the dynamical response of an ensemble of
classical magnetic moments. As a suitable probe of the intrinsic
dynamics of the system, we shall compute the linear dynamical
susceptibility, $\chi(\w)$, as a function of the temperature for
various frequencies and orientations of an external probing field
$\dB(t)=\dB\cos(\w t)$.

We compute the dynamical response for ensembles of $1000$ magnetic
moments. We integrate numerically the stochastic
Landau--Lifshitz--Gilbert equation of each spin by means of the
stochastic Heun scheme (\ref{heun:scheme}), and analyze the time
evolution of the total magnetic moment of the ensemble; the results
for the dynamical susceptibility have typically been averaged over
$50$--$100$ cycles of the oscillating field. In addition, in order to
reduce the statistical error bars, we apply at each $T$ the largest
probing field without leaving the {\em equilibrium\/} linear response
range (specifically, we scale the amplitude of the probing field with
the temperature according to $\mm\dBo=0.3\T$).

The damping coefficient, $\la$, the magnetic-anisotropy potential
barrier, $\dU=Kv$, and the magnitude of the magnetic moment, $\mm$,
are assumed to be the same for each spin. For non-interacting entities
the effects of a distribution in these parameters, as typically occurs
in nanoparticle ensembles, could be taken into account by an
appropriate rescaling and summation of the so-obtained results.

In all the figures which follow, the linear susceptibilities are
measured in units of $\mu_{0}\mm/\BK=\mu_{0}\mm^{2}/2Kv$ [the
transverse equilibrium susceptibility per spin at zero temperature in
the absence of a bias field; see Eq.\
(\ref{X:para:perp:approx})]. Furthermore, when the statistical error
bars of the numerical results are not shown, their size is, at most,
that of the plotted symbols.  Finally, in order to have a reference of
the discussed time scales, we can use the values of Eq.\
(\ref{tauK:estimation}), so that
$\tK^{-1}\sim10^{8}$--$10^{10}$\,s$^{-1}$ and the frequencies employed
below ($\w\tK/2\pi\sim10^{-3}$--$10^{-2}$) are then in the MHz range.

\subsubsection{Dynamical response in the absence of a bias field}

We shall first study the response of the spin ensemble in the absence
of a constant external field.

\paragraph{Longitudinal response.}

\begin{figure}[t!]
\vspace{-3.ex}
\eps{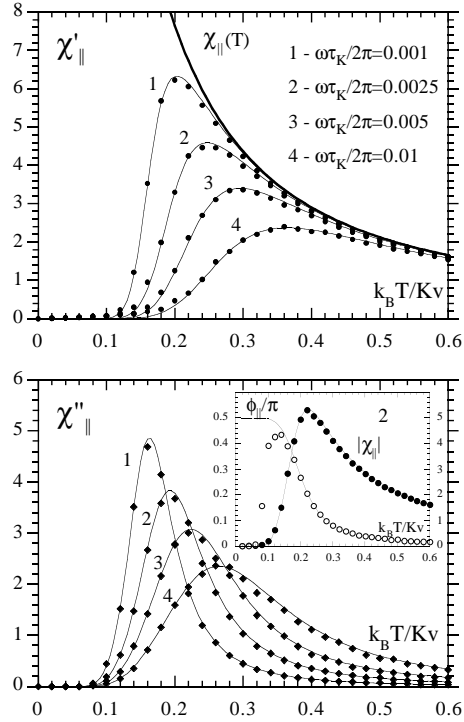}{0.7}
\vspace{-3.ex}
\caption[]
{
Longitudinal linear dynamical susceptibility $\chi_{\|}$ vs.\ $T$ in
the absence of a bias field. The symbols are for the numerically
computed $\chi_{\|}(\w,T)$ and the thin solid lines are Eq.\
(\ref{shschi:bias}) with $\tlo$ defined as integral relaxation time
[Eq.\ (\ref{tauint})]. The heavy solid line in the upper panel is the
thermal-equilibrium susceptibility [Eq.\ (\ref{X:para:perp})]. Inset:
Modulus and phase shift $\phase=\arctan(\chi''/\chi')$ for
$\w\tK/2\pi=0.0025$.
\label{chipara:plot}
}
\end{figure}

Figure \ref{chipara:plot} displays the results for the longitudinal
linear dynamical susceptibility vs.\ the temperature for an ensemble
of magnetic moments with parallel anisotropy axes
($\dB\parallel\hat{z}$).
No bias field has been applied and a damping
coefficient $\la=0.1$ has been used.%
\footnote{
Recall that, because of the axial symmetry considered, the effect of
$\la$ on the averaged quantities merely enters via the N\'{e}el time
$\tN=\s\tK$ [see the discussion after Eq.\
(\ref{brownfpe:axi:2})]. Because we measure the time in units of
$\tK$, the results presented for the {\em longitudinal\/} response are
independent of the $\la$ used.
}

At low temperatures, the longitudinal relaxation time obeys the
condition $\tlo\gg2\pi/\w$ [$t_{{\rm m}}(\w)=2\pi/\w$ is the {\em
dynamical\/} measurement time]. Consequently, during a large number of
cycles of the probing field, the probability of over-barrier rotations
is almost zero; the response consists of the rotations of the magnetic
moments close to the bottom of the potential wells (see the panel
$\T/\dU=0.12$ of Fig.\ \ref{jumps:plot}), whose averaged (over the
ensemble) projection onto the probing-field direction is quite small
(but non zero; see the enlargement of the low-temperature range in
Fig.\ \ref{chipara_teo_amp:plot}).  Moreover, as these
intra-potential-well relaxation modes are very fast ($\sim\tK$), this
small response is in phase with the probing field [see the low-$T$
part of the phase shift $\phase=\arctan(\chi''/\chi')$ in the inset of
Fig.\ \ref{chipara:plot}].

As the temperature is increased the magnetic moments can depart from
the potential minima by means of the energy gained from the heat
bath. Consequently, at a $\w$-dependent temperature
($\T/Kv\sim0.1$--$0.2$ for the frequencies employed), it emerges a
small probability of surmounting the magnetic-anisotropy potential
barrier during a number of cycles of the probing field (this
corresponds to the panel $\T/\dU=0.18$ of Fig.\
\ref{jumps:plot}). Accordingly, the averaged response starts to
increase steeply with $T$. However, as the thermally activated
response mechanism via over-barrier rotations is not efficient enough
at these temperatures, the signal exhibits a considerable lag behind
the probing field (see the inset of Fig.\
\ref{chipara:plot}). This is also reflected by the occurrence of a sizable out-of-phase
component of the response $\chi_{\|}''(T)$ (in fact, the response is mainly ``out of
phase").

At higher temperatures, the mechanism of over-barrier rotations
becomes increasingly efficient (panel $\T/\dU=0.28$ of Fig.\
\ref{jumps:plot}). Consequently, after exhibiting a maximum, the phase
shift starts to {\em decrease}, whereas the magnitude of the response
still {\em increases\/} steeply with $T$ (see the inset of Fig.\
\ref{chipara:plot}). However, if the temperature is further increased,
the very thermal agitation, which up to these temperatures was
responsible for the growth in the magnitude of the response, reaches a
level that: (i) efficiently produces over-barrier rotations, allowing
the magnetic moments to approximately redistribute according to the
instantaneous probing field, but, simultaneously, (ii) disturbs
sizably the alignment of the magnetic moments in the probing-field
direction. Consequently, at a temperature above that of the phase
maximum ($\T/Kv\sim0.2$--$0.3$ for the frequencies considered), the
magnitude of the response has a maximum and starts to decrease with
increasing $T$. The frequency-dependent temperature at which this
maximum occurs is called the {\em blocking\/} temperature.

Finally, at still higher temperatures ($\T/Kv\ge0.3$--$0.5$ for the
frequencies considered) the inequality $\tlo\ll2\pi/\w$ holds. Thus,
in comparison with $\tlo^{-1}$, the rate of change of the probing
field is quasi-stationary. Consequently, the magnetic moments can
quickly redistribute according to the conditions set by the
instantaneous probing field, almost being in the thermal-equilibrium
state associated with it (panel $\T/\dU=0.4$ of Fig.\
\ref{jumps:plot}). Then, the $\chi_{\|}'(T)$ curves corresponding to
different frequencies sequentially superimpose on the linear
equilibrium susceptibility, $\chi_{\|}(T)$, and, correspondingly,
$\chi_{\|}''(T)$ goes to zero.

The occurrence of a frequency-dependent maximum in the response of a
noisy non-linear multi-stable system to a periodic stimulus as a
function of the noise intensity, is one of the features usually
accompanying {\em stochastic resonance}. In this spin-dynamics case,
the maximum in the magnitude of the dynamical response as a function
of $T$ can be understood in terms of the quoted two-fold r\^{o}le
played by the temperature: (i) activating the dynamics of over-barrier
rotations, enabling the spins to (statistically) follow the
instantaneous field, but, (ii) provoking the thermal misalignment of
the spins from the driving-field direction. The maximum in the
response as a function of $T$ emerges as a result of the competition
between these two effects.

\paragraph{Transverse response.}

\begin{figure}[t!]
\vspace{-3.ex}
\eps{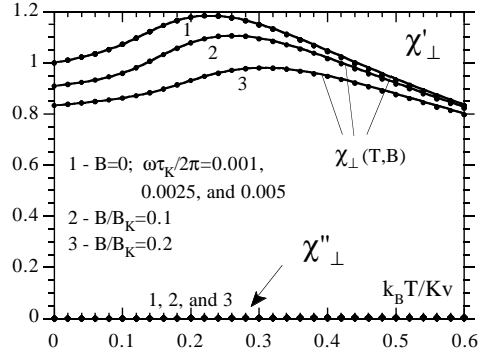}{0.7}
\vspace{-3.ex}
\caption[]
{
Transverse linear dynamical susceptibility $\chi_{\perp}$
vs.\ $T$ for the frequencies $\w\tK/2\pi=0.001$, $0.0025$, and
$0.005$. The damping coefficient is $\la=0.1$. Results in the unbiased
case ($B=0$) and in the presence of the longitudinal bias fields
$\Bred=0.1$ and $0.2$ are shown (for $\w\tK/2\pi=0.005$ only). The
heavy solid lines are the equilibrium susceptibilities [Eq.\
(\ref{X:para:perp:bias})].  $\chi_{\perp}'$ (circles) and
$\chi_{\perp}''$ (rhombi) have intentionally been plotted with the
same scale to show the relative smallness of the latter.
\label{chiperp:plot}
}
\end{figure}

We shall now study the {\em transverse\/} dynamical response of an ensemble of
magnetic moments with parallel anisotropy axes ($\dB\perp\hat{z}$). Figure
\ref{chiperp:plot} displays the transverse dynamical susceptibility for various
frequencies of the probing field (curves labelled 1; results in the
presence of a bias field, to be discussed below, are also shown).

For this transverse probing-field geometry, the mechanism of
inter-potential-well rotations plays a secondary dynamical r\^{o}le,
since it mainly pertains to the components of the magnetic moments
perpendicular to the probing field, whereas the response in the
probing-field direction is the one analyzed. This consists of
intra-potential-well rotations, which are very fast ($\sim\tK$) in
comparison with $\tm=2\pi/\w$ (see the panel $\mx$ vs.\ $t$ of Fig.\
\ref{jumps:ac:plot}).  Consequently, the dynamical susceptibilities
obtained are close to the equilibrium susceptibility in the whole
temperature range. Indeed, the $\chi_{\perp}'(T)$ curves corresponding
to different frequencies are very close to one another (they visually
coincide) and almost describe the equilibrium susceptibility
$\chi_{\perp}(T)$ (heavy solid line), while the out-of-phase component
$\chi_{\perp}''(T)$ is small.  Furthermore, $\chi_{\perp}''$ is not
only small in comparison with $\chi_{\perp}'$ but it is also much
smaller than the out-of-phase longitudinal susceptibility
$\chi_{\|}''$ (cf.\ Fig.\ \ref{chipara:plot}). Nevertheless,
$\chi_{\perp}''$ provides an interesting information concerning the
dynamics of $\m$, which will be discussed below.

For the transverse response, the maximum of $\chi_{\perp}'$ vs.\ $T$
is due to the crossover from the free-rotator regime ($\s=Kv/\T\ll 1$)
to the discrete-orientation regime ($\s\gg 1$), induced by the
bistable magnetic-anisotropy potential. This is essentially a {\em
thermal-equilibrium\/} effect (see Subsec.\ \ref{subsect:X}), with a
markedly different character from the {\em dynamical\/} maxima
exhibited by the longitudinal susceptibility $\chi_{\|}(\w,T)$.

\paragraph{Response for anisotropy axes distributed at random.}

Owing to the linearity of the response, when a distribution in
anisotropy axis orientations occurs, $\chi(\w)$ {\em in the absence of
a bias field\/} is merely given by the weighted sum of the
longitudinal and transverse dynamical susceptibilities, the weight
factors being $\avcosqal$ and $\avsenqal$, respectively. Here,
$\alpha$ is the angle between the anisotropy axis and the probing
field, and the angular brackets enclosing functions of $\alpha$ or
susceptibilities, stand for average over the anisotropy axis
distribution of an ensemble with the same parameters $\la$, $\dU=Kv$,
and $\mm$.
\begin{figure}[t!]
\vspace{-3.ex}
\eps{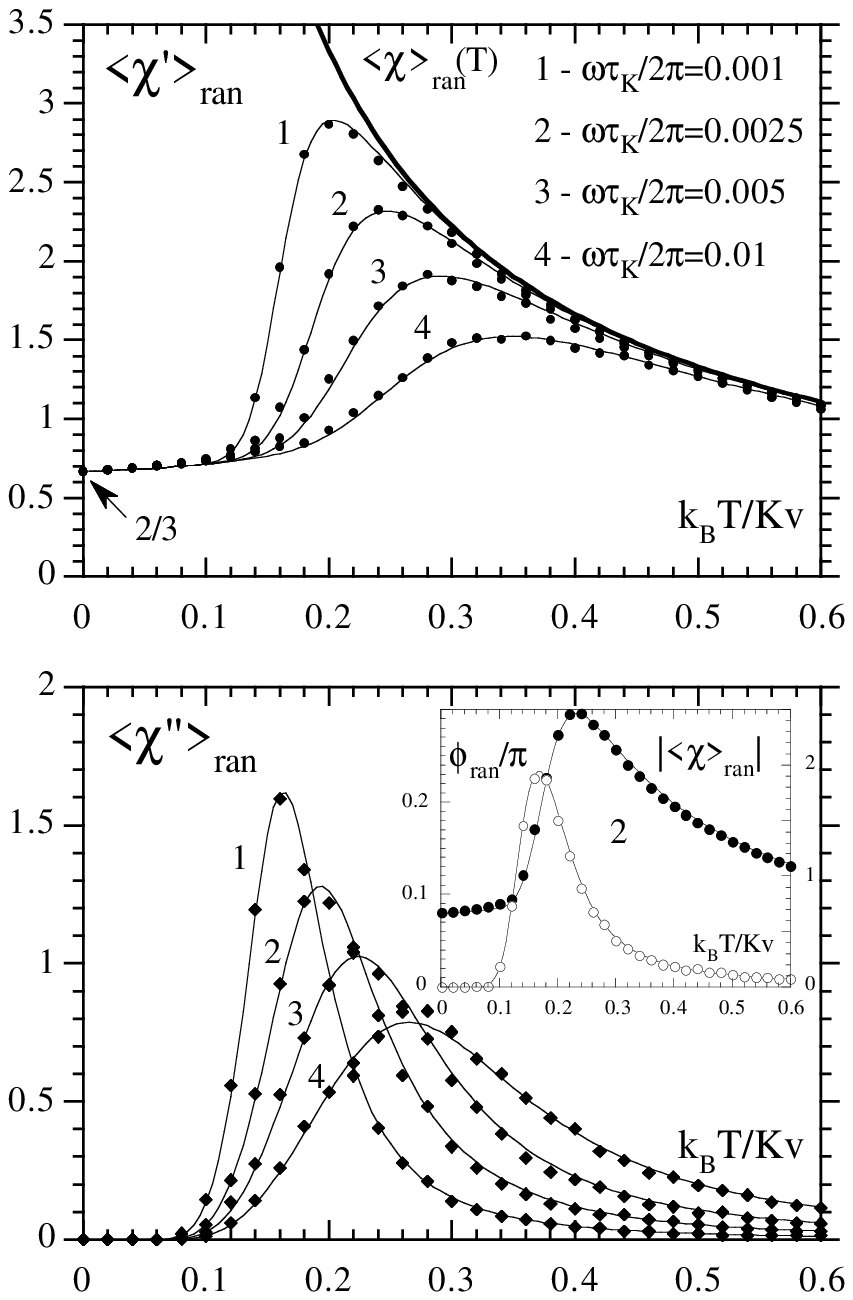}{0.7}
\vspace{-3.ex}
\caption[]
{
Linear dynamical susceptibility vs.\ $T$ for anisotropy axes
distributed at random, $B=0$, and $\la=0.1$. The symbols are for the
numerically computed $\avX$ and the thin solid lines are Eq.\
(\ref{shschi:bias}) with $\tlo$ defined as integral relaxation time
[Eq.\ (\ref{tauint})], and $\ttr$ given by the effective transverse
relaxation time (\ref{effeigentaumod}). The heavy solid line in the
upper panel is the thermal-equilibrium susceptibility [Eq.\
(\ref{X:ran})]. Inset: Modulus and phase shift
$\phase=\arctan(\chi''/\chi')$ for $\w\tK/2\pi=0.0025$.
\label{chiran:plot}
}
\end{figure}

The linear dynamical susceptibility for anisotropy axes distributed at
random ($\avcosqal=\avsenqal/2=1/3$) is displayed on Fig.\
\ref{chiran:plot}. The out-of-phase component, $\avXi$, is
overwhelmingly dominated by the responses to the components of the
probing field {\em along\/} the different anisotropy axes, and it is
almost $\frac{1}{3}\chi_{\|}''(\w,T)$ (cf.\ Fig.\
\ref{chipara:plot}). On the other hand, the in-phase component,
$\avXr$, is approximately $\frac{1}{3}\chi_{\|}'(\w,T)$ plus a
non-uniform upwards shift of magnitude $\frac{2}{3}\chi_{\perp}(T)$,
where $\chi_{\perp}(T)$ is the {\em equilibrium\/} transverse
susceptibility. This occurs in such a way that: (i) at high
temperatures, the Curie law $\avX|_{B=0}=\mu_{0}\mm^{2}/3\T$ is obeyed
(see Subsec.\ \ref{subsect:X}) and, (ii) at temperatures well below
the blocking temperatures, the response consists mainly of the
projection in the probing field direction of the rotations of the
magnetic moments close to the bottom of the potential wells towards
the transverse components of the probing field
($\frac{2}{3}\chi_{\perp}|_{T\simeq0}$). Due to the short
characteristic time of these intra-potential-well motions ($\sim\tK$;
see Fig.\ \ref{jumps:ac:plot}), this low-temperature response is
nearly instantaneous and in phase with the probing field (see the
inset of Fig.\ \ref{chiran:plot}).

Note that the large value of the effective $\tau_{0}$
($\sim10^{-8}$--$10^{-7}$\,s) in the Arrhenius law
$\tlo\simeq\tau_{0}\exp(\dU/\T)$, encountered in molecular magnetic
clusters having high spin in their ground state, entails that
experimental conditions with $\w/2\pi\sim10^{3}$--$10^{4}$\,Hz already
correspond to the frequency range considered here (the MHz range if
$\tK\sim10^{-10}$--$10^{-8}$\,s).  Indeed, these systems clearly
exhibit the qualitative features of the linear dynamical
susceptibility found at ``high" (but below ferromagnetic resonance)
frequencies: wide maxima in $\chi(\w,T)$ vs.\ $T$ for only one
potential barrier (relaxation time), sizable $\chi'(T)$ at
temperatures well below the blocking temperatures, and flattening of
the peak of $\chi''(T)$ with increasing $\w$ (Barra et~al., 1996;
Gomes et~al., 1998).

\subsubsection{Dynamical response in a longitudinal bias field}

We shall now study the effects of a constant magnetic field, $\B$,
applied along the common anisotropy axis direction of a spin ensemble
with parallel anisotropy axes ($\B\parallel\hat{z}$).

\paragraph{Longitudinal response.}

\begin{figure}[t!]
\vspace{-3.ex}
\eps{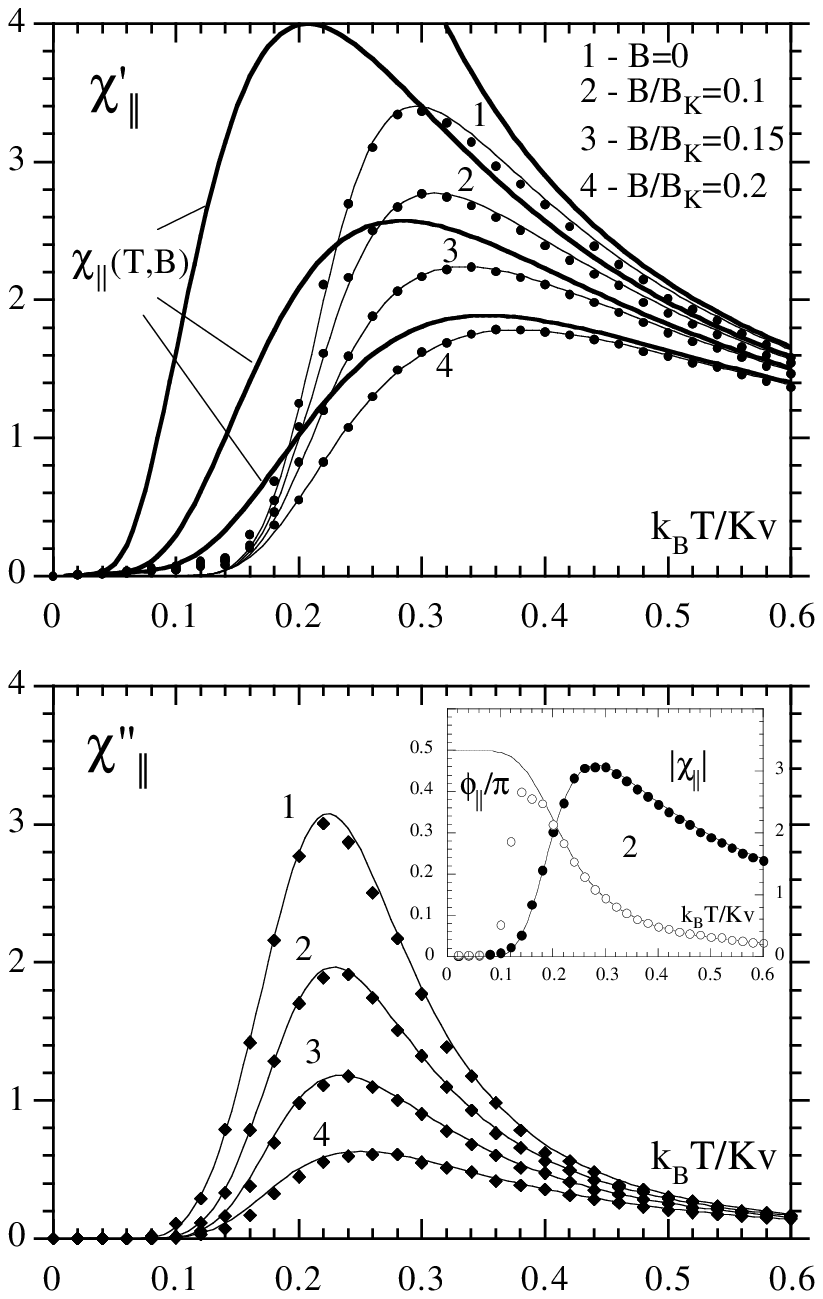}{0.7}
\vspace{-3.ex}
\caption[]
{
Longitudinal dynamical susceptibility $\chi_{\|}$ vs.\ $T$, for
$\la=0.1$, $\w\tK/2\pi=0.005$, and various values of the longitudinal
bias field. The symbols are for the numerically computed
$\chi_{\|}(\w,T,B)$ and the thin solid lines are Eq.\
(\ref{shschi:bias}) with $\tlo$ defined as integral relaxation time
[Eq.\ (\ref{tauint})]. The heavy solid lines in the upper panel are
the corresponding equilibrium susceptibilities [Eq.\
(\ref{X:para:perp:bias})]. Inset: Modulus and phase shift
$\phase=\arctan(\chi''/\chi')$ for $\Bred=0.1$.
\label{chiparabias:plot}
}
\end{figure}

Figure \ref{chiparabias:plot} displays the longitudinal
($\dB\parallel\hat{z}\parallel\B$) linear dynamical susceptibility of
the system for various values of the bias field.
The qualitative features of the susceptibility curves are similar to
those encountered in the unbiased case ($B=0$), and can be interpreted
in terms of the same processes:
\begin{enumerate}
\item
At low temperatures the response consists of the fast rotations of the
magnetic moments close to the bottom of the potential wells, with the
over-barrier relaxation mechanism being blocked. \item As $T$ is
increased the magnetic moments can depart from the potential minima by
means of the energy gained from the heat bath, and the response starts
to increase steeply with $T$ (with a sizable lag behind the probing
field). \item If $T$ is further increased the system reaches the
regime dominated by inter-potential-well rotations, exhibiting
dynamical maxima first in the phase shift and subsequently in the
magnitude of the response. \item In the high-temperature range, the
magnetic moments are almost in the thermal-equilibrium state
associated with the instantaneous probing field and, hence,
$\chi_{\|}'(T,B)$ approaches to the linear equilibrium susceptibility
while $\chi_{\|}''(T,B)$ tends to zero.
\end{enumerate}
Thus, the dynamics is qualitatively similar to the dynamics in the
unbiased case, the main difference being that the system now consists
of bistable {\em non-symmetrical\/} entities (recall the panel
$B/\BK=0.1$ of Fig.\ \ref{jumps:ac:plot}).

We remark in passing that the simple idea that the application of a
constant magnetic field reduces the potential barriers, so that the
relaxation rate increases and the blocking temperatures shift to lower
temperatures, should be viewed with caution.  The location of the
maximum of the dynamical response do depend on the potential-barrier
heights, but also on the form of the {\it equilibrium\/} response,
which is markedly different from that of the unbiased case.%
\footnote{
In a bias field, because $\chi_{\|}(T,B)$ is the slope of the
magnetization vs.\ field curve at $B$, instead of the initial slope of
the unbiased case, the equilibrium response already exhibits a maximum
as a function of $T$ (see Subsec.\ \ref{subsect:X}).
}
Indeed, for the frequencies and bias fields considered, the location
of the maxima of $\chi_{\|}''(T)$ is not very sensitive to the bias
field, while the maxima of $\chi_{\|}'(T)$ shift slightly to higher
temperatures as $B$ increases.

\paragraph{Transverse response.}

We shall finally consider the {\em transverse\/} dynamical response in
the presence of a {\em longitudinal\/} bias field
($\dB\perp\hat{z}\parallel\B$). Figure \ref{chiperp:plot} also
displays $\chi_{\perp}$ vs.\ $T$ for various values of the bias field
at $\w\tK/2\pi=0.005$ (curves labelled 2 and 3). The qualitative
features of the response are again similar to those encountered in the
unbiased case:
\begin{enumerate}
\item
The mechanism of inter-potential-well rotations plays a minor dynamical
r\^{o}le, with the response being dominated by the fast intra-potential-well rotations.
\item
The $\chi_{\perp}'(T,B)$ curves obtained are rather close to the corresponding
equilibrium susceptibilities (heavy solid lines).
\item
$\chi_{\perp}''(T,B)$ is small in comparison with both $\chi_{\perp}'(T,B)$ and
$\chi_{\|}''(T,B)$.
\end{enumerate}

\subsubsection{Comparison with different analytical expressions}

We shall now compare the linear dynamical susceptibility, obtained by
numerical integration of the stochastic Landau--Lifshitz--Gilbert
equation, with the heuristic models discussed in Subsec.\
\ref{subsect:models} and rigorous expressions. In this comparison {\em
no adjustable parameter\/} will be employed.

We shall sometimes use the word {\em exact\/} when referring to the
numerically computed quantities. Along with the feasible diminishing
of the statistical error bars of the computed quantities by averaging
over a sufficiently large number of trajectories, we also implicitly
mean that the numerical results are {\em exact\/} in the context of
the Brown--Kubo--Hashitsume stochastic model.

\paragraph{Longitudinal response.}

\begin{figure}[t!]
\vspace{-3.ex}
\eps{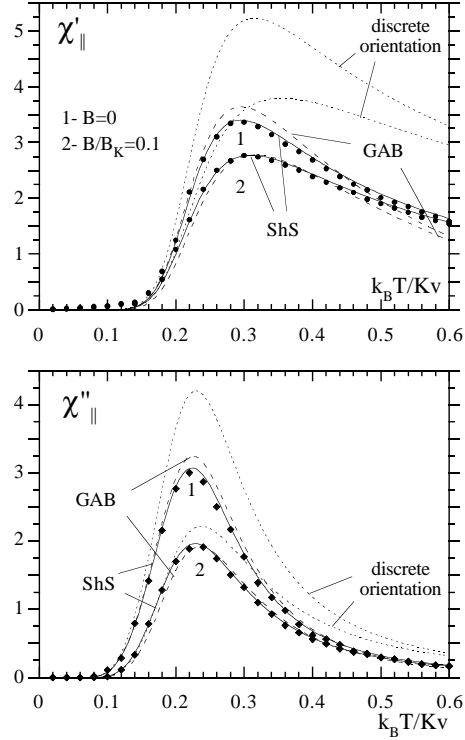}{0.7}
\vspace{-3.ex}
\caption[]
{
$\chi_{\|}$ vs.\ $T$ for $B=0$ and $\Bred=0.1$ with $\w\tK/2\pi=0.005$ (symbols).
The small dashing is for Eq.\ (\ref{isingchi:bias}), the medium dashing for Eq.\
(\ref{shschi:bias}) with the approximate Eq.\ (\ref{X:para:bias:approx}), and the solid
lines for Eq.\ (\ref{shschi:bias}). $\tlo$ defined as integral relaxation time [Eq.\
(\ref{tauint})] has been incorporated in the three equations.
\label{chipara_teo:plot}
}
\end{figure}

Figure \ref{chipara_teo:plot} shows the computed $\chi_{\|}(\w)$ in
the unbiased case and in the bias field $\Bred=0.1$.
The results of the heuristic discrete-orientation equation
(\ref{isingchi:bias}); Gittleman, Abeles, and Bozowski model [Eq.\
(\ref{shschi:bias}) with the approximate Eq.\
(\ref{X:para:bias:approx})]; and Shliomis and Stepanov equation
(\ref{shschi:bias}) are also shown. The longitudinal relaxation time,
$\tlo$, defined as the {\em integral relaxation time\/} $\tint$, has
been used in the three equations.

It is apparent that Eq.\ (\ref{isingchi:bias}) fails to describe the
numerical results; neither is the equilibrium (high-temperature)
susceptibility properly described.  Actually, the overall failure of
this expression could mainly be attributed to the rough approximation
used for its equilibrium part [Eq.\
(\ref{X:para:perp:bias:ising})]. The probability that $\m$ makes a
finite angle with the anisotropy axis is completely neglected in such
a discrete-orientation equation.

Concerning the Gittleman, Abeles, and Bozowski equation, it is more
suitable than the discrete-orientation equation, especially for the
matching of $\chi_{\|}''(T,B)$, although it fails to describe
$\chi_{\|}'(T,B)$. Again, not even the equilibrium susceptibility is
correctly described; the high-barrier approximation for
$\chi_{\|}(T,B)$ occurring in this model [Eq.\
(\ref{X:para:bias:approx})], although better than the
discrete-orientation approximation, is still not accurate enough at
the relevant temperatures. Furthermore, for bias fields
$\Bred\gsim0.15$, the divergence of this model from the exact results
becomes dramatic (results not shown).

In contrast, Eq.\ (\ref{shschi:bias}) approximates the numerical results reasonably.
This is in agreement with the comparison carried out by Ra{\u{\i}}kher et~al.\ (1997)
of the exact $\chi_{\|}(\w)$ with what they called the ``effective time
approximation" [which is indeed equivalent to the use of the longitudinal component of
Eq.\ (\ref{shschi:bias}) with $\tlo=\tint$]. Nevertheless, the exact analytical expression
for $\chi_{\|}(\w)$ comprises an infinite number of Debye-type relaxation
mechanisms, namely (see Appendix \ref{app:taus})
\begin{equation}
\label{chidyn:2}
\chi_{\|}(\w,T,B)
=
\chi_{\|}(T,B)
\sum_{k=1}^{\infty}\frac{a_{k}(T,B)}{1+i\omega/\Lambda_{k}(T,B)}
\;,
\end{equation}
where $a_{k}$ is the amplitude corresponding to the eigenvalue $\Lambda_{k}$ of the
Sturm--Liou\-ville equation associated with the Fokker--Planck equation. (Recall that
the first non-vanishing eigenvalue, $\Lamone$, is associated with the
inter-potential-well dynamics, whereas the higher-order eigenvalues, $\Lambda_{k}$,
$k\geq2$ are related with the intra-potential-well relaxation modes.) However, the
mentioned agreement could be expected in the unbiased case since, as was shown
numerically by Coffey et~al.\ (1994):
(i)
$a_{1}(B=0)\gg a_{k}(B=0),\,\forall k\geq2$
and
(ii)
$\Lamone^{-1}(B=0)\simeq\tint(B=0)$.
Indeed, Coffey, Crothers, Kalmykov and Waldron (1995{\em b}) shown that an
expression equivalent to the longitudinal component of Eq.\ (\ref{shschi:bias}),
together with the interpolation formula (\ref{creggtau}) for $\Lamone^{-1}$, well
describes the longitudinal dynamical polarisability of the congeneric nematic liquid
crystal with (unbiased) Meier-Saupe potential. (The {\em longitudinal\/} relaxation in
this system is mathematically identical with that of classical magnetic moments.) On
the other hand, in a constant longitudinal field the higher-order modes can make a
substantial contribution {\em in the low-temperature region ($\s\gg1$)}, and then
$\Lamone^{-1}$ largely deviates from $\tint$ while $a_{1}\gg a_{k}$ no longer holds
(Coffey, Crothers, Kalmykov and Waldron, 1995{\em a}; Garanin, 1996). Nevertheless,
for the frequencies employed here, the relevant dynamical phenomena occur in the
range $\s\sim2$--$10$, so that in the bias fields applied
$a_{1}\gg a_{k}$ and $\Lamone^{-1}\simeq\tint$ still hold approximately, and hence
Eq.\ (\ref{shschi:bias}) describes the exact results reasonably.

However, one could expect, even for $B=0$, a significant contribution
of the intra-potential-well relaxation modes to the longitudinal
response when the over-barrier dynamics is {\em blocked\/} at low $T$
($\w/\Lamone\gg1$). Indeed, on scrutinizing Figs.\ \ref{chipara:plot}
and \ref{chiparabias:plot}, one sees that Eq.\ (\ref{shschi:bias})
predicts, both for $B=0$ and $B\neq0$, a smaller $\chi_{\|}'$ when
departing from zero at temperatures well below the blocking
temperatures than the exact $\chi_{\|}'$. In contrast, because the
intra-potential-well modes are very fast ($\sim\tK$), their
contribution to the out-of-phase susceptibility is comparatively
smaller, so that $\chi_{\|}''$ is still described reasonably by the
Debye-type term associated with the inter-potential-well dynamics
($\chi_{\|}''\simeq\chi_{\|}(\w/\Lambda_{1})/[1+(\w/\Lambda_{1})^{2}]$).

\begin{figure}[b!]
\vspace{-3.ex}
\eps{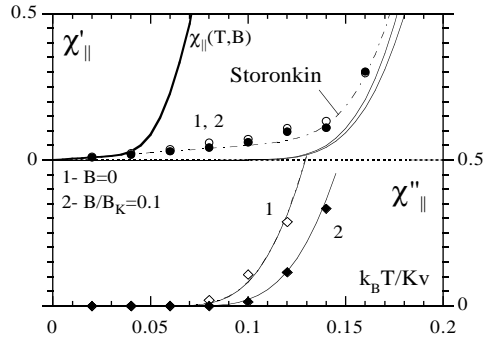}{0.7}
\vspace{-3.ex}
\caption[]
{
Detail of the low-temperature part of Fig.\ \ref{chipara_teo:plot}
showing the effect of the intra-potential-well relaxation modes. The
heavy solid line is the equilibrium susceptibility for $\Bred=0.1$,
the thin solid lines are for Eq.\ (\ref{shschi:bias}), and the
dashed-dotted lines for the asymptotic result (\ref{sto_chi}) by
Storonkin (for $B=0$ only). The out-of-phase components of Eqs.\
(\ref{shschi:bias}) and (\ref{sto_chi}) visually coincide.
\label{chipara_teo_amp:plot}
}
\end{figure}

These considerations are substantiated by comparing the numerical
results with the asymptotic ($\s\gg1$) expression for the longitudinal
dynamical susceptibility of the nematic liquid crystal derived by
Storonkin (1985), namely
\begin{equation}
\label{sto_chi}
\chi_{\|}
\simeq
\Xo
\bigg[
\underbrace{
\bigg( 1-\frac{1}{\s}-\frac{3}{4\s^{2}}
\bigg)
\frac{1}{1+i\w/\Lambda_{1}}
}_{\mbox{\small inter-potential-well mode}}
{}+
\underbrace{
\frac{1}{8\s^{2}}
\bigg(
\frac{1}{1+i\w/\Lambda_{3}}+\frac{1}{1+i\w/\Lambda_{5}}
\bigg)
}_{\mbox{\small intra-potential-well modes}}
\bigg]
\;,
\end{equation}
where [cf.\ Eq.\ (\ref{browntau}) at $B=0$]
\begin{eqnarray}
\label{browntaucorr}
\Lambda_{1}^{-1}
&
\simeq
&
\tau_{N}\frac{\sqrt{\pi}}{2}\s^{-3/2}\exp(\s)
\Big(1+\frac{1}{\s}+\frac{7}{4\s^{2}}\Big)
\;,
\\
\label{tau3-5}
\Lambda_{3}^{-1}
\simeq
\Lambda_{5}^{-1}
&
\simeq
&
\frac{1}{2}\frac{\tN}{\s}
\Big( 1+\frac{5}{2\s}+\frac{41}{4\s^{2}}
\Big)
\;.
\end{eqnarray}
Note that
$(\mu_{0}\mm^{2}/\T)(1-1/\s-3/4\s^{2})
\simeq\chi_{\|}(T)+{\cal O}(1/\s^{2})$
[see Eqs.\ (\ref{X:para:perp}) and (\ref{F-der:approx2b})], while the
correction terms in $\Lambda_{1}^{-1}$ agree with those derived by
Brown (1979) (see also Coffey et~al., 1994). Figure
\ref{chipara_teo_amp:plot} shows that Eq.\ (\ref{sto_chi}) remarkably
describes the $B=0$ numerical results at low temperatures.
Note that, because $\Lambda_{3,5}\sim\tN/\s=\tK$ [Eq.\
(\ref{neeltime-tauwell})] and $\w\tK\ll1$ for the frequencies
considered, it follows that
$1/(1+i\w/\Lambda_{3,5})\simeq1-i\w/\Lambda_{3,5}$. Therefore, since
$(\mu_{0}\mm^{2}/\T)\times(1/8\s^{2})\propto\T$, Storonkin formula
(\ref{sto_chi}) yields the low-temperature linear increase of
$\chi_{\|}'$ with $T$ due to the intra-potential-well relaxation
modes, whereas their contribution to $\chi_{\|}''$ is smaller by a
factor $\w/\Lambda_{3,5}\sim\w\tK$.

Furthermore, the intra-potential-well relaxation modes take a dramatic
reflection in the phase shifts (Ra{\u{\i}}kher and Stepanov, 1995{\em
b}). As any expression of the form $\chi(\w)=\chi/(1+i\w\tau)$
(Debye-type), the {\em longitudinal\/} component of Eq.\
(\ref{shschi:bias}) yields a phase shift
\begin{equation}
\label{phase:shift:debye}
\phase_{\|}
=
\arctan(\w\tlo)
\;,
\end{equation}
which increases monotonically with decreasing $T$ and, eventually,
reaches $\pi/2$ since at low temperatures $\w\tlo\gg1$ (see the insets
of Figs.\ \ref{chipara:plot} and \ref{chiparabias:plot}). However,
owing to the fact that the fast intra-potential-well relaxation modes
yield an almost instantaneous contribution to the response,
$\chi_{\|}'$ decreases with $T$ less steeply than
$\chi_{\|}/[1+(\w/\Lambda_{1})^{2}]$ at low temperatures, whereas
$\chi_{\|}''$ is still approximately given by
$\chi_{\|}(\w/\Lambda_{1})/[1+(\w/\Lambda_{1})^{2}]$. Consequently,
the actual phase shift (insets of Figs.\ \ref{chipara:plot} and
\ref{chiparabias:plot}), also increases monotonically with decreasing $T$ but, at a
temperature close to that of the peak of $\chi_{\|}''(T)$,
$\phase_{\|}(T)$ {\em exhibits a maximum\/} and then decreases to
zero, since at low $T$ the response is again ``in phase" with the
probing field due to the fast intra-potential-well modes. This
behavior of the phase shift is qualitatively similar to that
encountered in one-dimensional bistable systems (Morillo and
G{\'{o}}mez-Ord{\'{o}}{\~{n}}ez, 1993) and ascribed to the crossover
from the ``high-noise" regime, dominated by inter-potential-well
jumps, to the ``low-noise" regime, dominated by the fast
intra-potential-well motions.

Concerning the phase behavior for non-collinear situations, we must
bear in mind that the intra-potential-well motions make a relative
contribution to the transverse response much larger than to the
longitudinal response. Therefore, as the former contribution is
somehow taken into account by Eq.\ (\ref{shschi:bias}), via the {\em
equilibrium\/} transverse susceptibility, we find that, inasmuch as
$\avcosqal$ departs from unity, the Shliomis and Stepanov equation
describes the low-temperature phase shifts reasonably well (cf. the
inset of Fig.\ \ref{chipara:plot} with that of Fig.\
\ref{chiran:plot}). We finally remark that, because the
intra-potential-well relaxation modes are very fast and, thus,
$\chi_{\|}''$ is reasonably described by Eq.\ (\ref{shschi:bias}),
while $\chi_{\perp}''$ is relatively small, the theoretical background
of the methods of determination of the energy-barrier distribution of
Section \ref{sect:heuristic} that are based on the use of the {\em
out-of-phase\/} component of the low-frequency equation
(\ref{shschi:lowfrec}), result to be supported in the context of the
Brown--Kubo--Hashitsume stochastic model.

\paragraph{Transverse response.}

\begin{figure}[t!]
\vspace{-3.ex}
\eps{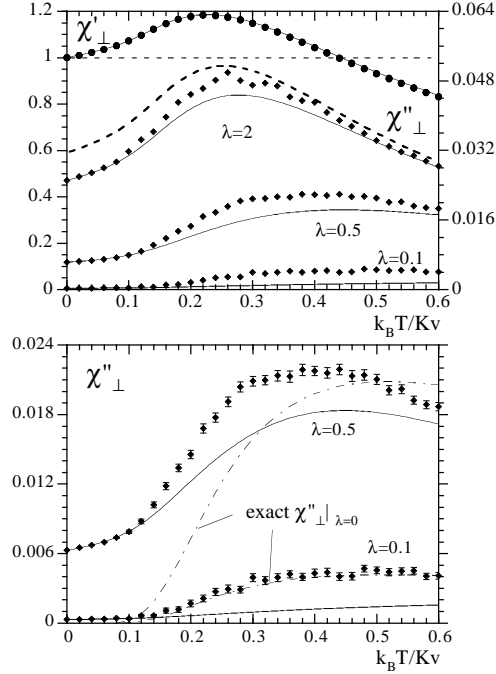}{0.56}
\vspace{-3.ex}
\caption[]
{
Upper panel: $\chi_{\perp}$ vs.\ $T$ for $B=0$, $\w\tK/2\pi=0.005$,
and various values of the damping coefficient $\la$. The circles are
for $\chi_{\perp}'$, and the rhombi for $\chi_{\perp}''$. The medium
dashed line corresponds to the constant $\chi_{\perp}'$ given by Eq.\
(\ref{gabchi}) and the solid lines to Eq.\ (\ref{shschi:bias}) with
$\ttr$ given by the effective transverse relaxation time
(\ref{effeigentaumod}). The heavy dashed curve is $\chi_{\perp}''$
with $\ttr$ given by the $\la\gg1$ result (\ref{effeigentau}). Lower
panel: Detail of $\chi_{\perp}''$ in the intermediate-to-weak damping
range together with the exact zero-damping formula (\ref{gip_chi})
(dashed-dotted lines).
\label{chiperp_teo:plot}
}
\end{figure}

Figure \ref{chiperp_teo:plot} displays the corresponding comparison
for $\chi_{\perp}(\w)$ in the unbiased case for various values of the
damping coefficient.%
\footnote{
In the cases with larger damping coefficients, $\la=0.5$ and $2$, we
have used a discretization time interval $\Delt=0.0025\tK$ in the
numerical integration of the stochastic Landau--Lifshitz--Gilbert
equation, instead of the value $\Delt=0.01\tK$ used in the rest of
this Section. 
}
For the transverse relaxation time, $\ttr$, we have employed the
effective relaxation time (\ref{effeigentaumod}), which has been
derived (Appendix \ref{app:taus}) from the low-frequency expansion of
the equation for $\chi_{\perp}(\w)$ of Ra\u{\i}kher and Shliomis
(1975; 1994).

For the transverse probing-field geometry, the discrete-orientation formula
(\ref{isingchi:bias}) predicts obviously an identically zero response, while the
Gittleman, Abeles, and Bozowski formula yields a constant $\chi_{\perp}'(T)$ and a
zero $\chi_{\perp}''(T)$. In contrast, the exact $\chi_{\perp}'(T)$ is well described by
Eq.\ (\ref{shschi:bias}), although, because $\w\ttr\ll1$ holds in the considered
frequency range, $\chi_{\perp}'(T)$ almost coincides with the equilibrium
susceptibility $\chi_{\perp}(T)$. Concerning $\chi_{\perp}''(T)$, Eq.\ (\ref{shschi:bias})
with the effective expression (\ref{effeigentaumod}) for $\ttr$ only matches the
out-of-phase response in the low-temperature range for the smallest damping
coefficient considered ($\la=0.1$). Nevertheless, Fig.\ \ref{chiperp_teo:plot} shows
that, as the damping coefficient is enlarged, the matching between the numerical
results and the simple Eq.\ (\ref{shschi:bias}) improves if one uses the effective $\ttr$
proposed [Eq.\ (\ref{effeigentaumod})]. This constitutes an advance over the usual
approach, where one employs the $\ttr$ derived by the effective-eigenvalue method
[Eq.\ (\ref{effeigentau:bias})], which yields the heavy dashed curve of Fig.\
\ref{chiperp_teo:plot} {\em irrespective of\/} $\la$.

The above comparison is in agreement with that made by Kalmykov and Coffey
(1997) of their numerical results, obtained by continued-fraction techniques, with the
complete (but approximate) expression for $\chi_{\perp}(\w)$ of Ra\u{\i}kher and
Shliomis (1975; 1994).%
\footnote{
In the frequency range below the ferromagnetic resonance range, this formula is
indistinguishable from the low-$\w$ expansion used here.
}
The failure of this expression for weak damping was explained in terms of the effects
of the gyromagnetic terms of the dynamical equation. When these terms dominate
($\la\ll1$), due to the occurrence of a spread of the precession frequencies of $\m$ in
the anisotropy field at intermediate temperatures (these frequencies are
$\propto\gBK\mz$), the response is not well described by a simple relaxation
mechanism. Then, only at low temperatures, where the magnetic moments are
concentrated close to the bottom of the potential wells (so the spread in precession
frequencies is reduced), the exact results are well described by the
$\chi_{\perp}(\w)$ of Ra\u{\i}kher and Shliomis.

The effects of the spread of the precession frequencies of $\m$ in the anisotropy field
had already been investigated by Gekht (1983) and independently by Garanin,
Ishchenko and Panina (1990). They  derived  the {\em exact\/} expression for
$\chi_{\perp}''(\w,T,B)$ in the $\la\to0$ limit, which accounts for the effects of
the phenomenon discussed (the former author employed a Liouville approach while
the latter ones started from the Fokker--Planck equation). Their formula can be
written as
\begin{equation}
\label{gip_chi}
\chi_{\perp}''|_{\la=0}
=
\Xo
\frac{\pi}{2}
\frac{\tilde{\w}}{(2|\s|)^{3}}
\frac{(2\s)^{2}-(\tilde{\w}-\xi)^{2}}{\Z}
\exp\Big(\frac{\tilde{\w}^{2}-\xi^{2}}{4\s}\Big)
\;,
\end{equation}
where $\tilde{\w}=\w(\mm/\gmr\T)$, $\xi=\mm B/\T$, $\Z$ is the longitudinal
partition function (\ref{Zpara}), and $\chi_{\perp}''(\w)$ is non-zero in the interval
$(\tilde{\w}-\xi)^{2}\leq(2\s)^{2}$. In order to compare the zero-damping formula
(\ref{gip_chi}) with the numerical results, we just write $\tilde{\w}=\w(2\la\tK\s)$,
which for fixed $\w\tK$ (as occurs in the plots) is a ``function" of $\la$.

The lower panel of Fig.\ \ref{chiperp_teo:plot} shows that, for $\la=0.5$, the
dampingless Eq.\ (\ref{gip_chi}) gives correctly the order of magnitude of the
numerical results at intermediate-to-high temperatures, while for $\la=0.1$ a good
agreement extending down to quite low temperatures can be seen. Since Eq.\
(\ref{gip_chi}) is the exact $\la=0$ result, this comparison indicates that, in the
intermediate-to-weak damping regime, the contribution of the spread of the
precession frequencies of the magnetic moment to $\chi_{\perp}''(\w)$ is sizable in
comparison with the effects of the damping. Therefore, by omitting this zero-damping
effect one could erroneously extract values of $\la$ from the $\chi_{\perp}''(\w)$ data
that overestimate the actual $\la$ and, for example, infer that the damping in
superparamagnets is stronger than it is in fact.

Another important manifestation of this effect was studied by
Ra{\u{\i}}kher and Stepanov (1995{\em a}). The contribution of the
damping to the absorption line in intrinsic ferromagnetic resonance
provokes a (unbounded) monotonic increase of the {\em linewidth\/}
with the temperature, whereas the linewidths experimentally observed
in certain magnetic nanoparticle systems are almost independent of the
temperature (Hennion et al., 1994). However, the spread of precession
frequencies in the anisotropy field also yields a contribution to the
linewidths, which in addition saturates at high temperatures. Thus,
the combination of both contributions leads to the appearance of an
intermediate temperature regime, fairly wide for systems with low
damping, in which the linewidth is quasiconstant.

%% file: garcms06.tex
\section[Foundation of the stochastic dynamical equations]
{Foundation of the stochastic dynamical\\ equations}
\label{sect:gle}

\subsection{Introduction}

In this Section we shall examine various topics related with the foundation of the
Brown--Kubo--Hashitsume stochastic model and possible extensions of this model
(Garc{\'{\i}}a-Palacios, 1999).

\subsubsection{Phenomenological equations}

The Brown--Kubo--Hashitsume model is phenomenological inasmuch as is
constructed by augmenting known phenomenological equations (Gilbert or
Landau--Lifshitz) by fluctuating fields. For subsequent reference, let us first rewrite
the basic equations of this model (see Subsec.\ \ref{subsect:brown}):
\begin{itemize}
\item
{\em Stochastic Gilbert equation}
\begin{equation}
\label{eqmot_M_Gil}
\frac{\D\m}{\D t}
=
\gmr\m\vecpro 
\bigg[
\Beff+\bfl(t)-(\gmr\mm)^{-1}\la\frac{\D\m}{\D t}
\bigg]
\;.
\end{equation}
This equation is equivalent to the stochastic Landau--Lifshitz--Gilbert equation
(\ref{stollgeq}), except for a ``renormalization" of the gyromagnetic ratio.
\item
{\em Stochastic Landau--Lifshitz equation}
\begin{equation}
\label{eqmot_M_LL}
\frac{\D\m}{\D t}
=
\gmr\m\vecpro 
\left[
\Beff+\bfl(t)
\right]
-\la\frac{\gmr}{\mm}
\m\vecpro 
\left(\m\vecpro\Beff\right)
\;.
\end{equation}
This equation may be regarded as the weak damping case ($\la\ll1$) of Eqs.\
(\ref{eqmot_M_Gil}) or (\ref{stollgeq}), although it can be considered as an
independent model as well. On the other hand, this is a Langevin equation more
archetypal those equations, in the sense that the fluctuating and relaxation (damping)
terms are not entangled.
\end{itemize}
In these dynamical equations $\la$ is a dimensionless damping coefficient and
$\Beff=-\partial\Hs/\partial\m$ is the (deterministic) effective field associated with
the Hamiltonian of the spin $\Hs(\m)$. This typically includes Zeeman and
magnetic-anisotropy energy terms, e.g., for uniaxial anisotropy with
symmetry axis $\vec{n}$
\[
\Hs
=
-\m\cdot\B-\half\Kuni(\m\cdot\vec{n})^{2}
\quad
\Longrightarrow
\quad
\Beff
=
\B+\hat{K}\m
\;,
\]
where $\hat{K}$ is a second-rank tensor with elements $K_{ij}=\Kuni n_{i}n_{j}$ [cf.\
Eq.\ (\ref{Beff})]. On the other hand, $\bfl(t)$ is a fluctuating field, the statistical
properties of which are
\begin{equation}
\label{bcorr:2}
\llangle b_{\fl,i}(t)\rrangle
=
0
\;,
\qquad
\llangle b_{\fl,i}(t)b_{\fl,j}(\tp)\rrangle
=
\frac{2\la\delta_{ij}}{\gmr\mm}\,\T
\delta(t-\tp)
\;,
\end{equation}
where we have taken into account that when one starts from the Gilbert equation one
must replace $\gmr\to\gmr/(1+\la^{2})$ in the results of Section \ref{sect:stochastic}
associated with the stochastic Landau--Lifshitz--Gilbert equation, so that $D_{{\rm
LLG}}$ is then identical with $D_{{\rm LL}}$ [see Eq.\ (\ref{coeffdif:ll-llg})]. Finally, on
introducing Eq.\ (\ref{stationary_condition}) into Eq.\ (\ref{brownfpe:2}), the
Fokker--Planck equation governing the time evolution of the non-equilibrium
probability distribution of spin orientations, associated with the above Langevin
equations, can be written as
\begin{equation}
\label{brownfpe:Gil-LL}
\frac{\partial\W}{\partial t}
=
-\frac{\partial {}}{\partial\m}\cdot
\left\{
\gmr\m\vecpro\Beff\W
-\la\frac{\gmr}{\mm}
\m\vecpro
\left[
\m\vecpro\left(\Beff-\T\frac{\partial {}}{\partial\m}\right)\W
\right]
\right\}
\;,
\end{equation}
where
$(\partial/\partial\m)\cdot\vec{J}
=\sum_{i} (\partial J_{i}/\partial\mi)$
and for the Gilbert case one must replace $\gmr$ by $\gmr/(1+\la^{2})$.

The Brown--Kubo--Hashitsume stochastic model has been the basis of significant
studies of the dynamics of classical magnetic moments. Nonetheless, there exist
important microscopic relaxation mechanisms that cannot be accommodated in the
context of this model, inasmuch as they do not produce a field-type perturbation on
the spin (``field-type" fluctuations). An important example is the coupling of the spin
to the lattice vibrations, which modulate the crystal-field and the exchange and
dipole-dipole interactions, and can produce fluctuations of the magnetic-anisotropy
potential of the spin (``anisotropy-type" fluctuations).

In order to take this phenomenon into account, Garanin, Ishchenko, and Panina (1990)
generalized the above Langevin equations to $\D\m/\D
t=\gmr\m\vecpro[\Beff+\vec{\bt}(t)+\hat{\kt}(t)\m]-\R$. Here, $\R$ is a relaxation
term to be determined and, in analogy with the expression $\Beff=\B+\hat{K}\m$ for
the effective field, $\vec{\bt}(t)$ is a stochastic {\em vector\/} that introduces the
field-type part of the thermal fluctuations, while $\hat{\kt}(t)$ is a stochastic {\em
second-rank tensor}, so that $\hat{\kt}(t)\m$ incorporates anisotropy-type
fluctuations into the dynamical equation.

On assuming the correlation properties
\begin{eqnarray}
\label{stats:gip}
\llangle\bt_{i}(t)\bt_{j}(\tp)\rrangle
&
=
&
\frac{2\la_{ij}}{\gmr\mm}\,\T\delta(t-\tp)
\;,
\nonumber
\\
\llangle\bt_{i}(t)\kt_{jk}(\tp)\rrangle
&
=
&
\frac{2\la_{i,jk}}{\gmr\mm}\,\T\delta(t-\tp)
\;,
\\
\llangle\kt_{ik}(t)\kt_{j\ell}(\tp)\rrangle
&
=
&
\frac{2\la_{ik,j\ell}}{\gmr\mm}\,\T\delta(t-\tp)
\;,
\nonumber
\end{eqnarray}
they constructed the associated Fokker--Planck equation
\begin{equation}
\label{fpe_gip}
\frac{\partial\W}{\partial t}
=
-\frac{\partial {}}{\partial\m}\cdot
\bigg\{
\gmr\m\vecpro\Beff
\W
-\bigg[
\R
-\frac{\gmr}{\mm}\T\,
\m\vecpro\TGIP
\bigg(
\m\vecpro 
\frac{\partial{}}{\partial\m}
\bigg)
\bigg]
\W
\bigg\}
\;,
\end{equation}
where $\TGIP$ is a symmetrical second-rank tensor related with the
correlation coefficients of the fluctuating terms by
\begin{equation}
\label{tensor:gip}
\tGIP_{ij}
=
\la_{ij}
+\sum_{k}(\la_{i,jk}+\la_{j,ik})\mk
+\sum_{k\ell}\la_{ik,j\ell}\mk\ml
\;.
\end{equation}
The relaxation term $\R$ was then determined by merely assuming that the
Boltzmann distribution $\Weq(\m)\propto\exp[-\Hs(\m)/\T]$ is a stationary solution
of the Fokker--Planck equation (\ref{fpe_gip}). This yields
$\R=(\gmr/\mm)\,\m\vecpro\TGIP(\m\vecpro\Beff)$, so the starting Langevin
equation finally reads [cf.\ Eq.\ (\ref{eqmot_M_LL})]
\begin{equation}
\label{eqmot_M_gip}
\frac{\D\m}{\D t}
=
\gmr\m\vecpro 
\left[
\Beff+\vec{\bt}(t)+\hat{\kt}(t)\m
\right]
-\frac{\gmr}{\mm}\m\vecpro\TGIP
\left(\m\vecpro\Beff\right)
\;.
\end{equation}
For an arbitrary form of $\TGIP$ the relaxation term in this equation deviates from
the form proposed by Landau and Lifshitz (1935). Only for $\tGIP_{ij}=\la\delta_{ij}$,
which for instance occurs when both the field-type and the anisotropy-type
fluctuations are isotropic ($\la_{ij}\propto\delta_{ij}$ and
$\la_{ik,j\ell}\propto\delta_{ij}\delta_{k\ell}$) and there is not interference between
them ($\la_{i,jk}\equiv0$), that archetypal relaxation term is recovered and the
Fokker--Planck equation of Garanin, Ishchenko, and Panina [Eq.\ (\ref{fpe_gip})]
reduces to Eq.\ (\ref{brownfpe:Gil-LL}).

\subsubsection{Dynamical approaches to the phenomenological equations}

There have been several attempts to justify, starting from dynamical descriptions of a
spin coupled to its surroundings, the phenomenological equations for the stochastic
spin dynamics.

Smith and De~Rozario (1976) considered a classical magnetic moment $\m$ coupled to
a field $\vec{b}(\ePm,\eQm)$ depending on the canonical momenta and coordinates
$(\ePm,\eQm)$ of the environment. They derived a master equation for $\m$ by
``projecting out" the environment variables, which, when the modulation due to the
surroundings is fast in comparison with the precession period of $\m$, reduces to the
Fokker--Planck equation (\ref{brownfpe:Gil-LL}).

Seshadri and Lindenberg (1982) studied a test spin interacting through a
Heisenberg-type Hamiltonian with an environment consisting of other spins. The
interaction among the latter was treated by a mean field approach, and a dynamical
equation for the test spin was obtained to second order in the spin-environment
coupling. The equation derived has the form of a generalized (i.e., containing
``memory" terms) Langevin equation, whose fluctuating and relaxation terms
naturally obey fluctuation-dissipation relations.

Jayannavar (1991) employed the {\em oscillator-bath\/} representation of the
environment (Magalinski{\u{\i}}, 1959; Ullersma, 1966; Zwanzig, 1973; Caldeira and
Leggett, 1983; Ford, Lewis and O'Connell, 1988), and assumed a coupling linear in both
the spin variables and the oscillator coordinates ({\em bilinear coupling}). A
generalized Langevin equation for the spin was derived, which, in the Markovian
approach (no memory) and for isotropic fluctuations, formally reduces to the
stochastic Gilbert equation (\ref{eqmot_M_Gil}). (A similar treatment was presented
by Klik, 1992.) Equations of Landau--Lifshitz form, akin to those derived by Seshadri
and Lindenberg, were also obtained in the weak-coupling regime.

Nevertheless, since spin-environment interactions linear in $\m$ produce a
field-type perturbation on the spin (see below), the treatments mentioned do not
account for fluctuations of the magnetic anisotropy of the spin. In this Section, in order
to incorporate this phenomenon, we shall extend the bilinear-coupling treatment of
Jayannavar by considering general dependences of the spin-environment coupling on
the spin variables. Furthermore, we shall also include interactions quadratic in the
oscillator variables (the classical analogue of, for example, two-phonon or two-photon
relaxation processes), which are essential at sufficiently high temperatures. Because
the ordinary formalism of the environment of independent oscillators is not directly
applicable when such quadratic couplings are included, we shall resort to a
perturbational expansion in the spin-environment coupling, which is inspired on that
of Cort{\'{e}}s, West and Lindenberg (1985).

We shall obtain dynamical equations for the spin that have the structure of
generalized Langevin equations with fluctuating terms $\gmr\m\vecpro\bfl(\m,t)$
and concomitant relaxation terms. These will have the form of a vector product of
$\m(t)$ with a memory integral, which includes $(\D\m/\D t)(\tp)$ or
$(\m\vecpro\Beff)(\tp)$ for weak coupling, taken along the past history of the spin
($\tp\leq t$). In the Markovian approach, the equations derived will reduce to the
form
\[
\frac{\D\m}{\D t}
=
\gmr\m\vecpro[\Beff+\bfl(\m,t)]
-\R
\;,
\]
where for couplings {\em linear\/} in the environmental coordinates the relaxation
term reads
$\R=(1/\mm)\,\m\vecpro\TL(\D\m/\D t)$
or
$\R=(\gmr/\mm)\,\m\vecpro\TL(\m\vecpro\Beff)$ for weak coupling, $\TL$
being a second-rank tensor depending on the structure of the coupling. In addition,
when interactions {\em quadratic\/} in the environment variables are also taken into
account, the relaxation term will depend explicitly on the temperature and, in the
Markovian approach, $\R$ will take the form
$\R=(\gmr/\mm)\,\m\vecpro\TLQ(\m\vecpro\Beff)$, with $\TLQ=\TL+\T\,\TQ$,
where the additional tensor $\TQ$ is determined by the quadratic portion of the
coupling.

Since the fluctuating effective field $\bfl(\m,t)$ will depend in general on $\m$, it can
incorporate fluctuations of the magnetic anisotropy of the spin. For instance, when the
spin-environment interaction includes terms up to quadratic {\em in the spin
variables}, $\bfl(\m,t)$ can be written as $\vec{\bt}(t)+\hat{\kt}(t)\m$, with the
correlation coefficients of the fluctuating terms being related with the tensors $\TLQ$
by expressions identical with Eq.\ (\ref{tensor:gip}). In this way, the generalization of
the classic Brown--Kubo--Hashitsume model effected by Garanin, Ishchenko, and
Panina will formally be obtained.

\subsection{Free dynamics and canonical variables}
\label{subsect:free_dynamics}

The dynamical equation for an isolated classical spin with Hamiltonian $\Hs(\m)$ is
\begin{equation}
\label{eqmot_free}
\frac{\D\m}{\D t}
=
\gmr\m\vecpro\Beff
\;,
\qquad
\Beff
=
-\frac{\partial\Hs}{\partial\m}
\;.
\end{equation}
By means of the formula (\ref{gradient}) for the gradient operator in spherical
coordinates, these vectorial equations, which merely express the precession of $\m$
about the instantaneous effective field, can be written as
\begin{equation}
\label{eqmot_free_polar}
\frac{\D\varphi}{\D t}
=
-\frac{\gmr}{\mm}\frac{1}{\sin\vartheta}
\frac{\partial\Hs}{\partial\vartheta}
\;,
\qquad
\frac{\D\vartheta}{\D t}
=
\frac{\gmr}{\mm}\frac{1}{\sin\vartheta}
\frac{\partial\Hs}{\partial\varphi}
\;,
\end{equation}
where $\varphi$ and $\vartheta$ are, respectively, the azimuthal and polar angles of
$\m$. Furthermore, these formulae are equivalent to the Hamilton equations
\[
\frac{\D\sQ}{\D t}
=
\frac{\partial\Hs}{\partial\sP}
\;,
\qquad
\frac{\D\sP}{\D t}
=
-\frac{\partial\Hs}{\partial\sQ}
\;,
\]
with the conjugate canonical variables%
\footnote{
The alternative choice $\tilde{\sQ}=\mz/\gmr$ and $\tilde{\sP}=-\varphi$ is
equivalent to the one used here through the {\em canonical\/} transformation
$\sQ=-\tilde{\sP}$ and $\sP=\tilde{\sQ}$.
}
\begin{equation}
\label{canonical_variables}
\sQ
=
\varphi
\;,
\qquad
\sP
=
\mz/\gmr
\;.
\end{equation}

In terms of the variables (\ref{canonical_variables}) the Cartesian components of
the magnetic moment are given by
\begin{equation}
\label{m(p,q)}
\mx
=
\sqrt{\mm^{2}-(\gmr\sP)^{2}}\cos\sQ
\;,
\quad
\my
=
\sqrt{\mm^{2}-(\gmr\sP)^{2}}\sin\sQ
\;,
\quad
\mz
=
\gmr\sP
\;.
\nonumber
\end{equation}
From these expressions for $\mi(\sP,\sQ)$ and the definition of the Poisson bracket of
two arbitrary dynamical variables
\[
\pbra A, B\pket
\equiv
\frac{\partial A}{\partial\sQ}
\frac{\partial B}{\partial\sP}
-\frac{\partial A}{\partial\sP}
\frac{\partial B}{\partial\sQ}
\;,
\]
one can readily obtain the customary Poisson-bracket (``commutation") relations
among the spin variables
\[
\pbra\mi,\mj\pket
=
\gmr\sum_{k}\epsilon_{ijk}\mk
\;,
\]
where $\epsilon_{ijk}$ is the Levi--Civita symbol.%
\footnote{
To illustrate, from
\[
\begin{array}{rclrclr}
\partial\mx/\partial \sQ
&
=
&
-\left[\mm^{2}-(\gmr\sP)^{2}\right]^{1/2}\sin\sQ
\;,
&
\quad
\partial\mx/\partial \sP
&
=
&
-\gmr^{2}\sP\left[\mm^{2}-(\gmr\sP)^{2}\right]^{-1/2}\cos\sQ
\;,
\\
\partial\my/\partial \sQ
&
=
&
\hspace{0.75em}
\left[\mm^{2}-(\gmr\sP)^{2}\right]^{1/2}\cos\sQ
\;,
&
\quad
\partial\my/\partial \sP
&
=
&
-\gmr^{2}\sP\left[\mm^{2}-(\gmr\sP)^{2}\right]^{-1/2}\sin\sQ
\;,
\end{array}
\]
one gets
$\pbra\mx,\my\pket
=\gmr^{2}\sP\sin^{2}\!\sQ+\gmr^{2}\sP\cos^{2}\!\sQ
=\gmr\mz$.
\qed
}
In addition, on using the {\em chain rule\/} of the Poisson bracket, namely
\[
\pbra f,g\pket
=
\sum_{i,k}
\frac{\partial f}{\partial x_{i}}
\frac{\partial g}{\partial x_{k}}
\pbra x_{i},x_{k}\pket
\;,
\qquad
x_{i}=x_{i}(\sP,\sQ)
\;,
\]
one gets the useful relation (cf.\ Eq.\ (13) by Smith and De~Rozario, 1976)
\begin{equation}
\label{MpoissonV}
\pbra\mi,\Vint(\m)\pket
=
-\gmr 
\bigg(
\m\vecpro\frac{\partial\Vint}{\partial\m}
\bigg)_{i}
\;,
\end{equation}
which is valid for any function of the spin variables $\Vint(\m)$.

Note finally that one can conversely {\em postulate\/} the relations
$\{\mi,\mj\}=\gmr\sum_{k}\epsilon_{ijk}\mk$ and then {\em derive\/} Eq.\
(\ref{eqmot_free}) starting from the basic Hamiltonian evolution equation $\D\mi/\D
t=\{\mi,\Hs\}$ and using Eq.\ (\ref{MpoissonV}). This can be considered as a
justification of the presence of the expression
$\Beff=-\partial\Hs/\partial\m$ in the dynamical equations for a classical spin.

\subsection{Dynamical equations for couplings linear in the environment variables}
\label{subsect:linear_coupling}

We shall now study a classical spin surrounded by an environment that can be
represented by a set of independent classical harmonic oscillators. In spite of its
academic appearance, those oscillators can correspond to the {\em normal modes\/} of
an electromagnetic field, the lattice vibrations (in the harmonic approximation), or
they can be an effective low-energy description of a more general surrounding
medium (Caldeira and Leggett, 1983). We shall assume that the spin-environment
interaction is linear in the coordinates of the oscillators but otherwise arbitrary in the
spin variables. In this way, fluctuations of the magnetic anisotropy of the spin will be
incorporated in the dynamical equations.

\subsubsection{The spin-environment Hamiltonian}

The total system consisting of the spin (the ``system of interest") plus the oscillators
representing the environment forms a {\em closed\/} dynamical system that we shall
describe by augmenting the isolated-spin Hamiltonian as follows
\begin{equation}
\label{hamiltonian:L:1}
\Ham_{\rm T}
=
\Hs(\m)+\sum_{\alpha}
\half
\Big\{
\eP_{\alpha}^{2}\overMa
+\Ma\w_{\alpha}^{2}
\Big[
\eQ_{\alpha}+\frac{\coupling}{\Ma\w_{\alpha}^{2}}\Fint_{\alpha}(\m)
\Big]^2
\Big\}
\;.
\end{equation}
Here, $\alpha$ is an oscillator index [e.g., the pair $(\vec{k},s)$ formed by the
wave-vector and branch index of a normal mode of the environment], and the
coupling terms $\Fint_{\alpha}(\m)$ are arbitrary functions of the spin variables
(typically polynomials in $\m$). These terms may depend on the parameters of the
oscillators $\w_{\alpha},\Ma$ but not on their dynamical variables
$\eP_{\alpha},\eQ_{\alpha}$. On the other hand, for the sake of convenience in
keeping track of the various orders, we have introduced a spin-environment coupling
constant $\coupling$, which in the weak-coupling approximation will be considered
small.

The terms proportional to $\Fint_{\alpha}^2$, which emerge when squaring
$\eQ_{\alpha}+(\coupling/\Ma\w_{\alpha}^{2})\Fint_{\alpha}$, are ``counter-terms"
introduced to balance the coupling-induced renormalization of the Hamiltonian of the
spin. The formalism takes as previously considered whether such a renormalization
actually occurs for a given interaction (Caldeira and Leggett, 1983), so that $\Hs$
would already include it (whenever exists). An advantage of this convention is that
one deals with the experimentally accessible energy of the spin, instead of the ``bare"
one, which might be difficult to determine.

The introduction of {\em non-linear\/} coupling terms $\Fint_{\alpha}(\m)$, as
otherwise occur in various relevant situations ($\Fint_{\alpha}\propto\sum\mk\ml$
for the magneto-elastic coupling of $\m$ to the lattice vibrations), will be essential to
get fluctuations of the magnetic anisotropy of the spin. The starting Hamiltonian in the
work of Jayannavar (1991) was similar to (\ref{hamiltonian:L:1}) with a special type
of {\em linear\/} $\Fint_{\alpha}(\m)$: the component $\mi$ of the magnetic moment
was coupled to the $i$th Cartesian component $\eQ_{\alpha,i}$ of certain
three-dimensional oscillators. This specific {\em bilinear\/} interaction yielded, not
only field-type fluctuations, but also uncorrelated ones. [Klik (1992) also considered
couplings non-linear in $\m$, but in that work the focus was on the existence of
thermal equilibrium in the Markovian limit.]

\subsubsection{Dynamical equations: general case}

For the sake of simplicity in notation but also of generality, we cast the Hamiltonian
(\ref{hamiltonian:L:1}) into the form
\begin{equation}
\label{hamiltonian:L:2}
\Ham_{\rm T}
=
\Hsm(\sP,\sQ)
+\sum_{\alpha}
\half
\left(
\eP_{\alpha}^{2}\overMa
+\Ma\w_{\alpha}^{2}\eQ_{\alpha}^{2}
\right)
+\coupling\sum_{\alpha}\eQ_{\alpha}\Fint_{\alpha}(\sP,\sQ)
\;,
\end{equation}
where $\sQ$ and $\sP$ are the canonical coordinate and conjugate momentum of a
system with Hamiltonian $\Hs(\sP,\sQ)$ [in the spin-dynamics case $\sP$ and $\sQ$
are given by Eqs.\ (\ref{canonical_variables})], and the ``modified" system
Hamiltonian $\Hsm$ augments $\Hs$ by the aforementioned counter-terms
\begin{equation}
\label{modified_hamiltonian:L}
\Hsm
=
\Hs
+\frac{\coupling^{2}}{2}
\sum_{\alpha}\frac{\Fint_{\alpha}^{2}}{\Ma\w_{\alpha}^{2}}
\;.
\end{equation}

The equation of motion for any dynamical variable $C$ without explicit dependence on
the time, $\partial C/\partial t\equiv0$, is given by the basic Hamiltonian evolution
equation
\[
\frac{\D C}{\D t}
=
\pbra C,\Ham_{\rm T}\pket
\;,
\]
where the whole Poisson bracket is given by
\[
\pbra A, B\pket
\equiv
\frac{\partial A}{\partial\sQ}
\frac{\partial B}{\partial\sP}
-\frac{\partial A}{\partial\sP}
\frac{\partial B}{\partial\sQ}
+\sum_{\alpha}
\frac{\partial A}{\partial\eQ_{\alpha}}
\frac{\partial B}{\partial\eP_{\alpha}}
-\frac{\partial A}{\partial\eP_{\alpha}}
\frac{\partial B}{\partial\eQ_{\alpha}}
\;.
\]
Therefore, the (coupled) equations of motion for {\em any dynamical variable
(observable) of  the system $A(\sP,\sQ)$\/} and the environment variables read
($C=A,\eP_{\alpha}$, and $\eQ_{\alpha}$)
\begin{eqnarray}
\label{eqmot_A:L:1}
\frac{\D A}{\D t}
&
=
&
\pbra A, \Hsm\pket+\coupling
\sum_{\alpha}\eQ_{\alpha}\pbra A, \Fint_{\alpha}\pket
\;,
\\
\label{eqmot_bath:L}
\frac{\D\eQ_{\alpha}}{\D t}
&
=
&
\eP_{\alpha}\overMa
\;,
\qquad
\frac{\D\eP_{\alpha}}{\D t}
=
-\Ma\w_{\alpha}^{2}\eQ_{\alpha}
-\coupling\Fint_{\alpha}
\;.
\end{eqnarray}
The goal is to derive a dynamical equation for $A(\sP,\sQ)$ involving the system
variables only ({\em reduced\/} dynamical equation). Then, the corresponding
equation for the spin will be obtained by replacing $A(\sP,\sQ)$ in that equation by
the Cartesian components of $\m$ [Eq.\ (\ref{m(p,q)})].

On considering that in Eqs.\ (\ref{eqmot_bath:L}) the term
$-\coupling\Fint_{\alpha}(t)=-\coupling\Fint_{\alpha}[\sP(t),\sQ(t)]$ plays the r\^{o}le
of a time-dependent forcing on the oscillators, those equations can be explicitly
integrated, yielding
\begin{equation}
\label{bath_coord:L:1}
\eQ_{\alpha}(t)
=
\eQ_{\alpha}^{\h}(t)
-\frac{\coupling}{\Ma\w_{\alpha}}
\int_{t_{0}}^{t}\!\!\D{\tp}\,
\sin[\w_{\alpha}(t-\tp)]
\Fint_{\alpha}(\tp)
\;,
\end{equation}
where
\begin{equation}
\label{qh}
\eQ_{\alpha}^{\h}(t)
=
\eQ_{\alpha}(t_{0})\cos[\w_{\alpha}(t-t_{0})]
+[\eP_{\alpha}(t_{0})/\Ma\w_{\alpha}]\sin[\w_{\alpha}(t-t_{0})]
\;,
\end{equation}
are the solutions of the {\em homogeneous\/} system of equations for the oscillators in
the absence of the system-environment interaction (proper modes of the
environment). Then, on integrating by parts in Eq.\ (\ref{bath_coord:L:1}) one gets for
the combination $\coupling\eQ_{\alpha}$ that appears in Eq.\ (\ref{eqmot_A:L:1})
\begin{equation}
\label{bath_coord:L:2}
\coupling\eQ_{\alpha}(t)
=
\ffl_{\alpha}(t)
-\left[
\K_{\alpha}(t-\tp)\Fint_{\alpha}(\tp)
\right]_{\tp=t_{0}}^{\tp=t}
+\int_{t_{0}}^{t}\!\!\D{\tp}\, 
\K_{\alpha}(t-\tp)\frac{\D\Fint_{\alpha}}{\D t}(\tp)
\;,
\end{equation}
where
\begin{equation}
\label{fluct-kernel:L:precursor}
\ffl_{\alpha}(t)
=
\coupling \eQ_{\alpha}^{\h}(t)
\;,
\qquad
\K_{\alpha}(\tau)
=
\frac{\coupling^{2}}{\Ma\w_{\alpha}^{2}}\cos(\w_{\alpha}\tau)
\;.
\end{equation}
Next, in order to eliminate the environment variables from the equation for
$A(\sP,\sQ)$, one substitutes Eq.\ (\ref{bath_coord:L:2}) back into Eq.\
(\ref{eqmot_A:L:1}). This yields a term $\sum_{\alpha}\pbra
A,\Fint_{\alpha}\pket\K_{\alpha}(t-t_{0})\Fint_{\alpha}(t_{0})$ that depends on the
initial state of the system $(\sP(t_{0}),\sQ(t_{0}))$ and produces a transient response
that can be ignored in the long-time dynamics (we shall however return to this
question below).%
\footnote{
In the ordinary independent oscillator model, one considers
$\Fint_{\alpha}(\sP,\sQ)\propto\sQ$ and the corresponding terms can formally be
removed from the dynamical equations by choosing the origin of the ``coordinate
frame" to lay at the ``position" of the system at $t=t_{0}$, that is,
$\Fint_{\alpha}(t_{0})\propto\sQ(t_{0})=0$.
However, this frame-dependent procedure cannot be employed if the system
comprises different entities. In addition, in the spin-dynamics case with, for instance,
$\Fint_{\alpha}(\m)$ linear in $\m$, one cannot set $\m(t_{0})=\vec{0}$ due to the
conservation of the magnitude of the spin.
}
The parallel term $-\sum_{\alpha}\pbra
A,\Fint_{\alpha}\pket\K_{\alpha}(0)\Fint_{\alpha}(t)$, which is derivable from a
Hamiltonian, is exactly balanced by the term emerging from the counter-terms in
$\pbra A, \Hsm\pket$. This can be shown by using $-\sum_{\alpha}
\pbra A,\Fint_{\alpha}\pket
\K_{\alpha}(0)\Fint_{\alpha}
=\pbra A,
-\half\sum_{\alpha}\K_{\alpha}(0)
\Fint_{\alpha}^{2}
\pket$, which follows from the {\em product rule\/} of the Poisson bracket
\begin{equation}
\label{productrule}
\pbra A,
BC\pket =\pbra A,
B\pket C+\pbra A, C\pket B
\;,
\end{equation}
and then using
$\K_{\alpha}(0)=\coupling^{2}/\Ma\w_{\alpha}^{2}$
[see Eq.\ (\ref{fluct-kernel:L:precursor})].

Therefore, one is finally left with the {\em reduced\/} dynamical equation
\begin{equation}
\label{eqmot_A:L}
\frac{\D A}{\D t}
=
\pbra A,\Hs\pket
+\sum_{\alpha}\pbra A, \Fint_{\alpha}\pket
\bigg[
\ffl_{\alpha}(t)
+\int_{t_{0}}^{t}\!\!\D{\tp}\,
\K_{\alpha}(t-\tp)
\frac{\D\Fint_{\alpha}}{\D t}(\tp)
\bigg]
\;,
\end{equation}
where the first term yields the free (conservative) time evolution of the system,
whereas the second term incorporates the effects of the interaction of the system with
its environment. The terms $\ffl_{\alpha}(t)$ are customarily interpreted as
{\em fluctuating\/} ``forces" (or ``fields"), while the integral term, which keeps in
general memory of the previous history of the system, provides the {\em
relaxation\/} due to the interaction with the surrounding medium. [Note that without
the integration by parts yielding Eq.\ (\ref{bath_coord:L:2}), the Hamiltonian
(renormalization) terms would occur inconveniently mixed in the integral term.]

The origin of both types of terms can be traced back as follows. Recall that in Eq.\
(\ref{bath_coord:L:1}) the time evolution of the oscillators has formally been written
as if they were driven by (time-dependent) forces
$-\coupling\Fint_{\alpha}[\sP(\tp),\sQ(\tp)]$ depending on the state of the system.
Therefore, $\eQ_{\alpha}(t)$ consists of the sum of the proper (free) mode
$\eQ_{\alpha}^{\h}(t)$ and the driven-type term, which naturally depends on the
``forcing" (state of the system) at previous times. Then, the replacement of
$\eQ_{\alpha}$ in the equation for the system variables by the driven-oscillator
solution incorporates:
\begin{enumerate}
\item
The time-dependent modulation due to the proper modes of the environment.
\item
The ``back-reaction" on the system of its preceding action on the surrounding medium.
\end{enumerate}
Thus, the formalism leads to a description in terms of a reduced
number of dynamical variables at the expense of both explicitly time-dependent
(fluctuating) terms and history-dependent (relaxation) terms (see Table
\ref{terms_interpretation:table}).
\begin{table}
\caption[]
{
Terms incorporating the effects of the interaction of the system
with the surrounding medium in the reduced dynamical equation
(\ref{eqmot_A:L}).
\label{terms_interpretation:table}
}
\begin{center}
\begin{tabular}{|c|c|c|}
{\bf term}
&
{\bf mechanism}
&
\quad{\bf interpretation}
\cr
\hline
\hline
$\ffl_{\alpha}(t)$
&
\quad
\parbox{12.5em}{\centering
time-dependent modulation\\ due to the proper modes\\ of the environment }
\quad
&
\quad
\parbox{5em}
{\centering
fluctuating
\\
term
}
\quad
\cr
\hline
\parbox{4em}{\centering
integral\\ term
}
&
\quad
\parbox{12.5em}{\centering
back-reaction on the system\\ of its preceding action\\ on the environment
}
\quad
&
\quad
\parbox{5em}{\centering
relaxation\\ term
}
\quad
\end{tabular}
\end{center}
\end{table}

\paragraph*{Archetypal example: the Brownian particle.}

In order to particularize these general expressions to definite situations, the structure
of the coupling terms $\Fint_{\alpha}$ needs to be specified. For instance, on setting
$\Fint_{\alpha}(\sP,\sQ)=-\Cint_{\alpha}\sQ$ (bilinear coupling), where the
$\Cint_{\alpha}=\Cint_{\alpha}(\w_{\alpha})$ are coupling constants, and writing
down Eq.\ (\ref{eqmot_A:L}) for $A=\sQ$ and $A=\sP$ with help from
$\pbra\sP,B\pket=-\partial B/\partial\sQ$
and
$\pbra\sQ,B\pket=\partial B/\partial\sP$,
one gets the celebrated generalized Langevin equation for a ``Brownian" particle
(Zwanzig, 1973)
\begin{equation}
\label{brownian_particle}
\frac{\D\sQ}{\D t}
=
\frac{\partial\Hs}{\partial\sP}
\;,
\qquad
\frac{\D\sP}{\D t}
=
-\frac{\partial\Hs}{\partial\sQ}
+f(t)
-\int_{t_{0}}^{t}\!\!\D{\tp}\,
\K(t-\tp)\frac{\D\sQ}{\D t}(\tp)
\;.
\end{equation}
Here, $f(t)=\sum_{\alpha}\Cint_{\alpha}\ffl_{\alpha}(t)$ is the fluctuating force and
$\K(\tau)=\sum_{\alpha}\Cint_{\alpha}^{2}\K_{\alpha}(\tau)$ is the memory kernel,
the relaxation term associated with which comprises minus the velocity $-(\D\sQ/\D
t)(\tp)$ of the particle ({\em viscous damping}).

In general, when $\pbra A,\Fint_{\alpha}\pket$ in Eq.\ (\ref{eqmot_A:L}) is not
constant, the fluctuating terms $\ffl_{\alpha}(t)$ enter multiplying the system
variables ({\em multiplicative\/} fluctuations). In this example, owing to the fact that
$\pbra\sQ,-\Cint_{\alpha}\sQ\pket=0$ and
$\pbra\sP,-\Cint_{\alpha}\sQ\pket=\Cint_{\alpha}$, the fluctuations are {\em
additive}.

\subsubsection{Dynamical equations: the spin-dynamics case}

Let us now particularize the above results to the dynamics of a classical spin. Here, we
introduce the coupling functions
\begin{equation}
\label{W:L}
\Fint_{\alpha}(\m)
=
\sum_{\ila}\Cint_{\alpha}^{\ila}\Vint_{\ila}(\m)
\;,
\end{equation}
where $\ila$ stands for a general index depending on the type of interaction, the
coefficients $\Cint_{\alpha}^{\ila}$ are spin-environment coupling constants, and the
terms $\Vint_{\ila}(\m)$ are certain functions of the spin variables. In order to
motivate this expression, consider, for example, the magneto-elastic coupling of $\m$
to the lattice vibrations. The index $\ila$ then stands for a pair of Cartesian indices
$(ij)$ and $\Vint_{\ila}\to\Vint_{ij}=\sum_{k\ell}a_{ij,k\ell}\mk\ml$, where the
$a_{ij,k\ell}$ are magneto-elastic coefficients.

In order to derive the reduced dynamical equation for the spin, we merely put
$A=\mi$, $i=x,y,z$, in Eq.\ (\ref{eqmot_A:L}), and then use Eq.\ (\ref{MpoissonV}) to
calculate the Poisson brackets required. On gathering the results so-obtained in
vectorial form and using $\Beff=-\partial\Hs/\partial\m$ and $\D\Vint_{\ilb}/\D
t=(\partial\Vint_{\ilb}/\partial\m)\cdot(\D\m/\D t)$, we arrive at
\begin{equation}
\label{eqmot_M_Giltyp}
\frac{\D\m}{\D t}
=
\gmr\m\vecpro 
\bigg\{
\Beff+\bfl(\m,t)
-
\int_{t_{0}}^{t}\!\!\D{\tp}\,
\TLpre(\m;t,\tp)
\frac{\D\m}{\D t}(\tp)
\bigg\}
\;.
\end{equation}
In this equation the {\em fluctuating magnetic field\/} is given by
\begin{equation}
\label{bfl:L}
\bfl(\m,t)
=
-\sum_{\ila}\ffl_{\ila}(t)
\frac{\partial\Vint_{\ila}}{\partial\m}
\;,
\end{equation}
which involves the environmental proper modes via the fluctuating sources
\begin{equation}
\label{fluct:L}
\ffl_{\ila}(t)
=
\coupling\sum_{\alpha}\Cint_{\alpha}^{\ila}\eQ_{\alpha}^{\h}(t)
\;.
\end{equation}
On the other hand, the relaxation tensor in Eq.\ (\ref{eqmot_M_Giltyp}) reads%
\footnote{
Although we omit the symbol of scalar product, the action of a dyadic
$\vec{A}\,\vec{B}$ on a vector $\vec{C}$ is the standard one:
$(\vec{A}\,\vec{B})\vec{C}\equiv\vec{A}(\vec{B}\cdot\vec{C})$.
}
\begin{equation}
\label{kernel:L:tensor}
\TLpre(\m;t,\tp)
=
\sum_{\ila,\ilb}\K_{\ila\ilb}(t-\tp)
\frac{\partial\Vint_{\ila}}{\partial\m}(t)
\frac{\partial\Vint_{\ilb}}{\partial\m}(\tp)
\;,
\end{equation}
where the {\em memory kernel\/} is given by%
\footnote{
Note that $\ffl_{\ila}(t)=\sum_{\alpha}\Cint_{\alpha}^{\ila}\ffl_{\alpha}(t)$
and
$\K_{\ila\ilb}(\tau)=\sum_{\alpha}\Cint_{\alpha}^{\ila}\Cint_{\alpha}^{\ilb}\K_{\alpha}(\tau)$,
where $\ffl_{\alpha}(t)$ and $\K_{\alpha}(\tau)$ are given by Eq.\
(\ref{fluct-kernel:L:precursor}).
}
\begin{equation}
\label{kernel:L}
\K_{\ila\ilb}(\tau)
=
\coupling^{2}
\sum_{\alpha}
\frac{\Cint_{\alpha}^{\ila}\Cint_{\alpha}^{\ilb}}{\Ma\w_{\alpha}^{2}}
\cos(\w_{\alpha}\tau)
\;.
\end{equation}

Equation (\ref{eqmot_M_Giltyp}) contains $\D\m/\D t$ on its right-hand side, so it
will be referred to as a {\em Gilbert-type\/} equation [cf.\ Eq.\ (\ref{eqmot_M_Gil})].
For $\coupling\ll1$, on replacing perturbatively that derivative by its conservative
part, $\D\m/\D t\simeq\gmr\m\vecpro\Beff$, one gets the weak-coupling {\em
Landau--Lifshitz-type\/} equation
\begin{equation}
\label{eqmot_M_LLtyp:L}
\frac{\D\m}{\D t}
=
\gmr\m
\vecpro
\left[
\Beff+\bfl(\m,t)
\right]
-\gmr\m
\vecpro
\bigg\{
\int_{t_{0}}^{t}\!\!\D{\tp}\,
\gmr
\TLpre(\m;t,\tp)
\left(\m\vecpro\Beff\right)(\tp)
\bigg\}
\;,
\end{equation}
which describes weakly damped precession.

For spin-environment interactions {\em linear\/} in the environment variables but
being otherwise {\em arbitrary\/} functions of $\m$, Eqs.\ (\ref{eqmot_M_Giltyp})
and (\ref{eqmot_M_LLtyp:L}) are the desired reduced dynamical equations for the
spin. They have the structure of generalized Langevin equations with {\em
fluctuating\/} terms $\gmr\m\vecpro\bfl(\m,t)$ (associated with the modulation by
the proper modes of the environment) and history-dependent {\em relaxation\/}
terms (corresponding to the back-reaction on the spin of its previous action on the
surrounding medium).

Note that $\ffl_{\ila}(t)$ [Eq.\ (\ref{fluct:L})] is a sum of a large number of sinusoidal
terms with different frequencies and phases; this can give to $\ffl_{\ila}(t)$ the form
of a highly irregular function of $t$ that is expected for a fluctuating term. However,
for a general form of the coupling functions $\Vint_{\ila}(\m)$, the term $\bfl(\m,t)$
{\em cannot\/} be interpreted as a fluctuating {\em ordinary\/} field, since it may
depend on $\m$, but it is rather a fluctuating {\em effective\/} field to be added to
the deterministic effective field $\Beff=-\partial\Hs/\partial\m$ [Eq.\
(\ref{eqmot_free})]. This can be illustrated by phrasing the discussion in terms of the
{\em fluctuating part\/} of the energy of the spin, namely [see Hamiltonian
(\ref{hamiltonian:L:2})]:
$\Hfl=\coupling\sum_{\alpha}\eQ_{\alpha}^{\h}(t)\Fint_{\alpha}(\m)$. From this
definition one first gets
\begin{equation}
\label{ufl:L}
\Hfl(\m,t)
=
\sum_{\ila}\ffl_{\ila}(t)\Vint_{\ila}(\m)
\;,
\qquad
\bfl(\m,t)
=
-\frac{\partial\Hfl}{\partial\m}
\;,
\end{equation}
so that $\bfl$ can be derived from $\Hfl$ in the same way as
$\Beff$ is obtained from $\Hs$. Next, recall that the non-linear part of
$\Hs(\m)$ carries the anisotropy-energy terms, e.g.,
$\Hs=-\m\cdot\B-\half\Kuni(\m\cdot\vec{n})^{2}$ in a uniaxial crystal. Analogously,
$\Hfl$ has the form $\Hfl(\m,t)=-\m\cdot\bfl(t)$, with $\bfl$ independent of $\m$,
only for linear $\Vint_{\ila}(\m)$ (bilinear coupling), so that {\em the non-linear part
of $\Vint_{\ila}(\m)$ incorporates fluctuations of the magnetic anisotropy of the spin}.

To illustrate, if the spin-environment interaction includes up to quadratic terms in
$\m$, one can write the coupling functions $\Vint_{\ila}(\m)$ as
\begin{equation}
\label{F:l_spin}
\Vint_{\ila}(\m)
=
\sum_{i}v_{\ila,i}\mi
+\half\sum_{ij}w_{\ila,ij}\mi\mj
\;,
\end{equation}
where the constants $v_{\ila,i}$ and $w_{\ila,ij}$ incorporate the symmetry of the
interaction. In this case, the fluctuating effective field (\ref{bfl:L}) can be cast into  the
form [cf.\ Eq.\ (\ref{eqmot_M_gip})]
\begin{equation}
\label{bfl:l_spin}
\bfl(\m,t)
=
\vec{\bt}(t)+\hat{\kt}(t)\m
\;,
\end{equation}
with the following expressions for the fluctuating sources $\vec{\bt}(t)$ and
$\hat{\kt}(t)$ in terms of the coupling constants
\[
\bt_{i}(t)
=
-\sum_{\ila}\ffl_{\ila}(t)v_{\ila,i}
\;,
\qquad
\kt_{ij}(t)
=
-\sum_{\ila}\ffl_{\ila}(t)w_{\ila,ij}
\;.
\]
As $\vec{\bt}(t)$ does not depend on $\m$, it can be interpreted as a fluctuating {\em
ordinary\/} field. The fluctuations of $\hat{\kt}(t)$, however, do not enter in this way,
since they occur via $\sum_{j}\kt_{ij}(t)\mj$, but they produce fluctuations of the
magnetic-anisotropy potential of the spin, both of the direction of the anisotropy axes
and of the magnitudes of the anisotropy constants. This is clearly perceived on
considering that the fluctuating part of the energy of the spin (\ref{ufl:L}) takes in this
case the form
\[
\Hfl(\m,t)
=
-\m\cdot\vec{\bt}(t)-\half\m\cdot\hat{\kt}(t)\m
\;.
\]
This resembles the scenario encountered for a mechanical oscillator
(Lindenberg and Seshadri, 1981), where the portion of the oscillator-environment
coupling quadratic in the coordinate of the test oscillator yields, instead of a
fluctuating force, a fluctuating contribution to its harmonic potential ({\em
frequency-type\/} fluctuations). Finally, if $\Vint_{\ila}(\m)$ only comprises
non-linear terms, such as those occurring in the magneto-elastic coupling mentioned
($\Vint_{\ila}\propto\sum\mk\ml$), no field-type fluctuating terms emerge and only
anisotropy-type fluctuations remain.

We remark in closing that, even for couplings linear in the spin variables, and hence
for $\bfl(t)$ independent of $\m$, the occurrence of the vector {\em product\/}
$\m\vecpro\bfl$ in the dynamical equations entails that the fluctuating terms enter in
a {\em multiplicative\/} way. This is at variance with the situation encountered in
ordinary mechanical systems (Lindenberg and Seshadri, 1981), where couplings linear
in the system variables lead to additive fluctuations [see
Eq.\ (\ref{brownian_particle})], whereas multiplicative fluctuating terms only emerge
for couplings non-linear in the system variables. To illustrate, for the mentioned
mechanical oscillator, the force- and frequency-type fluctuations provided by
$\Fint_{\alpha}
=-v_{\alpha}\sQ
-w_{\alpha}\sQ^{2}$
are, respectively, additive and multiplicative, whereas in the gyromagnetic case the
field-type fluctuations are already multiplicative.
Indeed, in the spin-dynamics case, in analogy with the results obtained for mechanical
rigid rotators (Lindenberg, Mohanty and Seshadri, 1983), the multiplicative character
of the fluctuations is a consequence of the Poisson bracket relations
$\pbra\mi,\mj\pket=\gmr\sum_{k}\epsilon_{ijk}\mk$ for angular-momentum-type
dynamical variables, which, even for $\Fint_{\alpha}$ linear in $\m$, lead to
non-constant $\pbra A,\Fint_{\alpha}\pket$ in Eq.\ (\ref{eqmot_A:L}). In our
derivation, this can straightly be traced back by virtue of the Poisson-bracket
formalism employed.

\subsubsection{Statistical properties of the fluctuating terms}

In order to determine the statistical properties of the fluctuating sources
$\ffl_{\ila}(t)$, one usually assumes that the environment was in thermodynamical 
equilibrium at the {\em initial\/} time (recall that no statistical assumption has been
explicitly introduced until this point). This initial state is customarily chosen in two
different ways.

\paragraph{Decoupled initial conditions.}

The environment variables are distributed at $t=t_{0}$ according to the Boltzmann law
associated with the environment Hamiltonian alone
\begin{eqnarray}
\label{gibbs:oscillators:dic}
\Weq(\ePm(t_{0}),\eQm(t_{0}))
&
\propto
&
\exp[-\Ham_{{\rm E}}(t_{0})/\T]
\;,
\\
\Ham_{{\rm E}}(t_{0})
&
=
&
\sum_{\alpha}
\half
\left[
\eP_{\alpha}(t_{0})^{2}\overMa
+\Ma\omega_{\alpha}^{2}\eQ_{\alpha}(t_{0})^{2}
\right]
\nonumber
\;,
\end{eqnarray}
where $(\ePm,\eQm)$ stands for the set of canonical variables of the environment. The
initial distribution is therefore Gaussian and one has for the first two moments of the
environmental variables
\begin{eqnarray*}
\llangle\eQ_{\alpha}(t_{0})\rrangle
=
0
\;,
&
&
\llangle\eP_{\alpha}(t_{0})\rrangle
=
0
\;,
\\
\llangle\eQ_{\alpha}(t_{0})\eQ_{\beta}(t_{0})\rrangle
=
\delta_{\alpha\beta}\frac{\T}{\Ma\w_{\alpha}^{2}}
\;,
&
\llangle\eQ_{\alpha}(t_{0})\eP_{\beta}(t_{0})\rrangle
=
0
\;,
&
\llangle\eP_{\alpha}(t_{0})\eP_{\beta}(t_{0})\rrangle
=
\delta_{\alpha\beta}\Ma\T
\;.
\end{eqnarray*}
From these results one readily gets the averages of the proper modes over initial
states of the environment (ensemble averages):
\begin{eqnarray*}
\llangle\eQ_{\alpha}^{\h}(t)\rrangle
&
=
&
\underbrace{
\llangle\eQ_{\alpha}(t_{0})\rrangle
}_{0}
\cos[\w_{\alpha}(t-t_{0})]
+\underbrace{
\llangle\eP_{\alpha}(t_{0})\rrangle
}_{0}
\frac{1}{\Ma\w_{\alpha}}\sin[\w_{\alpha}(t-t_{0})]
\;,
\\
\llangle\eQ_{\alpha}^{\h}(t)\eQ_{\beta}^{\h}(\tp)\rrangle
&
=
&
\underbrace{
\llangle\eQ_{\alpha}(t_{0})\eQ_{\beta}(t_{0})\rrangle
}_{\delta_{\alpha\beta}\T/\Ma\w_{\alpha}^{2}}
\cos[\w_{\alpha}(t-t_{0})]\cos[\w_{\beta}(\tp-t_{0})]
\\
&
& {}+
\underbrace{
\llangle\eQ_{\alpha}(t_{0})\eP_{\beta}(t_{0})\rrangle
}_{0}
\frac{1}{\Mb\w_{\beta}}
\cos[\w_{\alpha}(t-t_{0})]\sin[\w_{\beta}(\tp-t_{0})]
\\
&
& {}+
\underbrace{
\llangle\eP_{\alpha}(t_{0})\eQ_{\beta}(t_{0})\rrangle
}_{0}
\frac{1}{\Ma\w_{\alpha}}
\sin[\w_{\alpha}(t-t_{0})]\cos[\w_{\beta}(\tp-t_{0})]
\\
&
& {}+
\underbrace{
\llangle\eP_{\alpha}(t_{0})\eP_{\beta}(t_{0})\rrangle
}_{\delta_{\alpha\beta}\Ma\T}
\frac{1}{\Ma\w_{\alpha}\Mb\w_{\beta}}
\sin[\w_{\alpha}(t-t_{0})]\sin[\w_{\beta}(\tp-t_{0})]
\\
&
=
&
\T
\frac{\delta_{\alpha\beta}}{\Ma\w_{\alpha}^{2}}
\{
\cos[\w_{\alpha}(t-t_{0})]\cos[\w_{\alpha}(\tp-t_{0})]
\\
&
&
\hspace{4em}
+\sin[\w_{\alpha}(t-t_{0})]\sin[\w_{\alpha}(\tp-t_{0})]
\}
\;,
\end{eqnarray*}
so that
\begin{equation}
\label{stats:qh}
\llangle\eQ_{\alpha}^{\h}(t)\rrangle
=
0
\;,
\qquad
\llangle\eQ_{\alpha}^{\h}(t)\eQ_{\beta}^{\h}(\tp)\rrangle
=
\T\frac{\delta_{\alpha\beta}}{\Ma\w_{\alpha}^{2}}
\cos[\w_{\alpha}(t-\tp)]
\;.
\end{equation}

Thus, the fluctuating terms $\ffl_{\ila}(t)$ [Eq.\ (\ref{fluct:L})] are Gaussian stochastic
processes and the relevant averages over initial states of the environment are given by
\begin{eqnarray}
\label{stats:L:1st_moment}
\llangle\ffl_{\ila}(t)\rrangle
&
=
&
0
\;,
\\
\label{stats:L:2nd_moment}
\llangle\ffl_{\ila}(t)\ffl_{\ilb}(\tp)\rrangle
&
=
&
\T\,\K_{\ila\ilb}(t-\tp)
\;.
\end{eqnarray}
Equation (\ref{stats:L:2nd_moment}) relates the statistical time correlation of the
fluctuating terms $\ffl_{\ila}(t)$ with the relaxation memory kernels
$\K_{\ila\ilb}(\tau)$ occurring in the dynamical equations ({\em
fluctuation-dissipation\/} relations). Short (long) correlation times of the fluctuating
terms entail short-range (long-range) memory effects in the relaxation term, and vice
versa. The emergence of this type of relations is not surprising in this context, since
fluctuations and relaxation arise as different manifestations of the {\em same\/}
interaction of the system with the surrounding medium.

\paragraph{Coupled initial conditions.}

The environment is assumed to be at $t=t_{0}$ in thermal equilibrium {\em in the
presence of the system}, which is however taken as {\em fastened\/} in its initial state
(Ford, Lewis and O'Connell, 1988). Therefore, the corresponding initial distribution of
the environment variables is
\begin{eqnarray*}
\Weq(\ePm(t_{0}),\eQm(t_{0}))
&
\propto
&
\exp[-\Ham_{{\rm SE}}(t_{0})/\T]
\;,
\\
\Ham_{{\rm SE}}(t_{0})
&
=
&
\sum_{\alpha}
\half
\Big\{
\eP_{\alpha}(t_{0})^{2}\overMa
+\Ma\w_{\alpha}^{2}
\Big[
\eQ_{\alpha}(t_{0})
+\frac{\coupling}{\Ma\w_{\alpha}^{2}}\Fint_{\alpha}(t_{0})
\Big]^2
\Big\}
\;,
\end{eqnarray*}
where the $\Fint_{\alpha}(t_{0})$ are taken as constants. In this case, the dropped
terms depending on the initial state of the system
$\K_{\alpha}(t-t_{0})\Fint_{\alpha}(t_{0})$ [recall the remarks before Eq.\
(\ref{eqmot_A:L})], which for
$\Fint_{\alpha}=\sum_{\ila}\Cint_{\alpha}^{\ila}\Vint_{\ila}$ lead to the terms
$\sum_{\ilb}\K_{\ila\ilb}(t-t_{0})\Vint_{\ilb}(t_{0})$, are not omitted but they are
included into an alternative definition of the fluctuating sources, namely
$\tilde{f}_{\ila}(t)=\ffl_{\ila}(t)+\sum_{\ilb}\K_{\ila\ilb}(t-t_{0})\Vint_{\ilb}(t_{0})$.
The statistical properties of these terms, as determined by the above distribution, are
given by expressions {\em identical\/} with Eqs.\ (\ref{stats:L:1st_moment}) and
(\ref{stats:L:2nd_moment}).

Notice that the recourse to the ``process" of initial fastening (and subsequent releasing)
of the system by an external agency can, to a certain extent, be circumvented on
noting that the concomitant initial statistical properties of the environment are
consistent with the notion of a time-scale separation between the system and the
surrounding medium, i.e., the latter adjust rapidly to the state of the former
(Lindenberg and West, 1984).

Note finally that the differences associated with assuming decoupled initial conditions
or the more physically motivated coupled initial conditions diminish as long as the
weak-coupling condition is met. Anyhow, with both types of initial conditions one
obtains the {\em same\/} Langevin equation after a time, measured from $t_{0}$, of
the order of the width of the memory kernels $\K_{\ila\ilb}(\tau)$, which is the
characteristic time for the ``transient" terms
$\sum_{\ilb}\K_{\ila\ilb}(t-t_{0})\Vint_{\ilb}(t_{0})$ to die out.

\subsection[Dynamical equations for couplings linear-plus-quadratic in the
environment variables]
{Dynamical equations for couplings\\linear-plus-quadratic in the environment variables}
\label{subsect:linear+quadratic_coupling}

The introduction of interactions non-linear in the environment
variables is mandatory when relaxation mechanisms involving more than
one environmental normal mode (e.g., multi-phonon or multi-photon
processes) become relevant, as occurs at sufficiently high
temperatures. When such non-linear couplings are taken into account,
one must resort to approximate methods to derive a reduced equation of
motion for the spin. Here, we shall tackle the important weak-coupling
case by a perturbational treatment.

\subsubsection{The spin-environment Hamiltonian}

Let us consider the following generalization of the Hamiltonian (\ref{hamiltonian:L:1})
\begin{eqnarray}
\label{hamiltonian:L+Q:1}
\Ham_{\rm T}
=
\Hs(\m) &+&
\sum_{\alpha}
\half
\Big\{
\eP_{\alpha}^{2}\overMa
+\Ma\w_{\alpha}^{2}
\Big[
\eQ_{\alpha}+\frac{\coupling}{\Ma\w_{\alpha}^{2}}\Fint_{\alpha}(\m)
\Big]^2
\Big\}
\nonumber
\\
&+&
\half\sum_{\alpha\beta}
\Big[
\coupling \eQ_{\alpha}\eQ_{\beta}\Fint_{\alpha\beta}(\m)
+\frac
{\T\coupling^{2}}
{2\Ma\w_{\alpha}^{2}\Mb\w_{\beta}^{2}}
\Fint_{\alpha\beta}(\m)^{2}
\Big]
\;,
\end{eqnarray}
where couplings quadratic in the coordinates of the oscillators representing the
environment have been included. The part of this interaction depending on the spin
variables is introduced via the functions $\Fint_{\alpha\beta}$. On the other hand,
embodying the additional counter-terms (those proportional to
$\Fint_{\alpha\beta}^{2}$), the coupling-induced renormalization of the energy of the
spin is balanced to order $\coupling^{2}$. This renormalization results to be explicitly
dependent on the temperature for interactions non-linear in the environment
variables (see below).

\subsubsection{Dynamical equations: general case}

Again, for the sake of simplicity and generality, we rewrite the Hamiltonian
(\ref{hamiltonian:L+Q:1}) as [cf.\ Eq.\ (\ref{hamiltonian:L:2})]
\begin{eqnarray}
\label{hamiltonian:L+Q:2}
\Ham_{\rm T}
&
=
&
\Hsm(\sP,\sQ)
+\sum_{\alpha}
\half
\left(
\eP_{\alpha}^{2}\overMa
+\Ma\w_{\alpha}^{2}\eQ_{\alpha}^{2}
\right)
\nonumber
\\
&
& {}+
\coupling
\Big[
\sum_{\alpha}\eQ_{\alpha}\Fint_{\alpha}(\sP,\sQ)
+\half
\sum_{\alpha\beta}\eQ_{\alpha}\eQ_{\beta}\Fint_{\alpha\beta}(\sP,\sQ)
\Big]
\;,
\end{eqnarray}
where $\Hsm$ augments the system Hamiltonian by the counter-terms [cf.\
Eq.\ (\ref{modified_hamiltonian:L})]
\begin{equation}
\label{modified_hamiltonian:L+Q}
\Hsm
=
\Hs
+\frac{\coupling^{2}}{2}
\bigg(
\sum_{\alpha}\frac{\Fint_{\alpha}^{2}}{\Ma\w_{\alpha}^{2}}
+\T\sum_{\alpha\beta}
\frac
{\Fint_{\alpha\beta}^{2}}
{2\Ma\w_{\alpha}^{2}\Mb\w_{\beta}^{2}}
\bigg)
\;.
\end{equation}
The ordinary formalism of the environment of {\em independent\/} oscillators
(Magalinski{\u{\i}}, 1959; Ullersma, 1966; Zwanzig, 1973; Caldeira and Leggett, 1983;
Ford, Lewis and O'Connell, 1988) is not directly applicable when couplings non-linear
in the environment variables are included. For instance,
$\Fint_{\alpha\beta}\eQ_{\alpha}\eQ_{\beta}$ brings about an indirect interaction
among the oscillators so that these are no longer independent. Because a reduced
equation of motion for a dynamical variable $A(\sP,\sQ)$ cannot easily be derived for
an arbitrary strength of the coupling, we shall perform a perturbational treatment in
the weak-coupling case by means of simple extensions of the treatment developed by
Cort{\'{e}}s, West and Lindenberg (1985).

In Appendix \ref{app:corweslin} the corresponding calculations are detailed for a class
of Hamiltonians with quite general non-linear couplings in both the system and the
environment variables. The results obtained permit the incorporation of relaxation
mechanisms involving any number of environmental normal modes into the
dynamical equations of the system variables (under the weak-coupling condition
mentioned). In the linear-plus-quadratic case considered here, we find the following
{\em reduced\/} dynamical equation for any observable of the system $A(\sP,\sQ)$
[cf.\ Eq.\ (\ref{eqmot_A:L})]
\begin{eqnarray}
\label{eqmot_A:L+Q}
\frac{\D A}{\D t}
=
\pbra A, \Hs\pket &+&
\sum_{\alpha}\pbra A, \Fint_{\alpha}\pket
\bigg[
\ffl_{\alpha}(t)
+\int_{t_{0}}^{t}\!\!\D{\tp}\,
\K_{\alpha}(t-\tp)
\frac{\D\Fint_{\alpha}}{\D t}(\tp)
\bigg]
\nonumber
\\
&+&
\sum_{\alpha\beta}\pbra A, \Fint_{\alpha\beta}\pket
\bigg[
\ffl_{\alpha\beta}(t)+
\int_{t_{0}}^{t}\!\!\D{\tp}\,
\K_{\alpha\beta}(t-\tp)
\frac{\D\Fint_{\alpha\beta}}{\D t}(\tp)
\bigg]
\;.
\qquad
\quad
\end{eqnarray}
In this equation, the fluctuating terms $\ffl_{\alpha}(t)$ and the
corresponding kernels $\K_{\alpha}(\tau)$ are again given by Eqs.\
(\ref{fluct-kernel:L:precursor}), whereas their counterparts for the
quadratic portion of the coupling read
\begin{eqnarray}
\label{fluct:Q:precursor}
\ffl_{\alpha\beta}(t)
&
=
&
\frac{\coupling}{2}
\eQ_{\alpha}^{\h}(t)\eQ_{\beta}^{\h}(t)
\;,
\\
\label{kernel:Q:precursor}
\K_{\alpha\beta}(\tau)
&
=
&
\frac{\coupling^{2}}{2}
\frac{\T}{2\Ma\w_{\alpha}^{2}\Mb\w_{\beta}^{2}}
\big\{
\cos[(\w_{\alpha}-\w_{\beta})\tau]
+\cos[(\w_{\alpha}+\w_{\beta})\tau]
\big\}
\;,
\end{eqnarray}
where the $\eQ_{\alpha}^{\h}(t)$ are the environmental proper modes (\ref{qh}).

The treatment leading to Eq.\ (\ref{eqmot_A:L+Q}) can be summarized in
terms of the driven-oscillator picture discussed in Subsec.\
\ref{subsect:linear_coupling}. One part of the driving from the system
now depends on the state of the oscillators [cf.\ Eqs.\
(\ref{eqmot_bath:L}) with (\ref{eqmot_bath:gral})]; this state is
perturbatively replaced by the free evolution terms
$\eQ_{\alpha}^{\h}(t)$, and the back-reaction on the system is
averaged over initial states of the oscillators. This averaging yields
the explicit dependence of the kernels $\K_{\alpha\beta}(\tau)$ on the
temperature (and that of the associated counter-term
$\half\sum_{\alpha\beta}
\K_{\alpha\beta}(0)\Fint_{\alpha\beta}^{2}$).

\subsubsection{Dynamical equations: the spin-dynamics case}

In order to particularize the result (\ref{eqmot_A:L+Q}) to the dynamics of a classical
spin, the additional coupling functions $\Fint_{\alpha\beta}$ are expressed as
\[
\Fint_{\alpha\beta}(\m)
=
\sum_{\iqa}\Cint_{\alpha\beta}^{\iqa}\Vint_{\iqa}(\m)
\;,
\]
where the general index $\iqa$ is analogous to that introduced in the
linear case [Eq.\ (\ref{W:L})], the coefficients
$\Cint_{\alpha\beta}^{\iqa}$ are the spin-environment coupling
constants for the quadratic part of the interaction, and the terms
$\Vint_{\iqa}(\m)$ are certain functions of the spin variables. To
illustrate, for the coupling of $\m$ to the lattice vibrations
including quadratic terms in the strain tensor (``two-phonon"
processes), $\iqa$ stands for {\em two\/} pairs of Cartesian indices
and, for example,
$\Vint_{\iqa}\to\Vint_{ij,k\ell}=\sum_{rs}b_{ijk\ell,rs}\mr\ms$, where
the $b_{ijk\ell,rs}$ are second-order magneto-elastic coefficients.

Then, on merely replacing $A(\sP,\sQ)$ in Eq.\ (\ref{eqmot_A:L+Q}) by
the Cartesian components of the magnetic moment and then using Eq.\
(\ref{MpoissonV}) to calculate the corresponding Poisson brackets, one
arrives at the following reduced equation of motion for $\m$ [cf.\
Eq.(\ref{eqmot_M_LLtyp:L})]
\begin{eqnarray}
\label{eqmot_M_LLtyp:L+Q}
\frac{\D\m}{\D t}
&
=
&
\gmr\m
\vecpro
\left[
\Beff+\bfl(\m,t)
\right]
\nonumber
\\
&
& {}-\gmr\m
\vecpro
\bigg\{
\int_{t_{0}}^{t}\!\!\D{\tp}\,
\gmr
\left[
\TLpre+\T\,\TQpre
\right]_{(\m;t,\tp)}
\left(\m\vecpro\Beff\right)(\tp)
\bigg\}
\;.
\qquad
\quad
\end{eqnarray}
Here, the fluctuating effective field generalizes the expression (\ref{bfl:L}) to
\begin{equation}
\label{bfl:L+Q}
\bfl(\m,t)
=
-\bigg[
\sum_{\ila}\ffl_{\ila}(t)\frac{\partial\Vint_{\ila}}{\partial\m}
+\sum_{\iqa}\ffl_{\iqa}(t)\frac{\partial\Vint_{\iqa}}{\partial\m}
\bigg]
\;,
\end{equation}
where the $\ffl_{\ila}(t)$ are given by Eq.\ (\ref{fluct:L}) and the
$\ffl_{\iqa}(t)=\sum_{\alpha\beta}\Cint_{\alpha\beta}^{\iqa}\ffl_{\alpha\beta}(t)$ are
additional fluctuating terms
\begin{equation}
\label{fluct:Q}
\ffl_{\iqa}(t)
=
\frac{\coupling}{2}
\sum_{\alpha\beta}
\Cint_{\alpha\beta}^{\iqa}
\,\eQ_{\alpha}^{\h}(t)\eQ_{\beta}^{\h}(t)
\;.
\end{equation}
Concerning the relaxation terms, $\TLpre$ is again given by Eq.\
(\ref{kernel:L:tensor}), while the part of the  relaxation tensor associated with the
quadratic part of the coupling in given by
\begin{equation}
\label{kernel:Q:tensor}
\T\,\TQpre(\m;t,\tp)
=
\sum_{\iqa,\iqb}\K_{\iqa\iqb}(t-\tp)
\frac{\partial\Vint_{\iqa}}{\partial\m}(t)
\frac{\partial\Vint_{\iqb}}{\partial\m}(\tp)
\;,
\end{equation}
where the kernel is given by $\K_{\iqa\iqb}(\tau)=\sum_{\alpha\beta}
\Cint_{\alpha\beta}^{\iqa}\Cint_{\alpha\beta}^{\iqb}\K_{\alpha\beta}(\tau)$ or,
explicitly
\begin{equation}
\label{kernel:Q}
\K_{\iqa\iqb}(\tau)
=
\T\frac{\coupling^{2}}{2}
\sum_{\alpha\beta}
\frac
{\Cint_{\alpha\beta}^{\iqa}\Cint_{\alpha\beta}^{\iqb}}
{2\Ma\w_{\alpha}^{2}\Mb\w_{\beta}^{2}}
\big\{
\cos[(\w_{\alpha}-\w_{\beta})\tau]
+\cos[(\w_{\alpha}+\w_{\beta})\tau]
\big\}
\;.
\end{equation}

Note that the equation (\ref{eqmot_M_LLtyp:L+Q}) is of Landau--Lifshitz type since
the derivative $\D\m/\D t$ that would appear in the relaxation term has been
replaced, within the approximation used ($\coupling\ll1$), by its free evolution part
$\D\m/\D t\simeq\gmr\m\vecpro\Beff$ [see the remarks after Eq.\
(\ref{eqmot_A:gral})]. Notice also that we have explicitly shown the temperature
dependence of the relaxation term, which is caused by the quadratic portion of the
coupling.

Equation (\ref{eqmot_M_LLtyp:L+Q}) is the desired dynamical equation for the spin
when its interaction with the environment is weak and embodies linear-plus-quadratic
terms in the variables of the oscillators representing the environment. Note that, in
the pictorial quantum-mechanical language, the term comprising
$\cos(\w_{\alpha}\tau)$ in the memory kernel (\ref{kernel:L}) would correspond to a
relaxation mechanism (transition) via the emission or absorption of a vibrational
quantum of energy
$\hbar\w_{\alpha}$. Similarly,
$\cos[(\w_{\alpha}+\w_{\beta})\tau]$ in the kernel (\ref{kernel:Q}) would be
associated with relaxation mechanisms with either the emission or the absorption of
two vibrational quanta, whereas $\cos[(\w_{\alpha}-\w_{\beta})\tau]$ would
correspond to the absorption of one quantum and the emission of a second one
(scattering processes).

Finally, the definition (\ref{ufl:L}) of the fluctuating part of the energy of the spin can
be generalized to
\begin{equation}
\label{ufl:L+Q}
\Hfl(\m,t)
=
\sum_{\ila}
\ffl_{\ila}(t)\Vint_{\ila}(\m)+\sum_{\iqa}\ffl_{\iqa}(t)\Vint_{\iqa}(\m)
\;,
\end{equation}
whence $\bfl=-\partial\Hfl/\partial\m$, in correspondence with
$\Beff=-\partial\Hs/\partial\m$. Remarks similar to those made after Eq.\
(\ref{ufl:L}) concerning the structure of $\Hfl(\m,t)$ for linear and non-linear (in the
spin variables) spin-environment interactions, and the corresponding nature of the
fluctuations (field- and/or anisotropy-type), are in order here.

\subsubsection{Statistical properties of the fluctuating terms}

The statistical properties of the $\ffl_{\ila}(t)$, as determined by
the initial distribution (\ref{gibbs:oscillators:dic}) of the environment variables ({\em
decoupled initial conditions}), are given by Eqs.\ (\ref{stats:L:1st_moment}) and
(\ref{stats:L:2nd_moment}), whereas the statistical properties of the $\ffl_{\iqa}(t)$
and their cross-correlations read
\begin{eqnarray}
\label{stats:Q:1st_moment}
\llangle\ffl_{\iqa}(t)\rrangle
&
=
&
0
\;,
\\
\label{stats:L_cross_Q}
\llangle\ffl_{\ila}(t)\ffl_{\iqa}(\tp)\rrangle
&
=
&
0
\;,
\\
\label{stats:Q:2nd_moment}
\llangle\ffl_{\iqa}(t)\ffl_{\iqb}(\tp)\rrangle
&
=
&
\T\,\K_{\iqa\iqb}(t-\tp)
\;.
\end{eqnarray}
In order to obtain Eq.\ (\ref{stats:Q:1st_moment}), i.e., centered fluctuating  sources, as
well as Eq.\ (\ref{stats:Q:2nd_moment}), we have assumed that
$\Cint_{\alpha\beta}^{\iqa}\equiv0$ for $\alpha=\beta$. If such a restriction is not
applied, one has, for example, $\langle\ffl_{\iqa}(t)\rangle\neq0$, which represents a
non-vanishing average forcing of the spin. Note however that to retain those terms
must cause no harm since, when the double sums over oscillators
$\sum_{\alpha\beta}(\cdot)$ are transformed into double integrals for (quasi-)
continuous distributions of oscillators, such $\alpha=\beta$ terms constitute a
zero-measure set whose contribution can therefore be ignored.

The Gaussian property of the $\ffl_{\iqa}(t)$ can then be established on the basis that
these terms are sums over a large number of contributions
$\Cint_{\alpha\beta}^{\iqa}\,\eQ_{\alpha}^{\h}(t)\eQ_{\beta}^{\h}(t)$ with mean zero
and equivalent statistical properties (Central Limit Theorem). On the other hand, Eq.\
(\ref{stats:Q:2nd_moment}) expresses that the fluctuating sources $\ffl_{\iqa}(t)$ and
the relaxation memory kernels
$\K_{\iqa\iqb}(\tau)$ associated with the quadratic portion of the coupling also obey
fluctuation-dissipation relations. In addition, the zero cross-correlations of Eq.\
(\ref{stats:L_cross_Q}) are also fluctuation-dissipation relations involving null kernels
[see Eq.\ (\ref{Ksb})].

We finally remark that on assuming {\em coupled initial conditions}, without
modifying the definitions of the fluctuating terms, the corrections to Eqs.\
(\ref{stats:L:1st_moment}) and (\ref{stats:Q:1st_moment}), and to the relations
(\ref{stats:L:2nd_moment}), (\ref{stats:L_cross_Q}), and (\ref{stats:Q:2nd_moment})
are, respectively, of order $\coupling^{2}$ and $\coupling^3$; these corrections are of
order higher than the terms retained in the weak-coupling approximation used (see
Appendix \ref{app:corweslin}).

\subsection{Markovian regime and phenomenological equations}

We shall now study the form that the dynamical equations derived exhibit in the
absence of memory effects. Then, we shall consider some specific spin-environment
interactions, formally obtaining the Langevin equations mentioned at the beginning
of this section.

\subsubsection{Markovian regime}

The Markovian regime arises when the relaxation memory kernels are sharply peaked
at $\tau=0$, the remainder terms in the memory integrals change slowly enough
in the relevant range, and the kernels enclose a finite non-zero algebraic area. Under
these conditions, one can replace the kernels by Dirac deltas and no memory effects
occur.

\paragraph{Langevin equations.}

Let us assume that the memory kernel (\ref{kernel:L}) can be replaced by a Dirac
delta
\begin{equation}
\label{kernel:L:markov}
\K_{\ila\ilb}(\tau)
=
2(\la_{\ila\ilb}/\gmr\mm)\delta(\tau)
\;,
\end{equation}
where the $\la_{\ila\ilb}$ are {\em damping coefficients\/} related with the strength
and characteristics of the coupling (see below). Then, on using
$\int_{0}^{\infty}\!\D\tau\,\delta(\tau)h(\tau)=h(0)/2$, equation
(\ref{eqmot_M_Giltyp}) reduces to the {\em Gilbert-type\/} equation [cf.\ Eq.\
(\ref{eqmot_M_Gil})]
\begin{equation}
\label{eqmot_M_Giltyp_markov}
\frac{\D\m}{\D t}
=
\gmr\m\vecpro 
\bigg[
\Beff+\bfl(\m,t)-(\gmr\mm)^{-1}\TL\frac{\D\m}{\D t}
\bigg]
\;,
\end{equation}
where $\TL(\m)$ is a dimensionless second-rank tensor with elements
\begin{equation}
\label{tensor:L}
\tL_{ij}(\m)
=
\sum_{\ila,\ilb}\la_{\ila\ilb}
\frac{\partial\Vint_{\ila}}{\partial\mi}
\frac{\partial\Vint_{\ilb}}{\partial\mj}
\;.
\end{equation}
Likewise, on inserting Eq.\ (\ref{kernel:L:markov}) in the weak-coupling Eq.\
(\ref{eqmot_M_LLtyp:L}) we get the following {\em Landau--Lifshitz-type\/}
equation [cf.\ Eq.\ (\ref{eqmot_M_LL})]
\begin{equation}
\label{eqmot_M_LLtyp_markov:L}
\frac{\D\m}{\D t}
=
\gmr\m\vecpro 
\left[
\Beff+\bfl(\m,t)
\right]
-\frac{\gmr}{\mm}\m\vecpro\TL
\left(\m\vecpro\Beff\right)
\;.
\end{equation}
Note that the tensor $\TL$, the precursor of which is the tensor $\TLpre$ [Eq.\
(\ref{kernel:L:tensor})] occurring in the memory integrals, is symmetrical since
$\la_{\ila\ilb}$ is so [see Eq.\ (\ref{lambda_ab:oscillators}) below].

On the other hand, the Markovian case of the dynamical equation for couplings
linear-plus-quadratic in the environment coordinates emerges when the additional
memory kernel can also be replaced by a Dirac delta, namely
\begin{equation}
\label{kernel:Q:markov}
\K_{\iqa\iqb}(\tau)
=
2(\la_{\iqa\iqb}\T/\gmr\mm)\delta(\tau)
\;,
\end{equation}
where we have explicitly shown the temperature dependence arising from the
kernel (\ref{kernel:Q}). Under these conditions, Eq.\ (\ref{eqmot_M_LLtyp:L+Q})
reduces to the {\em Landau--Lifshitz-type\/} equation
\begin{equation}
\label{eqmot_M_LLtyp_markov:L+Q}
\frac{\D\m}{\D t}
=
\gmr\m\vecpro 
\left[
\Beff+\bfl(\m,t)
\right]
-\frac{\gmr}{\mm}\m\vecpro\TLQ
\left(\m\vecpro\Beff\right)
\;,
\end{equation}
where $\bfl(\m,t)$ is now given by Eq.\ (\ref{bfl:L+Q}). In this equation the
relaxation tensor
\begin{equation}
\label{tensor:L+Q}
\TLQ
=
\TL+\T\,\TQ
\;,
\end{equation}
where
\begin{equation}
\label{tensor:Q}
\tQ_{ij}(\m)
=
\sum_{\iqa,\iqb}\la_{\iqa\iqb}
\frac{\partial\Vint_{\iqa}}{\partial\mi}
\frac{\partial\Vint_{\iqb}}{\partial\mj}
\;,
\end{equation}
introduces an explicit dependence on the temperature rooted in the quadratic portion
of the coupling.

For a general form of the spin-environment interaction, due to the occurrence of the
tensors $\TLQ$ the structure of the relaxation terms in the above equations
deviates from the forms proposed by Gilbert and Lan\-dau and Lif\-shitz. Such
deviations can be produced by couplings non-linear in $\m$, for which $\TL_{ij}$ and
$\TQ_{ij}$ depend in general on the spin variables, but they also emerge when these
tensors are independent of $\m$ (for example, for couplings linear in $\m$) but they
are not proportional to $\delta_{ij}$. The relaxation is then anisotropic because, for
instance, $-\m\vecpro\TLQ(\m\vecpro\Beff)$ no longer points from $\m$ to the
direction of $\Beff$.

Finally, owing to the fluctuation-dissipation relations (\ref{stats:L:2nd_moment}) and
(\ref{stats:Q:2nd_moment}), the fluctuating terms corresponding to the Markovian
memory kernels are delta-correlated in time. Consequently, the statistical
properties of the fluctuating terms take the form
\begin{eqnarray}
\label{stats:L:1st_moment:markov}
\llangle\ffl_{\ila}(t)\rrangle
&
=
&
0
\;,
\\
\label{stats:L:2nd_moment:markov}
\llangle\ffl_{\ila}(t)\ffl_{\ilb}(\tp)\rrangle
&
=
&
\frac{2\la_{\ila\ilb}}{\gmr\mm}\,\T\delta(t-\tp)
\;,
\end{eqnarray}
and
\begin{eqnarray}
\label{stats:Q:1st_moment:markov}
\llangle\ffl_{\iqa}(t)\rrangle
&
=
&
0
\;,
\\
\label{stats:L_cross_Q:markov}
\llangle\ffl_{\ila}(t)\ffl_{\iqa}(\tp)\rrangle
&
=
&
0
\;,
\\
\label{stats:Q:2nd_moment:markov}
\llangle\ffl_{\iqa}(t)\ffl_{\iqb}(\tp)\rrangle
&
=
&
\frac{2(\la_{\iqa\iqb}\T)}{\gmr\mm}\,\T\delta(t-\tp)
\;.
\end{eqnarray}
Notice the double occurrence of $\T$ in the last relation.

\paragraph{Damping coefficients.}

On taking Eqs.\ (\ref{kernel:L:markov}) and (\ref{kernel:Q:markov}) into account, one
can calculate the damping coefficients from the area enclosed by the memory kernels,
namely
\begin{equation}
\label{lambdas:kernel}
\frac{\la_{\ila\ilb}}{\gmr\mm}
=
\int_{0}^{\infty}\!\!\D\tau\,\K_{\ila\ilb}(\tau)
\;,
\qquad
\label{lambda_abcd:kernel}
\frac{\la_{\iqa\iqb}\T}{\gmr\mm}
=
\int_{0}^{\infty}\!\!\D\tau\,\K_{\iqa\iqb}(\tau)
\;.
\end{equation}
These areas must be: (i) {\em finite\/} and (ii) {\em different from zero}, for the
Markovian approximation to work.

On the other hand, since it could be difficult to find the kernels exactly in some cases,
it is convenient to have alternative means for calculating the areas required only.
Thus, on inserting the definitions of the kernels (\ref{kernel:L}) and (\ref{kernel:Q})
into the above integrals and using
$\int_{0}^{\infty}\!\D\tau\,\cos(\w\tau)=\pi\delta(\w)$,
we arrive at the following expressions for the damping coefficients in terms of the
distribution of normal modes and spin-environment coupling constants
\begin{eqnarray}
\label{lambda_ab:oscillators}
\frac{\la_{\ila\ilb}}{\gmr\mm}
&
=
&
\pi\coupling^{2}
\sum_{\alpha}
\frac
{\Cint_{\alpha}^{\ila}\Cint_{\alpha}^{\ilb}}
{\Ma\w_{\alpha}^{2}}\delta(\w_{\alpha})
\;,
\\
\label{lambda_abcd:oscillators}
\frac{\la_{\iqa\iqb}}{\gmr\mm}
&
=
&
\pi\frac{\coupling^{2}}{2}
\sum_{\alpha\beta}
\frac
{\Cint_{\alpha\beta}^{\iqa}\Cint_{\alpha\beta}^{\iqb}}
{2\Ma\w_{\alpha}^{2}\Mb\w_{\beta}^{2}}
\big[
\delta(\w_{\alpha}-\w_{\beta})
+\delta(\w_{\alpha}+\w_{\beta})
\big]
\;.
\end{eqnarray}
Note that the Dirac deltas in these formulae make sense under integral signs for
(quasi-) continuous distributions of environmental modes. Recall in this connection
that the coupling constants can depend on the frequencies of these normal modes.

\paragraph{Fokker--Planck equations.}

The Markovian Langevin equations can be employed to construct the corresponding
Fokker--Planck equations governing the time evolution of the non-equilibrium
probability distribution of spin orientations $\W(\m,t)$. On examining the
statistical properties (\ref{stats:L:2nd_moment:markov}) and
(\ref{stats:Q:2nd_moment:markov}), one realizes that, to do so, Langevin
equations where the noise terms {\em are not\/} statistically independent need to be
considered.

Let us then consider the general system of Langevin equations
\begin{equation}
\label{langevinequation:n-dim:corr}
\frac{\D y_{i}}{\D t}
=
\drift_{i}(\ym,t)
+\sum_{k}\diff_{ik}(\ym,t)\Lan_{k}(t)
\;,
\end{equation}
where $\ym=(y_{1},\ldots,y_{n})$, $k$ runs over a given set of indices, and the
Langevin sources $\Lan_{k}(t)$ are Gaussian stochastic processes satisfying
\begin{equation}
\label{langevin:n-dim:corr}
\llangle\Lan_{k}(t)\rrangle
=
0
\;,
\qquad
\llangle\Lan_{k}(t)\Lan_{\ell}(\tp)\rrangle
=
2D_{k\ell}\delta(t-\tp)
\;.
\end{equation}
The constant (symmetrical) matrix $D_{k\ell}$ accounts for the possible correlations
among the $\Lan_{k}(t)$ [cf. Eq.\ (\ref{langevin:n-dim})].

The time evolution of $\W(\ym,t)$, the non-equilibrium probability distribution of
$\ym$ at time $t$, is given by the following generalization of the (Stratonovich)
Fokker--Planck equation (\ref{fokkerplanck:langevin:n-dim})
\begin{eqnarray*}
\frac{\partial\W}{\partial t}
&
=
&
-\sum_{i}\frac{\partial {}}{\partial y_{i}}
\bigg[
\bigg(
\drift_{i}
+\sum_{jk\ell}\,\diff_{j\ell}D_{\ell k}
\frac{\partial\diff_{ik}}{\partial y_{j}}
\bigg)\W
\bigg]
\\
&
& {}+\sum_{ij}
\frac{\partial^{2}}{\partial y_{i}\partial y_{j}}
\bigg[
\bigg(
\sum_{k\ell}\,\diff_{ik}D_{k\ell}\diff_{j\ell}
\bigg)
\W
\bigg]
\;.
\end{eqnarray*}
As in Subsec.\ \ref{subsect:brown}, we take the $y_{j}$-derivatives of the
diffusion term in order to cast the Fokker--Planck equation into the form of a
continuity equation for the probability distribution
\begin{equation}
\label{fokkerplanck:langevin:n-dim:cont:corr}
\frac{\partial\W}{\partial t}
=
-\sum_{i}\frac{\partial {}}{\partial y_{i}}
\bigg\{
\bigg[
\drift_{i}
-\sum_{k\ell}\diff_{ik}D_{k\ell}
\bigg(
\sum_{j}\frac{\partial\diff_{j\ell}}{\partial y_{j}}
\bigg)
-\sum_{jk\ell}\,\diff_{ik}D_{k\ell}\diff_{j\ell}
\frac{\partial {}}{\partial y_{j}}\,
\bigg]
\W
\bigg\}
\;.
\end{equation}
Note that, for uncorrelated fluctuations, $D_{k\ell}=D\delta_{k\ell}$, these equations
duly reduce to Eqs.\ (\ref{fokkerplanck:langevin:n-dim}) and
(\ref{fokkerplanck:langevin:n-dim:cont}).

Now, on considering the {\em Landau--Lifshitz-type equation
(\ref{eqmot_M_LLtyp_markov:L})}, supplemented by the statistical properties
(\ref{stats:L:1st_moment:markov}) and (\ref{stats:L:2nd_moment:markov}), the
substitutions [cf.\ Eqs.\ (\ref{F:ll-llg}) and (\ref{G:ll-llg})]
\begin{eqnarray*}
(k,\ell)
&
=
&
(\ila,\ilb)
\;,
\qquad (y_{1},y_{2},y_{3})
=
(\mx,\my,\mz)
\;,
\\
\Lan_{\ila}(t)
&
=
&
\ffl_{\ila}(t)
\;,
\qquad
\qquad
\quad D_{\ila\ilb}
=
\frac{\la_{\ila\ilb}}{\gmr\mm}\T
\;,
\\
\drift_{i}
&
=
&
\left[
\gmr\m\vecpro\Beff-\frac{\gmr}{\mm}\m\vecpro\TL
\left(\m\vecpro\Beff\right)
\right]_{i}
\;,
\\
\diff_{i\ila}
&
=
&
-\gmr 
\sum_{rs}
\epsilon_{irs}\mr\frac{\partial\Vint_{\ila}}{\partial\ms}
\;,
\end{eqnarray*}
cast those equations into the form of the general system of Langevin equations
(\ref{langevinequation:n-dim:corr}) supplemented by Eqs.\
(\ref{langevin:n-dim:corr}). Therefore, on using [cf.\ Eq.\ (\ref{Gij:derivative:ll-llg})] 
\[
\frac{\partial\diff_{i\ila}}{\partial\mj}
=
-\gmr 
\bigg(
\sum_{s}
\epsilon_{ijs}\frac{\partial\Vint_{\ila}}{\partial\ms}
+\sum_{rs}
\epsilon_{irs}\mr
\frac{\partial^{2} \Vint_{\ila}}{\partial\mj\partial\ms}
\bigg)
\;,
\]
one finds that $\sum_{j}\partial\diff_{j\ila}/\partial\mj\equiv0,\,\forall \ila$ due to
the fact that $\epsilon_{jjs}=0$ and the vanishing of the contraction of symmetrical
tensors with antisymmetrical ones. Consequently, the second term on the
right-hand side of the general Fokker--Planck equation
(\ref{fokkerplanck:langevin:n-dim:cont:corr}) also vanishes in this case. For the third
term, by repeated use of
$(\vec{J}\vecpro\vec{J}')_{i}=\sum_{rs}\epsilon_{irs}J_{r}J_{s}'$ and recalling the
definition (\ref{tensor:L}), we obtain
\[
-\sum_{j\ila\ilb}\,
\diff_{i\ila}D_{\ila\ilb}\diff_{j\ilb}
\frac{\partial\W}{\partial\mj}
=
\frac{\gmr}{\mm}\T
\bigg[
\m\vecpro
\TL
\bigg(
\m\vecpro\frac{\partial\W}{\partial\m}
\bigg)
\bigg]_{i}
\;.
\]
On introducing these results into Eq.\ (\ref{fokkerplanck:langevin:n-dim:cont:corr})
one eventually arrives at the Fokker--Planck equation [cf.\ Eqs.\
(\ref{brownfpe:Gil-LL}) and (\ref{fpe_gip})]
\begin{equation}
\label{fpe}
\frac{\partial\W}{\partial t}
=
-\frac{\partial {}}{\partial\m}\cdot
\bigg\{
\gmr\m\vecpro\Beff
\W
-\frac{\gmr}{\mm}\m\vecpro\TL
\bigg[
\m\vecpro 
\bigg(
\Beff-\T\frac{\partial {}}{\partial\m}
\bigg)\W
\bigg]
\bigg\}
\;,
\end{equation}
where
$(\partial/\partial\m)\cdot\vec{J}
=\sum_{i} (\partial J_{i}/\partial\mi)$.
In addition, by means of similar considerations and allowing the index in the
Langevin sources $\Lan_{k}(t)$ to run also over the indices $\iqa$, the
Landau--Lifshitz-type equation (\ref{eqmot_M_LLtyp_markov:L+Q}) leads to a
Fokker--Planck equation analogous to the above one with $\TL$ augmented to
$\TLQ=\TL+\T\,\TQ$, namely
\begin{equation}
\label{fpe:L+Q}
\frac{\partial\W}{\partial t}
=
-\frac{\partial {}}{\partial\m}\cdot
\bigg\{
\gmr\m\vecpro\Beff
\W
-\frac{\gmr}{\mm}\m\vecpro\TLQ
\bigg[
\m\vecpro 
\bigg(
\Beff-\T\frac{\partial {}}{\partial\m}
\bigg)\W
\bigg]
\bigg\}
\;.
\end{equation}

Concerning the stationary solution of these Fokker--Planck equations, one can use
$\Beff=-\partial\Hs/\partial\m$ and
$(\partial/\partial\m)
\cdot
\big(\m\vecpro\Beff\Weq\big)=0$
(see Subsec.\ \ref{subsect:brown}), to demonstrate that the Boltzmann
distribution, $\Weq(\m)\propto\exp[-\Hs(\m)/\T]$, is indeed a stationary solution of
Eqs.\ (\ref{fpe}) and (\ref{fpe:L+Q}). This entails that under external stationary
conditions
$\W(\m,t)\stackrel{t\to\infty}{\longrightarrow}\Weq(\m)$,
that is, the spin eventually reaches the thermal equilibrium distribution of
orientations. Note that this is a consequence of the formalism employed, instead of a
constrain imposed separately, as is done in the phenomenological approaches (see
Subsec.\ \ref{subsect:brown}).

Note nevertheless that we have only proved the thermal equilibration for Eqs.\
(\ref{eqmot_M_LLtyp_markov:L}) and (\ref{eqmot_M_LLtyp_markov:L+Q}), i.e., in the
weak-coupling case. In this connection, it is to be recalled that, inasmuch as the
spin-environment coupling Hamiltonians themselves are commonly obtained via
perturbation theory (so they are ``small" in some sense), the study of the
arbitrary-coupling case of such Hamiltonians is mainly of an academic interest.

\subsubsection{Brown--Kubo--Hashitsume model}

When the spin-environment interaction is linear in the spin variables, the obtained
Markovian equations formally reduce to the equations occurring in the
Brown--Kubo--Hashitsume model. To illustrate, let us consider the simpler case of
couplings linear in the environment coordinates. Then, if the $\Vint_{\ila}(\m)$ are
linear in $\m$, both the relaxation tensor $\TL$ and the fluctuating field $\bfl$ are
independent of $\m$ [see Eqs.\ (\ref{tensor:L}) and (\ref{bfl:L}), respectively]. From
the statistical properties (\ref{stats:L:1st_moment:markov}) and
(\ref{stats:L:2nd_moment:markov}) of the fluctuating sources $\ffl_{\ila}(t)$, one then
gets [cf.\ Eqs.\ (\ref{bcorr:2})]
\begin{equation}
\label{stats:bfl:L}
\llangle b_{{\rm fl},i}(t)\rrangle
=
0
\;,
\quad
\llangle b_{{\rm fl},i}(t)b_{{\rm fl},j}(\tp)\rrangle
=
\frac{2\tL_{ij}}{\gmr\mm}\T\delta(t-\tp)
\;,
\end{equation}
where the last result establishes the relation between the structure of the correlations
among the components of $\bfl(t)$ and the form of the relaxation tensor $\TL$.%
\footnote{
Note that for $\bfl(\m,t)$ depending on $\m$, one cannot merely employ
Eqs.\ (\ref{stats:L:1st_moment:markov}) and (\ref{stats:L:2nd_moment:markov}) to
derive the statistical properties of $\bfl(\m,t)$, since $\m(t)$ and $\ffl_{\ila}(t)$ {\em
are not\/} independent.
}
The corresponding result by Jayannavar (1991) comprised an
uncorrelated $\bfl(t)$ (a diagonal $\tL_{ij}$ in our formulation) due to special bilinear
interaction that he considered [recall the discussion after Eq.\ (\ref{hamiltonian:L:1})].

On the other hand, if the spin-environment interaction yields uncorrelated {\em
and\/} isotropic fluctuations ($\tL_{ij}=\la\delta_{ij}$), one finds that: (i) the statistical
properties (\ref{stats:bfl:L}) reduce to (\ref{bcorr:2}), (ii) the Langevin equations
(\ref{eqmot_M_Giltyp_markov}) and (\ref{eqmot_M_LLtyp_markov:L}) reduce,
respectively, to the stochastic Gilbert [Eq.\ (\ref{eqmot_M_Gil})] and Landau--Lifshitz
[Eq.\ (\ref{eqmot_M_LL})] equations, and (iii) the Fokker--Planck equation (\ref{fpe})
reduces to (\ref{brownfpe:Gil-LL}). Thus, the phenomenological
Brown--Kubo--Hashitsume model is formally obtained.

Note that these results also hold when couplings quadratic in the environment
variables are included [Eq.\ (\ref{eqmot_M_LLtyp_markov:L+Q})], with the difference
that the relaxation terms (effective damping coefficients) are then explicitly dependent
on the temperature.

\subsubsection{Garanin, Ishchenko, and Panina model}

We shall now show that the weak-coupling Landau--Lifshitz-type equations
(\ref{eqmot_M_LLtyp_markov:L}) and (\ref{eqmot_M_LLtyp_markov:L+Q}), formally
reduce to the Langevin equation (\ref{eqmot_M_gip}) of Garanin, Ishchenko, and
Panina, when the spin-environment interaction includes up to quadratic terms in the
{\em spin variables}. In this case, the coupling functions $\Vint_{\ila}$ and
$\Vint_{\iqa}$ can be written as the natural extension of Eq.\ (\ref{F:l_spin}), namely
\begin{eqnarray}
\label{F:l+q_spin}
\Vint_{\ila}(\m)
&
=
&
\sum_{i}v_{\ila,i}\mi
+\half\sum_{ij}w_{\ila,ij}\mi\mj
\;,
\\
\label{G:l+q_spin}
\Vint_{\iqa}(\m)
&
=
&
\sum_{i}v_{\iqa,i}\mi
+\half\sum_{ij}w_{\iqa,ij}\mi\mj
\;,
\end{eqnarray}
where the $v_{\ila,i}$, $w_{\ila,ij}$, $v_{\iqa,i}$, and $w_{\iqa,ij}$ are coupling
constants incorporating the symmetry of the interaction. As in Subsec.\
\ref{subsect:linear_coupling}, the fluctuating effective field (\ref{bfl:L+Q}) can be
separated in an ordinary-field part and an anisotropy-field part
\begin{equation}
\label{bfl:l+q_spin}
\bfl(\m,t)
=
\vec{\bt}(t)+\hat{\kt}(t)\m
\;,
\end{equation}
while, in this case, the expressions for the fluctuating sources in terms of the
coupling constants are generalized to
\begin{eqnarray*}
\bt_{i}(t)
&
=
&
-\Big[\sum_{\ila}\ffl_{\ila}(t)v_{\ila,i}
+\sum_{\iqa}\ffl_{\iqa}(t)v_{\iqa,i}\Big]
\;,
\\
\kt_{ij}(t)
&
=
&
-\Big[\sum_{\ila}\ffl_{\ila}(t)w_{\ila,ij}
+\sum_{\iqa}\ffl_{\iqa}(t)w_{\iqa,ij}\Big]
\;.
\end{eqnarray*}
Naturally, the fluctuating part of the energy of the spin (\ref{ufl:L+Q}), which gives
$\bfl=-\partial\Hfl/\partial\m$, also takes in this case the form
$\Hfl
=-\m\cdot\vec{\bt}(t)
-\half\m\cdot\hat{\kt}(t)\m$.

In the Markovian regime, the auto- and cross-correlations of $\vec{\bt}(t)$ and
$\hat{\kt}(t)$ can be obtained by dint of Eqs.\ (\ref{stats:L:2nd_moment:markov}),
(\ref{stats:L_cross_Q:markov}), and (\ref{stats:Q:2nd_moment:markov}).  Such
correlations can be cast into the form proposed by Garanin, Ishchenko, and Panina
[Eq.\ (\ref{stats:gip})], with the following expressions for the correlation coefficients
\begin{eqnarray}
\label{lambdas:L+Q}
\la_{ij}
&
=
&
\sum_{\ila,\ilb}\la_{\ila\ilb}v_{\ila,i}v_{\ilb,j}
+\T\sum_{\iqa,\iqb}\la_{\iqa\iqb}v_{\iqa,i}v_{\iqb,j}
\;,
\nonumber
\\
\la_{i,jk}
&
=
&
\sum_{\ila,\ilb}\la_{\ila\ilb}v_{\ila,i}w_{\ilb,jk}
+\T\sum_{\iqa,\iqb}\la_{\iqa\iqb}v_{\iqa,i}w_{\iqb,jk}
\;,
\\
\la_{ik,j\ell}
&
=
&
\sum_{\ila,\ilb}\la_{\ila\ilb}w_{\ila,ik}w_{\ilb,j\ell}
+\T\sum_{\iqa,\iqb}\la_{\iqa\iqb}w_{\iqa,ik}w_{\iqb,j\ell}
\;.
\nonumber
\end{eqnarray}
Concerning the relaxation term, the tensor $\TLQ=\TL+\T\,\TQ$ [Eq.\
(\ref{tensor:L+Q})] associated with the coupling functions (\ref{F:l+q_spin}) and
(\ref{G:l+q_spin}), is given by
\begin{eqnarray*}
\tLQ_{ij}
&
=
&
\sum_{\ila,\ilb}\la_{\ila\ilb}
\Big( v_{\ila,i}
+\sum_{k}w_{\ila,ik}\mk
\Big)
\Big( v_{\ilb,j}
+\sum_{\ell}w_{\ilb,j\ell}\ml
\Big)
\\
&
& {}+
\T
\sum_{\iqa,\iqb}\la_{\iqa\iqb}
\Big( v_{\iqa,i}
+\sum_{k}w_{\iqa,ik}\mk
\Big)
\Big( v_{\iqb,j}
+\sum_{\ell}w_{\iqb,j\ell}\ml
\Big)
\;.
\end{eqnarray*}
However, this expression can be written in terms of the correlation coefficients
(\ref{lambdas:L+Q}) as
\begin{equation}
\label{tensor:L+Q-lambdas}
\tLQ_{ij}
=
\la_{ij}
+\sum_{k}(\la_{i,jk}+\la_{j,ik})\mk
+\sum_{k\ell}\la_{ik,j\ell}\mk\ml
\;,
\end{equation}
which is identical with the relation (\ref{tensor:gip}) between the tensor $\TGIP$ in
Eq.\ (\ref{eqmot_M_gip}) and the correlation coefficients in Eq.\ (\ref{stats:gip}).

Therefore, we find that when the spin-environment coupling includes up to quadratic
terms in the spin variables, the structures of the fluctuating effective field
$\bfl(\m,t)$
and of the relaxation term
$\R=(\gmr/\mm)\,\m\vecpro\TLQ(\m\vecpro\Beff)$
in the Landau--Lifshitz-type equation (\ref{eqmot_M_LLtyp_markov:L+Q}), as well as
the relation between them, are identical with the structures and mutual relations of
the corresponding terms in the Langevin equation (\ref{eqmot_M_gip}) of Garanin,
Ishchenko, and Panina. Naturally, the Fokker--Planck equation (\ref{fpe:L+Q}) then
reduces to Eq.\ (\ref{fpe_gip}).

\subsection{Discussion}

Starting from a Hamiltonian description of a classical spin interacting with the
surrounding medium, we have derived generalized Langevin equations, which, in the
Markovian approach, reduce to known stochastic equations of motion for classical
magnetic moments.

Note however that the presented derivation of the equations of Garanin, Ishchenko,
and Panina and, similarly, the previous derivations of the equations occurring in the
Brown--Kubo--Hashitsume model (Smith and De~Rozario, 1976;
Seshadri and Lindenberg, 1982; Jayannavar, 1991; Klik, 1992), are formal in the sense that one must
still investigate specific realizations of the spin-plus-environment whole system, and
then prove that the assumptions employed (mainly that of Markovian behavior) are at
least approximately met. A paradigmatic case in which the Markovian approach breaks
down, is the case of the magneto-elastic coupling of the spin to the lattice vibrations (in
two or three dimensions) {\em linear\/} in the corresponding normal modes (Garg and
Kim, 1991). The associated memory kernel crosses zero, changes it sign, and tends to
zero from negative values as $\tau\to\infty$, {\em enclosing a zero algebraic area}.
One then gets identically zero $\la_{\ila\ilb}$ by Eq.\ (\ref{lambdas:kernel}) and
hence a zero tensor $\TL$ by Eq.\ (\ref{tensor:L}). Therefore, on replacing such a
kernel by a Dirac delta, one looses the relaxational effects associated with the portion
of the coupling {\em linear\/} in the environment variables (``one-phonon" processes),
which are dominant at sufficiently low temperatures.

On the other hand, we have considered the classical regime of the environment and
the spin. A classical description of the environment is adequate, for example, for the
coupling to low-frequency ($\hbar\w_{\alpha}/\T\ll1$) normal modes, while, for
instance, the magnetic moment of a nanometric particle
($\mm\sim10^{3}$--$10^{5}\,\mu_{{\rm B}}$) behaves, except for very low
temperatures, as a classical spin. In addition, the equations derived might also serve
as a limit description of the semi-classical dynamics of molecular magnetic clusters
with high spin ($S\gsim10$) in their ground state.

%% file: garcms07.tex
\section{Summary and conclusions}

To conclude, let us summarize the most important results presented in
this Chapter:

Approximate and exact results for a number of thermal equilibrium
quantities for non-interacting classical magnetic moments with a
simple axially symmetric anisotropy potential, have been derived and
analyzed. The results obtained also apply to systems described as
assemblies of classical dipole moments with Hamiltonians comprising a
coupling term to an (electric or magnetic) external field plus an
axially symmetric orientational potential. Concerning their
application to superparamagnetic systems, it has been shown the
fundamental r\^{o}le of the magnetic anisotropy in the
thermal-equilibrium properties of magnetic nanoparticles and,
consequently, the inadequacy of the approaches that ignore these
effects on the basis of a restrictive ascription of superparamagnetism
to the temperature range where the anisotropy energy is smaller than
the thermal energy.

In the study of the dynamics of individual magnetic moments by the
Langevin dynamics approach, interesting phenomena in the over-barrier
rotation process have been found, such as crossing-back and multiple
crossing of the potential barrier, which can be explained in terms of
the gyromagnetic nature of the system.

The results for the linear dynamical susceptibility, $\chi(\w)$,
obtained from the stochastic Landau--Lifshitz--Gilbert equation, have
been compared with different analytical expressions used to model the
relaxation of nanoparticle ensembles, assessing their accuracy. It has
been found that, among a number of heuristic expressions for
$\chi(\w)$, only the simple formula proposed by Shliomis and Stepanov
matches the coarse features of the susceptibility reasonably. On the
other hand, we have investigated the effects of the
intra-potential-well relaxation modes on the low-temperature
longitudinal dynamical response, showing their relatively small
reflection in the $\chi_{\|}(\w,T)$ curves (remarkably small in
$\chi_{\|}''$) but their dramatic influence on the phase
shifts. Concerning the transverse response, the sizable relative
contribution to $\chi_{\perp}''(\w)$ of the spread of the precession
frequencies of the magnetic moment in the anisotropy field at
intermediate-to-high temperatures, has been demonstrated by comparing
the numerical results with the exact zero-damping expression for
$\chi_{\perp}''(\w)$. Taking this effect into account may be relevant
to properly assess the strength of the damping is superparamagnetic
systems.

Dynamical equations for a classical spin interacting with the
surrounding medium have been derived by means of the formalism of the
oscillator-bath environment. The customary bilinear-coupling treatment
has been extended to couplings that depend arbitrarily on the spin
variables and are linear or linear-plus-quadratic in the environment
dynamical variables. The equations obtained have the structure of
generalized Langevin equations, which, in the Markovian approach,
formally reduce to known semi-phenomenological equations of motion for
classical magnetic moments. Specifically, the generalization of the
stochastic Landau--Lifshitz equation effected by Garanin, Ishchenko,
and Panina in order to incorporate fluctuations of the magnetic
anisotropy of the spin, has been obtained for spin-environment
interactions including up to quadratic terms in the spin variables.
On the other hand, the portion of the coupling quadratic in the
environment variables introduces an explicit dependence of the
effective damping coefficients on the temperature.

%% file: garcms08.tex
\appendix

\setcounter{equation}{0}
\section[The functions $\F^{(\ell)}(\s)$]{The functions $\F^{(\ell)}(\s)$}
\label{app:F}

In this appendix, we shall summarize some properties of the function $\F(\s)$ and its
derivatives:
\[
\F^{(\ell)}(\s)
=
\int_{0}^{1}\!\!\D{z}\,z^{2\ell}\exp(\s z^{2})
\;,
\qquad
\ell
=
0,1,2,\ldots
\;.
\]
These functions, which were introduced by Ra{\u{\i}}kher and Shliomis (1975), play an
important r\^{o}le in the study of the equilibrium and dynamical properties of
classical magnetic moments with the simplest axially symmetric anisotropy potential.

We shall also derive approximate expressions for the most familiar combinations of
the type $\F^{(\ell)}/\F$, which will be valid in the ranges $|\s|\ll1$ and $|\s|\gg1$.
These approximate formulae can be employed to derive the corresponding
approximate expressions for a number of quantities.

\subsection{Relations with known special functions}

The functions $\F^{(\ell)}(\s)$ are related with certain special functions, e.g.,
{\em the Kummer functions, error functions, and the Dawson integral}.

The definition of the confluent hypergeometric (Kummer) functions is (Arfken, 1985,
p.~753)
\begin{eqnarray}
\label{kummer:series}
&
& M(a,c\,;x)
=
\sum_{n=0}^{\infty}
\frac{(a)_{n}}{(c)_{n}}
\frac{x^{n}}{n!}
\;,
\qquad c\neq 0,-1,-2,\ldots
\;,
\\
&
& (a)_{n}
=
a(a+1)\cdots(a+n-1)
=
(a+n-1)!/(a-1)!
\;,
\quad (a)_{0}
=
1
\;,
\nonumber
\end{eqnarray}
where $(a)_{n}$ is the Pochhammer symbol. For non-integer argument the
factorial signs are to be interpreted as gamma functions $a!\stackrel{{\rm
def}}{=}\Gamma(a+1)$ with
\begin{equation}
\label{gamma:integral}
\Gamma(z)
=
\int_{0}^{\infty}\!\!\D{t}\, t^{z-1}\,e^{-t}
\;,
\qquad
\Re(z)>0
\;,
\end{equation}
where $\Re(\cdot)$ denotes real part. The relation between the functions
$\F^{(\ell)}(\s)$ and Kummer functions reads
\begin{equation}
\label{F-kummer}
\F^{(\ell)}(\s)
=
\frac{M(\ell+\half,\ell+\threehalfs;\s)}{2\ell+1}
\;,
\quad
\ell
=
0,1,2,\ldots
\;.
\end{equation}
On using $M(a,c\,;x=0)=1$ [see Eq.\  (\ref{kummer:series})], one gets from Eq.\
(\ref{F-kummer}) as a corollary the derivatives of $\F(\s)$ at the origin
\begin{equation}
\label{F:zero}
\F^{(\ell)}(0)
=
\frac{1}{2\ell+1}
\;,
\quad
\ell
=
0,1,2,\ldots
\;.
\end{equation}
The relations (\ref{F-kummer}) can easily be derived from the following integral
representation of the Kummer function
\begin{equation}
\label{kummer:integral} M(a,c\,;x)
=
\frac{2\Gamma(c)}{\Gamma(a)\Gamma(c-a)}
\int_{0}^{1}\!\!\D{z}\,e^{x\,z^{2}}z^{2a-1}(1-z^{2})^{c-a-1}
\;,
\quad
\Re(c)>\Re(a)>0
\;,
\end{equation}
which follows from the more familiar one (Arfken, 1985, p.~754)
\begin{equation}
\label{kummer:integral:2} M(a,c\,;x)
=
\frac{\Gamma(c)}{\Gamma(a)\Gamma(c-a)}
\int_{0}^{1}\!\!\D{t}\, e^{xt}t^{a-1}(1-t)^{c-a-1}
\;,
\quad
\Re(c)>\Re(a)>0
\;,
\end{equation}
by dint of the substitution $t=z^{2}$. For $a=\ell+\half$ and $c=\ell+\threehalfs$,
one has $c-a=1$, so that
\[
\frac{2\Gamma(c)}{\Gamma(a)\Gamma(c-a)}
=
\frac
{2\Gamma(\ell+\threehalfs)}
{\Gamma(\ell+\half)\Gamma(1)}
=
2\ell+1
\;,
\]
where $\Gamma(z+1)=z\Gamma(z)$ and $\Gamma(1)=1$ have been employed. Then,
on using $c-a-1=0$ and $2a-1=2\ell$, the right-hand side of Eq.\ (\ref{F-kummer})
can be written by means of the integral representation (\ref{kummer:integral}) as
\begin{eqnarray*}
\frac{M(\ell+\half,\ell+\threehalfs;\s)}{2\ell+1}
=
\frac{1}{2\ell+1}
\times(2\ell+1)
\int_{0}^{1}\!\!\D{z}\,e^{\s\,z^{2}}z^{2\ell}
\stackrel{{\rm def}}{=}
\F^{(\ell)}(\s)
\;.
\qed
\end{eqnarray*}

On introducing the {\em error\/} functions of real and ``imaginary" argument, namely
\begin{equation}
\label{erf-erfi}
{\rm erf}(x)
=
\sqrt{4/\pi}
\int_{0}^{x}\!\!\D{t}\,
\exp(-t^{2})
\;,
\qquad {\rm erfi}(x)
=
\sqrt{4/\pi}
\int_{0}^{x}\!\!\D{t}\,
\exp(t^{2})
\;,
\end{equation}
one can alternatively write $\F(\s)$ as
\begin{equation}
\label{F-error}
\F(\s)
=
\left\{
\begin{array}{ll}
{\displaystyle \sqrt{\pi/4\s}\,{\rm erfi}(\s^{1/2})}
&
\mbox{ for }\s>0
\\
{\displaystyle \sqrt{\pi/4|\s|}\,{\rm erf}(|\s|^{1/2})}
&
\mbox{ for }\s<0
\end{array}
\;.
\right.
\end{equation}
The less familiar ${\rm erfi}(x)$ is directly related with the Dawson  integral
\begin{equation}
\label{dawson} D(x)
=
\exp(-x^{2})
\int_{0}^{x}\!\!\D{t}\,
\exp(t^{2})
\;,
\end{equation}
which is a tabulated function also available in certain mathematical libraries of
computers. Consequently, the first equation in (\ref{F-error}) is essentially the known
relation between $\F(\s)$ and the Dawson integral (see Coffey, Cregg and Kalmykov,
1993, p.~368)
\begin{equation}
\label{F-dawson}
\F(\s)
=
\frac{\exp(\s)}{\sqrt{\s}}D(\sqrt{\s})
\;,
\qquad
\s>0
\;,
\end{equation}
which, as is indicated, only holds for positive argument.
\\
Proofs:
\begin{enumerate}
\item
By means of the substitution $t=\sqrt{\pm\s}\,z$, where the upper and
lower signs correspond, respectively, to $\s>0$ and $\s<0$, one finds
\begin{eqnarray*}
\sqrt{\pi/4(\pm\s)}
\times
\left\{
\begin{array}{l}
{\rm erfi}(\sqrt{\s})
\\
{\rm erf}(\sqrt{-\s})
\end{array}
\right.
&
=
&
\sqrt{\pi/4(\pm\s)}
\sqrt{4/\pi}
\int_{0}^{\sqrt{\pm\s}}\!\!\D{t}\,
\exp(\pm t^{2})
\\
&
=
&
\underbrace{
\sqrt{1/(\pm\s)}\sqrt{\pm\s}
}_{1}
\int_{0}^{1}\!\!\D{z}\,\exp(\s z^{2})
\;,
\end{eqnarray*}
from which Eqs.\ (\ref{F-error}) follow.\qed
\item
On the other hand, Eqs.\ (\ref{erf-erfi}) and (\ref{dawson}) immediately yield
\[
{\rm erfi}(x)
=
\sqrt{4/\pi}
\int_{0}^{x}\!\!\D{t}\,
\exp(t^{2})
=
\sqrt{4/\pi}
\exp(x^{2})D(x)
\;,
\]
from which one gets Eq.\ (\ref{F-dawson}) through the already demonstrated Eq.\
(\ref{F-error}).\qed
\end{enumerate}

\subsection{Recurrence relations}

The functions $\F^{(\ell)}$ satisfy the following recurrence relations:
\begin{eqnarray}
\label{F:recurrence:1}
\F^{(\ell+1)}
=
\frac{e^{\s}-(2\ell+1)\F^{(\ell)}}{2\s}
\;,
\qquad
\F^{(\ell)}
=
\frac{e^{\s}-2\s \F^{(\ell+1)}}{2\ell+1}
\;,
\end{eqnarray}
which can readily be obtained by integrating by parts the definition of $\F^{(\ell)}$.
The $\ell=0$ particular case of these relations is frequently employed. It can be
written in the following equivalent forms
\begin{equation}
\label{F-Fp:1}
\F'
=
\frac{e^{\s}-\F}{2\s}
\quad\Leftrightarrow\quad
\F
=
e^{\s}-2\s \F'
\quad\Leftrightarrow\quad
\frac{e^{\s}}{\F}
=
1+2\s\frac{\F'}{\F}
\;,
\end{equation}
where the prime denotes derivative with respect to $\s$.

One can also derive recurrence relations among the combinations $\F^{(\ell)}/\F$,
which occur in the expressions for a number of quantities (e.g., the linear and
non-linear susceptibilities). On dividing both sides of the first Eq.\
(\ref{F:recurrence:1}) by $\F$ and using Eq.\ (\ref{F-Fp:1}) to eliminate
$e^{\s}/(2\s\F)$, one gets the following relation between quotients of the form
$\F^{(\ell)}/\F$:
\begin{equation}
\label{F:recurrence:3}
\frac{\F^{(\ell+1)}}{\F}
=
\frac{\F^{(1)}}{\F}
+\frac{1}{2\s}
\left[ 1-(2\ell+1)\frac{\F^{(\ell)}}{\F}
\right]
\;.
\end{equation}
The following particular case
\begin{eqnarray}
\label{F-Fp:2}
\frac{\F''}{\F}
=
\frac{\F'}{\F}-\frac{1}{2\s}\left(3\frac{\F'}{\F}-1\right)
\;,
\end{eqnarray}
is especially useful. For instance, it can be employed to calculate $\F''/\F$ from
$\F'/\F$.

\subsection{Series expansions}

Series expansions for $\F(\s)$ and its derivatives can easily be obtained from the
corresponding expansions of the Kummer functions.

\subsubsection{Power series}

From the relations (\ref{F-kummer}) between $\F^{(\ell)}(\s)$ and Kummer functions,
one can construct the Taylor expansion of the former through the power series
(\ref{kummer:series}) for the latter. For the quotient of Pochhammer symbols
required one gets
\[
\frac{1}{2\ell+1}
\frac{(\ell+\half)_{n}}{(\ell+\threehalfs)_{n}}
=
\frac{1}{2\ell+1}
\frac
{(\ell+n-\half)!/(\ell-\half)!}
{(\ell+n+\half)!/(\ell+\half)!}
=
\frac{1}{2(\ell+n)+1}
\;,
\]
from which we obtain the desired power series of $\F^{(\ell)}(\s)$
\[
\F^{(\ell)}(\s)
=
\sum_{n=0}^{\infty}
\frac{1}{n!}
\;
\frac{\s^{n}}{2(\ell+n)+1}
\;.
\]

\subsubsection{Asymptotic formula for large positive argument}

For $x\gg1$, the Kummer functions are approximately given by (Arfken, 1985, p.~757)
\begin{eqnarray}
\label{kummer:asympt1}
\lefteqn{ M(a,c\,;x)
=
\frac{\Gamma(c)}{\Gamma(a)}
\frac{e^{x}}{x^{c-a}} }
\nonumber
\\
&
&
\times
\left[
  1+\frac{(1-a)(c-a)}{x}
+\frac{(1-a)(2-a)(c-a)(c-a+1)}{2x^{2}}+\cdots
\right]
\;.
\qquad
\quad
\end{eqnarray}
Then, on using the relations (\ref{F-kummer}) and noting that in this case
$1-a=-(2\ell-1)/2$ and $c-a=1$, we obtain the following asymptotic expansion of
$\F^{(\ell)}(\s)$
\begin{equation}
\label{Fderivatives:asympt1}
\F^{(\ell)}(\s)
=
\frac{e^{\s}}{2\s}
\left\{ 1-\frac{(2\ell-1)}{2\s}+\frac{(2\ell-1)(2\ell-3)}{4\s^{2}}+\cdots
\right\}
\;,\quad
\s\gg1
\;.
\end{equation}
This expansion generalizes for an arbitrary $\ell$ the results derived by Ra{\u{\i}}kher
and Shliomis (1975) for $\ell=0,1,2$, and $3$. Note finally that, one can use Eq.\
(\ref{Fderivatives:asympt1}) to take the $\s\to\infty$ limit of the quotient
$\F^{(\ell)}/\F$, getting
\begin{eqnarray}
\label{lim1}
\frac{\F^{(\ell)}}{\F}
\simeq
\frac
{1-(2\ell-1)/2\s+(2\ell-1)(2\ell-3)/4\s^{2}+\cdots}
{1+1/2\s+3/4\s^{2}+\cdots}
\stackrel{\s\to\infty}{\longrightarrow}1
\;,
\quad
\forall\ell
\;.
\end{eqnarray}

\subsubsection{Asymptotic formula for large negative argument}

Asymptotic expressions for $\F^{(\ell)}(\s\ll-1)$ can be derived from the asymptotic
expansion of the Kummer functions for large negative argument. The latter is easily
obtained from the expansion (\ref{kummer:asympt1}) for large positive
argument by dint of {\em Kummer's first formula\/}
$M(a,c\,;x)=e^{x}M(c-a,c\,;-x)$
(Arfken, 1985, p.~754)
\begin{eqnarray}
\label{kummer:asympt2}
M(a,c\,;x)
&
\simeq
&
\frac{\Gamma(c)}{\Gamma(c-a)}
\frac{1}{(-x)^{a}}
\nonumber
\\
&
& {}\times
\left[ 1+\frac{(c-a-1)a}{x}
+\frac{(c-a-2)(c-a-1)a(a+1)}{2x^{2}}+\cdots
\right]
\;.
\nonumber
\\
\end{eqnarray}
Then, taking once more the relations (\ref{F-kummer}) into account, one obtains the
approximate expression
\begin{equation}
\label{Fderivatives:asympt2}
\F^{(\ell)}(\s)\simeq
\frac{\pi^{1/2}}{2^{2\ell+1}}
\frac{(2\ell)!}{\ell!}
\frac{1}{(-\s)^{\ell+1/2}}
\;,
\qquad
\s\ll-1
\;,
\end{equation}
for the derivation of which we have also employed the following useful result for the
gamma function of half-odd-integer argument
\begin{equation}
\label{gamma:halfinteger}
\Gamma(\ell+\half)
=
\frac{\pi^{1/2}}{2^{2\ell}}\frac{(2\ell)!}{\ell!}
\;.
\end{equation}
Note that the next terms in the asymptotic expansion (\ref{Fderivatives:asympt2})
vanish identically, since $c-a-1=0$ in this case [see Eq.\ (\ref{kummer:asympt2})].
Finally, for the quotient $\F^{(\ell)}/\F$ one gets the limit
\begin{eqnarray}
\label{lim2}
\frac{\F^{(\ell)}}{\F}
\simeq
\frac{1}{2^{2\ell}}\frac{(2\ell)!}{\ell!}\frac{1}{(-\s)^{\ell}}
\stackrel{\s\to-\infty}{\longrightarrow}0
\;,
\quad
\forall\ell\geq1
\;.
\end{eqnarray}

To conclude, as an exercise of consistency, one can obtain from the derived $|\s|\ll1$
and $\s\gg1$ expansions of $\F(\s)$, via the relation (\ref{F-dawson}), the known
power and asymptotic series of the Dawson integral (see, for example, Coffey, Cregg
and Kalmykov, 1993, p.~368):
\[
D(x)
=
\left\{
\begin{array}{ll}
{\displaystyle
x-\frac{2}{3}x^{3}+\frac{4}{15}x^{5}+\cdots
}
\;,
&
x\ll1
\\
{\displaystyle
\frac{1}{2x}\left(1+\frac{1}{2x^{2}}+\cdots\right)
}
\;,
&
x\gg 1
\end{array}
\right.
\;.
\]

\subsection{Approximate formulae for $\F'/\F$ and $\F''/\F$}

We shall now derive approximate expressions for $\F'/\F$ valid in the $|\s|\ll1$ and
$|\s|\gg1$ ranges. These expressions, along with the recurrence relations
(\ref{F:recurrence:3}) between consecutive $\F^{(\ell)}/\F$, would provide
approximate expressions for $\F^{(\ell)}/\F$ with $\ell\geq2$. We shall explicitly give
these approximate formulae for $\F''/\F$.

The following approximate expressions will be obtained by constructing approximate
solutions of the differential equation that the function $G=\F'/\F$ satisfies, namely
\begin{equation}
\label{odeG}
\frac{\D G}{\D\s}
=
\frac{1}{2\s}(1-3G)+G(1-G)
\;,
\end{equation}
which can easily be derived from Eq.\ (\ref{F-Fp:2}).

\subsubsection{Power series}

To obtain $G|_{|\s|\ll 1}$, we shall seek for a solution of the differential equation
(\ref{odeG}) in the form of a power series
$G=\sum_{n=0}^{\infty}a_{n}\s^{n}$. Prior to do that, however, in order to remove the
singularities in the coefficients in that differential equation, these are multiplied by
$2\s$, yielding the equivalent equation
$2\s(\D G/\D\s)=(1-3G)+2\s G(1-G)$.
This is a non-homogeneous non-linear differential equation, and  these
features will take reflection in the form of the constructed  solution.

On inserting $G=\sum_{n=0}^{\infty}a_{n}\s^{n}$ into the above differential equation,
redefining the summation indices in order to obtain the same exponent for $\s$ under
each summation symbol, and equating coefficients, one gets for the $a_{n}$:
\[
a_{0}
=
1/3
\;,
\qquad
\Big(n+\frac{3}{2}\Big)
a_{n}
=
a_{n-1}-\sum_{k=0}^{n-1}a_{k}a_{n-1-k}
\;,
\quad
\mbox{ for } n\geq 1
\;.
\]
The fact that $a_{0}$ is not a free parameter results from the non-homogeneous
character of the differential equation. On the other hand, the above recurrence
relation among the $a_{n}$ shows that, as a consequence of the  non-linearity of the
differential equation, the computation of each coefficient involves all the previous
ones. Finally, on computing the first few coefficients, $G=\F'/\F$ emerges in the
approximate form
\begin{equation}
\label{Gapprox1} G\simeq\frac{1}{3}
\left(
  1+\frac{4}{15}\s+\frac{8}{315}\s^{2}
-\frac{16}{4725}\s^{3}-\frac{32}{31185}\s^{4}
\right)
\;.
\end{equation}
We have carried out the expansion up to the fourth order in $\s$ because some
quantities are approximately obtained up to terms of order $\s^{3}$ and, for example,
$\F''/\F$ involves $G'$ [see Eq.\ (\ref{Fpp-G}) below].

The formulae required to derive some approximate expressions in the main text are
\begin{eqnarray}
\label{F-der:approx1}
\frac{\F'}{\F}
&
\simeq
&
\frac{1}{3}
\left(1+\frac{4}{15}\s+\frac{8}{315}\s^{2}
-\frac{16}{4725}\s^{3}\right)
\;,
\nonumber
\\
\bigg(\frac{\F'}{\F}\bigg)^{2}
&
\simeq
&
\frac{1}{9}
\left(1+\frac{8}{15}\s+\frac{64}{525}\s^{2}
+\frac{32}{4725}\s^{3}\right)
\;,
\\
\frac{\F''}{\F}
&
\simeq
&
\frac{1}{5}
\left(1+\frac{8}{21}\s+\frac{16}{315}\s^{2}
-\frac{32}{10395}\s^{3}\right)
\;.
\nonumber
\end{eqnarray}
For instance, the combinations entering in the equations for the non-linear
susceptibility read
\begin{eqnarray}
\label{F-der:approx1:comb}
\frac{1}{2}
\left[
\frac{1}{3}\frac{\F''}{\F}-\bigg(\frac{\F'}{\F}\bigg)^{2}
\right]
&
\simeq
&
-\frac{1}{45}
\left(1+\frac{16}{21}\s+\frac{8}{35}\s^{2}+\frac{32}{1485}\s^{3}\right)
\;,
\nonumber
\\
\frac{1}{2}
\left[
\bigg(\frac{\F'}{\F}\bigg)^{2}-\frac{\F''}{\F}
\right]
&
\simeq
&
-\frac{2}{45}
\left(1+\frac{4}{21}\s-\frac{4}{105}\s^{2}-\frac{32}{2079}\s^{3}\right)
\;,
\nonumber
\\
\frac{1}{16}
\left[
-1
+2\frac{\F'}{\F}
-2\bigg(\frac{\F'}{\F}\bigg)^{2}
+\frac{\F''}{\F}
\right]
&
\simeq
&
-\frac{1}{45}
\left(1-\frac{8}{21}\s+\frac{128}{10395}\s^{3}\right)
\;.
\end{eqnarray}
The expression for $\F''/\F$ in Eq.\ (\ref{F-der:approx1}) has been obtained from
$\F'/\F$, through the relation
\begin{equation}
\label{Fpp-G}
\frac{\F''}{\F}
=
G'+G^{2}
\;,
\end{equation}
which is directly demonstrated by taking the derivative
$G'=(\F'/\F)'=(\F''/\F)-(\F'/\F)^{2}$.

\subsubsection{Asymptotic formulae}

We shall now derive approximate expressions for $G=\F'/\F$ valid in the $|\s|\gg1$
ranges. To this end we make in Eq.\ (\ref{odeG}) the substitution
$\varrho=1/\s$, which casts it into the form
\[
-\varrho^{2}\frac{\D G}{\D\varrho}
=
\frac{\varrho}{2}(1-3G)+G(1-G)
\;.
\]
Let us seek for solutions of this differential equation in the form of a series of powers
of $\varrho$.%
\footnote{
A similar method was employed by Ra{\u{\i}}kher and Shliomis (1975) to
derive the aforementioned asymptotic series of $\F^{(\ell)}(\s)$.
}
On inserting $G=\sum_{n=0}^{\infty}b_{n}\varrho^{n}$ into the
above equation, redefining conveniently the summation indices, and
equating coefficients, one gets for the $b_{n}$:
\begin{eqnarray*}
b_{0}(1-b_{0})
&
=
&
0
\;,
\qquad b_{1}
=
\frac{1}{2}\frac{3b_{0}-1}{1-2b_{0}}
\;,
\\
(1-2b_{0})b_{n}
&
=
&
\Big(
\frac{5}{2}-n
\Big) b_{n-1}+\sum_{k=1}^{n-1}b_{k}b_{n-k}
\;,
\quad
\mbox{ for } n \geq 2
\;.
\end{eqnarray*}
Again, the first coefficient is not a free parameter and the above recurrence  relation
involves all the coefficients preceding a given one.

As could be expected from the fact that we are searching for solutions in two
different asymptotic ranges ($\s\to\pm\infty$), we obtain two different solutions. The
one that corresponds to the choice $b_{0}=0$ (denoted $G_{1}$), when expressed in
terms of the original variable $\s=1/\varrho$, takes the simple form
\[
G_{1}
=
-\frac{1}{2\s}
\;,
\quad (b_{0}=0)
\;,
\]
where all the remainder terms vanish identically. (As can be readily seen,
$G=-1/2\s$ is an exact solution of the original differential equation (\ref{odeG}),
although, since it diverges at $\s=0$, it is not the selfsame $\F'/\F$.) On the other hand,
the solution that  corresponds to the choice $b_{0}=1$ (denoted $G_{2}$) is given by
\[
G_{2}
=
1-\frac{1}{\s}-\frac{1}{2\s^{2}}
-\frac{5}{4\s^{3}}+\cdots
\;,
\quad (b_{0}=1)
\;.
\]
We must now ascribe each solution to one of the two asymptotic
ranges. On recalling Eqs.\ (\ref{lim1}) and (\ref{lim2}), we conclude
that $G_{1}$ and $G_{2}$ correspond, respectively, to the $\s\ll-1$
and $\s\gg1$ ranges. Note anyway that the $\s\ll-1$ result can {\em
directly\/} be obtained from the asymptotic results
(\ref{Fderivatives:asympt2}).
\begin{figure}[b!]
\vspace{-3.ex}
\eps{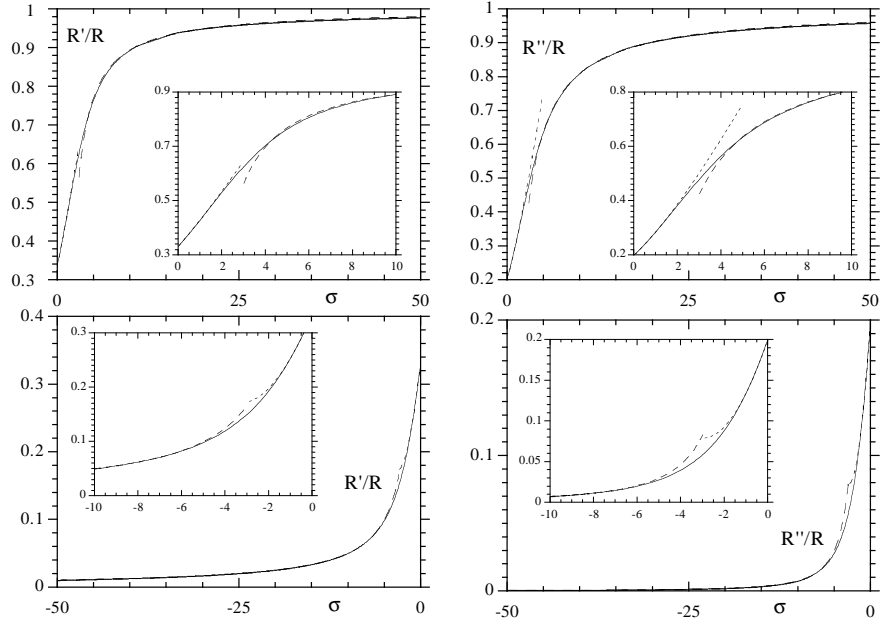}{1}
\vspace{-3.ex}
\caption[]
{
The functions $\F'/\F$ and $\F''/\F$ together with their small and
large $\s$ approximations. The continuous lines represent the exact
functions. Long dashes: large $|\s|$ approximations
(\ref{F-der:approx2a}) and (\ref{F-der:approx2b}). Short dashes: small
$\s$ approximations (\ref{F-der:approx1}). The insets show the details
of the zones where the small $\s$ approximation might be swapped by
the corresponding large $\s$ approximation, without a significant loss
of accuracy.
\label{Fp-Fpp_approx:plot}
}
\end{figure}

We can now derive the combinations of $\F(\s)$ and its derivatives that are
required in the main text to construct approximate formulae for
various quantities. For $\s\ll-1$, these are:
\begin{equation}
\label{F-der:approx2a}
\frac{\F'}{\F}\simeq-\frac{1}{2\s}
\;,
\qquad
\bigg(\frac{\F'}{\F}\bigg)^{2}\simeq\frac{1}{4\s^{2}}
\;,
\qquad
\frac{\F''}{\F}\simeq\frac{3}{4\s^{2}}
\;,
\end{equation}
and the combinations
\begin{eqnarray}
\label{F-der:approx2a:comb}
\frac{1}{2}
\left[
\frac{1}{3}\frac{\F''}{\F}-\bigg(\frac{\F'}{\F}\bigg)^{2}
\right]
&
\simeq
&
0
\;,
\nonumber
\\
\frac{1}{2}
\left[
\bigg(\frac{\F'}{\F}\bigg)^{2}-\frac{\F''}{\F}
\right]
&
\simeq
&
-\frac{1}{4\s^{2}}
\;,
\\
\frac{1}{16}
\left[
-1+2\frac{\F'}{\F}-2\bigg(\frac{\F'}{\F}\bigg)^{2}+\frac{\F''}{\F}
\right]
&
\simeq
&
-\frac{1}{16}
\left(1+\frac{1}{\s}-\frac{1}{4\s^{2}}\right)
\;.
\nonumber
\end{eqnarray}
Similarly, for $\s\gg1$ we find
\begin{eqnarray}
\label{F-der:approx2b}
\frac{\F'}{\F}
&
\simeq
&
1-\frac{1}{\s}-\frac{1}{2\s^{2}}-\frac{5}{4\s^{3}}
\;,
\nonumber
\\
\bigg(\frac{\F'}{\F}\bigg)^{2}
&
\simeq
&
1-\frac{2}{\s}-\frac{3}{2\s^{3}}
\;,
\\
\frac{\F''}{\F}
&
\simeq
&
1-\frac{2}{\s}+\frac{1}{\s^{2}}-\frac{1}{2\s^{3}}
\;,
\nonumber
\end{eqnarray}
and their combinations
\begin{eqnarray}
\label{F-der:approx2b:comb}
\frac{1}{2}
\left[
\frac{1}{3}\frac{\F''}{\F}-\bigg(\frac{\F'}{\F}\bigg)^{2}
\right]
&
\simeq
&
-\frac{1}{3}
\left(1-\frac{2}{\s}-\frac{1}{2\s^{2}}-\frac{2}{\s^{3}}\right)
\;,
\nonumber
\\
\frac{1}{2}
\left[
\bigg(\frac{\F'}{\F}\bigg)^{2}-\frac{\F''}{\F}
\right]
&
\simeq
&
-\left(\frac{1}{2\s^{2}}+\frac{1}{2\s^{3}}\right)
\;,
\\
\frac{1}{16}
\left[
-1+2\frac{\F'}{\F}-2\bigg(\frac{\F'}{\F}\bigg)^{2}+\frac{\F''}{\F}
\right]
&
\simeq
&
0
\;.
\nonumber
\end{eqnarray}
We also write down the leading terms in the $|\s|\ll1$ and $|\s|\gg1$ expansions of the
combination $\F''/\F-(\F'/\F)^{2}$, which occurs in some expressions studied in
Sections \ref{sect:quantities} and \ref{sect:heuristic},
\begin{equation}
\label{comb:F:leadingterm}
\frac{\F''}{\F}
-\left(\frac{\F'}{\F}\right)^{2}
\simeq\left\{
\begin{array}{cl}
1/2\s^2
&
\mbox{ for } \s\ll-1
\\
4/45
&
\mbox{ for } |\s|\ll1
\\
1/\s^2
&
\mbox{ for } \s\gg1
\end{array}
\right.
\;.
\end{equation}

The appropriate combination of Eqs.\ (\ref{F-der:approx1}),
(\ref{F-der:approx2a}), and (\ref{F-der:approx2b}) almost patch the
corresponding exact curves over the entire $\s$-range. This is shown
in Fig.\ \ref{Fp-Fpp_approx:plot}, where it can be seen that the use
of the $|\s|\ll1$ results, swapped at some point between $|\s|=2$ and
$|\s|=5$ by the corresponding $|\s|\gg1$ formulae, is a reasonable
approximation of the exact functions.

\setcounter{equation}{0}
\section{Derivation of the formulae for the relaxation times}
\label{app:taus}

In this appendix we shall give demonstrations of the formulae (\ref{tauint}) and
(\ref{effeigentaumod}) for the relaxation times.

\subsection{Integral relaxation time}

\subsubsection{The integral relaxation time and the low-frequency dynamical 
susceptibility}

The integral relaxation time defined by Eq.\ (\ref{tauint:def}) is expressible in
terms of the eigenvalues, $\Lambda_{k}$, and amplitudes, $a_{k}$, of the
Sturm--Liou\-ville problem associated with the axially symmetric Fokker--Planck
equation [Eq.\ (\ref{tauint-lambda})]. In addition, $\tint$ can also be written in terms
of the low-frequency dynamical susceptibility (Garanin, Ishchenko, and Panina, 1990;
Garanin, 1996). In order to show this, let us first write down the general result from
linear-response theory
\begin{equation}
\label{chi-relax}
\chi(\w)
=
\chi
-i\w
\int_{0}^{\infty}\!\!\D{t}\, e^{-i\w t}
\frac
{\llangle\mm(\infty)\rrangle-\llangle\mm(t)\rrangle}
{\dBo}
\;,
\end{equation}
where $\mm(t)$ is the relaxing quantity, $\chi(\w)$ is its
susceptibility counterpart ($\chi$ being the equilibrium
susceptibility), and $\dBo$ is the infinitesimal change in the
external control parameter. On applying this result to
$\langle\mz(\infty)\rangle-\langle\mz(t)\rangle$ from Eq.\
(\ref{mrelax}) and using the sum rule $\sum_{k\geq1}a_{k}=1$, one
finds
\[
\chi_{\|}(\w)
=
\chi_{\|}\sum_{k\geq1}a_{k}
\bigg\{ 1-i\w\int_{0}^{\infty}\!\!\D{t}\,
\exp
\left[
-(i\w+\Lambda_{k})t
\right]
\bigg\}
\;,
\]
from which it follows
\begin{equation}
\label{chidyn}
\chi_{\|}(\w)
=
\chi_{\|}\sum_{k\geq1}\frac{a_{k}}{1+i\w\Lambda_{k}^{-1}}
\;.
\end{equation}
Thus, each exponential mode in the relaxation curve (\ref{mrelax}) gives a Debye-type
factor in $\chi_{\|}(\w)$ weighted by $a_{k}$ and with characteristic time
$\Lambda_{k}^{-1}$. Finally, on expanding $\chi_{\|}(\w)$ for low frequencies by dint
of the binomial formula, one gets
\begin{equation}
\label{chidyn:approx}
\chi_{\|}(\w)
\simeq\chi_{\|}\sum_{k\geq1}a_{k}\left(1-i\w\Lambda_{k}^{-1}+\cdots\right)
=
\chi_{\|}\left(1-i\w\tint+\cdots\right)
\;,
\end{equation}
where we have again used $\sum_{k\geq1}a_{k}=1$ and taken Eq.\
(\ref{tauint-lambda}) into account. Equation (\ref{chidyn:approx}) demonstrates that
the calculation of $\tint$ can effectively be reduced to the calculation of the
low-frequency dynamical susceptibility.

\subsubsection{Perturbational solution of the Fokker--Planck equation in the presence
of a low sinusoidal field}

In order to calculate $\chi_{\|}(\w)$ one applies a low sinusoidal field,
$\Delta\xi(t)=\Delta\xi\exp(i\w t)$ where $\Delta\xi=\mm\dBo/\T$, along
the $z$ (symmetry) axis and then calculates the solution of the axially symmetric
Fokker--Planck equation (\ref{brownfpe:axi:2}) in such a situation. Since
$-\beta\Hs(z,t)=-\beta\Hs_{0}(z)+z\Delta\xi(t)$, where $\Hs_{0}$ is the unperturbed
Hamiltonian, we shall seek for a solution for the probability distribution in the
stationary regime of the form
\begin{equation}
\label{PDF:ansatz}
\W(z,t)
=
\Weq(z)[1+q(z)\Delta\xi(t)]
\;,
\end{equation}
where $\Weq=\Z_{0}^{-1}\exp(-\beta\Hs_{0})$ is the equilibrium probability
distribution in the absence of the oscillating field.

On introducing the above $\W(z,t)$ into the Fokker--Planck equation
(\ref{brownfpe:axi:2}) one gets, to first order in
$\Delta\xi$ (linear response regime), the following second-order differential
equation for $q(z)$
\begin{equation}
\label{odeq}
\left(
-\beta\Hs_{0}'+\frac{\D {}}{\D z}
\right)
\left[
\Omega(z)\frac{\D q}{\D z}
\right]
-i\w2\tN q
=
-\Omega(z)\beta\Hs_{0}'+\Omega'
\;,
\end{equation}
where $\Omega(z)=1-z^{2}$ and the primes denote differentiation with respect to $z$.
On the other hand, by taking the introduced form of $\W(z,t)$ into account, the
non-equilibrium average of the $z$ component of magnetic moment can be written as
\[
\llangle\mz(t)\rrangle
=
\int_{-1}^{1}\!\!\D{z}\,\W(z,t)\mz
=
\mm\llangle z\rrangle_{\eq}+\mm\Delta\xi(t)\int_{-1}^{1}\!\!\D{z}\,\Weq(z)q(z)z
\;,
\]
where $\langle z\rangle_{\eq}=\int_{-1}^{1}\!\!\D{z}\,\Weq(z)z$ is the equilibrium
average in the unperturbed case. Next, since
$\Delta\xi(t)=(\mm\dBo/\T)\exp(i\w t)$, the dynamical susceptibility, which is the
coefficient of $\mu_{0}^{-1}\dBo e^{i\w t}$ in $\langle\mz(t)\rangle$, is given by
\begin{equation}
\label{chidynint}
\chi_{\|}(\w)
=
\Xo\int_{-1}^{1}\!\!\D{z}\,\Weq(z)q(z)z
\;.
\end{equation}
Comparison of this equation with Eq.\ (\ref{chidyn:approx}) reveals that only the
low-frequency part of $q(z)$ is required to calculate $\tint$. This is important
since Eq.\ (\ref{odeq}) cannot be solved analytically in the general case. In contrast, it
can be solved perturbatively for low $\w$ because, for $\w=0$, only $q'(z)$ and
$q''(z)$ occur in that equation. This enables one to low the order of the differential
equation (\ref{odeq}) by introducing an auxiliary function $g(z)=q'(z)$, and solving
successively the system of first-order differential equations for $q(z)$ and $g(z)$.

Let us accomplish this. First, one introduces the perturbational expansion
\[
q(z)
=
q_{0}(z)-(i\w)q_{1}(z)+(i\w)^{2}q_{2}(z)-\cdots
\;,
\]
into Eq.\ (\ref{chidynint}), getting
\[
\chi_{\|}(\w)
=
\Xo\int_{-1}^{1}\!\!\D{z}\,\Weq(z)q_{0}(z)z
-i\w\Xo\int_{-1}^{1}\!\!\D{z}\,\Weq(z)q_{1}(z)z
+\cdots
\]
Then, on comparing this result with Eq.\ (\ref{chidyn:approx}), one obtains the
following integral representation of $\tint$:
\begin{equation}
\label{tauint:q1}
\tint
=
\frac{1}{\partial\llangle z\rrangle_{\eq}/\partial\xi}
\int_{-1}^{1}\!\!\D{z}\,\Weq(z)q_{1}(z)z
\;,
\end{equation}
where we have used $\chi_{\|}=(\mu_{0}\mm^{2}/\T)\partial\llangle
z\rrangle_{\eq}/\partial\xi$ [cf.\ Eqs.\ (\ref{FDR:equilibrium}) and (\ref{X:FDR}); we
can differentiate with respect to $B$ since this is parallel to the probing field
$\dBo$]. Equation (\ref{tauint:q1}) shows that the calculation of $\tint$ effectively
reduces to that of $q_{1}(z)$. In order to obtain this quantity, we introduce the
above perturbational expansion of $q(z)$, along with $g_{i}\equiv\D q_{i}/\D z$, into
Eq.\ (\ref{odeq}), getting
\begin{eqnarray}
\label{odeqper}
&
&
\left(-\beta\Hs_{0}'+\frac{\D {}}{\D z}\right)
\left\{\Omega(z)[g_{0}-(i\w)g_{1}+(i\w)^{2}g_{2}-\cdots]\right\}
\nonumber
\\
&
&
-i\w2\tN[q_{0}-(i\w)q_{1}+\cdots]
=
-\Omega(z)\beta\Hs_{0}'+\Omega'
\;.
\end{eqnarray}
The zeroth-order equation has the thermal equilibrium solution
\begin{equation}
\label{q0} q_{0}
=
z-\llangle z\rrangle_{\eq}
\;,
\end{equation}
as can be shown by using the definition of $q(z)$ and expanding the
equilibrium probability distribution associated with
$\beta\Hs=\beta\Hs_{0}-z\Delta\xi$ [i.e., the $\w=0$ limit of $\beta\Hs(t)$] in
powers of
$\Delta\xi$.

The $(i\w)$-order term of Eq.\ (\ref{odeqper}) reads
\[
\left(-\beta\Hs_{0}'+\frac{\D {}}{\D z}\right)
\left[\Omega(z)g_{1}\right]
+2\tN q_{0}
=
0
\;.
\]
This differential equation can be integrated by quadratures yielding
\begin{equation}
\label{g1}
g_{1}(z)
=
\frac{2\tN}{\Omega(z)}
\exp[\beta\Hs_{0}(z)]
\left[c_{1}+\Z_{0}\Phi(z)\right]
\;,
\end{equation}
where $\Z_{0}$ is the (unperturbed) equilibrium partition function and $\Phi(z)$ is
given by
\begin{equation}
\label{Phi:2}
\Phi(z)
=
\int_{-1}^{z} \D z_{1}\Weq(z_{1})
\underbrace{
(\llangle z\rrangle_{\eq}-z_{1})
}_{-q_{0}(z_{1})}
\;.
\end{equation}
On using the condition $J_{z}|_{z=\pm1}=0$ (which follows from the tangency of the
current of probability to the unit sphere) and $\Phi(-1)=\Phi(1)=0$ (which
immediately follow from the above definition), one gets for the integration constant
$c_{1}=0$. Consequently, $q_{1}(z)=\int^{z}\!\D{z_{2}}\,g_{1}(z_{2})$ is given by
\begin{equation}
\label{q1}
q_{1}(z)
=
c_{2}+2\tN\int_{-1}^{z}
\frac{\D z_{2}}{\Omega(z_{2})}\Phi(z_{2})/\Weq(z_{2})
\equiv c_{2}+\tilde{q}_{1}
\;,
\end{equation}
where we have written $\Z_{0}\exp[\beta\Hs_{0}(z_{2})]=1/\Weq(z_{2})$. The new
integration constant, $c_{2}$, can be obtained by solving the $(i\w)^{2}$-order
equation and imposing anew the aforementioned condition on the current of
probability at the boundaries. On doing so, one finds
$c_{2}=-\llangle\tilde{q}_{1}\rrangle_{\eq}
=-\int_{-1}^{1}\!\!\D{z}\,\Weq(z)\tilde{q}_{1}(z)$, where $\tilde{q}_{1}(z)$ is the
integral term in Eq.\ (\ref{q1}).

\subsubsection{The Garanin, Ishchenko, and Panina formula}

We can already do the integral involving $q_{1}(z)$ in the formula
(\ref{tauint:q1}) for the integral relaxation time:
\begin{eqnarray*}
\int_{-1}^{1}\!\!\D{z}\,\Weq(z)q_{1}(z)z
&
=
&
\int_{-1}^{1}\!\!\D{z}\,\Weq(z)
\left[
\tilde{q}_{1}(z)-\llangle\tilde{q}_{1}\rrangle_{\eq}
\right] z
\\
&
=
&
\int_{-1}^{1}\!\!\D{z}\,
\Weq(z)\tilde{q}_{1}(z)z
-\llangle z\rrangle_{\eq}
\underbrace{
\int_{-1}^{1}\!\!\D{z}\,
\Weq(z)\tilde{q}_{1}(z)
}_{\llangle\tilde{q}_{1}\rrangle_{\eq}}
\\
&
=
&
-\int_{-1}^{1}
\underbrace{
\D z\Weq(z)\left(\llangle z\rrangle_{\eq}-z\right)
}_{\D\Phi(z){\rm
~by~Eq.~(\ref{Phi:2})}}
\tilde{q}_{1}(z)
\\
&
=
&
-\underbrace{
\left[
\Phi(z)\tilde{q}_{1}(z)
\right]_{-1}^{1}
}_{0{\rm ~by~}\Phi(-1)
=
\Phi(1)=0}
+\int_{-1}^{1}\!\!\D{z}\,\Phi(z)
\underbrace{
\frac{2\tN}{\Omega(z)}\Phi(z)/\Weq(z)
}_{\tilde{q}_{1}'(z){\rm~by~Eq.~(\ref{q1})}}
\;.
\end{eqnarray*}
Then, on introducing this result into Eq.\ (\ref{tauint:q1}) one obtains
\begin{equation}
\label{tauint:gral}
\tint
=
\frac{2\tN}{\partial\llangle z\rrangle_{\eq}/ \partial\xi}
\int_{-1}^{1}\frac{\D z}{\Omega(z)}\Phi(z)^{2}/\Weq(z)
\;,
\end{equation}
whence, on recalling that $\Omega(z)$ is a shorthand for $1-z^{2}$, one finally gets the
result (\ref{tauint}) of Garanin, Ishchenko, and Panina (1990).

\subsubsection{Explicit expressions for $\Phi(z)$}

Let us conclude with the calculation of explicit expressions for $\Phi(z)$ for
particular forms of the Hamiltonian. Let us assume that $\Hs_{0}$ comprises a uniaxial
anisotropy term, $-Kv z^{2}$, plus a Zeeman term, $-\mm B z$, i.e.,
$-\beta\Hs_{0}=\sigma z^{2}+\xi z$ [see Eq.\ (\ref{sigma-xi})].

\paragraph{Isotropic case.}

When $\sigma=0$, the equilibrium probability distribution is given by Eq.\
(\ref{pdfboltzmann:eff:langevin}). Thus, one of the contributions to $\Phi(z)$ is
\[
\llangle z\rrangle_{\eq}
\int_{-1}^{z}\!\!\D{z_{1}}\,
\Weq(z_{1})
=
L(\xi)
\frac{e^{\xi z}-e^{-\xi}}{2\sinh\xi}
\;,
\]
where we have used $\langle z\rangle_{\eq}=L(\xi)$, $L(\xi)$ being the Langevin
function. The remainder contribution to $\Phi(z)$ is
\begin{eqnarray*}
-\int_{-1}^{z}\!\!\D{z_{1}}\,
\Weq(z_{1})z_{1}
&
=
&
-\frac{\xi}{2\sinh\xi}
\,
\frac{\partial {}}{\partial\xi}
\int_{-1}^{z}\!\!\D{z_{1}}\,
\exp(\xi z_{1})
\\
&
=
&
-\frac{e^{\xi z}}{2\sinh\xi}
\left[
\left(z-\frac{1}{3}\right)+e^{-\xi(1+z)}\left(1+\frac{1}{\xi}\right)
\right]
\;.
\end{eqnarray*}
On adding these two contributions and recalling the definition (\ref{langevin:function})
of the Langevin function, one finally gets the explicit result
\begin{equation}
\label{Phiisotropic}
\Phi_{\lan}(z)
=
\frac{\Weq(z)}{\xi}
\left[
\coth\xi-z-\frac{\exp(-\xi z)}{\sinh\xi}
\right]
\;.
\end{equation}

\paragraph{Zero-field case.}

For $\xi=0$ the equilibrium probability distribution is given by Eq.\
(\ref{pdfboltzmann:eff:unb}). Therefore, $\langle z\rangle_{\eq}=0$ and
\[
\Phi(z)
=
-\frac{1}{2\F(\s)}
\int_{-1}^{z}\!\!\D{z_{1}}\,
\exp(\s z_{1}^{2})z_{1}
\;.
\]
Then, on expressing the result of the integral in terms of the probability distribution
(\ref{pdfboltzmann:eff:unb}), one gets [note that $\Weq(-1)=\Weq(1)$]
\begin{equation}
\label{Phiunbiased}
\Phi_{\unb}(z)
=
\frac{1}{2\sigma}
\left[\Weq(1)-\Weq(z)\right]
\;.
\end{equation}

\subsection{Effective transverse relaxation time}

We shall now derive Eq.\ (\ref{effeigentaumod}) for the effective transverse
relaxation time by performing the low-frequency expansion of the formula for
$\chi_{\perp}(\w)$ of Ra\u{\i}kher and Shliomis (1975; 1994).

\subsubsection{The Ra\u{\i}kher and Shliomis formula for the transverse dynamical
susceptibility}

The expression for $\chi_{\perp}(\w)$ derived by these authors can be written as
\begin{equation}
\label{chiperp:RSh}
\chi_{\perp}(\w,T)
=
\chi_{\perp}(T)
\frac
{\lambda_{a}(\lambda_{b}+i\w2\tN)+\lambda_{c}}
{(\lambda_{1}+i\w2\tN)(\lambda_{2}+i\w2\tN)}
\;,
\end{equation}
where $\chi_{\perp}(T)$ is the equilibrium transverse susceptibility
(\ref{X:para:perp}). The coefficients $\lambda_{a}$, $\lambda_{b}$, and
$\lambda_{c}$ are given, in terms of the functions $\F^{(\ell)}(\s)$ [Eq.\
(\ref{F-Fderivatives})] and the dimensionless damping coefficient $\la$ in the
Landau--Lifshitz equation, by
\[
\lambda_{a}
=
\frac{\F+\F'}{\F-\F'}
\;,
\qquad
\lambda_{b}
=
\frac{\F-3\F'+4\F''}{\F'-\F''}
\;,
\qquad
\lambda_{c}
=
\frac{2\s}{\la^{2}}\frac{3\F'-\F}{\F-\F'}
\;,
\]
whereas $\lambda_{1}$ and $\lambda_{2}$ are the roots of the second-degree
equation
$x^{2}-(\lambda_{a}+\lambda_{b})x+(\lambda_{a}\lambda_{b}+\lambda_{c})=0$. On
using that the roots $x_{1}$ and $x_{2}$ of $ax^{2}+bx+c=0$ obey $x_{1}+x_{2}=-(b/a)$
and $x_{1}x_{2}=c/a$, we can write the expression in denominator of
$\chi_{\perp}(\w)$ in terms of $\lambda_{a}$, $\lambda_{b}$, and $\lambda_{c}$ as
\[
(\lambda_{1}+i\w2\tN)(\lambda_{2}+i\w2\tN)
=
(\lambda_{a}\lambda_{b}+\lambda_{c})
-4\w^{2}\tN^{2}+i\w2\tN(\lambda_{a}+\lambda_{b})
\;.
\]
Accordingly, the transverse susceptibility (\ref{chiperp:RSh}) can equivalently be
written as
\begin{equation}
\label{chiperp:RSh:red}
\chi_{\perp}(\w,T)
=
\chi_{\perp}(T)
\frac
{
{\displaystyle
1+i\w2\tN
\frac{\lambda_{a}}{\lambda_{a}\lambda_{b}+\lambda_{c}}
}
}
{
{\displaystyle
1-4\w^{2}\tN^{2}\frac{1}{\lambda_{a}\lambda_{b}+\lambda_{c}}
+i\w2\tN\frac{\lambda_{a}+\lambda_{b}}{\lambda_{a}\lambda_{b}+\lambda_{c}}
}
}
\;.
\end{equation}

\subsubsection{Low-frequency expansion of $\chi_{\perp}(\w)$ and the
effective transverse relaxation time}

On expanding $\chi_{\perp}(\w)$ from (\ref{chiperp:RSh:red}) in powers of
$\w\tN$ to first order, we get the simple result
\begin{equation}
\label{chiperp:RSh:red:expansion}
\chi_{\perp}(\w,T)/\chi_{\perp}(T)
\simeq 1-i\w2\tN\frac{\lambda_{b}}{\lambda_{a}\lambda_{b}+\lambda_{c}}
\simeq
\frac
{1}
{
{\displaystyle
1+i\w2\tN\frac{\lambda_{b}}{\lambda_{a}\lambda_{b}+\lambda_{c}}
}
}
\;,
\end{equation}
where the last approximate equality has been obtained by means of the binomial
expansion $(1+x)^{\epsilon}=1+\epsilon x+\cdots$. Therefore, in the low-frequency
range $\chi_{\perp}(\w)$ has a Debye-type form, so that the quantity multiplying
$i\w$ defines an effective relaxation time, namely
\begin{equation}
\label{effeigentaumod:1}
\ttr|_{\xi=0}
=
2\tN
\frac{1}{\lambda_{a}}
\frac{1}{1+\lambda_{c}/\lambda_{a}\lambda_{b}}
\;.
\end{equation}
To conclude, with help from the results of Appendix \ref{app:F}, let us write the
coefficients $\lambda_{a}$, $\lambda_{b}$, and $\lambda_{c}$ in terms of $\legunb$
[the average of the second Legendre polynomial (\ref{S2}) at zero field]
\[
\lambda_{a}
=
\frac{2+\legunb}{1-\legunb}
\;,
\qquad
\lambda_{b}
=
\frac{2\s}{3}\frac{2+\legunb(1-6/\s)}{\legunb}
\;,
\qquad
\lambda_{c}
=
\frac{1}{\la^{2}}\frac{6\s \legunb}{1-\legunb}
\;,
\]
From these equations we get
\[
\frac{1}{\lambda_{a}}
=
\frac{1-\legunb}{2+\legunb}
\;,
\qquad
\frac{\lambda_{c}}{\lambda_{a}\lambda_{b}}
=
\frac{1}{\la^{2}}
\frac
{(3\legunb)^{2}}
{(2+\legunb)\big[2+\legunb(1-6/\s)\big] }
\;,
\]
which when inserted in Eq.\ (\ref{effeigentaumod:1}) yield the effective transverse
relaxation time (\ref{effeigentaumod}).

Note finally that, as introduced, the effective transverse relaxation time is a sort of
transverse {\em integral\/} relaxation time $\tau_{{\rm int},\perp}$ [compare the first
approximate equality of Eq.\ (\ref{chiperp:RSh:red:expansion}) with Eq.\
(\ref{chidyn:approx})]. However, its usefulness is questionable in the transverse-field
case as the magnetization relaxation curve then comprises {\em oscillating\/} terms,
so that the area under such a curve may largely overestimate the relaxation rate.

\setcounter{equation}{0}
\section{Reduced equations of motion for non-linear system-environment couplings}
\label{app:corweslin}

In this appendix we shall derive a reduced equation of motion for any  dynamical
variable $A(\sP,\sQ)$ whose time evolution is determined by the Hamiltonian
(\ref{hamiltonian:L+Q:2}). This will be carried out by means of a perturbational
expansion in the coupling parameter $\coupling$. Nevertheless, we shall first study the
weak-coupling dynamics associated with a larger class of Hamiltonians of the form
\begin{equation}
\label{hamiltonian:gral}
\Ham_{\rm T}
=
\Hsm(\sP,\sQ)
+\sum_{\alpha}
\half
\left(
\eP_{\alpha}^{2}\overMa
+\Ma\w_{\alpha}^{2}\eQ_{\alpha}^{2}
\right)
+\coupling\sum_{N}\G^{N}(\eQm)\Fint_{N}(\sP,\sQ)
\;,
\end{equation}
where the coupling terms $\G^{N}(\eQm)$ are {\em arbitrary\/} functions of the
environment coordinates $\eQm$ and $N$ stands for a general index, which can run,
for example, over single oscillator indices, pairs, triplets, etc.\
(${\alpha},{\alpha\beta},{\alpha\beta\gamma},\ldots$). On the other hand, the
modified system Hamiltonian $\Hsm$ augments the system Hamiltonian $\Hs$ by
appropriate counter-terms, which will be determined below.

We shall first derive the reduced dynamical equations associated with the Hamiltonian
(\ref{hamiltonian:gral}), so that one could incorporate relaxation mechanisms
involving any number of environmental normal modes into the dynamical equations
of the system variables. This will be done by a perturbational treatment that is an
extension of the treatment developed by Cort{\'{e}}s, West and Lindenberg (1985) to
deal with a system-environment coupling {\em linear\/} in the system coordinate [the
case $\Fint_{N}(\sP,\sQ)\propto\sQ$ of the Hamiltonian (\ref{hamiltonian:gral})], but
otherwise arbitrary in the environment coordinates.%
\footnote{
Brun (1993) also treated rather general non-bilinear interactions by perturbation
theory.
}
Eventually, we shall particularize the results obtained to the
Hamiltonian (\ref{hamiltonian:L+Q:2}), which is recovered when:
\begin{enumerate}
\item
$N$ only runs over single oscillator indices $\alpha$ and pairs $\alpha\beta$.
\item
The corresponding coupling terms are $\G^{\alpha}(\eQm)=\eQ_{\alpha}$ and
$\G^{\alpha\beta}(\eQm)=\half\eQ_{\alpha}\eQ_{\beta}$.
\end{enumerate}

The coupled dynamical equations for $A(\sP,\sQ)$ and the environment variables
associated with the Hamiltonian (\ref{hamiltonian:gral}) are [cf.\ Eqs.\
(\ref{eqmot_A:L:1}) and (\ref{eqmot_bath:L})]
\begin{eqnarray}
\label{eqmot_A:gral:1}
\frac{\D A}{\D t}
&
=
&
\pbra A, \Hsm\pket
+
\coupling\sum_{N}\G^{N}(\eQm)\pbra A, \Fint_{N}\pket
\;,
\\
\label{eqmot_bath:gral}
\frac{\D\eQ_{\alpha}}{\D t}
&
=
&
\eP_{\alpha}\overMa
\;,
\quad
\frac{\D\eP_{\alpha}}{\D t}
=
-\Ma\w_{\alpha}^{2}\eQ_{\alpha}
-\coupling\sum_{N}\G_{\alpha}^{N}(\eQm)\Fint_{N}
\;,
\end{eqnarray}
where we have used the shorthand
\[
\G_{\alpha}^{N}
=
\partial\G^{N}/\partial\eQ_{\alpha}
\;.
\]
Equations (\ref{eqmot_bath:gral}) can {\em formally\/} be integrated, yielding an
equation akin to Eq.\ (\ref{bath_coord:L:1}) with
$\Fint_{\alpha}(\tp)\to\sum_{N}\G_{\alpha}^{N}[\eQm(\tp)]\Fint_{N}(\tp)$, namely
\[
\eQ_{\alpha}(t)
=
\eQ_{\alpha}^{\h}(t)
-\frac{\coupling}{\Ma\w_{\alpha}}
\int_{t_{0}}^{t}\!\!\D{\tp}\,
\sin[\w_{\alpha}(t-\tp)]
\sum_{N}\G_{\alpha}^{N}[\eQm(\tp)]\Fint_{N}(\tp)
\;,
\]
where the $\eQ_{\alpha}^{\h}(t)$ are the solutions (\ref{qh}) for the free oscillators
and $\Fint_{N}(\tp)=\Fint_{N}[\sP(\tp),\sQ(\tp)]$. On integrating by parts in this
equation one gets [cf.\ Eq.\ (\ref{bath_coord:L:2})]
\begin{equation}
\label{bath_coord:gral}
\eQ_{\alpha}(t)
=
\eQ_{\alpha}^{\h}(t)
-\coupling\sum_{N}
\left[ D_{\alpha}^{N}(\eQm;t,\tp)\Fint_{N}(\tp)
\right]_{\tp=t_{0}}^{\tp=t}
+\coupling
\int_{t_{0}}^{t}\!\!\D{\tp}\, 
\sum_{N}D_{\alpha}^{N}(\eQm;t,\tp)\frac{\D\Fint_{N}}{\D t}(\tp)
\;,
\end{equation}
where we have introduced the indefinite integral
\begin{equation}
\label{D} D_{\alpha}^{N}(\eQm;t,\tp)
=
\frac{1}{\Ma\w_{\alpha}}\int^{\tp}\!\!\D{\tp'}\,\sin[\w_{\alpha}(t-\tp')]
\G_{\alpha}^{N}[\eQm(\tp')]
\;.
\end{equation}

Recall that writing $\eQ_{\alpha}(t)$ in the form (\ref{bath_coord:gral}) by an
integration by parts, enables one to separate the Hamiltonian (renormalization)
and relaxational terms (Subsec.\ \ref{subsect:linear_coupling}). However, Eq.\
(\ref{bath_coord:gral}) gives $\eQ_{\alpha}(t)$ in implicit form, since
$\eQ_{\alpha}(t)$ also appears on the right-hand side via $\G_{\alpha}^{N}(\eQm)$.
Thus, Eq.\ (\ref{bath_coord:gral}) is an explicit solution only in the linear
$\G^{N}(\eQm)$ case of the Hamiltonian (\ref{hamiltonian:L:2}).

For weak system-environment interactions, we shall solve Eq.\ (\ref{bath_coord:gral})
for $\eQ_{\alpha}(t)$ perturbatively in $\coupling$. However, as pointed out by
Cort{\'{e}}s, West and Lindenberg (1985), in order to get eventually a
thermodynamically consistent description, the expansion cannot be uniform in
$\coupling$. If one keeps fluctuating terms up to order $\coupling^{k}$, the relaxation
terms must be retained up to order $\coupling^{2k}$, in order to obtain proper
fluctuation-dissipation relations [see, for example, Eqs.\ (\ref{fluct:L}), (\ref{kernel:L})
and (\ref{stats:L:2nd_moment})].

The $\coupling$-expansion of $\eQ_{\alpha}(t)$ reads
\[
\eQ_{\alpha}(t)
=
\eQ_{\alpha}^{\h}(t)
+\coupling\,\delta\eQ_{\alpha}(t)+\ldots
\;,
\]
where $\coupling\,\delta \eQ_{\alpha}(t)$ is given by the second plus third terms on
the right-hand side of Eq.\ (\ref{bath_coord:gral}) when $\eQm^{\h}$ (the
zeroth-order term) is substituted for $\eQm$ in $D_{\alpha}^{N}(\eQm;t,\tp)$, namely
\[
\coupling\,\delta\eQ_{\alpha}(t)
=
-\coupling\sum_{N}
\left[ D_{\alpha}^{N}(\eQm^{\h};t,\tp)\Fint_{N}(\tp)
\right]_{\tp=t_{0}}^{\tp=t}
+\coupling\int_{t_{0}}^{t}\!\!\D{\tp}\, 
\sum_{N}D_{\alpha}^{N}(\eQm^{\h};t,\tp)\frac{\D\Fint_{N}}{\D t}(\tp)
\]
[that is, we iterate Eq.\ (\ref{bath_coord:gral}) into itself]. The corresponding
expansion of the coupling functions is given by
\begin{equation}
\label{B:expansion}
\coupling\G^{N}(\eQm)
=
\coupling\G^{N}(\eQm^{\h})
+{\coupling}^{2}
\sum_{\alpha}\G_{\alpha}^{N}(\eQm^{\h})\delta \eQ_{\alpha}
+\ldots
\;,
\end{equation}
which enters into Eq.\ (\ref{eqmot_A:gral:1}). The term
\begin{equation}
\label{fluct:gral}
\ffl_{N}(t)
\equiv
\coupling\G^{N}[\eQm^{\h}(t)]
\;,
\end{equation}
per analogy with $\ffl_{\alpha}(t)=\coupling \eQ_{\alpha}^{\h}(t)$ [Eq.\
(\ref{fluct-kernel:L:precursor})], is interpreted as the lowest order fluctuation.
Following the programme of Cort{\'{e}}s, West and Lindenberg (1985), we shall retain
fluctuations only to this order.%
\footnote{
In order to ensure $\langle\ffl_{N}(t)\rangle=0$, where the angular
brackets denote average over initial states of the oscillators, one could assume that,
for instance, at least one coordinate enters in $\G^{N}(\eQm)$ an odd number of times.
Nevertheless, as discussed after Eqs.\ (\ref{stats:Q:1st_moment}),
(\ref{stats:L_cross_Q}), and (\ref{stats:Q:2nd_moment}), such a restriction is not
actually needed when the frequency spectrum of the oscillators is sufficiently dense.
}

Concerning the back-reaction part, one first introduces the quantities
\begin{eqnarray}
\label{kernel:gral}
\K^{N,M}(t,\tp)
&
=
&
{\coupling}^{2}
\Big\langle\sum_{\alpha}
\G_{\alpha}^{N}[\eQm^{\h}(t)]D_{\alpha}^{M}(\eQm^{\h};t,\tp)
\Big\rangle
\;,
\\
\label{delta_kernel:gral}
\delta\K^{N,M}(t,\tp)
&
=
&
{\coupling}^{2}
\sum_{\alpha}
\G_{\alpha}^{N}[\eQm^{\h}(t)]D_{\alpha}^{M}(\eQm^{\h};t,\tp)
-\K^{N,M}(t,\tp)
\;,
\end{eqnarray}
so that the second term in the expansion (\ref{B:expansion}) can be
decomposed as
\begin{eqnarray*}
{\coupling}^{2}
\sum_{\alpha}
\G_{\alpha}^{N}(\eQm^{\h})
\delta\eQ_{\alpha}
&
=
&
-\Big[
\sum_{M}
\left[
\K^{N,M}(t,\tp)+\delta\K^{N,M}(t,\tp)
\right]
\Fint_{M}(\tp)
\Big]_{\tp=t_{0}}^{\tp=t}
\\
&
& {}+
\int_{t_{0}}^{t}\!\!\D{\tp}\,
\sum_{M}
\left[
\K^{N,M}(t,\tp)+\delta\K^{N,M}(t,\tp)
\right]
\frac{\D\Fint_{M}}{\D t}(\tp)
\;.
\end{eqnarray*}
Each kernel $\K^{N,M}$ gives a different type of contribution whereas the contribution
of $\delta\K^{N,M}$ can be interpreted as fluctuations around the former (Cort{\'{e}}s,
West and Lindenberg, 1985). As these fluctuations are of order
higher ($\coupling^{2}$) than the fluctuations that we are retaining in the present
treatment, the terms $\delta\K^{N,M}$ will henceforth be omitted. On the other hand,
the terms
$\sum_{M}\K^{N,M}(t,t_{0})\Fint_{M}(t_{0})$
in
${\coupling}^{2}\sum_{\alpha}\G_{\alpha}^{N}\delta\eQ_{\alpha}$ will also be ignored
as they are the generalization of those terms that give a transient in the response
(see Subsec.\ \ref{subsect:linear_coupling}; recall however that they could be
incorporated into an alternative definition of the fluctuating sources but, as they are
of order $\coupling^{2}$, they would anyhow be ignored).
Finally, the parallel terms
$-\sum_{M}\K^{N,M}(t,t)\Fint_{M}(t)$ give the Hamiltonian contributions. In order to
prove this, note first that, since $\K^{N,M}(t,\tp)$ comprises equilibrium averages, it
depends on $(t-\tp)$ and, hence, $\K^{N,M}(t,t)$ is independent of $t$. By the same
reasoning one can demonstrate the symmetry property
$\K^{N,M}=\K^{M,N}$.%
\footnote{
We shall anyway verify explicitly these two results for the Hamiltonian
(\ref{hamiltonian:L+Q:2}).
}
Then, by using the product rule of the Poisson bracket (\ref{productrule}), one finds
that the contribution originating from $-\sum_{M}\K^{N,M}(t,t)\Fint_{M}(t)$ in the
equation for $A(\sP,\sQ)$ is given by
\[
-\sum_{NM}\K^{N,M}(0)\pbra A, \Fint_{N}\pket\Fint_{M}
=
\Big\{ A,-\half\sum_{NM}\K^{N,M}(0)\Fint_{N}\Fint_{M}
\Big\}
\;,
\]
which is indeed derivable from a (time-independent) Hamiltonian. This term is
associated with the coupling-induced renormalization of the energy of the system and
is balanced by the counter-terms incorporated into $\Hsm$, now explicitly identified
as [cf.\ Eq.\ (\ref{modified_hamiltonian:L})]
\begin{equation}
\label{modified_hamiltonian:gral}
\Hsm
=
\Hs+\half\sum_{NM}\K^{N,M}(0)\Fint_{N}\Fint_{M}
\;.
\end{equation}

On collecting the terms whose retention has hitherto been argued and introducing
them into Eq.\ (\ref{eqmot_A:gral:1}), one finally gets the approximate reduced
equation of motion for any dynamical variable $A(\sP,\sQ)$ [cf.\ Eq.\
(\ref{eqmot_A:L})]
\begin{equation}
\label{eqmot_A:gral}
\frac{\D A}{\D t}
=
\pbra A, \Hs\pket
+\sum_{N}\pbra A, \Fint_{N}\pket
\bigg[
\ffl_{N}(t)
+\int_{t_{0}}^{t}\!\!\D{\tp}\,
\sum_{M}\K^{N,M}(t-\tp)\frac{\D\Fint_{M}}{\D t}(\tp)
\bigg]
\;.
\end{equation}
In addition, within the approximation used (fluctuating and relaxation terms to order
$\coupling$ and $\coupling^{2}$, respectively), one can replace $\D\Fint_{M}/\D t$
in the memory integral by its  conservative part $\D\Fint_{M}/\D
t\simeq\pbra\Fint_{M},\Hs\pket$. On the other hand, one can establish
fluctuation-dissipation relations by means of arguments parallel to those presented by
Cort{\'{e}}s, West and Lindenberg (1985).

To conclude, we shall particularize these results to the linear-plus-quadratic couplings
of the Hamiltonian (\ref{hamiltonian:L+Q:2}). This is recovered when $N$ runs over
single oscillator indices $\alpha$, with $\G^{\alpha}=\eQ_{\alpha}$, and pairs
$\alpha\beta$, with $\G^{\alpha\beta}=\half\eQ_{\alpha}\eQ_{\beta}$. Then, the
fluctuating terms $\ffl_{N}(t)=\coupling\G^{N}[\eQm^{\h}(t)]$ are given by
$\ffl_{\alpha}(t)=\coupling\eQ_{\alpha}^{\h}(t)$ [Eq.\ (\ref{fluct-kernel:L:precursor})]
and
$\ffl_{\alpha\beta}(t)=(\coupling/2)\eQ_{\alpha}^{\h}(t)\eQ_{\beta}^{\h}(t)$
[Eq.\ (\ref{fluct:Q:precursor})]. On the other hand, by inserting the derivatives
\begin{eqnarray}
\label{B_alpha}
\G_{\gamma}^{\alpha}
&
=
&
\partial\G^{\alpha}/\partial\eQ_{\gamma}
=
\delta_{\alpha\gamma}
\;,
\\
\label{B_alpha_beta}
\G_{\gamma}^{\alpha\beta}
&
=
&
\partial\G^{\alpha\beta}/\partial\eQ_{\gamma}
=
\half(\delta_{\alpha\gamma}\eQ_{\beta}
+\delta_{\beta\gamma}\eQ_{\alpha})
\;,
\end{eqnarray}
into Eq.\ (\ref{D}), the functions $D_{\gamma}^{N}(\eQm;t,\tp)$ emerge in the form
($N=\alpha,\,\alpha\beta$)
\begin{eqnarray}
\label{D_alpha} D_{\gamma}^{\alpha}(\eQm;t,\tp)
&
=
&
\frac{\delta_{\alpha\gamma}}{\Ma\w_{\alpha}^{2}}
\cos[\w_{\alpha}(t-\tp)]
\;,
\\
\label{D_alpha_beta} D_{\gamma}^{\alpha\beta}(\eQm;t,\tp)
&
=
&
\frac{1}{\Mg\w_{\gamma}}\int^{\tp}\!\!\D{\tp'}\,\sin[\w_{\gamma}(t-\tp')]
\half[\delta_{\alpha\gamma}\eQ_{\beta}(\tp')
+\delta_{\beta\gamma}\eQ_{\alpha}(\tp')]
\;.
\qquad
\quad
\end{eqnarray}
Therefore, on taking the averages in Eq.\ (\ref{kernel:gral}) with
respect to the distribution (\ref{gibbs:oscillators:dic}) (decoupled
initial conditions) by means of Eqs.\ (\ref{stats:qh}), we get for the
kernels $\K^{N,M}$ (see proofs below)
\begin{eqnarray}
\label{Ksa}
\K^{\alpha,\beta}(\tau)
&
=
&
\delta_{\alpha\beta}\frac{\coupling^{2}}{\Ma\w_{\alpha}^{2}}
\cos(\w_{\alpha}\tau)
\;,
\\
\label{Ksb}
\K^{\alpha,\beta\gamma}(\tau)
&
=
&
\K^{\alpha\beta,\gamma}(\tau)
=
0
\;,
\\
\K^{\alpha\beta,\gamma\delta}(\tau)
&
=
&
\half (\delta_{\alpha\gamma}\delta_{\beta\delta}
+\delta_{\alpha\delta}\delta_{\beta\gamma})
\nonumber
\\
\label{Ksc}
&
& {}\times\frac{\coupling^{2}}{2}
\frac{\T}{2\Ma\w_{\alpha}^{2}\Mb\w_{\beta}^{2}}
\big\{
\cos[(\w_{\alpha}-\w_{\beta})\tau]
+\cos[(\w_{\alpha}+\w_{\beta})\tau]
\big\}
\;.
\qquad
\end{eqnarray}
These kernels satisfy the properties mentioned above: they depend on
$\tau=t-\tp$ and are symmetrical with respect to the indices separated
by commas, which correspond to the general indices $N,M$.

On introducing all these results in Eq.\ (\ref{eqmot_A:gral}), the
resulting dynamical equation for $A(\sP,\sQ)$ is given by Eq.\
(\ref{eqmot_A:L+Q}). For the sake of simplicity, we have introduced in
that equation the kernels $\K_{\alpha}(\tau)$ and
$\K_{\alpha\beta}(\tau)$, which are defined in terms of the above
kernels by
\begin{eqnarray*}
\K^{\alpha,\beta}(\tau)
&
=
&
\delta_{\alpha\beta}\K_{\alpha}(\tau)
\;,
\\
\K^{\alpha\beta,\gamma\delta}(\tau)
&
=
&
\half(\delta_{\alpha\gamma}\delta_{\beta\delta}
+\delta_{\alpha\delta}\delta_{\beta\gamma})
\K_{\alpha\beta}(\tau)
\;.
\end{eqnarray*}
Besides, on explicitly writing the counter-term of Eq.\
(\ref{modified_hamiltonian:gral}) in this linear-plus-quadratic case, one arrives at Eq.\
(\ref{modified_hamiltonian:L+Q}).

Note finally that, owing to the fact that
$\G_{\gamma}^{\alpha}(\eQm^{\h}) D_{\gamma}^{\beta}(\eQm^{\h};t,\tp)$
does not depend on $\eQm^{\h}$, the kernel $\K_{\alpha}(\tau)$ is not
affected by the averaging procedure, whereas this renders
$\K_{\alpha\beta}(\tau)$ explicitly dependent on the temperature (see
below). In this connection, we remark that the modifications of this
last kernel obtained when one assumes coupled initial conditions,
begin at order $\coupling^{3}$.

\subsection*{Derivation of the kernels}

\paragraph{Derivation of $\K^{\alpha,\beta}(\tau)$.}

From Eqs.\ (\ref{B_alpha}) and (\ref{D_alpha}) and the general definition
$\K^{N,M}(t,\tp)
=\coupling^{2}\langle\sum_{\rho}\G_{\rho}^{N}D_{\rho}^{M}\rangle$,
one gets
\begin{eqnarray*}
\K^{\alpha,\beta}(t,\tp)
&
=
&
\coupling^{2}
\Big\langle\sum_{\rho}
\G_{\rho}^{\alpha}D_{\rho}^{\beta}
\Big\rangle
\\
&
=
&
\coupling^{2}
\Big\langle\sum_{\rho}
\delta_{\alpha\rho}
\frac{\delta_{\beta\rho}}{\Mb\w_{\beta}^{2}}
\cos[\w_{\beta}(t-\tp)]
\Big\rangle
=
\delta_{\alpha\beta}\frac{\coupling^{2}}{\Ma\w_{\alpha}^{2}}
\cos[\w_{\alpha}(t-\tp)]
\;,
\end{eqnarray*}
where the average has played no r\^{o}le.\qed

\paragraph{Derivation of $\K^{\alpha,\beta\gamma}(\tau)$.}

From Eqs.\ (\ref{B_alpha}) and (\ref{D_alpha_beta}) we obtain
\[
\llangle\G_{\rho}^{\alpha}D_{\rho}^{\beta\gamma}
\rrangle
=
\frac{\delta_{\alpha\rho}}{\Mr\w_{\rho}}
\int^{\tp}\!\!\D{\tp'}\,\sin[\w_{\rho}(t-\tp')]
\half
\Big[\delta_{\beta\rho}
\underbrace{
\llangle\eQ_{\gamma}^{\h}(\tp')\rrangle
}_{0}
+\delta_{\gamma\rho}
\underbrace{
\llangle\eQ_{\beta}^{\h}(\tp')\rrangle
}_{0}
\Big]
=
0
\;,
\]
where Eqs.\ (\ref{stats:qh}) have been employed. Therefore, from this result
and the general definition (\ref{kernel:gral}) it follows that
$\K^{\alpha,\beta\gamma}(t,\tp)=0$.\qed

\paragraph{Derivation of $\K^{\alpha\beta,\gamma}(\tau)$.}

The average of the product of Eqs.\ (\ref{B_alpha_beta}) and (\ref{D_alpha})
evaluated at $\eQm^{\h}$ is zero as well. Indeed,
\[
\llangle\G_{\rho}^{\alpha\beta}D_{\rho}^{\gamma}
\rrangle
=
\half
\Big[
\delta_{\alpha\rho}
\underbrace{
\llangle\eQ_{\beta}^{\h}(t)\rrangle
}_{0}
+\delta_{\beta\rho}
\underbrace{
\llangle\eQ_{\alpha}^{\h}(t)\rrangle
}_{0}
\Big]
\frac{\delta_{\gamma\rho}}{\Mg\w_{\gamma}^{2}}
\cos[\w_{\gamma}(t-\tp)]
=
0
\;,
\]
whence one gets the stated result $\K^{\alpha\beta,\gamma}(t,\tp)=0$.\qed

\paragraph{Derivation of $\K^{\alpha\beta,\gamma\delta}(\tau)$.}

Finally, for the average of the product of Eqs.\ (\ref{B_alpha_beta}) and
(\ref{D_alpha_beta}) evaluated at $\eQm^{\h}$ one has
\begin{eqnarray*}
\lefteqn{
\llangle\G_{\rho}^{\alpha\beta} D_{\rho}^{\gamma\delta}
\rrangle }
&
&
\\
&
&
=
\Big\langle
\half
\big[
\delta_{\alpha\rho}
\eQ_{\beta}^{\h}(t)
+\delta_{\beta\rho}
\eQ_{\alpha}^{\h}(t)
\big]
\\
&
&
\qquad
{}\times
\frac{1}{\Mr\w_{\rho}}
\int^{\tp}\!\!\D{\tp'}\,\sin[\w_{\rho}(t-\tp')]
\half
\big[\delta_{\gamma\rho}
\eQ_{\delta}^{\h}(\tp')
+\delta_{\delta\rho}
\eQ_{\gamma}^{\h}(\tp')
\big]
\Big\rangle
\;.
\end{eqnarray*}
Therefore, we need to calculate the following average
\begin{eqnarray*}
\lefteqn{
\Big\langle\big[
\delta_{\alpha\rho}
\eQ_{\beta}^{\h}(t)
+\delta_{\beta\rho}
\eQ_{\alpha}^{\h}(t)
\big]
\big[\delta_{\gamma\rho}
\eQ_{\delta}^{\h}(\tp')
+\delta_{\delta\rho}
\eQ_{\gamma}^{\h}(\tp')
\big]
\Big\rangle }
\qquad
\quad
&
&
\\
&
=
&
\T
\bigg\{
\delta_{\alpha\rho}\delta_{\gamma\rho}
\frac{\delta_{\beta\delta}}{\Mb\w_{\beta}^{2}}\cos[\w_{\beta}(t-\tp')]
+\delta_{\alpha\rho}\delta_{\delta\rho}
\frac{\delta_{\beta\gamma}}{\Mb\w_{\beta}^{2}}\cos[\w_{\beta}(t-\tp')]
\\
&
&
\qquad
\quad {}+\delta_{\beta\rho}\delta_{\gamma\rho}
\frac{\delta_{\alpha\delta}}{\Ma\w_{\alpha}^{2}}\cos[\w_{\alpha}(t-\tp')]
+\delta_{\beta\rho}\delta_{\delta\rho}
\frac{\delta_{\alpha\gamma}}{\Ma\w_{\alpha}^{2}}\cos[\w_{\alpha}(t-\tp')]
\bigg\}
\\
&
=
&
\frac{\T}{\Ma\w_{\alpha}^{2}\Mb\w_{\beta}^{2}}
\Big\{
\delta_{\alpha\rho} (\delta_{\gamma\rho}\delta_{\beta\delta}
+\delta_{\delta\rho}\delta_{\beta\gamma})
\Ma\w_{\alpha}^{2}\cos[\w_{\beta}(t-\tp')]
\\
&
&
\qquad
\quad {}+\delta_{\beta\rho} (\delta_{\gamma\rho}\delta_{\alpha\delta}
+\delta_{\delta\rho}\delta_{\alpha\gamma})
\Mb\w_{\beta}^{2}\cos[\w_{\alpha}(t-\tp')]
\Big\}
\;,
\end{eqnarray*}
where we have used Eqs.\ (\ref{stats:qh}). Next, on multiplying this expression by
$\sin[\w_{\rho}(t-\tp')]/\Mr\w_{\rho}$, and summing over $\rho$ we obtain
\begin{eqnarray*}
\lefteqn{
\sum_{\rho}
\Big\langle
\big[
\delta_{\alpha\rho}
\eQ_{\beta}^{\h}(t)
+\delta_{\beta\rho}
\eQ_{\alpha}^{\h}(t)
\big]
\big[\delta_{\gamma\rho}
\eQ_{\delta}^{\h}(\tp')
+\delta_{\delta\rho}
\eQ_{\gamma}^{\h}(\tp')
\big]
\Big\rangle\sin[\w_{\rho}(t-\tp')]/\Mr\w_{\rho}
}
&
&
\\
&
=
&
\frac{\T}{\Ma\w_{\alpha}^{2}\Mb\w_{\beta}^{2}}
(\delta_{\alpha\gamma}\delta_{\beta\delta}
+\delta_{\alpha\delta}\delta_{\beta\gamma})
\\
&
&
{}\times
\Big\{
\w_{\alpha}\cos[\w_{\beta}(t-\tp')]\sin[\w_{\alpha}(t-\tp')]
+\w_{\beta}\cos[\w_{\alpha}(t-\tp')]\sin[\w_{\beta}(t-\tp')]
\Big\}
\;.
\end{eqnarray*}
Then, on taking into account that
$(\D/\D{\tp'})\{\cos[\w_{\alpha}(t-\tp')]\cos[\w_{\beta}(t-\tp')]\}$
is equal to the term within the above curly brackets when calculating the integral
occurring in
$\K^{\alpha\beta,\gamma\delta}(t,\tp)
=\coupling^{2}
\langle\sum_{\rho}
\G_{\rho}^{\alpha\beta} D_{\rho}^{\gamma\delta}
\rangle$, we arrive at
\[
\K^{\alpha\beta,\gamma\delta}(t,\tp)
=
\half(\delta_{\alpha\gamma}\delta_{\beta\delta}
+\delta_{\alpha\delta}\delta_{\beta\gamma})
\frac{\coupling^{2}}{2}
\frac{\T}{\Ma\w_{\alpha}^{2}\Mb\w_{\beta}^{2}}
\cos[\w_{\alpha}(t-\tp)]\cos[\w_{\beta}(t-\tp)]
\;,
\]
whence one immediately obtains Eq.\ (\ref{Ksc}).\qed

\section*{Acknowledgments}
\addcontentsline{toc}{section}{Acknowledgments}

{\small
Helpful discussions with Prof.\ W.\ T.\ Coffey, Dr.\ D.\ A.\ Garanin,
and Dr.\ P.~Svedlindh are gratefully acknowledged. I also wish to
express my gratitude to the colleagues of {\em Instituto de Ciencia de
Materiales de Arag\'{o}n} (Consejo Superior de Investigaciones
Cient\'{\i}ficas--Universidad de Zaragoza), as well as to those of the
{\em Department of Materials Science - Division of Solid State
Physics\/} (Uppsala University) with which I have worked during these
years. This work has partially been supported by Diputaci\'{o}n
General de Arag\'{o}n and the Swedish Science Research Council (NFR).
}